\documentclass[english,aps,reprint,10pt,superscriptaddress,amssymb,amsfonts,nofootinbib]{revtex4-1}
\usepackage[T1]{fontenc}
\usepackage[latin9]{inputenc}
\usepackage{xcolor}
\usepackage{units}
\usepackage{mathtools}
\usepackage{amsmath}
\usepackage{amsthm}
\usepackage{amssymb}
\usepackage{stmaryrd}
\usepackage{graphicx}
\usepackage{nicefrac}
\usepackage{comment}

\makeatletter

\providecommand{\tabularnewline}{\\}

  \theoremstyle{plain}
  
 \ifx\proof\undefined\
   \newenvironment{proof}[1][\proofname]{\par
     \normalfont\topsep6\p@\@plus6\p@\relax
     \trivlist
     \itemindent\parindent
     \item[\hskip\labelsep
           \scshape
       #1]\ignorespaces
   }{%
     \endtrivlist\@endpefalse
   }
   \providecommand{\proofname}{Proof}
 \fi

\@ifundefined{date}{}{\date{}}
\usepackage[colorlinks,bookmarks=false,citecolor=blue,linkcolor=red,urlcolor=blue]{hyperref}
\usepackage{breqn}

\usepackage{color}
\usepackage{amsmath,amssymb,amsthm,amscd,latexsym,epsfig,bm,subfigure}
\usepackage{mathtools}
\usepackage{leftidx}

\usepackage{tikz}
\usetikzlibrary{shapes}
\usetikzlibrary{arrows, bending}
\usetikzlibrary{knots}

\usetikzlibrary{decorations.markings}
\tikzset{->-/.style={decoration={
  markings,
  mark=at position #1 with {\arrow{stealth'}}},
  postaction={decorate}}}
\tikzset{-<-/.style={decoration={
  markings,
  mark=at position #1 with {\arrow{stealth' reversed}}},
  postaction={decorate}}}

\makeatother

\usepackage{babel}

\usepackage{babel}
\providecommand{\lemmaname}{Lemma}

\makeatletter
\let\cat@comma@active\@empty
\makeatother

\begin{document}

\title{Twisted Fracton Models in Three Dimensions}

\author{Hao Song}
\email{haosong@ucm.es}
\affiliation{Departamento de F\'isica Te\'orica, Universidad Complutense, 28040 Madrid, Spain}
\author{Abhinav Prem}
\affiliation{Department of Physics and Center for Theory of Quantum Matter,
University of Colorado, Boulder, Colorado 80309, USA}
\author{Sheng-Jie Huang}
\affiliation{Department of Physics and Center for Theory of Quantum Matter,
University of Colorado, Boulder, Colorado 80309, USA}
\author{M.A. Martin-Delgado}
\affiliation{Departamento de F\'isica Te\'orica, Universidad Complutense, 28040 Madrid, Spain}

\date{\today}
\begin{abstract}
We study novel three-dimensional gapped quantum phases of matter which support quasiparticles with restricted mobility, including immobile ``fracton'' excitations. So far, most existing fracton models may be instructively viewed as generalized Abelian lattice gauge theories. Here, by analogy with  Dijkgraaf-Witten topological gauge theories, we discover a natural generalization of fracton models, obtained by twisting the gauge symmetries. Introducing generalized gauge transformation operators carrying an extra phase factor depending on local configurations, we construct a plethora of exactly solvable three-dimensional models, which we dub ``twisted fracton models.'' A key result of our approach is to demonstrate the existence of rich non-Abelian fracton phases of distinct varieties in a three-dimensional system with finite-range interactions. For an accurate characterization of these novel phases, the notion of being  \emph{inextricably non-Abelian} is introduced for fractons and quasiparticles with one-dimensional mobility, referring to their new behavior of displaying braiding statistics that is, and remains, non-Abelian regardless of which quasiparticles with higher mobility are added to or removed from them. 
We also analyze these models by embedding them on a three-torus and computing their ground state degeneracies, which exhibit a surprising and novel dependence on the system size in the non-Abelian fracton phases. Moreover, as an important advance in the study of fracton order, we develop a general mathematical framework which systematically captures the fusion and braiding properties of fractons and other quasiparticles with restricted mobility.
\end{abstract}
\maketitle

\tableofcontents{}

\section{Introduction}
\label{intro}

The study of topological quantum phases of matter  has led to remarkable new discoveries, both theoretically and experimentally, and has profoundly influenced our understanding of quantum many-body physics. Starting with the discovery of the fractional quantum Hall effect~\cite{Tsui,Laughlin}, it was realized that there exist quantum phases of matter which lie outside Landau's symmetry breaking paradigm. 
One such class of phases are those with intrinsic topological order, which are gapped quantum phases of matter distinguished by patterns of long-range entanglement in their ground states~\cite{Levin2006,preskill,chen2010}. Nontrivial topological orders, examples of which include quantum Hall states and gapped spin liquids, may exhibit striking phenomena such as excitations with fractionalized statistics, locally indistinguishable degenerate ground states, and robust gapless edge states~\cite{Laughlin,mooreread,wenniu,wen2002,moessner,balents2002,sondhi,RG,K,KITAEV20062}. 
The potential application of topological states for
fault-tolerant quantum computation \cite{KITAEV20032,ColorCode2D,ColorCode3D,RMP3} has provided another main motivation for current intensive study on topological orders.

The landscape of topological quantum phases becomes much richer in the presence of global symmetries. Even in the absence of intrinsic topological order, distinct phases protected by some unbroken symmetry are possible, leading to the modern notion of symmetry protected topological (SPT) phases~\cite{CGLW,chen2013}, of which the 1D Haldane phase for spin-1 chain~\cite{haldane,AKLT,VMC,Pollmann,GuWen2009} and topological insulators~\cite{kanemele,BernevigTI,fukanemele,rroyTI,moorebalentsTI,RMP1,RMP2,VS,MKF} are paradigmatic examples. Further considering the interplay between non-trivial topological order and global symmetries leads to the concept of symmetry enriched topological (SET) phases, which have been of much recent interest \cite{moroz,mesaros13ying,essin,barkeshli,lu,hsongSGSET,hsongThesis,hsongPGSPT}. 

The topological nature of these phases is reflected in the fact that their low-energy behavior is governed by a topological quantum field theory (TQFT), which in turn allows one to develop general mathematical frameworks for understanding their physics. In particular, the language of tensor category theory has proven hugely successful in analyzing intrinsic topological orders in $d=2$ spatial dimensions. It is now well-understood that the fusion and braiding properties of quasiparticles---anyons---in a topologically ordered spin system are described by a unitary modular tensor category (UMTC)~\cite{KITAEV20062,rowell}. For instance, the UMTC describing the anyons in Kitaev's quantum double model (\emph{i.e.}, a lattice realization of gauge theory for $d=2$) based on a finite group $G$ is given by the representation theory of the quantum double algebra $\mathcal{D}(G)$ \cite{KITAEV20032,bais2002,BM1,BM2}.
Rich topological orders also exist in $d=3$ spatial dimensions \cite{HZW,CC3D,Wang3loop, Jiang2014} and they may provide fault-tolerant quantum computing schemes with advantages over their $d=2$ cousins as exemplified by the color codes \cite{ColorCode2D,ColorCode3D,3DcolorThresholds}.
Since challenges vary with dimensions as seen in classifying manifolds~\cite{W,P1,P2,P3}, the theory of topological orders for $d=3$ is less developed compared to $d=2$ and remains an active research topic~\cite{Lan3DtopB,Lan3DtopF}.

Recently, a new class of models have brought to light novel gapped quantum phases of matter which  lie beyond the conventional framework of topological order. These phases, which are said to possess ``fracton order,'' were originally discovered in exactly solvable $d=3$ lattice models and exhibit a rich phenomenology, including a locally stable ground state degeneracy on the 3-torus which depends sub-extensively (hence non-topologically) on the system size and quasiparticles with restricted mobility~\cite{chamon,haah,fracton1,vijay16xcube}. In particular, these models strikingly host quasiparticles---fractons---which are intrinsically immobile (\emph{i.e.}, cannot be moved by string operators). This peculiar and striking feature serves as a defining characteristic of fracton phases and has recently led to a flurry of theoretical interest in understanding these phases from a variety of perspectives~\cite{chamon,bravyi,castelnovo,haah,haah2,yoshida,fracton1,vijay16xcube,sub,genem,williamson,prem,han,sagar,hsieh,slagle1,nonabelian,decipher,balents,slagle2,chiral,prem2,regnault,valbert,devakul,regnault2,han2,albert,leomichael,gromov,shirley,slagle3,pai,yizhi1,cagenet,pinch}. A recent review on current progress in this field can be found in Ref.~\cite{fractonreview}. 

The gapped fracton models discovered and studied thus far can be be broadly separated into type-I and type-II fracton phases, in the taxonomy of Ref.~\cite{vijay16xcube}. In type-I (resp. type-II) phases, fractons appear at the corners of membrane-like (fractal-like) operators. A further distinguishing feature of type-I phases is the presence of topologically non-trivial excitations which are mobile along sub-dimensional manifolds (lines or planes) of the three-dimensional system, while all topologically non-trivial excitations in type-II phases are strictly immobile. Well-known examples of type-I phases are Chamon's model~\cite{chamon,bravyi}, the X-cube model~\cite{vijay16xcube}, and the checkerboard model~\cite{vijay16xcube}, while Haah's code~\cite{haah} remains the paradigmatic model for type-II fracton phases. In this paper, we will restrict our attention to type-I fracton phases.

A natural question to pose is whether existing models exhaust the possible kinds of quasiparticles which a fracton phase may harbor. In particular, is it possible for fractons or excitations with restricted mobility to have a multi-channel fusion rule, \emph{i.e.}, be non-Abelian? In type-I fracton phases, certain topological excitations can move only along sub-dimensional manifolds and may thus braid non-trivially with each other, allowing some notion of non-trivial statistics to survive even though the system is three-dimensional. Thus, while in principle there appears to be no obstruction to realizing non-Abelian statistics in type-I fracton phases, a new framework is clearly needed to capture this more general class of systems, which is a primary motivation for this work.

Constructing and studying exactly solvable models has proven a fruitful approach in exploring the landscape of gapped quantum phases of matter~\cite{KITAEV20032,KITAEV20062,levinwen,HZW,GuWen,hungwan,LevinGu,BM1,BM2,BSW,CCW,CC3D,ColorCode2D,ColorCode3D,kitaevkong,kong2014,linlevin2014,lan2014,CW,lin2017,wang2017}.
For conventional topological orders, a key insight for constructing new exactly solvable models was provided by a gauging procedure relating (short-range entangled) SPT states to (long-range entangled) topological orders described by twisted gauge theories~\cite{LevinGu}. Specifically, this gauging procedure relates lattice non-linear $\sigma$-models for SPT phases~\cite{chen2013} to (lattice realizations of)  Dijkgraaf-Witten topological gauge theories~\cite{dijkgraaf1990} describing topological orders which host quasiparticle (resp. loop) excitations in $d=2$ (resp. $d=3$) spatial dimensions with rich statistical properties~\cite{hu13twisted,Wang3loop, Jiang2014,WangWen2015,wan3dtwisted,Gu2015}.  Dijkgraaf-Witten topological gauge theory (also referred to as twisted gauge theory) generalizes standard lattice gauge theory
by ``twisting'' its gauge transformations, \emph{i.e.}, by allowing them to carry an extra phase factor specified by a $(d+1)$-cocycle $\omega\in Z^{d+1}(G,U(1))$ and local field configurations, where $G$ is the gauge group. 

Similarly to the duality between SPT states and topological orders, it has been realized that certain (Abelian) type-I fracton models, such as the X-cube, can be related through a generalized gauging procedure to short-range entangled states with subsystem symmetries~\cite{vijay16xcube,yizhi1,yizhi2}. Based on this observation, most exactly solvable fracton models can be naturally interpreted as generalized lattice gauge theories~\cite{vijay16xcube}. Motivated by this interpretation of fracton models, and by the twisting procedure for obtaining Dijkgraaf-Witten theories from standard gauge theories, here we consider twisting certain type-I fracton models along planes by 3-cocycles. This allows us to systematically generate a rich family of type-I fracton models---dubbed ``twisted fracton models''---which realize non-Abelian excitations with restricted mobility, such as non-Abelian fractons. In this paper we extensively explore the properties of twisted fracton models, which form a natural platform for realizing a wide variety of novel quasiparticles, and elucidate the related notion of braiding excitations with restricted mobility.

Given the length of this paper, we now highlight our procedure and main results.

\subsection{Summary of main results}\label{subsec:intro_sum}

In this paper, we develop a general procedure for systematically constructing exactly solvable models, which we dub \emph{twisted fracton models}, thereby greatly expanding the set of type-I fracton phases and establishing a general mathematical framework within which to study non-Abelian fracton orders. We start by observing that the X-cube and checkerboard models~\cite{vijay16xcube}, as originally defined, can both be viewed as generalized Abelian lattice gauge theories. Then, in analogy with  Dijkgraaf-Witten topological gauge theory, we observe that the generalized gauge transformations can be twisted as well. This leads us to a plethora of exactly solvable three-dimensional models exhibiting a landscape of rich and hitherto undiscovered behaviors, of which we present the twisted X-cube and twisted checkerboard models as paradigmatic examples.

Importantly, these exactly solvable models establish the existence of novel type-I fracton phases hosting \emph{inextricably non-Abelian} fractons, which 
we will define shortly in this section before providing examples based on concrete models
in later sections. Moreover, in contrast to other approaches for generalizing type-I fracton orders~\cite{nonabelian,cagenet}, which are based on coupling stacks of $d=2$ topological phases, our construction here has a cleaner connection to TQFTs (explicitly, Dijkgraaf-Witten topological gauge theories) realizing similar braiding properties, which enables us to thoroughly analyze the resulting fracton models.  
For instance, we compute their ground state degeneracy (GSD) on a three-torus $\mathtt{T}^{3}$ explicitly, revealing the novel dependence of this GSD on system size in non-Abelian fracton phases for the first time.

In our analysis of the spectrum of twisted fracton models, there emerges a systematic route for describing the braiding and fusion properties of quasiparticles, including those with restricted mobility. Some key definitions which intuitively reveal the structure of these phases are as follows. A 0d (resp. 1d, 2d) mobile quasiparticle is an excited finite region which can move as a whole in 0 (resp. 1, 2) dimensions. We refer to such a  quasiparticle as an \emph{intrinsic} 0d (resp. 1d, 2d) mobile quasiparticle if it is not a fusion result of quasiparticles with higher mobility. For instance, an intrinsic 1d mobile quasiparticle cannot be obtained by fusing quasiparticles mobile in 2 dimensions. A \emph{fracton} is thus simply understood as an intrinsic 0d mobile quasiparticle. 

Assuming no non-trivial quasiparticles mobile in 3 dimensions, which is the case in all type-I gapped fracton models, the $x$ (resp. $y$, $z$) \emph{topological charge} of a quasiparticle can be detected by braiding 2d mobile quasiparticles around it in the $yz$ (resp. $zx$, $xy$) planes. The particle type of an excitation is then specified by its $x,y,z$ topological charges, which may be subject to some constraints. In addition, the \emph{quantum dimension} of a quasiparticle equals the product of quantum dimensions associated with its topological charges in the three directions. 

We now define what it means for quasiparticles with restricted mobility to be \emph{inextricably non-Abelian}. A quasiparticle is \emph{Abelian} (resp. \emph{non-Abelian}) if its quantum dimension is 1 (resp. greater than 1). An \emph{inextricably} non-Abelian fracton is one which is \emph{not} a fusion result of an Abelian fracton with some mobile quasiparticles. Similarly, an \emph{inextricably} non-Abelian 1d mobile quasiparticle is one which cannot be obtained by fusing an Abelian 1d mobile quasiparticle with some 2d mobile quasiparticles. Significantly, this implies that a fracton model hosting either an inextricably non-Abelian fracton or an inextricably non-Abelian 1d mobile quasiparticle \emph{cannot} be understood as some Abelian fracton order weakly coupled to layers of conventional two-dimensional topological states. This is one of the central results of our work, as it demonstrates the existence of a fundamentally new class of fracton orders.

Studying the excitations of twisted fracton models, we show that both inextricably non-Abelian fractons and inextricably non-Abelian 1d mobile quasiparticles may be realized within twisted checkerboard models. On the other hand, twisted X-cube models host only inextricably non-Abelian 1d mobile quasiparticles. Thus, we find two basic types of non-Abelian fracton orders: one which allows fractons (and 1d mobile quasiparticles simultaneously) inextricably non-Abelian and one which only hosts inextricably non-Abelian 1d mobile quasiparticles. Actually, in our twisted fracton models, quasiparticles may have inextricably non-Abelian topological charges in one, two or three directions, which reveals a further distinction between varieties of fracton phases.

As a further technical contribution, we provide a detailed derivation of the categorical description for anyons in twisted discrete gauge theories directly from their lattice models in two spatial dimensions, which is absent in the literature. Necessary mathematical details are included in appendices to make our derivation and discussion self-contained. This treatment applies straightforwardly to studying twisted fracton models as well.

\subsection{Outline}

We now outline the remainder of this paper. In Sec.~\ref{sec:TwistedGauge}, we treat lattice models of twisted gauge theories in two spatial dimensions. While the results contained in the section may be familiar to readers, we emphasize that our treatment differs from previous approaches and is directly applicable to the twisted fracton models introduced later. We characterize conventional topological orders by deriving properties such as their ground state degeneracy on a torus and the braiding and fusion properties of anyons. The braiding of anyons is especially transparent in our treatment, wherein anyons are represented as punctures on a disk. 

In Sec.~\ref{sec:twistedfracton}, we introduce new families of exactly solvable twisted fracton models. In particular, we introduce twisted versions of two paradigmatic examples of three-dimensional fracton order: twisted X-cube models and twisted checkerboard models. Rather than reviewing the untwisted $\mathbb{Z}_2$ X-cube and checkerboard models, for which the reader is referred to Refs.~\cite{fracton1,vijay16xcube}, we first define these models based on arbitrary finite Abelian groups $G$. We then twist the gauge symmetry by non-trivial 3-cocycles to arrive at the twisted fracton models. 

Secs.~\ref{sec:GSDXcube} and~\ref{sec:GSDchecker} are devoted to calculating the non-trivial ground state degeneracies (GSD) of twisted fracton models with the system defined on a three-torus. The explicit calculations for both the twisted X-cube (Sec.~\ref{sec:GSDXcube}) and checkerboard (Sec.~\ref{sec:GSDchecker}) models serve three purposes. Firstly, the sub-extensive system size dependence of the GSD in all cases demonstrates clearly that the models under consideration are gapped phases. In fact, it becomes clear from our later analysis of quasiparticles that this GSD is stable against arbitrary local perturbations and hence reveals that the system is non-trivially long-range entangled. Secondly, the dependence of the GSD on the system size in all models under consideration establishes the geometric nature of fracton phases: they are sensitive not only to the global topology but also to geometry. This provides a clear distinction between conventional topological order and fracton order. Thirdly, the new exotic  dependence of GSD on  system size (e.g. Eqs.~\eqref{eq:GSD_Xcube_Z}, \eqref{eq:GSD_Xcube_2}, \eqref{eq:GSD_Xcube_3}, and \eqref{eq:GSD_checker3}) strongly hints at the existence of novel non-Abelian fracton phases. 

In Secs.~\ref{sec:QPXcube} and~\ref{sec:QPchecker}, the quasiparticle spectra of the twisted X-cube and checkerboard models are analyzed respectively; here, we classify all particle types and study their braiding and fusion properties. Importantly, our analysis uncovers a systematic route for describing quasiparticles in type-I fracton phases. First, we explain how the particle type of an excited spot is labelled by its $x,y,z$ \emph{topological charges}, which can be detected by braiding 2d mobile quasiparticles around it in the $yz,zx,xy$ planes respectively. Further, we elucidate the notions of \emph{mobility} and \emph{quantum dimension} for quasiparticles and determine them through the topological charge data. We also discuss certain fusion and braiding processes in general. In order to illustrate the variety of novel fracton phases which may be accessed through our framework, we examine certain examples explicitly. We find that semionic or inextricably non-Abelian 1d mobile quasiparticles are allowed in some twisted X-cube models (Sec.~\ref{sec:QPXcube}). On the other hand, the twisted checkerboard models (Sec.~\ref{sec:QPchecker}) are shown to realize a broader variety of excitations. Specifically, we show the existence of inextricably non-Abelian fractons in a twisted checkerboard model based on the group $G=\mathbb{Z}_{2}\times\mathbb{Z}_{2}\times\mathbb{Z}_{2}$.

The paper concludes in Sec.~\ref{cncls} with a discussion of avenues for future investigation and of open questions raised by the present results. To keep this paper self-contained, necessary  mathematical materials are provided in the appendices. Specifically, Appendix~\ref{sec:GCDW} contains the definitions of group cohomology, triangulated manifolds, and the associated  Dijkgraaf-Witten weight and partition function. Appendix~\ref{sec:AlgebraDG} reviews the quasi-triangular quasi-Hopf algebra structure of a twisted quantum double $\mathcal{D}^{\omega}(G)$ and the tensor category of its representations.


\section{2D Lattice models of twisted gauge theories}
\label{sec:TwistedGauge}

\subsection{Description of lattice models}

In the following, we will describe exactly solvable lattice models
motivated by the gauge theory in two spatial dimensions based on a finite group $G$ and its
twisted versions. Here $G$ may be non-Abelian and its identity element
is denoted as $e$.

\subsubsection{Untwisted models}

As a warm-up, let us first recall the standard lattice model of a
gauge theory in two spatial dimensions ~\cite{KITAEV20032}, based on a finite group $G$
with identity element denoted by $e$. To be concrete, we work on a square
lattice (\emph{i.e.}, a two-dimensional manifold composed of square plaquettes) $\Sigma$. The discussion in this section actually applies to any other planar lattice.

\begin{figure}
	\noindent\begin{minipage}[t]{1\columnwidth}%
		\includegraphics[width=0.75\columnwidth]{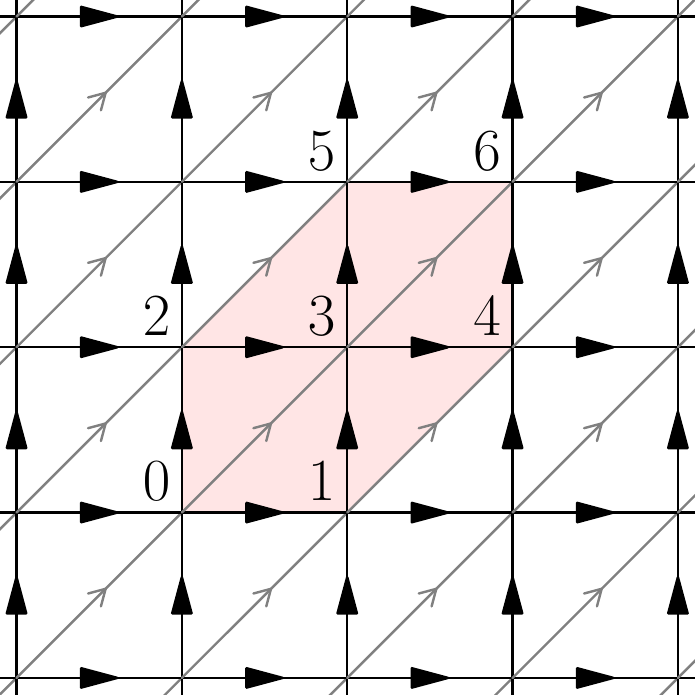}
		
		(a)%
	\end{minipage}
	
	\bigskip{}
	
	\noindent\begin{minipage}[t]{1\columnwidth}%
		\includegraphics[width=0.85\columnwidth]{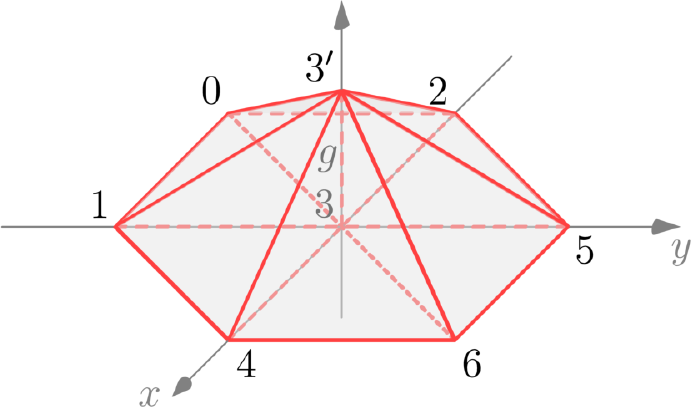}
		
		(b)%
	\end{minipage}
	
	\caption{Lattice model of gauge theory in 2+1 dimensions. (a) The physical
		degrees of freedom are on the black oriented edges of the square lattice.
		Auxiliary grey edges are added to give a complete triangulation. (b)
		$P_{v}^{g}$ for $v=3$ is presented by a triangulated pyramid with $\left[33'\right]$ colored by $g\in G$, where
		the edge orientations are picked according to the local ordering of
		vertices $0<1<2<3<3'<4<5<6$. For each tetrahedron, $\text{sgn}\left(\left[v_{0}v_{1}v_{2}v_{3}\right]\right)$
		equals the sign of the triple product $\protect\overrightarrow{v_{0}v_{1}}\cdot\left(\protect\overrightarrow{v_{0}v_{2}}\times\protect\overrightarrow{v_{0}v_{3}}\right)$.
		For example, $\text{sgn}\left(\left[0133'\right]\right)=+1$ and $\text{sgn}\left(\left[233'5\right]\right)=-1$.}
	
	\label{fig:twisted_QD}
\end{figure}

Let $E\left(\Sigma\right)$ be the set of its edges with a chosen
orientation each as shown by the black arrows in Fig.~\ref{fig:twisted_QD}(a).
In addition,  the sets of vertices and plaquettes of $\Sigma$ are
denoted by $V\left(\Sigma\right)$ and $F\left(\Sigma\right)$ respectively.
Technically, all the edges and plaquettes are thought to be closed,
\emph{i.e.}, they include their boundaries. In particular, each plaquette
contains all its edges. 
Moreover, for any region (\emph{i.e.}, subspace) $\Gamma$ of $\Sigma$, let $V\left(\Gamma\right)$,
$E\left(\Gamma\right)$ and $F\left(\Gamma\right)$ denote the subsets
of $V\left(\Sigma\right)$, $E\left(\Sigma\right)$ and $F\left(\Sigma\right)$
that collect all the vertices, edges and plaquettes inside $\Gamma$
respectively.

A local Hilbert space (also called a \emph{spin} for short) with an orthonormal basis $\left\{ \left|\ell,\sigma\right\rangle \right\} _{\sigma\in G}$ is assigned to each edge $\ell\in E(\Sigma)$. 
Thus, the Hilbert space $\mathcal{H}\left(E\left(\Gamma\right),G\right)$
associated with any region $\Gamma$ of $\Sigma$ is spanned by the vectors
\begin{equation}
\left|\zeta\right\rangle \coloneqq\bigotimes_{\ell\in E\left(\Gamma\right)}\left|\ell,\zeta\left(\ell\right)\right\rangle 
\end{equation}
labeled by $\zeta\in G^{E\left(\Gamma\right)}$, where $G^{E\left(\Gamma\right)}\coloneqq\text{Fun}\left(E\left(\Gamma\right),G\right)$
is the set of functions from $E\left(\Gamma\right)$ to $G$. Each element of $G^{E\left(\Gamma\right)}$ specifies a \emph{spin configuration} on $\Gamma$. On the whole lattice $\Sigma$, the total Hilbert space is ${\cal H}\left(E\left(\Sigma\right),G\right)$. 

Suppose that $\mathcal{O}$ is an operator acting on ${\cal H}\left(E\left(\Sigma\right),G\right)$.
We say that $\mathcal{O}$ is supported on a region $\Gamma\subseteq\Sigma$
if it can be expressed as 
\begin{equation}
\mathcal{O}=\mathcal{O}_{\Gamma}\otimes1_{\Sigma\backslash\Gamma},
\end{equation}
where $\mathcal{O}_{\Gamma}$ is an operator acting on ${\cal H}\left(E\left(\Gamma\right),G\right)$
and $1_{\Sigma\backslash\Gamma}$ denotes the identity operator acting
on the rest of the spins. Usually, $1_{\Sigma\backslash\Gamma}$ is omitted
in notations and the operators acting on ${\cal H}\left(E\left(\Gamma\right),G\right)$
are automatically viewed as operators acting on ${\cal H}\left(E\left(\Sigma\right),G\right)$
as well.

On each vertex $v$, we have a \emph{gauge transformation operator}
\begin{equation}
A_{v}^{g}\coloneqq\bigotimes_{\ell\ni v}L_{v}^{g}\left(\ell\right)\label{eq:A_v}
\end{equation}
for each $g\in G$, where $\ell\ni v$ means that $\ell$ connects
to $v$. In addition, for $\ell=\left[v_{0}v_{1}\right]$,
\begin{equation}
L_{v}^{g}\left(\ell\right)\coloneqq\begin{cases}
\sum_{\sigma\in G}\left|\ell,g\sigma\right\rangle \left\langle \ell,\sigma\right|, & v_{0}=v,v_{1}\neq v,\\
\sum_{\sigma\in G}\left|\ell,\sigma\right\rangle \left\langle \ell,\sigma g\right|, & v_{0}\neq v,v_{1}=v,\\
\sum_{\sigma\in G}\left|\ell,g\sigma g^{-1}\right\rangle \left\langle \ell,\sigma\right|, & v_{0}=v,v_{1}=v,\\
1, & v\notin\ell.
\end{cases}\label{eq:L_v}
\end{equation}
The third line in this definition of $L_{v}^{g}$ takes care of the
possibility $\ell$ being a loop, which happens when the size of the
square lattice with periodic boundary condition reduces to 1 in one direction.

It is straightforward to check that $\forall v,v_{0},v_{1}\in V\left(\Sigma\right)$,
$\forall g,h\in G$,
\begin{align}
\left(A_{v}^{g}\right)^{\dagger} & =A_{v}^{g^{-1}},\label{eq:A1}\\
A_{v}^{g}A_{v}^{h} & =A_{v}^{gh},\label{eq:A2}\\
\left[A_{v_{0}}^{g},A_{v_{1}}^{h}\right] & =0,\text{ if }v_{0}\neq v_{1}.\label{eq:A3}
\end{align}
Thus, we have a set of mutually commuting Hermitian local projectors,
also known as \emph{stabilizers} in the quantum computation literature, 
\begin{equation}
A_{v}\coloneqq\frac{1}{\left|G\right|}\sum_{g\in G}A_{v}^{g}
\end{equation}
labeled by vertices.

On each plaquette $p$, we have a projector which requires the triviality
of flux 
\begin{equation}
B_{p}\coloneqq\sum_{\zeta\in G^{E\left(p\right)}}\delta_{\zeta\left(\left[v_{0}v_{1}\right]\right)\zeta\left(\left[v_{1}v_{2}\right]\right)\zeta\left(\left[v_{2}v_{3}\right]\right)\zeta\left(\left[v_{3}v_{0}\right]\right),e}\left|\zeta\right\rangle \left\langle \zeta\right|,
\end{equation}
where $e$ is the identity element of $G$ and $v_{0}v_{1}v_{2}v_{3}$
is a sequence of vertices around the boundary of $p$. If the orientation
of an edge $\ell=\left[v_{0}v_{1}\right]$ is inverse to what is picked
in Fig.~\ref{fig:twisted_QD}(a) (\emph{i.e.}, $\left[v_{1}v_{0}\right]\in E\left(\Sigma\right)$),
then $\zeta\left(\ell\right)\coloneqq\left(\zeta\left(\left[v_{1}v_{0}\right]\right)\right)^{-1}$. 

It can be checked that all the projectors $A_{v}$ and $B_{p}$ labeled
by vertices and plaquettes commute with each other. They form a set
of stabilizers that completely fix local degrees of freedom. In other
words, 
\begin{equation}
H=-\sum_{v}A_{v}-\sum_{p}B_{p}
\end{equation}
is a gapped Hamiltonian. In particular, it has a finite ground state
degeneracy when embedded in a torus (\emph{i.e.}, with periodic boundary
conditions), which is independent of system size and robust to any
local perturbations. 

\subsubsection{Twisted models}

Motivated by Dijkgraaf-Witten topological gauge theories~\cite{dijkgraaf1990},
the above lattice model based on a finite group $G$ can be twisted by a 3-cocycle $\omega\in Z^{3}\left(G,U\left(1\right)\right)$, resulting in generalizations classified by the corresponding group cohomology $\left[\omega\right]\in H^{3}\left(G,U\left(1\right)\right)$.
More details of  Dijkgraaf-Witten topological gauge theories and
group cohomology are summarized in Appendix~\ref{sec:GCDW}.

In the twisted model, we will keep $B_{p}$ unchanged. For any region
$\Gamma$ of $\Sigma$, let 
\begin{align}
G_{B}^{E\left(\Gamma\right)} & \coloneqq\left\{ \zeta\in G^{E\left(\Sigma\right)}\;|\;B_{p}\left|\zeta\right\rangle =\left|\zeta\right\rangle ,\forall p\in F\left(\Gamma\right)\right\} ,
\end{align}
whose element are called \emph{locally flat} spin configurations on $\Gamma$. 
Let $\mathcal{H}_{B}\left(E\left(\Gamma\right),G\right)$
denote the Hilbert subspace spanned by $\left|\zeta\right\rangle $
with $\zeta\in G_{B}^{E\left(\Gamma\right)}$. 

In order to define twisted versions of gauge transformation operators,
we pick a complete triangulation of $\Sigma$ by adding the grey oriented
edges shown in Fig.~\ref{fig:twisted_QD}(a). The orientations of
edges are picked such that there is no triangle whose three edges
form a closed walk; such a choice is called a \emph{branching structure}~\cite{chen2013}. Then every triangle $\tau$ is ordered and should
be labeled as $\left[\tau_{0}\tau_{1}\tau_{2}\right]$ with vertices
ordered such that the orientations of the edges $\left[\tau_{0}\tau_{1}\right]$,
$\left[\tau_{1}\tau_{2}\right]$ and $\left[\tau_{0}\tau_{2}\right]$
coincide with the branching structure. 

Technically, the branched triangulation makes $\Sigma$ into a $\Delta$-complex. 
The definition
of a $\Delta$-complex is given in Ref.~\cite{Hatcher}. For a general
$\Delta$-complex $\mathtt{X}$, we denote the set of $n$-simplices (\emph{i.e.},
vertices for $n=0$, edges for $n=1$, triangles for $n=2$, tetrahedrons for 
$n=3$ and so on) in $\mathtt{X}$ by $\Delta^{n}\left(\mathtt{X}\right)$. 

A function
$\xi\in G^{\Delta^{1}\left(\mathtt{X}\right)}$ is called a \emph{coloring
}~\cite{mesaros13ying} of $\mathtt{X}$ with $G$, if $\xi\left(\left[\tau_{0}\tau_{1}\right]\right)\xi\left(\left[\tau_{1}\tau_{2}\right]\right)=\xi\left(\left[\tau_{0}\tau_{2}\right]\right)$
on any triangle $\left[\tau_{0}\tau_{1}\tau_{2}\right]\in\Delta^{2}\left(\mathtt{X}\right)$.
The set of colorings of $\mathtt{X}$ with $G$ is denoted 
$\text{Col}\left(\mathtt{X};G\right)$ 
or simply $\text{Col}\left(\mathtt{X}\right)$ when $G$ does not need to be specified
explicitly.

For each $v\in V\left(\Sigma\right)$, let $\Sigma\left[v\right]$
be the region inside $\Sigma$ made of all plaquettes adjacent
to $v$. 
Take the vertex $v=3$ shown in Fig.~\ref{fig:twisted_QD}(a) for instance: $\Sigma\left[v\right]$ contains four plaquettes (around $v$)  
including their edges (twelve in total).
Then $\Sigma\left[v\right]$ is a $\Delta$-subcomplex of $\Sigma$ as well and each  $\zeta\in G_{B}^{E\left(\Sigma\left[v\right]\right)}$ determines a coloring of $\Sigma\left[v\right]$.
In particular, the group
element assigned to the edge $\left[03\right]$ is $\zeta\left(\left[01\right]\right)\zeta\left(\left[13\right]\right)=\zeta\left(\left[02\right]\right)\zeta\left(\left[23\right]\right)$.

Further, we construct a pyramid $\mathtt{P}_{v}$ over $v$, whose bottom is the union of all triangles adjacent to $v$. Let $v'$ denote the apex of $\mathtt{P}_{v}$. With $\left[vv'\right]=g$ (\emph{i.e.}, $\left[vv'\right]$ colored by $g\in G$), the pyramid presents an operator 
\begin{equation}
P_{v}^{g}\coloneqq\sum_{\zeta\in G_{B}^{E\left(\Sigma\left[v\right]\right)}}\left|\zeta\right\rangle \omega\left[\zeta,\mathtt{P}_{v}^{g}\right]\left\langle \zeta A_{v}^{g}\right|,\label{eq:P_v}
\end{equation}
where $\zeta A_{v}^{g}\in G_{B}^{E\left(\Sigma\left[v\right]\right)}$
is determined by $\left\langle \zeta A_{v}^{g}\right|=\left\langle \zeta\right|A_{v}^{g}$ and 
$\omega\left[\zeta,\mathtt{P}_{v}^{g}\right]$ is the
Dijkgraaf-Witten weight, defined by Eq.~\eqref{eq:DW_weight},
on $\mathtt{P}_{v}$ with the coloring specified by $\left[vv'\right]=g$ and $\zeta$ on the bottom.
Explicitly, for $v=3$ in Fig.~\ref{fig:twisted_QD}(b), 
\begin{equation}
\omega\left[\zeta,\mathtt{P}_{v}^{g}\right]=\frac{\left[0133'\right]\left[133'4\right]\left[33'46\right]}{\left[0233'\right]\left[233'5\right]\left[33'56\right]},
\end{equation}
where each tetrahedron $\left[v_{0}v_{1}v_{2}v_{3}\right]$ stands
for the phase factor $\omega\left(\left[v_{0}v_{1}\right],\left[v_{1}v_{2}\right],\left[v_{2}v_{3}\right]\right)$
with edges short for their associated group elements. For example, 
\begin{align}
\left[0133'\right] & =\omega\left(\zeta\left(\left[01\right]\right),\zeta\left(\left[13\right]\right),g\right),\\
\left[133'4\right] & =\omega\left(\zeta\left(\left[13\right]\right),g,g^{-1}\zeta\left(\left[34\right]\right)\right).
\end{align}
Clearly, $g=e$ implies $\omega\left[\zeta,\mathtt{P}_{v}^{g}\right]=1,\forall \zeta\in G_{B}^{E\left(\Sigma\left[v\right]\right)}$.

\begin{figure}
\includegraphics[width=0.85\columnwidth]{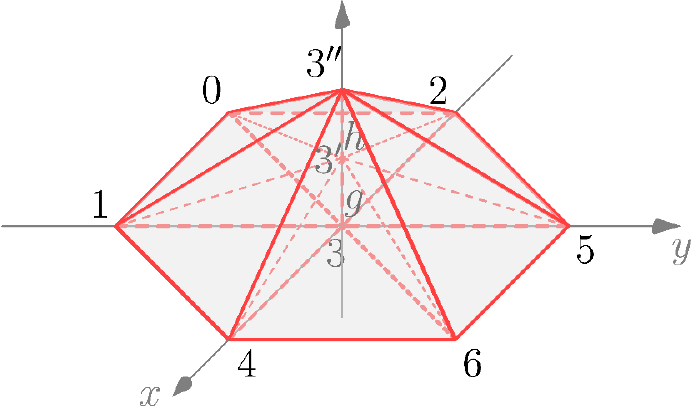}

\caption{Stacking two pyramids over a vertex as a graphic representation of $P_{v}^{g}P_{v}^{h}$.}

\label{fig:twisted_QD-1}
\end{figure}

\begin{figure}
\noindent\begin{minipage}[t]{1\columnwidth}%
\includegraphics[width=0.85\columnwidth]{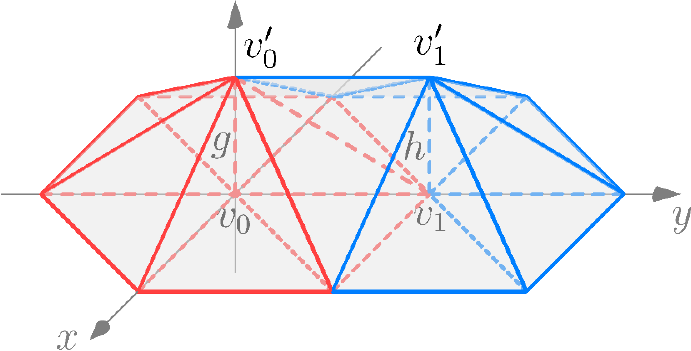}

(a)%
\end{minipage}

\bigskip{}

\noindent\begin{minipage}[t]{1\columnwidth}%
\includegraphics[width=0.85\columnwidth]{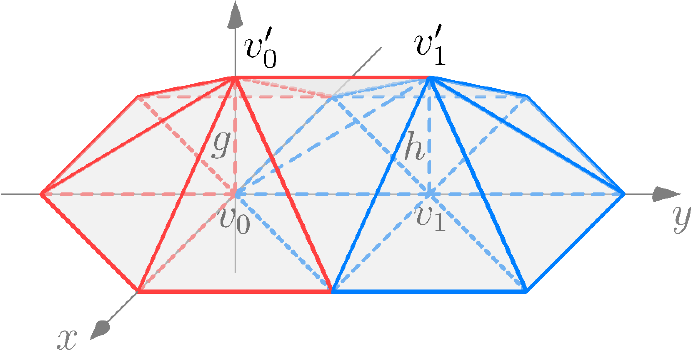}

(b)%
\end{minipage}

\caption{Two orders of stacking two pyramids over two adjacent vertices, with
(a) and (b) presenting $P_{v_{0}}^{g}P_{v_{1}}^{h}$ and $P_{v_{1}}^{h}P_{v_{0}}^{g}$
respectively.}

\label{fig:twisted_QD-2}
\end{figure}

With this graphic representation, we can demonstrate some crucial properties
of these operators. First, on a single vertex, $\omega\left[\zeta,\mathtt{P}_{v}^{g}\right]\omega\left[\zeta A_{v}^{g},\mathtt{P}_{v}^{h}\right]$
can be presented as a stack of pyramids colored by $\zeta$ on the
bottom and $\left[33'\right]=g$, $\left[3'3''\right]=h$ as shown in
Fig.~\ref{fig:twisted_QD-1}. Thus, it is the Dijkgraaf-Witten weight
on this particular coloring of the stack, which is a pyramid over
$v$ with a different bulk triangulation. Topologically, the pyramid
is just a ball with a particular surface triangulation. The cocycle
condition of $\omega$ implies that the Dijkgraaf-Witten weight assigned to a ball only
depends on its surface triangulation and coloring, which is discussed
in Appendix~\ref{subsec:DW} in a general setting. Therefore, $\omega\left[\zeta,\mathtt{P}_{v}^{g}\right]\omega\left[\zeta A_{v}^{g},\mathtt{P}_{v}^{h}\right]=\omega\left[\zeta,\mathtt{P}_{v}^{gh}\right]$
and hence
\begin{equation}
P_{v}^{g}P_{v}^{h}=P_{v}^{gh}.\label{eq:P_gh}
\end{equation}
Setting $h=g^{-1}$, we get $\omega\left[\zeta,\mathtt{P}_{v}^{g}\right]\omega\left[\zeta A_{v}^{g},\mathtt{P}_{v}^{g^{-1}}\right]=1$
and hence $\omega\left[\zeta A_{v}^{g},\mathtt{P}_{v}^{g^{-1}}\right]=\left(\omega\left[\zeta,\mathtt{P}_{v}^{g}\right]\right)^{*}$.
Thus,
\begin{equation}
\left(P_{v}^{g}\right)^{\dagger}=P_{v}^{g^{-1}}.\label{eq:P_dagger}
\end{equation}
Together, Eqs.~\eqref{eq:P_gh} and \eqref{eq:P_dagger} imply that
\begin{equation}
P_{v}\coloneqq\frac{1}{\left|G\right|}\sum_{g\in G}P_{v}^{g}
\end{equation}
is a Hermitian projector. 

In addition, on two distinct vertices $v_{0}$ and $v_{1}$, the operators
$P_{v_{0}}^{g}P_{v_{1}}^{h}$ and $P_{v_{1}}^{h}P_{v_{0}}^{g}$ are
supported on $\Sigma\left[v_{0}\right]\cup\Sigma\left[v_{1}\right]$.
If $v_{0}$ and $v_{1}$ are adjacent (\emph{i.e.}, $\left[v_{0}v_{1}\right]\in\Delta^{1}\left(\Sigma\right)$),
their matrix elements $\left\langle \zeta\right|P_{v_{0}}^{g}P_{v_{1}}^{h}\left|\zeta'\right\rangle $
and $\left\langle \zeta\right|P_{v_{1}}^{h}P_{v_{0}}^{g}\left|\zeta'\right\rangle $
are equal, because they are Dijkgraaf-Witten weight on the two stacks of
pyramids in Fig.~\ref{fig:twisted_QD-2}, which are the same topological space with identical surface triangulation and coloring. Therefore, $P_{v_{0}}^{g}P_{v_{1}}^{h}=P_{v_{1}}^{h}P_{v_{0}}^{g}$,
which becomes even more obvious if $v_{0}$ and $v_{1}$ are not adjacent.
In short,
\begin{equation}
\left[P_{v_{0}}^{g},P_{v_{1}}^{h}\right]=0\text{ if }v_{0}\neq v_{1}.\label{eq:Pv1_Pv2}
\end{equation}
As a result, the set of Hermitian projectors $\left\{ P_{v}\right\} _{v\in V\left(\Sigma\right)}$ labeled by vertices commute with each other.

When the 3-cocycle is completely trivial (\emph{i.e.}, $\omega\equiv1$), the operator $P_{v}^{g}$
reduces to $A_{v}^{g}\prod_{p\ni v}B_{p}$. So $P_{v}^{g}$ is the
twisted version of $A_{v}^{g}$ with the projector $\prod_{p\ni v}B_{p}$
included. The Hamiltonian can be simply expressed as 
\begin{equation}
H=-\sum_{v}P_{v},
\end{equation}
whose ground states are specified by $P_{v}=1$. When $\omega\equiv1$,
the ground states are the same as those specified by $A_{v}=B_{p}=1$.

\begin{figure}
	\includegraphics[width=0.7\columnwidth]{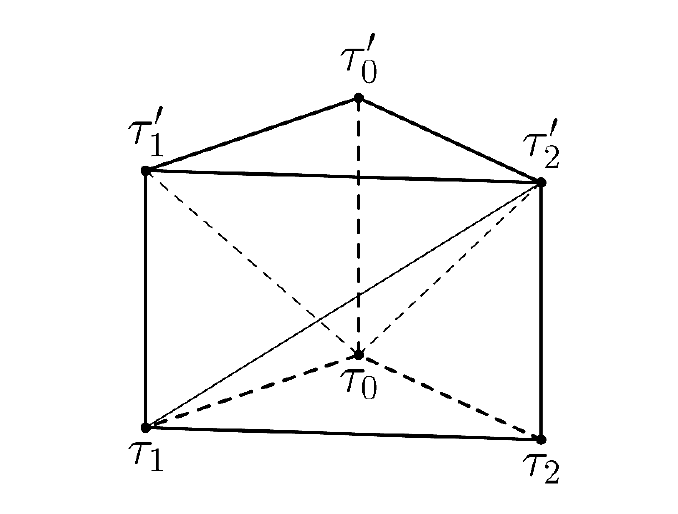}
	
	\caption{Default triangulation of a prism over $\tau\times \mathtt{I}$, where $\tau=\left[\tau_{0}\tau_{1}\tau_{2}\right]$
		and $\mathtt{I}=\left[0,1\right]$. The Dijkgraaf-Witten weight on this prism
		is $\left[\tau_{0}\tau_{1}\tau_{2}\tau_{2}^{\prime}\right]\left[\tau_{0}\tau_{0}^{\prime}\tau_{1}^{\prime}\tau_{2}^{\prime}\right]/\left[\tau_{0}\tau_{1}\tau_{1}^{\prime}\tau_{2}^{\prime}\right]$,
		where each tetrahedron $\left[v_{0}v_{1}v_{2}v_{3}\right]$ stands
		for $\omega\left(\left[v_{0}v_{1}\right],\left[v_{1}v_{2}\right],\left[v_{2}v_{3}\right]\right)$
		with edges short for the group elements assigned by the coloring.}
	
	\label{fig:prism-1}
\end{figure}

To conclude this subsection, we would like to generalize the above
definition of $P_{v}^{g}$ to take care of singular triangulations,
where vertices of a triangle may coincide. This is done by replacing
$\omega\left[\zeta,\mathtt{P}_{v}^{g}\right]$
in Eq.~\eqref{eq:P_v} with 
\begin{equation}
\omega\left[\Sigma,v;\zeta,g\right]\coloneqq\prod_{\tau\in\Delta^{2}\left(v,\Sigma\right)}\left(\frac{\left[\tau_{0}\tau_{1}\tau_{2}\tau_{2}^{\prime}\right]\left[\tau_{0}\tau_{0}^{\prime}\tau_{1}^{\prime}\tau_{2}^{\prime}\right]}{\left[\tau_{0}\tau_{1}\tau_{1}^{\prime}\tau_{2}^{\prime}\right]}\right)^{\text{sgn}\left(\tau\right)},\label{eq:w-1}
\end{equation}
where $\Sigma$ is a surface whose triangulation may be singular
and $\Delta^{2}\left(v,\Sigma\right)$ denotes the set of triangles
adjacent to the vertex $v$ in $\Sigma$. For each triangle $\tau=\left[\tau_{0}\tau_{1}\tau_{2}\right]$,
the sign $\text{sgn}\left(\tau\right)$ is $+1$ (resp. $-1$) if
the branching structure orders its vertices in the counterclockwise
(resp. clockwise) way. To define and compute
$\left[\tau_{0}\tau_{1}\tau_{2}\tau_{2}^{\prime}\right]\left[\tau_{0}\tau_{0}^{\prime}\tau_{1}^{\prime}\tau_{2}^{\prime}\right]/\left[\tau_{0}\tau_{1}\tau_{1}^{\prime}\tau_{2}^{\prime}\right]\in U\left(1\right)$,
we present it graphically as a prism in Fig.~\ref{fig:prism-1} with
bottom $\left[\tau_{0}\tau_{1}\tau_{2}\right]$ colored by $\zeta$.
Moreover, for $i=0,1,2$, we color $\left[\tau_{i}\tau_{i}^{\prime}\right]$
by $g$ if $\tau_{i}=v$ and $e$ otherwise. Then the coloring of
the rest of the edges is completely determined and each tetrahedron stands
for the phase factor assigned by $\omega$. For example, by $\left[\tau_{0}\tau_{1}\tau_{2}\tau_{2}^{\prime}\right]$
we mean $\omega\left(\left[\tau_{0}\tau_{1}\right],\left[\tau_{1}\tau_{2}\right],\left[\tau_{2}\tau_{2}^{\prime}\right]\right)$
with edges $\left[\tau_{0}\tau_{1}\right],\left[\tau_{1}\tau_{2}\right],\left[\tau_{2}\tau_{2}^{\prime}\right]$
short for the group elements assigned to them by this coloring. It
is easy to see that $\omega\left[\Sigma,v;\zeta,g\right]$ reduces
back to $\omega\left[\zeta,\mathtt{P}_{v}^{g}\right]$
if the triangulation is regular.

\begin{figure}
\includegraphics[width=0.6\columnwidth]{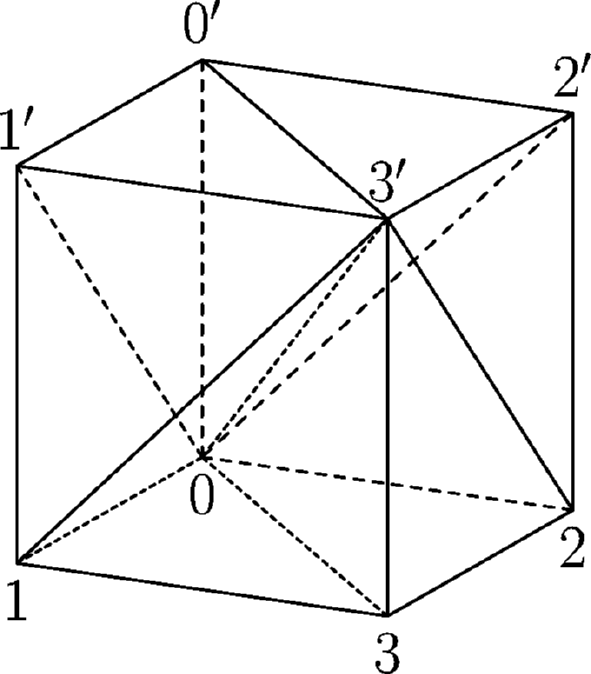}

\caption{A triangulation of a cube. The eight vertices are ordered as $0<0'<1<1'<2<2'<3<3'$;
their ordering assigns orientations to edges, triangles and tetrahedrons.
Gluing the three pairs of opposite faces of the cube gives a triangulated
three-dimensional torus $\mathtt{T}^{3}$.}

\label{fig:t3}
\end{figure}

\subsection{Ground state degeneracy on torus}

Suppose that the lattice model of a twisted gauge theory is defined
with periodic boundary conditions in both directions. In other words,
the lattice $\Sigma$ is embedded in a topological torus $\mathtt{T}^{2}$. 

Let us compute its ground state degeneracy. Technically, it equals
$\text{tr}P$, the trace of 
\begin{equation}
P\coloneqq\prod_{v\in V\left(\Sigma\right)}P_{v}
\end{equation}
over the physical Hilbert space $\mathcal{H}\left(E(\Sigma),G\right)$,
or equivalently over $\mathcal{H}_{B}\left(E(\Sigma),G\right)$. Hence,
\begin{align}
\text{tr}P & =\frac{1}{\left|G^{V\left(\Sigma\right)}\right|}\sum_{\zeta\in G_{B}^{E\left(\Sigma\right)}}\sum_{\eta\in G^{V\left(\Sigma\right)}}\left\langle \zeta\right|\prod_{v}P_{v}^{\eta\left(v\right)}\left|\zeta\right\rangle ,\label{eq:TrP}
\end{align}
where $V\left(\Sigma\right)$ is the set of vertices of $\Sigma$
and $G^{V\left(\Sigma\right)}$ is the set of functions from $V\left(\Sigma\right)$
to $G$. 

Pick a vertex $u\in V\left(\Sigma\right)$ and two non-contractible
loops $q_{x}$, $q_{y}$ based at $u$ along the two spatial directions.
For any $\zeta\in\text{Col}\left(\Sigma,G\right)$, let $h_1$ and
$h_2$ be the group elements assigned by $\zeta$ to $q_{x}$ and
$q_{y}$ respectively. The choices of $\eta$ such that $\left\langle \zeta\right|\prod_{v}A_{v}^{\eta\left(v\right)}\left|\zeta\right\rangle \neq0$
are labeled by $h_{3}\coloneqq\eta\left(u\right)\in Z_{G}\left(h_{1},h_{2}\right)$,
where $Z_{G}\left(h_{1},h_{2}\right)$ is the centralizer of $\left\{ h_{1},h_{2}\right\} $
in $G$. Actually, $\left\langle \zeta\right|\prod_{v}A_{v}^{\eta\left(v\right)}\left|\zeta\right\rangle $
can be thought of as the Dijkgraaf-Witten weight $\omega\left[\mathtt{T}^{2}\times \mathtt{I};\zeta,\zeta,h_{3}\right]$
on a triangulated space $\mathtt{T}^{2}\times \mathtt{I}$ with its bottom
$\mathtt{T}^{2}\times\left\{ 0\right\} $ and top $\mathtt{T}^{2}\times\left\{ 1\right\} $
both colored by $\zeta$ and an edge $\left[uu'\right]$ colored by
$h_{3}$. Here $\mathtt{I}=\left[0,1\right]$ and $u$, $u'$ stand for $u\times\left\{ 0\right\} $,
$u\times\left\{ 1\right\} $ respectively. Further, since the bottom
and top of $\mathtt{T}^{2}\times \mathtt{I}$ are identically triangulated
and colored, we can simply glue them together and think of $\left\langle \zeta\right|\prod_{v}A_{v}^{\eta\left(v\right)}\left|\zeta\right\rangle $
as the Dijkgraaf-Witten weight on the three dimensional torus $\mathtt{T}^{3}$,
which only depends on the group elements associated with the three
non-contractible loops based at a vertex. Thus, 
\begin{equation}
\left\langle \zeta\right|\prod_{v}A_{v}^{\eta\left(v\right)}\left|\zeta\right\rangle =\omega\left[\mathtt{T}^{3};h_{1},h_{2},h_{3}\right],
\end{equation}
when it is nonzero. Further, Eq.~\eqref{eq:TrP} reduces to
\begin{equation}
\text{tr}P=\mathcal{Z}_{\omega}\left(\mathtt{T}^{3}\right).
\end{equation}
In other words, the ground state degeneracy on $\mathtt{T}^{2}$ equals
the Dijkgraaf-Witten partition function on $\mathtt{T}^{3}$.

For the purpose of calculation, we can use the simplest triangulation of $\mathtt{T}^{3}$ shown in Fig.~\ref{fig:t3} and get 
\begin{align}
\omega\left[\mathtt{T}^{3};h_{1},h_{2},h_3\right] & =\frac{\left[0133'\right]\left[00'1'3'\right]}{\left[011'3'\right]}\cdot\frac{\left[022'3'\right]}{\left[0233'\right]\left[00'2'3'\right]}\nonumber \\
 & =\frac{\omega_{h_3}\left(h_{1},h_{2}\right)}{\omega_{h_3}\left(h_{2},h_{1}\right)},
\end{align}
where $\left[01\right]=h_{1}$, $\left[02\right]=h_{2}$, $\left[00'\right]=h_{3}$ and for $g,s,t\in G$,
\begin{equation}
\omega_{g}\left(s,t\right)\coloneqq\frac{\omega\left(g,s,t\right)\omega\left(s,t,\left(st\right)^{-1}gst\right)}{\omega\left(s,s^{-1}gs,t\right)}.
\end{equation}
Thus, we can compute the ground states degeneracy $\text{tr}P$ on
$\mathtt{T}^{2}$ explicitly by
\begin{multline}
\text{tr}P=\mathcal{Z}_{\omega}\left(\mathtt{T}^{3}\right)=
\frac{1}{\left|G\right|}\sum_{h_{1},h_{2},h_3\in G}
\delta_{h_{1}h_{2},h_{2}h_{1}}\\
\cdot\delta_{h_1 h_3,h_3 h_1}\cdot\delta_{h_2 h_{3},h_{3} h_2}\cdot
\frac{\omega_{h_3}\left(h_{1},h_{2}\right)}{\omega_{h_3}\left(h_{2},h_{1}\right)}.
\end{multline}
In particular, if $G$ is Abelian and $\omega\equiv1$, then $\text{tr}P=\left|G\right|^{2}$. 

\subsubsection{Example: $G=\mathbb{Z}_{2}$ twisted}
\label{subsec:DW_Z2}

As an example of a twisted model, we consider $G=\mathbb{Z}_{2}=\left\{ 0,1\right\} $
with a non-trivial 3-cocyle
\begin{equation}
\omega\left(f,g,h\right)=\begin{cases}
-1, & f=g=h=1,\\
1, & \text{otherwise}.
\end{cases}
\end{equation}
We will see soon that this gives rise to anyons with topological spins
$\pm i$ and hence this model is called a double semion model~\cite{levinwen,FREEDMAN2004428}.
Although $\left[\omega\right]$ is nontrivial in $H^{3}\left(G,U\left(1\right)\right)$,
we still have $\frac{\omega_{h_3}\left(h_{1},h_{2}\right)}{\omega_{h_3}\left(h_{2},h_{1}\right)} \equiv 1$
and hence $\text{tr}P=\left|G\right|^{2}$, the same ground state
degeneracy on $\mathtt{T}^{2}$ as in the untwisted model. 

\subsubsection{Example: $G=\mathbb{Z}_{m}^{3}$ with $\omega\left(f,g,h\right)=e^{i\frac{2\pi}{m}f^{\left(1\right)}g^{\left(2\right)}h^{\left(3\right)}}$}
\label{subsec:DW_Zm3}

Another interesting model can be constructed with $G=\mathbb{Z}_{m}\times\mathbb{Z}_{m}\times\mathbb{Z}_{m}\equiv\mathbb{Z}_{m}^{3}$
with a 3-cocyle
\begin{equation}
\omega\left(f,g,h\right)=e^{i\frac{2\pi}{m}f^{\left(1\right)}g^{\left(2\right)}h^{\left(3\right)}},
\end{equation}
for $f=\left(f^{\left(1\right)},f^{\left(2\right)},f^{\left(3\right)}\right),\;g=\left(g^{\left(1\right)},g^{\left(2\right)},g^{\left(3\right)}\right),\;h=\left(h^{\left(1\right)},h^{\left(2\right)},h^{\left(3\right)}\right)\in G$,
where the multiplication $f^{\left(1\right)}g^{\left(2\right)}h^{\left(3\right)}$
is well-defined from the ring structure of $\mathbb{Z}_{m}$. Often,
we also write the elements of $G$ simply as $000$, $100$, $110$
and so on for short. As examples, we have $\omega\left(100,010,001\right)=-1$
and $\omega\left(100,001,010\right)=1$ in such notations for $m=2$.
Now 
\begin{align}
\frac{\omega_{h_3}\left(h_{1},h_{2}\right)}{\omega_{h_3}\left(h_{2},h_{1}\right)} & =\exp\left(i\frac{2\pi}{m}\left|\begin{array}{ccc}
h_{1}^{\left(1\right)} & h_{2}^{\left(1\right)} & h_3^{\left(1\right)}\\
h_{1}^{\left(2\right)} & h_{2}^{\left(2\right)} & h_3^{\left(2\right)}\\
h_{1}^{\left(3\right)} & h_{2}^{\left(3\right)} & h_3^{\left(3\right)}
\end{array}\right|\right)\\
 & =\exp\left\{ i\frac{2\pi}{m}\left(h_{1}\times h_{2}\right)\cdot h_3\right\} 
\end{align}
is nontrivial, where we write 
\begin{align}
f\times g & \coloneqq\left(f^{\left(2\right)}g^{\left(3\right)}-f^{\left(3\right)}g^{\left(2\right)},\right.\nonumber \\
 & \left.f^{\left(3\right)}g^{\left(1\right)}-f^{\left(1\right)}g^{\left(3\right)},f^{\left(1\right)}g^{\left(2\right)}-f^{\left(2\right)}g^{\left(1\right)}\right),\label{eq:crossprod}\\
f\cdot g & \coloneqq f^{\left(1\right)}g^{\left(1\right)}+f^{\left(2\right)}g^{\left(2\right)}+f^{\left(3\right)}g^{\left(3\right)},\label{eq:dotprod}
\end{align}
for any $g,h\in G$. 

By noticing the identity
\begin{equation}
\frac{1}{\left|G\right|}\sum_{h_3\in G}\exp\left\{ i\frac{2\pi}{m}\left(h_{1}\times h_{2}\right)\cdot h_3\right\} =\delta_{h_{1}\times h_{2},0},
\end{equation}
we get an explicit formula for the ground state degeneracy on $\mathtt{T}^{2}$
\begin{equation}
\text{tr}P=\mathcal{Z}_{\omega}\left(\mathtt{T}^{3}\right)=\sum_{h_{1},h_{2}\in G}\delta_{h_{1}\times h_{2},0}.
\end{equation}
In particular, for $m=2$, we have 
\begin{equation}
\text{tr}P=\mathcal{Z}_{\omega}\left(\mathtt{T}^{3}\right)=22,
\end{equation}
which is quite different from the untwisted case whose ground state
degeneracy on $\mathtt{T}^{2}$ is $\left|G\right|^{2}=64$.

\subsection{Anyons and twisted quantum double algebra}

It is well-known that the quasiparticles in these two-dimensional models 
are anyons and that the total number of particle types equals the ground state
degeneracy on a torus $\mathtt{T}^2$.
Explicitly, the particle types of anyons can be labeled by irreducible
representations (up to isomorphism) of the twisted quantum double
algebra $\mathcal{D}^{\omega}\left(G\right)$. 
Actually, $\mathcal{D}^{\omega}\left(G\right)$
can be enhanced into a quasi-triangular quasi-Hopf algebra equipped with a coproduct
$\varDelta:\mathcal{D}^{\omega}\left(G\right)\otimes\mathcal{D}^{\omega}\left(G\right)\rightarrow\mathcal{D}^{\omega}\left(G\right)$
and a universal $R$-matrix $R\in\mathcal{D}^{\omega}\left(G\right)\otimes\mathcal{D}^{\omega}\left(G\right)$;
the extra structures encode the fusion and braiding properties of
anyons. If $\omega\equiv 1$, then  $\mathcal{D}^{\omega}\left(G\right)$ reduces to the normal quantum double $\mathcal{D}(G)$, which is a quasi-triangular Hopf algebra and used in studying the standard gauge theories in two spatial dimensions \cite{KITAEV20032,bais2002,BM1,BM2}. The lattices models of (twisted) gauge theories in two spatial dimensions are thus also often called (twisted) quantum double models.
The mathematical details of $\mathcal{D}^{\omega}\left(G\right)$
and its representations are summarized in Appendix~\ref{sec:AlgebraDG}.
Below, we will elucidate the notion of anyon and its connection to
the representation theory of $\mathcal{D}^{\omega}\left(G\right)$
in the concrete lattice models. 

\subsubsection{Topological charge and representation}
\label{subsec:Top_charge}

\begin{figure}
\noindent\begin{minipage}[t]{1\columnwidth}%
\includegraphics[width=0.75\columnwidth]{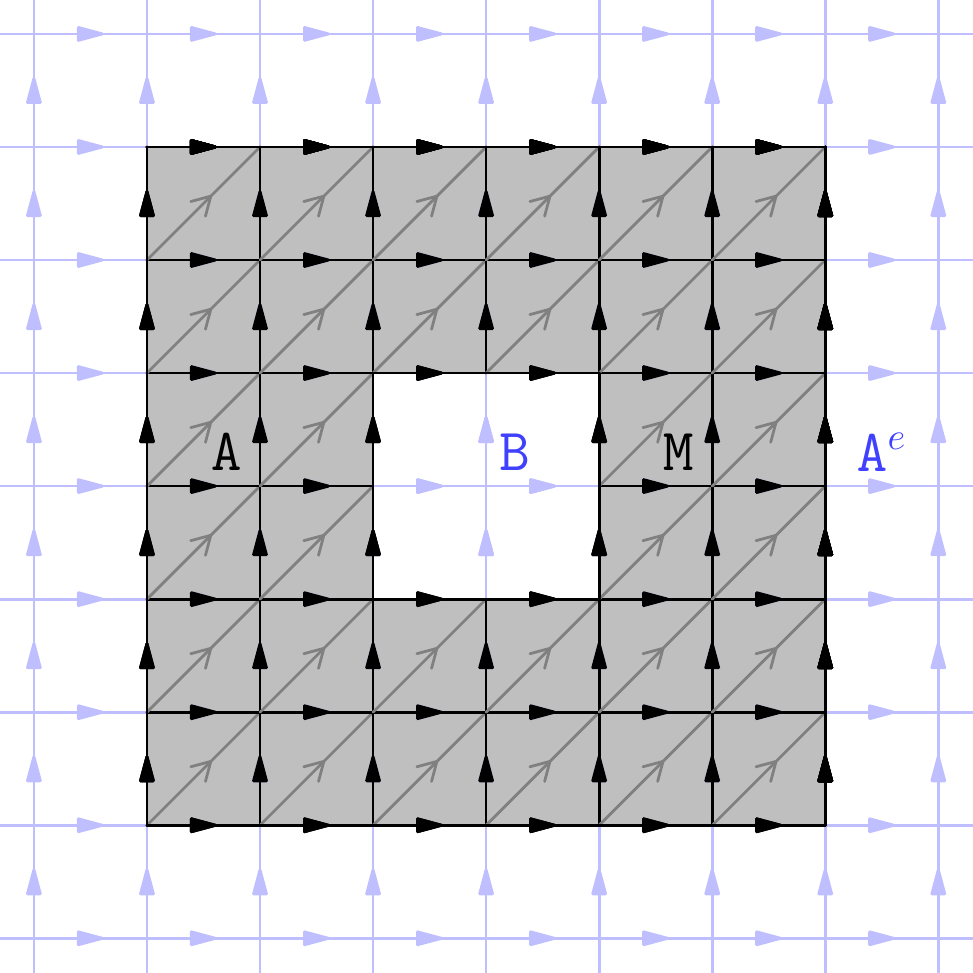}

(a)%
\end{minipage}

\bigskip{}

\noindent\begin{minipage}[t]{1\columnwidth}%
\includegraphics[width=0.75\columnwidth]{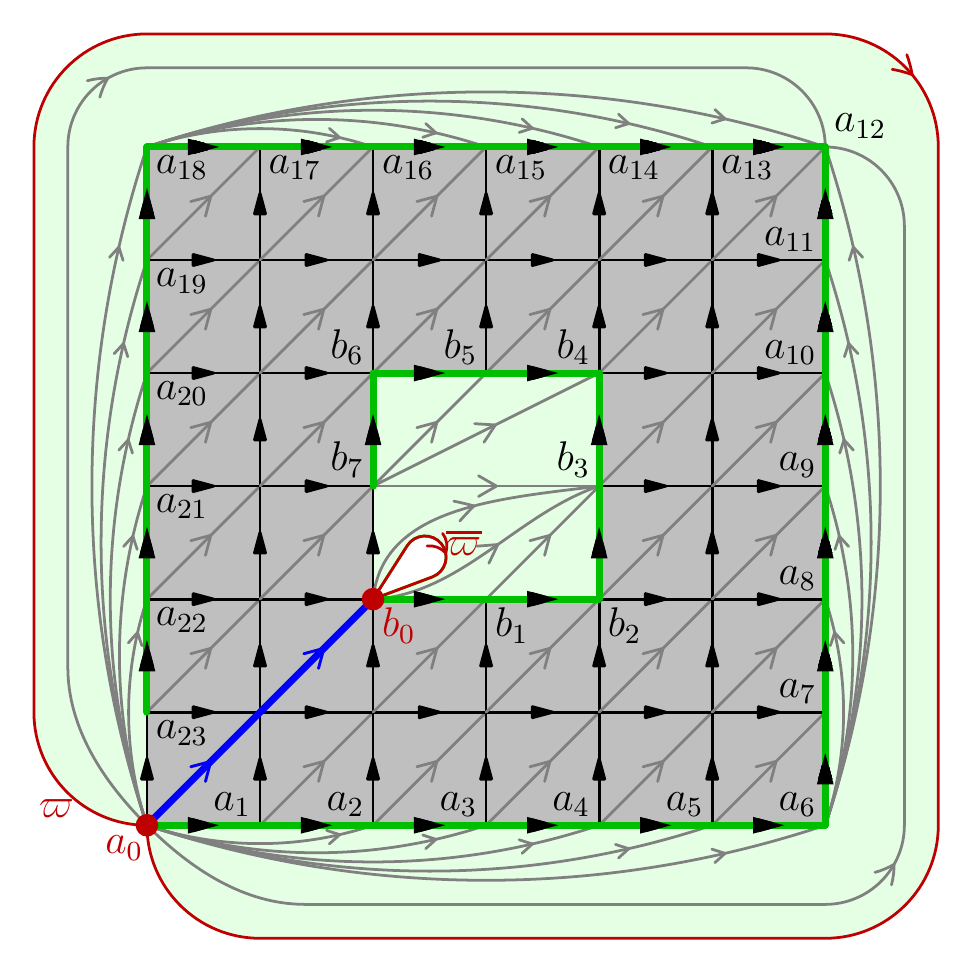}

(b)%
\end{minipage}

\caption{Hilbert space used for classifying excitations within a finite region
$\mathtt{B}$. (a) There is no excitation in the grey region 
$\mathtt{M}=\mathtt{A}-\mathtt{B}^{\circ}$, where $\mathtt{A}$ 
is a larger region containing $\mathtt{B}$ and $\mathtt{B}^\circ$
is the interior of $\mathtt{B}$.
(b) Extra $P_{v}^{g}$ operators for $v\in V\left(\partial\mathtt{M}\right)$
and $g\in G$ can be defined by embedding $\mathtt{M}$ into a topological
annulus $\overline{\mathtt{M}}$. The boundary of $\overline{\mathtt{M}}$
is the disjoint union of two loops; $\partial\overline{\mathtt{M}}=\left(-\varpi\right)\cup\overline{\varpi}$,
where the minus sign means that the orientation of $\varpi$ is opposite
to the one induced from $\overline{\mathtt{M}}$.}

\label{fig:tqd_annulus}
\end{figure}

First, let us classify the \emph{particle types} of 
an excited finite region (\emph{i.e.}, a localized quasiparticle), 
such as the small white square region $\mathtt{B}$ at the center 
of Fig.~\ref{fig:tqd_annulus}(a).
For topologically ordered systems in two spatial dimensions, 
we also use the terminology \emph{topological charge} as an synonym of particle type. 
To make it well-defined, we suppose that $\mathtt{B}$ is an isolated excitation
inside a much larger region $\mathtt{A}$. 
Then there is an excitation-free topological 
annulus $\mathtt{M}\coloneqq\mathtt{A}-\mathtt{B}^{\circ}$, such as the shaded
region in Fig.~\ref{fig:tqd_annulus}(a), 
separating $\mathtt{B}$ from the other excitations in $\mathtt{A}^{e}$.
Here $\mathtt{X}^e$ (resp. $\mathtt{X}^\circ$) denotes the exterior (resp. interior) 
of any topological space $\mathtt{X}$.
Such states lie in the Hilbert subspace 
${\cal H}\left(E\left(\mathtt{A}^{e}\right),G\right)\otimes{\cal H}_{0}\left(\mathtt{M}\right)\otimes{\cal H}\left(E\left(\mathtt{B}^{\circ}\right),G\right)$,
where ${\cal H}_{0}\left(\mathtt{M}\right)$ is the Hilbert subspace selected out of
the locally flat states ${\cal H}_{B}\left(E\left(\mathtt{M}\right),G\right)$ 
by the projector
\begin{equation}
P\left(\mathtt{M}\right)\coloneqq\prod_{v\in V\left(\mathtt{M}^{\circ}\right)}P_{v}.
\end{equation}
Since hopping between states of $\mathcal{H}\left(\mathtt{A}^{e}\right)$
(resp. $\mathcal{H}\left(\mathtt{B}^{\circ}\right)$) can be made by 
operators supported on $\mathtt{A}^e$ (resp. $\mathtt{B}^{\circ}$), 
they are irrelevant to the discussion of particle types. Below, we
only need to focus on $\mathcal{H}_{0}\left(\mathtt{M}\right)$. 

It is easy to see that the dimension of ${\cal H}_{0}\left(\mathtt{M}\right)$
grows with the number of spins along the boundary of $\mathtt{M}$.
To facilitate the analysis of $\mathcal{H}_{0}\left(\mathtt{M}\right)$, 
we embed $\mathtt{M}$ in a larger topological annulus $\overline{\mathtt{M}}$
which covers all shaded regions (grey and green online) and add edges
to finish a triangulation of $\overline{\mathtt{M}}$ as shown in
Fig.~\ref{fig:tqd_annulus}(b). Obviously, each coloring of $\mathtt{M}$
extends to $\overline{\mathtt{M}}$ uniquely. We label the outer (resp.
inner) boundary of $\overline{\mathtt{M}}$ by $\varpi$ (resp. $\overline{\varpi}$),
which is a loop with base point $a_{0}$ (resp. $b_{0}$). 
Let $\left\langle a_{0}b_{0}\right\rangle $
be the thick path (blue online) from $a_{0}$ to $b_{0}$ 
in Fig.~\ref{fig:tqd_annulus}(b). 
Let $T_{g}$, $\overline{T}_{h}$ and $T_{s}^{\left\langle a_{0}b_{0}\right\rangle }$
be the Hermitian projectors requiring the group elements associated with
paths $\left[a_0 a_1 a_2 \cdots a_{23} a_0\right]$, 
$\left[b_0 b_1 b_2 \cdots b_{7} b_0\right]$ and $\left\langle a_{0}b_{0}\right\rangle $
to be $g,h,s\in G$ respectively. 

As shown in Fig.~\ref{fig:tqd_annulus}(b), the vertices along the
outer and inner boundaries of $\mathtt{M}$ 
(\emph{i.e.}, $\partial\mathtt{A}$ and $\partial\mathtt{B}$)
are labeled as $a_{0},a_{1},\cdots,a_{23}$
and $b_{0},b_{1},\cdots,b_{7}$ respectively. Pick any two functions
\begin{align}
\chi: & \left\{ \left[a_{i}a_{i+1}\right]\right\} _{i=0,1,\cdots,22}\rightarrow G,\\
\overline{\chi}: & \left\{ \left[b_{i}b_{i+1}\right]\right\} _{i=0,1,\cdots,6}\rightarrow G.
\end{align}
Let $T\left[\chi\right]$ (resp. $T\left[\overline{\chi}\right]$)
be the Hermitian projector requiring $\zeta\left(\left[a_{i}a_{i+1}\right]\right)=\chi\left(\left[a_{i}a_{i+1}\right]\right)$
for $i=0,1,\cdots,22$ (resp. $\zeta\left(\left[a_{i}a_{i+1}\right]\right)=\overline{\chi}\left(\left[b_{i}b_{i+1}\right]\right)$
for $i=0,1,\cdots,6$). Obviously, $T\left[\chi\right]$ (resp.
$T\left[\overline{\chi}\right]$) is supported on the thick edges (green online) 
along the outer (resp. inner) boundary of $\mathtt{M}$. 

The Hermitian projectors $T_{g}$, $T_{s}^{\left\langle a_{0}b_{0}\right\rangle }$,
$T\left[\chi\right]$ and $T\left[\overline{\chi}\right]$ commute
with each other. It is a straightforward computation to show that,
on $\mathcal{H}\left(E\left(\mathtt{M}\right),G\right)$,
\begin{equation}
\text{tr}\left(T_{g}T_{s}^{\left\langle a_{0}b_{0}\right\rangle }T\left[\chi\right]T\left[\overline{\chi}\right]P\left(\mathtt{M}\right)\right)=1.
\end{equation}
Therefore, we can label a basis of $\mathcal{H}_{0}\left(\mathtt{M}\right)$
by $g,s,\chi$ and $\overline{\chi}$. 

\begin{figure}
\noindent\begin{minipage}[t]{1\columnwidth}%
\includegraphics[width=0.6\columnwidth]{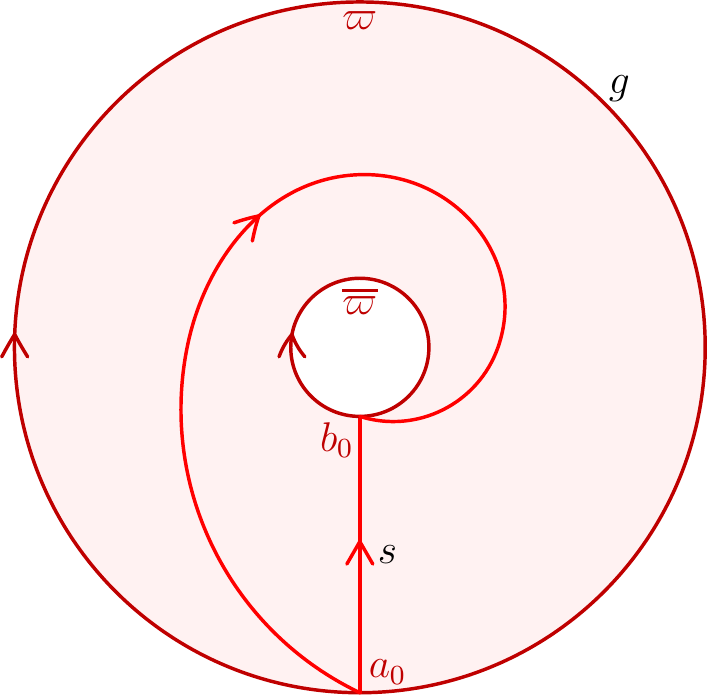}

(a)%
\end{minipage}

\bigskip{}

\noindent\begin{minipage}[t]{1\columnwidth}%
\includegraphics[width=0.7\columnwidth]{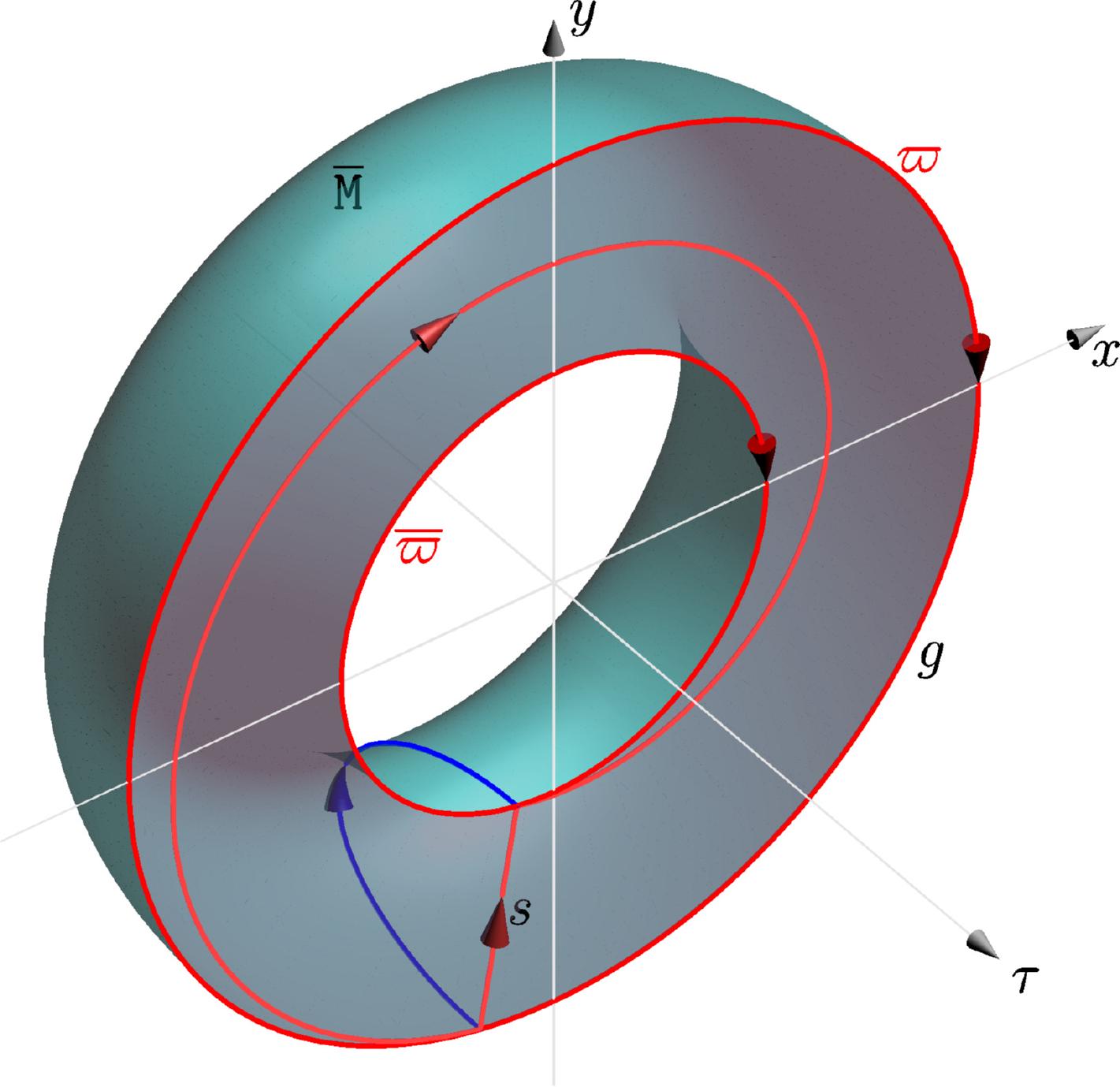}

(b)%
\end{minipage}

\caption{Graphic representation of 
$\left|\chi,\overline{\chi};\mathtt{D}_{g}^{s}\right\rangle $.
(a) An annulus (\emph{i.e.}, one-hole disk)
with colored triangulation $\mathtt{D}_{g}^{s}$. (b) A solid torus whose surface is 
$\left(-\overline{\mathtt{M}}\right)\cup_{\varpi\overline{\varpi}}\mathtt{D}_{g}^{s}$.
The minus sign emphasizes that the orientation of $\overline{\mathtt{M}}$
points towards the inside of the solid torus according to the right
hand rule. The two annuli $\overline{\mathtt{M}}$ and $\mathtt{D}_{g}^{s}$
are drawn curved and flat respectively; their shared boundary is the
disjoint union of the two loops $\varpi$ and $\overline{\varpi}$.
The state $\left|\chi,\overline{\chi};\mathtt{D}_{g}^{s}\right\rangle $ is specified by
$\left\langle \zeta|\chi,\overline{\chi};\mathtt{D}_{g}^{s}\right\rangle =
\mathcal{N} \mathcal{Z}_{\omega}\left(\zeta;\mathtt{D}_{g}^{s}\right)$
for $\zeta \in G^{E\left(\mathtt{M}\right)}\left(\chi,\overline{\chi}\right)$, 
where $\mathcal{Z}_{\omega}\left(\zeta;\mathtt{D}_{g}^{s}\right)$
is the Dijkgraaf-Witten partition function on this solid torus and 
$\mathcal{N}$ is a normalization factor. }

\label{fig:D_ht}
\end{figure}

To give a graphic representation of the basis vectors, let $\mathtt{D}_{g}^{s}$ be
an annulus with \emph{colored triangulation} (\emph{i.e.}, triangulation in which 
some edges carry fixed group elements)
as shown in Fig.~\ref{fig:D_ht}(a).
Gluing $\overline{\mathtt{M}}$ with $\mathtt{D}_{g}^{s}$ along the outer
and inner boundaries (\emph{i.e.}, loops $\varpi$ and $\overline{\varpi}$)
respectively, we get a triangulated torus, denoted 
$\left(-\overline{\mathtt{M}}\right)\cup_{\varpi\overline{\varpi}}\mathtt{D}_{g}^{s}$. 
Let $G^{E\left(\mathtt{M}\right)}\left(\chi,\overline{\chi}\right)$
be the set of spin configurations 
coinciding with $\chi,\overline{\chi}$ on the corresponding edges 
(green online in Fig.~\ref{fig:tqd_annulus}(b)).
Further, let $Z_{\omega}\left(\zeta;\mathtt{D}_{g}^{s}\right)$ be 
the Dijkgraaf-Witten partition function on a solid torus whose surface is 
$\left(-\overline{\mathtt{M}}\right)\cup_{\varpi\overline{\varpi}}\mathtt{D}_{g}^{s}$,
like the one in Fig.~\ref{fig:D_ht}(b),
with $E\left(\mathtt{M}\right)$ fixed to 
$\zeta \in G^{E\left(\mathtt{M}\right)}\left(\chi,\overline{\chi}\right)$.
Explicitly, $Z_{\omega}\left(\zeta;\mathtt{D}_{g}^{s}\right)$ is the sum of Dijkgraaf-Witten weight over colorings of the solid torus coinciding with $g$, $s$ and $\zeta$ on the corresponding edges.
Details of Dijkgraaf-Witten partition function are included
in Appendix~\ref{sec:GCDW}. Here, 
if $\zeta$ is locally flat (\emph{i.e.}, 
$\zeta\in G_B^{E\left(\mathtt{M}\right)}$) 
and assigns $g$, $s$ 
to paths $\left[a_0 a_1 a_2 \cdots a_{23} a_0\right]$, 
$\left\langle a_{0}b_{0}\right\rangle$ respectively, then
$\mathcal{Z}_{\omega}\left(\zeta;\mathtt{D}_{g}^{s}\right)\in U\left(1\right)$;
otherwise, there is no valid coloring on the solid torus and hence 
$\mathcal{Z}_{\omega}\left(\zeta;\mathtt{D}_{g}^{s}\right)=0$.
Then, the vectors 
\begin{equation}
\left|\chi,\overline{\chi};\mathtt{D}_{g}^{s}\right\rangle \coloneqq\sum_{\zeta\in
G_{B}^{E\left(\mathtt{M}\right)}\left(\chi,\overline{\chi}\right)
}
\frac{\mathcal{Z}_{\omega}\left(\zeta;\mathtt{D}_{g}^{s}\right)}{\left|G\right|^{\frac{\left|V\left(\mathtt{M}^{\circ}\right)\right|}{2}}}\left|\zeta\right\rangle \label{eq:wfn_D}
\end{equation}
labeled by $\chi,\overline{\chi},g,s$ form an orthonormal basis of 
$\mathcal{H}_{0}\left(\mathtt{M}\right)$, 
where $G_B^{E\left(\mathtt{M}\right)}\left(\chi,\overline{\chi}\right)\coloneqq G^{E\left(\mathtt{M}\right)}\left(\chi,\overline{\chi}\right)\cap
G_B^{E\left(\mathtt{M}\right)}$. 
For $v\in\mathtt{M}^{\circ}$, it is obvious 
$P_v=1$ (and hence  $P_v^g=1,\forall g\in G$) 
on these states by noticing
\begin{equation}
\left\langle \zeta\right|P_v\left|\chi,\overline{\chi};\mathtt{D}_{g}^{s}\right\rangle =\mathcal{Z}_{\omega}\left(\zeta;\mathtt{D}_{g}^{s}\right)=\left\langle \zeta|\chi,\overline{\chi};\mathtt{D}_{g}^{s}\right\rangle 
\end{equation}
using the graphic representation of $P_{v}^{g}$ given by Eq.~\eqref{eq:P_v} and
Fig.~\ref{fig:twisted_QD}(b).

\begin{figure}
\includegraphics[width=0.6\columnwidth]{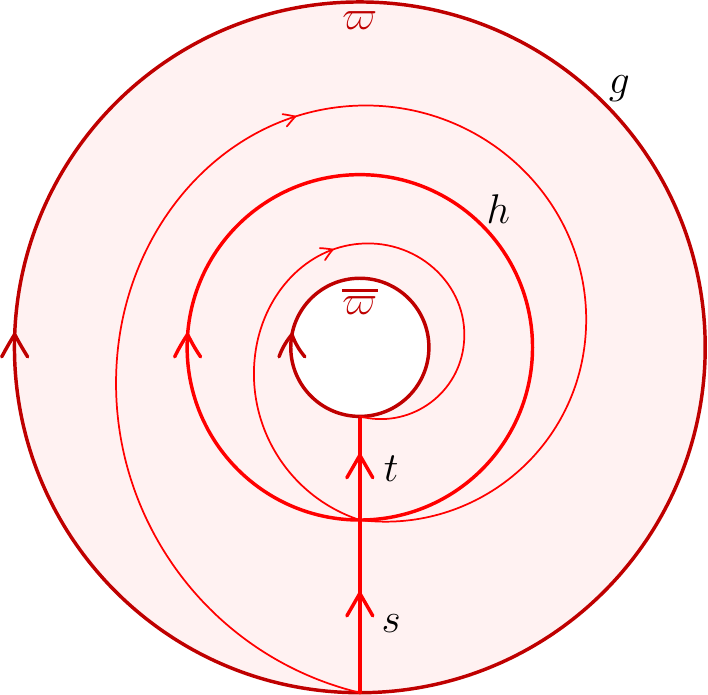}

\caption{An annulus (\emph{i.e.}, one-hole disk) with colored triangulation
$\mathtt{D}_{g}^{s}\mathtt{D}_{h}^{t}$. It determines a legitimate
coloring and corresponds to a nonzero state $\left|\mathtt{D}_{g}^{s}\mathtt{D}_{h}^{t}\right\rangle $
if and only if $g=shs^{-1}$. }

\label{fig:DD}
\end{figure}

In general, we can consider states presented by other triangulations
of the annulus. Let $\mathtt{D}_{g}^{s}\mathtt{D}_{h}^{t}$ denote an triangulated annulus carrying fixed group elements $g,h,s,t\in G$ on its four edges as shown in Fig.~\ref{fig:DD}; it is obtained by connecting
$\mathtt{D}_{g}^{s}$ and $\mathtt{D}_{h}^{t}$. Similarly, 
gluing $\overline{\mathtt{M}}$ and 
$\mathtt{D}_{g}^{s}\mathtt{D}_{h}^{t}$ along loops 
$\varpi$ and $\overline{\varpi}$ gives a torus 
$\left(-\overline{\mathtt{M}}\right)\cup_{\varpi\overline{\varpi}}\mathtt{D}_{g}^{s}\mathtt{D}_{h}^{t}$. 
Analogous to Eq.~\eqref{eq:wfn_D}, we can define 
\begin{equation}
\left|\chi,\overline{\chi};\mathtt{D}_{g}^{s}\mathtt{D}_{h}^{t}\right\rangle \coloneqq\sum_{\zeta\in
G_{B}^{E\left(\mathtt{M}\right)}\left(\chi,\overline{\chi}\right)}
\frac{\mathcal{Z}_{\omega}\left(\zeta;\mathtt{D}_{g}^{s}\mathtt{D}_{h}^{t}\right)}{\left|G\right|^{\frac{\left|V\left(\mathtt{M}^{\circ}\right)\right|}{2}}}\left|\zeta\right\rangle ,
\end{equation}
where $\mathcal{Z}_{\omega}\left(\zeta;\mathtt{D}_{g}^{s}\mathtt{D}_{h}^{t}\right)$ is 
the Dijkgraaf-Witten partition function on a solid torus whose surface is $\left(-\overline{\mathtt{M}}\right)\cup_{\varpi\overline{\varpi}}\mathtt{D}_{g}^{s}\mathtt{D}_{h}^{t}$. 

We notice 
\begin{align}
\mathcal{Z}_{\omega}\left(\zeta;\mathtt{D}_{g}^{s}\mathtt{D}_{h}^{t}\right) & =\mathcal{Z}_{\omega}\left(\zeta;\mathtt{D}_{g}^{st}\right)\mathcal{Z}_{\omega}\left(\mathtt{D}_{g}^{st};\mathtt{D}_{g}^{s}\mathtt{D}_{h}^{t}\right),\\
\mathcal{Z}_{\omega}\left(\mathtt{D}_{g}^{st};\mathtt{D}_{g}^{s}\mathtt{D}_{h}^{t}\right) & =\delta_{g,shs^{-1}}\omega_{g}\left(s,t\right),
\end{align}
where $\omega_{g}\left(s,t\right)$ is defined as
\begin{equation}
\omega_{g}\left(s,t\right)\coloneqq\frac{\omega\left(g,s,t\right)\omega\left(s,t,\left(st\right)^{-1}g\left(st\right)\right)}{\omega\left(s,s^{-1}gs,t\right)}.
\end{equation}
Therefore, $\forall g,h,s,t \in G$, we have
\begin{equation}
\left|\chi,\overline{\chi};\mathtt{D}_{g}^{s}\mathtt{D}_{h}^{t}\right\rangle =\delta_{g,shs^{-1}}\omega_{g}\left(s,t\right)
\left|\chi,\overline{\chi};\mathtt{D}_{g}^{st}\right\rangle ,
\end{equation}
which motivates the definition of an algebra $\mathcal{D}^{\omega}\left(G\right)$, 
called a \emph{twisted quantum double} of $G$. Formally, 
$\mathcal{D}^{\omega}\left(G\right)$ is spanned by
$\left\{ D_{g}^{s}\right\} _{g,s\in G}$ with multiplication rule
\begin{equation}
D_{g}^{s}D_{h}^{t}\coloneqq\delta_{g,shs^{-1}}\omega_{g}\left(s,t\right)D_{g}^{st},
\quad\forall g,h,s,t \in G.
\end{equation}
More details about $\mathcal{D}^{\omega}\left(G\right)$ 
are included in Appendix~\ref{sec:AlgebraDG}. 

We have seen that $\mathcal{H}_{0}\left(\mathtt{M}\right)$ factors into 
\begin{equation}
\mathcal{H}_{0}\left(\mathtt{M}\right)=
\mathcal{H}\left(\partial\mathtt{A}\right)
\otimes \mathcal{H}\left(\partial\mathtt{B}\right)
\otimes \mathcal{H}_{*}\left(\overline{\mathtt{M}}\right),
\end{equation}
where $\mathcal{H}\left(\partial\mathtt{A}\right)$, 
$\mathcal{H}\left(\partial\mathtt{B}\right)$ and
$\mathcal{H}_{*}\left(\overline{\mathtt{M}}\right)$ 
are the Hilbert spaces spanned by orthonormal bases 
$\left\{ \left|\chi\right\rangle \right\} $, 
$\left\{ \left|\overline{\chi}\right\rangle \right\} $ and 
$\left\{ \left|\mathtt{D}_{g}^{s}\right\rangle \right\} $ respectively.  
Using $\overline{\mathtt{M}}$ and 
Eq.~\eqref{eq:P_v} with $\omega\left[\zeta,\mathtt{P}_{v}^{g}\right]$ replaced by Eq.~\eqref{eq:w-1}, we extend
the definition of $P_v^g$ to include vertices on $\partial\mathtt{M}$ as well. Explicitly, 
\begin{equation}
P_{v}^{g}\coloneqq\sum_{\zeta\in G_{B}^{E\left(\overline{\mathtt{M}\left[v\right]}\right)}}\left|\zeta\right\rangle \omega\left[\overline{\mathtt{M}},v;\zeta,g\right]\left\langle \zeta A_{v}^{g}\right|,
\end{equation}
$\forall g\in G, \forall v \in V\left(\mathtt{M}\right)$, where $\overline{\mathtt{M}[v]}=\mathtt{M}[v]$ for $v\in\mathtt{M}^{\circ}$ while $\overline{\mathtt{M}[v]}=\mathtt{M}[v]\cup \partial\mathtt{A}$ (resp. $\overline{\mathtt{M}[v]}=\mathtt{M}[v]\cup \partial\mathtt{B}$) for  $v\in\partial{\mathtt{A}}$ (resp.  $v\in\partial{\mathtt{B}}$) with $\mathtt{M}[v]$ the region made of all plaquettes adjacent to $v$ inside $\mathtt{M}$.
They still satisfy Eq.~\eqref{eq:Pv1_Pv2}.
Except for $v=a_0$ and $b_0$, Eqs.~(\ref{eq:P_gh}) and (\ref{eq:P_dagger}) also hold. Hence 
$\{P_{v}\coloneqq\frac{1}{\left|G\right|}\sum_{g}P_{v}^{g}\}_{v\neq a_{0},b_{0}}$ are mutual 
commuting Hermitian projectors. Let
\begin{align}
P_{\partial\mathtt{A}} & \coloneqq\prod_{v\in V\left(\partial\mathtt{A}\right)\backslash\left\{ a_{0}\right\} }P_{v},\\
P_{\partial\mathtt{B}} & \coloneqq\prod_{v\in V\left(\partial\mathtt{B}\right)\backslash\left\{ b_{0}\right\} }P_{v}.
\end{align}
Then $\left|G\right|^{\left|V\left(\partial\mathtt{A}\right)\right|-1} T\left[\chi\right]P_{\partial\mathtt{A}}T\left[\chi'\right]$ realizes a generic operator
$\left|\chi\right\rangle \left\langle \chi'\right|$ 
on $\mathcal{H}\left(\partial\mathtt{A}\right)$. Thus,
$\mathcal{H}\left(\partial\mathtt{A}\right)$
describes only degrees of freedom near 
$\partial\mathtt{A}$ (\emph{i.e.}, the outer boundary
of $\mathtt{M}$). 
Similarly, $\mathcal{H}\left(\partial\mathtt{B}\right)$
describes only degrees of freedom near $\partial\mathtt{B}$.
Both $\mathcal{H}\left(\partial\mathtt{A}\right)$ and $\mathcal{H}\left(\partial\mathtt{B}\right)$ are irrelevant to classification of particle types.

Moreover, the operators
\begin{align}
\pi\left(D_{g}^{s}\right) & \coloneqq\left|G\right|^{\left|V\left(\partial\mathtt{A}\right)\right|-1}\sum_{\chi}T\left[\chi\right]T_{g}P_{a_{0}}^{s}P_{\partial\mathtt{A}}T\left[\chi\right], \label{eq:pi_D} \\
\overline{\pi}\left(D_{h}^{t}\right) & \coloneqq\left|G\right|^{\left|V\left(\partial\mathtt{B}\right)\right|-1}\sum_{\overline{\chi}}T\left[\overline{\chi}\right]\left(\overline{T}_{h}P_{b_{0}}^{t}\right)^{\dagger}P_{\partial\mathtt{B}}T\left[\overline{\chi}\right],
\end{align}
labeled by $D_{g}^{s},D_{h}^{t}\in\mathcal{D}^{\omega}\left(G\right)$ and supported near $\partial\mathtt{A}$, 
$\partial\mathtt{B}$ respectively, only act nontrivially on 
$\mathcal{H}_{*}\left(\overline{\mathtt{M}}\right)$. Explicitly,
\begin{align}
\pi\left(D_{g}^{s}\right)\left|\chi,\overline{\chi};\mathtt{D}_{h}^{t}\right\rangle  & =\left|\chi,\overline{\chi};\mathtt{D}_{g}^{s}\mathtt{D}_{h}^{t}\right\rangle ,\\
\overline{\pi}\left(D_{h}^{t}\right)\left|\chi,\overline{\chi};\mathtt{D}_{g}^{s}\right\rangle  & =\left|\chi,\overline{\chi};\mathtt{D}_{g}^{s}\mathtt{D}_{h}^{t}\right\rangle .
\end{align}
Thus, $\pi$ and $\overline{\pi}$ turn $\mathcal{H}_*\left(\overline{\mathtt{M}}\right)$ into a regular $\mathcal{D}^{\omega}\left(G\right)$-$\mathcal{D}^{\omega}\left(G\right)$-bimodule; 
\emph{i.e.}, the left and right actions of $\mathcal{D}^{\omega}\left(G\right)$ on $\mathcal{H}_*\left(\overline{\mathtt{M}}\right)$ specified by $\pi$ and $\overline{\pi}$ respectively are the same as how $\mathcal{D}^{\omega}\left(G\right)$ acts on itself via algebra multiplication.

In addition, an $*$-algebra structure on 
$\mathcal{D}^{\omega}\left(G\right)$ can be specified by
\begin{equation}
\left(D_{g}^{s}\right)^{\dagger}\coloneqq\omega_{g}^{*}\left(s,s^{-1}\right)D_{s^{-1}gs}^{s^{-1}},
\quad\forall g,s\in G,
\end{equation} 
where $\omega_{g}^{*}\left(s,s^{-1}\right)$ is the complex conjugate
of $\omega_{g}\left(s,s^{-1}\right)$. Then it is straightforward
to check that 
\begin{align}
\pi\left(\left(D_{g}^{s}\right)^{\dagger}\right) & =\left(\pi\left(D_{g}^{s}\right)\right)^{\dagger},\\
\overline{\pi}\left(\left(D_{h}^{t}\right)^{\dagger}\right) & =\left(\overline{\pi}\left(D_{h}^{t}\right)\right)^{\dagger},
\end{align}
by using the identity~\cite{1991NuPhS..18...60D}
\begin{equation}
\omega_{g}\left(s,t\right)\omega_{g}\left(st,u\right)=\omega_{g}\left(s,tu\right)\omega_{s^{-1}gs}\left(t,u\right),
\label{eq:D_mult}
\end{equation}
$\forall g,s,t,u\in G$. We also notice that setting $t=s^{-1}$ and $u=s$ in Eq.~\eqref{eq:D_mult} gives $\omega_{g}\left(s,s^{-1}\right)=\omega_{s^{-1}gs}\left(s^{-1},s\right)$, which ensures $((D_g^s)^{\dagger})^{\dagger}=D_g^s$.

Since $\pi$ is a left regular representation of a unital algebra, it is
faithful. So $\mathcal{D}^{\omega}\left(G\right)$ can be viewed
as a subalgebra, closed under the Hermitian conjugate, of $\mathcal{L}\left(\mathcal{H}_{*}\left(\overline{\mathtt{M}}\right)\right)$. Here $\mathcal{L}\left(V\right)$ denote the algebra of all linear operators on a vector space $V$.
Thus, $\mathcal{D}^{\omega}\left(G\right)$ is a finite dimensional
$C^{*}$-algebra and hence semisimple. Therefore, $\mathcal{D}^{\omega}\left(G\right)$
is isomorphic to a direct sum of matrix algebras
\begin{equation}
\mathcal{D}^{\omega}\left(G\right)\xrightarrow[*-\text{algebra }\cong]{\rho\coloneqq\bigoplus_{\mathfrak{a}\in\mathfrak{Q}}\rho_{\mathfrak{a}}}\bigoplus_{\mathfrak{a}\in\mathfrak{Q}}\mathcal{L}\left(\mathcal{V}_{\mathfrak{a}}\right),\label{eq:wd_3}
\end{equation}
where $\mathfrak{Q}$ labels the isomorphism classes of irreducible
representations of $\mathcal{D}^{\omega}\left(G\right)$ and $\mathcal{V}_{\mathfrak{a}}=\left(\rho_{\mathfrak{a}},V_{\mathfrak{a}}\right)$
is a finite dimensional Hilbert space carrying an representation $\rho_{\mathfrak{a}}$
corresponding to $\mathfrak{a}\in\mathfrak{Q}$. Moreover, $\mathcal{L}\left(\mathcal{V}_{\mathfrak{a}}\right)$
is the algebra of linear operators on $\mathcal{V}_{\mathfrak{a}}$;
it is isomorphic to the algebra of $n_{\mathfrak{a}}\times n_{\mathfrak{a}}$
square matrices, where $n_{\mathfrak{a}}\coloneqq\dim_{\mathbb{C}}\mathcal{V}_{\mathfrak{a}}$.
More explanations about this isomorphism $\rho$ are included in Appendix~\ref{subsec:Reps_DG}.
Let $\left\{ \left|\mathfrak{a};i\right\rangle \right\} _{i=1,\cdots,n_{\mathfrak{a}}}$
be an orthonormal basis of $\mathcal{V}_{\mathfrak{a}}$. Via $\rho$
in Eq.~\eqref{eq:wd_3}, we can view $\{\left|\mathfrak{a};i\right\rangle \left\langle \mathfrak{a};j\right|\}_{i,j=1,2,\cdots,n_{\mathfrak{a}}}^{\mathfrak{a}\in\mathfrak{Q}}$
as a basis of $\mathcal{D}^{\omega}\left(G\right)$. 

As a $\mathcal{D}^{\omega}\left(G\right)$-$\mathcal{D}^{\omega}\left(G\right)$-bimodule, $\mathcal{H}_{*}\left(\overline{\mathtt{M}}\right)$ can be identified with
$\mathcal{D}^{\omega}\left(G\right)$ and further get decomposed
\begin{align}
\mathcal{H}_{*}\left(\overline{\mathtt{M}}\right)\xrightarrow[\sim]{\left|\mathtt{D}_{g}^{s}\right\rangle \mapsto D_{g}^{s}} & \mathcal{D}^{\omega}\left(G\right)\\
\xrightarrow[\sim]{\tilde{\rho}\coloneqq\bigoplus_{\mathfrak{a}\in\mathfrak{Q}}\sqrt{\frac{n_{\mathfrak{a}}}{\left|G\right|}}\rho_{\mathfrak{a}}} & \bigoplus_{\mathfrak{a}\in\mathfrak{Q}}\mathcal{L}\left(\mathcal{V}_{\mathfrak{a}}\right)=\bigoplus_{\mathfrak{a}\in\mathfrak{Q}}\mathcal{V}_{\mathfrak{a}}\otimes\mathcal{V}_{\mathfrak{a}}^{*}. \label{eq:rho_DG}
\end{align}
The normalizations for $\tilde{\rho}$ on each sector are picked different
from $\rho$ such that inner product is also respected. As a Hilbert space,
it is convenient to write the basis vectors of $\mathcal{L}\left(\mathcal{V}_{\mathfrak{a}}\right)$
(resp. $\mathcal{V}_{\mathfrak{a}}^{*}$) as $\left|\mathfrak{a};i,\overline{j}\right\rangle \coloneqq\left|\mathfrak{a};i\right\rangle \left\langle \mathfrak{a};j\right|$
(resp. $\left|\mathfrak{a};\overline{j}\right\rangle \coloneqq\left\langle \mathfrak{a};j\right|$).
The default inner product on $\mathcal{V}_{\mathfrak{a}}^{*}$ is
given by $\left\langle \mathfrak{a};\overline{j'}|\mathfrak{a};\overline{j}\right\rangle \coloneqq\left\langle \mathfrak{a};j|\mathfrak{a};j'\right\rangle $.
The tensor product specifies the inner product on $\mathcal{L}\left(\mathcal{V}_{\mathfrak{a}}\right)$;
equivalently, $\left\langle \mathcal{O}_{1}|\mathcal{O}_{2}\right\rangle =\text{tr}\left(\mathcal{O}_{1}^{\dagger}\mathcal{O}_{2}\right)$,
$\forall\mathcal{O}_{1},\mathcal{O}_{2}\in\mathcal{L}\left(\mathcal{V}_{\mathfrak{a}}\right)$.

By construction, $\left|\mathfrak{q};k'\right\rangle \left\langle \mathfrak{q};k\right|\in \mathcal{D}^{\omega}\left(G\right)$
acts as 
\begin{align}
\pi\left(\left|\mathfrak{q};k'\right\rangle \left\langle \mathfrak{q};k\right|\right)\left|\mathfrak{a};i,\overline{j}\right\rangle  & =\delta_{\left(\mathfrak{q},k\right),\left(\mathfrak{a},i\right)}\left|\mathfrak{a};k',\overline{j}\right\rangle ,\label{eq:local_op}\\
\overline{\pi}\left(\left|\mathfrak{q};k'\right\rangle \left\langle \mathfrak{q};k\right|\right)\left|\mathfrak{a};i,\overline{j}\right\rangle  & =\delta_{\left(\mathfrak{a},j\right),\left(\mathfrak{q},k'\right)}\left|\mathfrak{a};i,\overline{k}\right\rangle .
\end{align}
Clearly, each $\mathfrak{a}\in\mathfrak{Q}$ labels a topological charge; it can be detected but cannot be changed by operators supported near either $\partial\mathtt{A}$ or $\partial\mathtt{B}$. Moreover, $i$ and $\overline{j}$ in $\left|\mathfrak{a};i,\overline{j}\right\rangle$, \emph{i.e.}, the two factors of 
$\mathcal{V}_{\mathfrak{a}}\otimes\mathcal{V}_{\mathfrak{a}}^{*}$ in Eq.~\eqref{eq:rho_DG}, 
describe the remaining degrees of freedom near $\partial\mathtt{A}$ and $\partial\mathtt{B}$ respectively.

Applying the above analysis of topological charges to the reduced situation with $\mathtt{B}=\emptyset$ and
$\mathtt{M}=\mathtt{A}-\mathtt{B}^{\circ}=\mathtt{A}$, 
we can prove that the ground states on any closed manifold are locally indistinguishable.
Now $\mathcal{H}_{0}\left(\mathtt{M}\right)$ has only the degrees of freedom labeled by $\chi$ along $\partial\mathtt{A}$. Further, suppose that $\mathcal{O}$ is any local operator inside $\mathtt{A}$ (away from $\partial\mathtt{A}$). 
Then $P\left(\mathtt{M}\right)\mathcal{O}P\left(\mathtt{M}\right)$ (equal to the action of $\mathcal{O}$ on the ground state subspace)
must be a scalar, because it commutes with $T\left[\chi\right]$ and hence cannot flip $\chi$. Therefore, no local operator can distinguish ground states.

To facilitate later discussions, let us describe $\mathcal{H}_0\left(\mathtt{M}\right)$ in more detail when $\mathtt{B}=\emptyset$. The $\Delta$-complex $\overline{\mathtt{M}}$ used for defining $P_v^g$ for $v\in\partial\mathtt{M}$ now reduces to a disk whose boundary is a loop $\varpi$. 
With no hole in $\overline{\mathtt{M}}$, the group element assigned to $\varpi$ must be trivial. Thus, we may also view $\overline{\mathtt{M}}$ as a sphere by identifying all points of $\varpi$
without affecting the definition of $P_v^g$. 
Then, analogous to Eq.~\eqref{eq:wfn_D}, a basis of $\mathcal{H}_0\left(\mathtt{M}\right)$ can be specified by the Dijkgraaf-Witten partition function on a ball with surface $\overline{\mathtt{M}}$.

\subsubsection{Fusion and coproduct}
\label{subsec:Fusions-and-braidings}

\begin{figure}
\includegraphics[width=0.6\columnwidth]{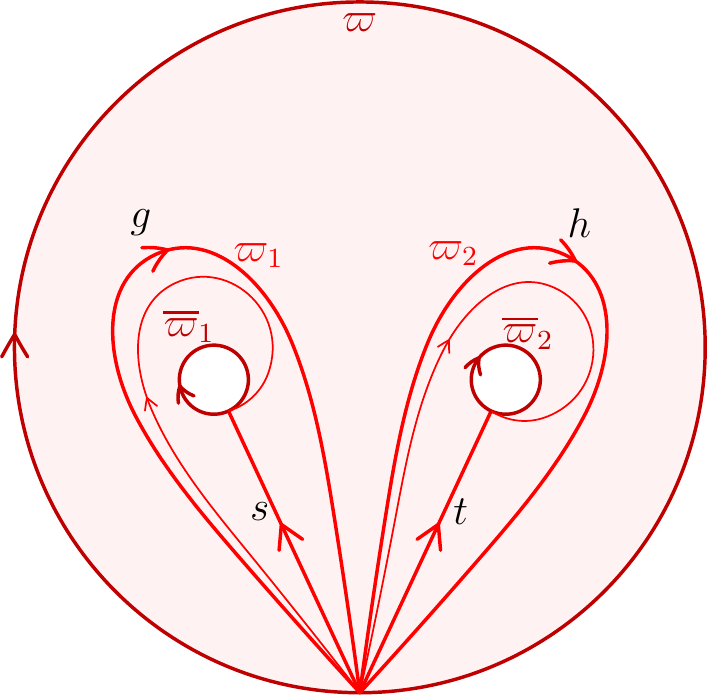}

\caption{A two-hole disk with colored triangulation $\mathtt{D}_{g}^{s}\otimes\mathtt{D}_{h}^{t}$.}

\label{fig:D2h}
\end{figure}

The setup is similar to Fig.~\ref{fig:tqd_annulus}(a), but now $\mathtt{A}$
contains two spatially separated excited spots $\mathtt{B}_{1}$,
$\mathtt{B}_{2}$ and we are going to analyze the Hilbert space on
the region 
$\mathtt{\mathtt{M}}\coloneqq\mathtt{A}-\mathtt{B}_{1}^{\circ}-\mathtt{B}_{2}^{\circ}$.
We embed $\mathtt{M}$ into a slightly bigger triangulated two-hole
disk $\overline{\mathtt{M}}$ with extra edges added along $\partial\mathtt{B}_{1}$,
$\partial\mathtt{B}_{2}$ and $\partial\mathtt{A}$, as we did for
$\mathtt{A}-\mathtt{B}^{\circ}$. The boundary of $\overline{\mathtt{M}}$ is
three disjoint loops, \emph{i.e.}, $\partial\mathtt{M}=\left(-\varpi\right)\cup\varpi_{1}\cup\varpi_{2}$,
where the minus sign means that the orientation of $\varpi$ is opposite
to the one induced from $\overline{\mathtt{M}}$ as shown in Fig.~\ref{fig:D2h}.

Let $\mathtt{D}_{g}^{s}\otimes\mathtt{D}_{h}^{t}$ denote a two-hole
disk with the colored triangulation shown in Fig.~\ref{fig:D2h}.
Analogous to the case of one-hole disk (\emph{i.e.}, annulus),
the Hilbert space $\mathcal{H}_{*}\left(\overline{\mathtt{M}}\right)$ 
relevant to topological charge analysis is spanned by 
spanned by $\left\{ \left|\mathtt{D}_{g}^{s}\otimes\mathtt{D}_{h}^{t}\right\rangle \right\} _{g,h,s,t\in G}$. 
It describes the states selected out of $\mathcal{H}\left(E(\mathtt{M}),G\right)$ 
by $P_v=1,\forall v\in V\left(\mathtt{M}^{\circ}\right)$
up to some compatible colorings $\chi$, $\overline{\chi}_1$ and $\overline{\chi}_2$ of
$\partial\mathtt{A}$, $\partial\mathtt{B}_1$ and $\partial\mathtt{B}_2$, 
via the the analogue of Eq.~\eqref{eq:wfn_D} on a three-dimensional
manifold with surface $\left(-\overline{\mathtt{M}}\right)\cup_{\varpi\overline{\varpi}_{1}\overline{\varpi}_{2}}\mathtt{D}_{g}^{s}\otimes\mathtt{D}_{h}^{t}$
(\emph{i.e.}, the genus-two surface obtained by gluing $\overline{\mathtt{M}}$
with $\mathtt{D}_{g}^{s}\otimes\mathtt{D}_{h}^{t}$ along loops
$\varpi$, $\overline{\varpi}_{1}$ and $\overline{\varpi}_{2}$).
The minus sign emphasizes that the orientation of $\overline{\mathtt{M}}$ points towards the inside of the three-dimensional manifold. In general, 
other colored triangulations
of a two-hole disk with boundary $\left(-\varpi\right)\cup\overline{\varpi}_{1}\cup\overline{\varpi}_{2}$
can be used to present states in $\mathcal{H}_{*}\left(\overline{\mathtt{M}}\right)$
as well. 

\begin{figure}
\noindent\begin{minipage}[t]{1\columnwidth}%
\includegraphics[width=0.6\columnwidth]{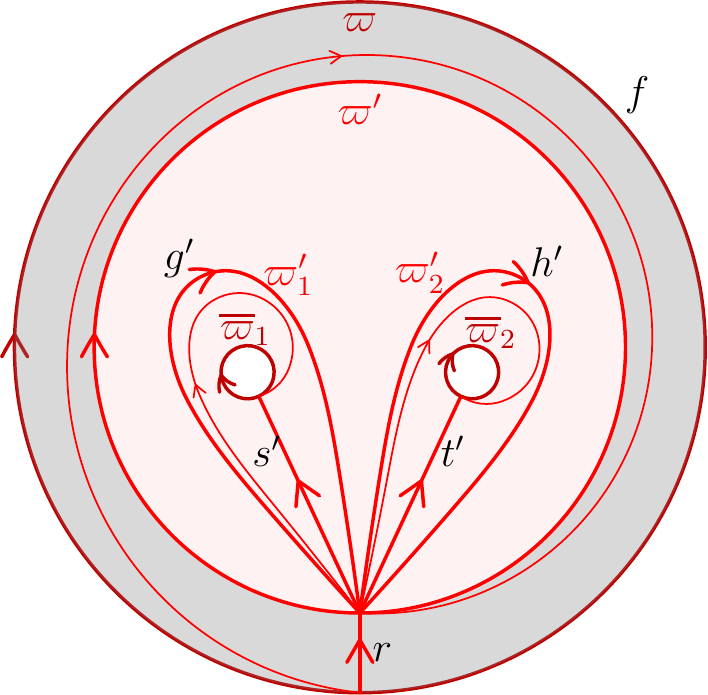}

(a)%
\end{minipage}

\bigskip{}

\noindent\begin{minipage}[t]{1\columnwidth}%
\includegraphics[width=0.6\columnwidth]{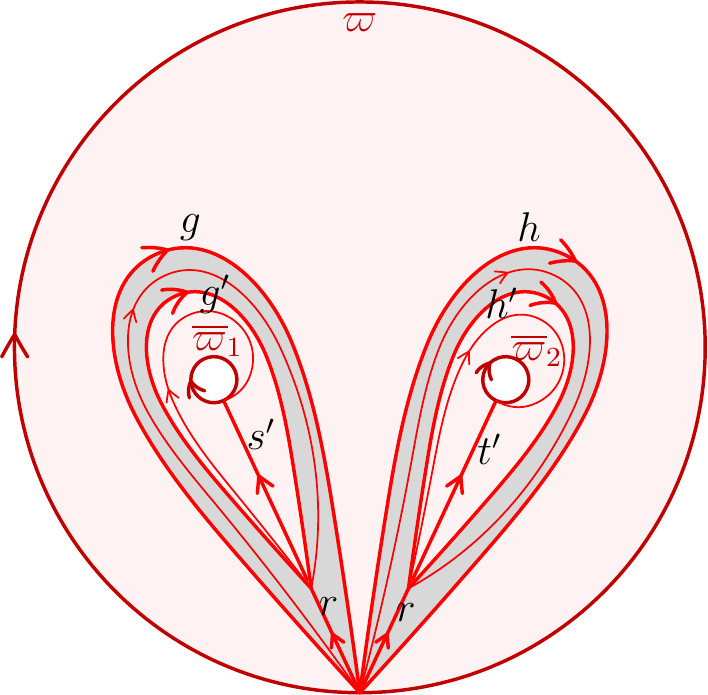}

(b)%
\end{minipage}

\caption{Graphic representations of (a) $\pi\left(D_{f}^{r}\right)\left|\mathtt{D}_{g'}^{s'}\otimes\mathtt{D}_{h'}^{t'}\right\rangle $
and (b) $\left|\mathtt{D}_{g}^{r}\mathtt{D}_{g'}^{s'}\otimes\mathtt{D}_{h}^{r}\mathtt{D}_{h'}^{t'}\right\rangle $.}

\label{fig:DD2h}
\end{figure}

Using the isomorphism $\tilde{\rho}$ in Eq.~\eqref{eq:rho_DG}, we have
\begin{align}
{\cal H}_{*}\left(\overline{\mathtt{M}}\right) & \xrightarrow[\sim]{\left|\mathtt{D}_{g}^{s}\otimes\mathtt{D}_{h}^{t}\right\rangle \mapsto D_{g}^{s}\otimes D_{h}^{t}}\mathcal{D}^{\omega}\left(G\right)^{\otimes2}\\
 & \xrightarrow[\sim]{\tilde{\rho}\otimes\tilde{\rho}}\bigotimes_{\mathfrak{a}_{1},\mathfrak{a}_{2}\in\mathfrak{Q}}
\mathcal{V}_{\mathfrak{a}_{1}}^{(1)}\otimes\mathcal{V}_{\mathfrak{a}_{1}}^{(1)*}\otimes\mathcal{V}_{\mathfrak{a}_{2}}^{(2)}\otimes\mathcal{V}_{\mathfrak{a}_{2}}^{(2)*},
\end{align}
where $\mathcal{V}_{\mathfrak{a}_1}^{(1)}$ and $\mathcal{V}_{\mathfrak{a}_2}^{(2)}$ are Hilbert spaces carrying irreducible representations corresponding to $\mathfrak{a}_1,\mathfrak{a}_2 \in \mathfrak{Q}$.
The degrees of freedom $\mathcal{V}_{\mathfrak{a}_{1}}^{(1)*}$ and $\mathcal{V}_{\mathfrak{a}_{2}}^{(2)*}$ 
(in particular, topological charges $\mathfrak{a}_1$ and $\mathfrak{a}_2$) 
can be pinned by operators supported near near $\partial\mathtt{B}_{1}$ and $\partial\mathtt{B}_{2}$ respectively. 
The operators $\pi (\mathtt{D}_f^r)$ for $f,r\in G$ defined by Eq.~\eqref{eq:pi_D}
specifies the total topological charge inside $\mathtt{A}$.
The action of $\pi (\mathtt{D}_f^r)$ is presented in Fig.~\ref{fig:DD2h}(a).
Explicitly, 
\begin{equation}
\pi\left(D_{f}^{r}\right)\left|\mathtt{D}_{g'}^{s'}\otimes\mathtt{D}_{h'}^{t'}\right\rangle =\sum_{gh=f}\omega^{r}\left(g,h\right)\left|\mathtt{D}_{g}^{r}\mathtt{D}_{g'}^{s'}\otimes\mathtt{D}_{h}^{r}\mathtt{D}_{h'}^{t'}\right\rangle ,\label{eq:fusion}
\end{equation}
for $f,g',h',r,s',t'\in G$, where 
\begin{equation}
\omega^{r}\left(g,h\right)\coloneqq\frac{\omega\left(g,h,r\right)\omega\left(r,r^{-1}gr,r^{-1}hr\right)}{\omega\left(g,r,r^{-1}hr\right)}
\end{equation}
and $\left|\mathtt{D}_{g}^{r}\mathtt{D}_{g'}^{s'}\otimes\mathtt{D}_{h}^{r}\mathtt{D}_{h'}^{t'}\right\rangle $
is presented by the colored triangulation in Fig.~\ref{fig:DD2h}(b).
A quick way to check Eq.~\eqref{eq:fusion} is to notice that $\left\langle \mathtt{D}_{g}^{s}\otimes\mathtt{D}_{h}^{t}\right|\pi(D_{f}^{r})\left|\mathtt{D}_{g'}^{s'}\otimes\mathtt{D}_{h'}^{t'}\right\rangle $
corresponds to a solid torus whose surface is the gluing result of
the two-hole disks in Fig.~\ref{fig:D2h} and Fig.~\ref{fig:DD2h}(a)
along $\varpi$, $\overline{\varpi}_{1}$ and $\overline{\varpi}_{2}$. 
The solid torus can be partitioned
into two solid tori relating $\mathtt{D}_{g}^{rs'}\sim\mathtt{D}_{g}^{r}\mathtt{D}_{g'}^{s'}$, 
$\mathtt{D}_{h}^{rt'}\sim \mathtt{D}_{h}^{r}\mathtt{D}_{h'}^{t'}$  and a prism over the triangle
with edges $\varpi,\varpi_{1},\varpi_{2}$. 
The prism gives the factor $\omega^{r}\left(g,h\right)$. 
Thus, $\pi(D_f^r)$ is specified by the coproduct 
\begin{align}
\varDelta:\mathcal{D}^{\omega}\left(G\right) & \rightarrow\mathcal{D}^{\omega}\left(G\right)\otimes\mathcal{D}^{\omega}\left(G\right);\nonumber \\
D_{f}^{r} & \mapsto\varDelta\left(D_{f}^{r}\right)=\sum_{gh=f}\omega^{r}\left(g,h\right)D_{g}^{r}\otimes D_{h}^{r}.
\end{align}

On $\mathcal{V}_{\mathfrak{a}_{1}}^{(1)}\otimes\mathcal{V}_{\mathfrak{a}_{2}}^{(2)}$
(\emph{i.e.}, a sector of $\mathcal{H}_{*}\left(\overline{\mathtt{M}}\right)$
with local degrees of freedom at $\partial\mathtt{B}_{1}$and $\partial\mathtt{B}_{2}$
pinned), the operator $\pi(D_{f}^{r})$ acts as $(\rho_{\mathfrak{a}_{1}}^{(1)}\otimes\rho_{\mathfrak{a}_{2}}^{(2)})\circ\varDelta$, making $\mathcal{V}_{\mathfrak{a}_{1}}^{(1)}\otimes\mathcal{V}_{\mathfrak{a}_{2}}^{(2)}$
a representation of $\mathcal{D}^{\omega}\left(G\right)$. In general, $\mathcal{V}_{\mathfrak{a}_{1}}^{(1)}\otimes\mathcal{V}_{\mathfrak{a}_{2}}^{(2)}$
is reducible
\begin{equation}
\mathcal{V}_{\mathfrak{a}_{1}}^{(1)}\otimes\mathcal{V}_{\mathfrak{a}_{2}}^{(2)}=\bigoplus_{\mathfrak{a}}V_{\mathfrak{a}}^{\mathfrak{a}_{1}\mathfrak{a}_{2}}\otimes\mathcal{V}_{\mathfrak{a}}
\end{equation}
with spaces of intertwiners $V_{\mathfrak{a}}^{\mathfrak{a}_{1}\mathfrak{a}_{2}}\coloneqq\text{Hom}(\mathcal{V}_{\mathfrak{a}},\mathcal{V}_{\mathfrak{a}_{1}}^{(1)}\otimes\mathcal{V}_{\mathfrak{a}_{2}}^{(2)})$,
where $\mathcal{V}_{\mathfrak{a}}$ is a Hilbert space carrying an irreducible representation of $\mathcal{D}^{\omega}(G)$ corresponding to the total topological charge $\mathfrak{a}\in\mathfrak{Q}$. The dimension $N_{\mathfrak{a}_{1}\mathfrak{a}_{2}}^{\mathfrak{a}}=\dim_{\mathbb{C}}V_{\mathfrak{a}}^{\mathfrak{a}_{1}\mathfrak{a}_{2}}$
counts the number of ways to fuse $\mathfrak{a}_{1}$, $\mathfrak{a}_{2}$
into $\mathfrak{a}$ and is called a \emph{fusion rule}. 

\begin{figure}
\includegraphics[width=0.6\columnwidth]{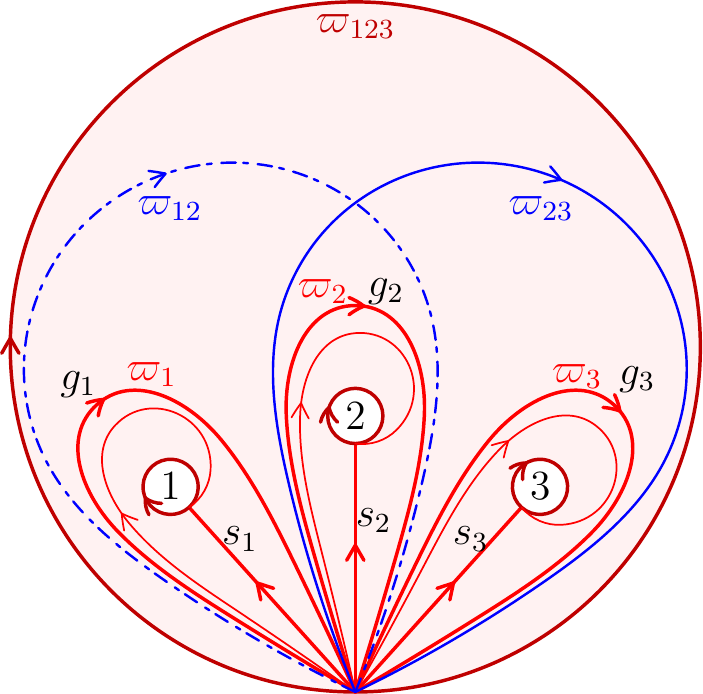}

\caption{A retriangulation of a three-hole disk is made by replacing loop $\varpi_{12}$
with loop $\varpi_{23}$. }

\label{fig:D3h}
\end{figure}

To describe more than two quasiparticles, we need to understand the associative property of any three topological charges.
As before, the topological degrees of freedom are encoded in the Hilbert space
$\mathcal{H}_{*}(\overline{\mathtt{M}})$ with a basis presented 
by colorings of a triangulated three-hole disk. 
However, there are two natural triangulations as shown in Fig.~\ref{fig:D3h};
we can either group holes $1,2$ together by loop $\varpi_{12}$ or
group holes $2,3$ together by loop $\varpi_{23}$. The two triangulations
lead to two different bases $\left\{ \left|\left(\mathtt{D}_{g_{1}}^{s_{1}}\otimes\mathtt{D}_{g_{2}}^{s_{2}}\right)\otimes\mathtt{D}_{g_{3}}^{s_{3}}\right\rangle \right\} $
and $\left\{ \left|\mathtt{D}_{g_{1}}^{s_{1}}\otimes\left(\mathtt{D}_{g_{2}}^{s_{2}}\otimes\mathtt{D}_{g_{3}}^{s_{3}}\right)\right\rangle \right\} $. 
Noticing that changing from the triangulation with $\varpi_{12}$ to
the one with $\varpi_{23}$ corresponds
to a tetrahedron whose edges are loops $\varpi_{1}$, $\varpi_{2}$,
$\varpi_{3}$, $\varpi_{12}$, $\varpi_{23}$ and $\varpi_{123}$,
we get the basis transformation 
\begin{equation}
\left|\left(\mathtt{D}_{g_{1}}^{s_{1}}\otimes\mathtt{D}_{g_{2}}^{s_{2}}\right)\otimes\mathtt{D}_{g_{3}}^{s_{3}}\right\rangle =\frac{\left|\mathtt{D}_{g_{1}}^{s_{1}}\otimes\left(\mathtt{D}_{g_{2}}^{s_{2}}\otimes\mathtt{D}_{g_{3}}^{s_{3}}\right)\right\rangle }{\omega\left(g_{1},g_{2},g_{3}\right)}.
\end{equation}
In other words, $\mathcal{H}_{*}{\left(\overline{\mathtt{M}}\right)}$ can be identified with
$\left(\mathcal{D}^{\omega}\left(G\right)\right)^{\otimes3}$ in two ways
\begin{align}
\varphi_{\left(12\right)3}:\mathcal{H}_{*}\left(\overline{\mathtt{M}}\right) & \xrightarrow{\sim} 
\left(\mathcal{D}^{\omega}\left(G\right)\right)^{\otimes3},\nonumber \\
\left|\left(\mathtt{D}_{g_{1}}^{s_{1}}\otimes\mathtt{D}_{g_{2}}^{s_{2}}\right)\otimes\mathtt{D}_{g_{3}}^{s_{3}}\right\rangle & \mapsto
D_{g_{1}}^{s_{1}}\otimes D_{g_{2}}^{s_{2}}\otimes D_{g_{3}}^{s_{3}} ,\\
\varphi_{1\left(23\right)}:\mathcal{H}_{*}\left(\overline{\mathtt{M}}\right) &
\xrightarrow{\sim}
\left(\mathcal{D}^{\omega}\left(G\right)\right)^{\otimes3},\nonumber \\
\left|\mathtt{D}_{g_{1}}^{s_{1}}\otimes\left(\mathtt{D}_{g_{2}}^{s_{2}}\otimes\mathtt{D}_{g_{3}}^{s_{3}}\right)\right\rangle & \mapsto
D_{g_{1}}^{s_{1}}\otimes D_{g_{2}}^{s_{2}}\otimes D_{g_{3}}^{s_{3}} ,
\end{align}
with the basis transformation encoded by the Drinfeld associator 
$\phi\coloneqq\sum_{f,g,h}\omega\left(f,g,h\right)^{-1}D_{f}^{e}\otimes D_{g}^{e}\otimes D_{h}^{e}$ (\emph{i.e.}, 
$\phi A=\varphi_{1(23)}\circ\varphi_{(12)3}^{-1}(A),\forall A\in \mathcal{D}^{\omega}(G)^{\otimes 3}$).

Three copies of Eq.~\eqref{eq:rho_DG} give 
\begin{equation}
\left(\mathcal{D}^{\omega}\left(G\right)\right)^{\otimes3}\xrightarrow[\sim]{\tilde{\rho}^{\otimes3}}\bigoplus_{\mathfrak{a}_{1},\mathfrak{a}_{2},\mathfrak{a}_{3}\in\mathfrak{Q}}\bigotimes_{n=1}^{3}\left(\mathcal{V}_{\mathfrak{a}_{n}}^{\left(n\right)}\otimes\mathcal{V}_{\mathfrak{a}_{n}}^{\left(n\right)*}\right),\label{eq:wd_3-1}
\end{equation}
where $\mathcal{V}_{\mathfrak{a}_{n}}^{\left(n\right)}=(\rho_{\mathfrak{a}_{n}}^{\left(n\right)},V_{\mathfrak{a}_{n}}^{\left(n\right)})$
is an irreducible representation of $\mathcal{D}^{\omega}\left(G\right)$ 
on a Hilbert space $V_{\mathfrak{a}_{n}}^{\left(n\right)}$. Since
$\phi$ does not act on local degrees of freedom $\mathcal{V}_{\mathfrak{a}_{n}}^{\left(n\right)*}$,
we can safely fixed a state of $\mathcal{V}_{\mathfrak{a}_{n}}^{\left(n\right)*}$
and just consider $\bigotimes_{n=1}^{3}\mathcal{V}_{\mathfrak{a}_{n}}^{\left(n\right)}$
for describing fusion and braiding processes. But to interpret the
states, we need to specify whether we are using $\varphi_{\left(12\right)3}$
or $\varphi_{1\left(23\right)}$ by writing $\bigotimes_{n=1}^{3}\mathcal{V}_{\mathfrak{a}_{n}}^{\left(n\right)}$
as either $(\mathcal{V}_{\mathfrak{a}_{1}}^{\left(1\right)}\otimes\mathcal{V}_{\mathfrak{a}_{2}}^{\left(2\right)})\otimes\mathcal{V}_{\mathfrak{a}_{3}}^{\left(3\right)}$
or $\mathcal{V}_{\mathfrak{a}_{1}}^{\left(1\right)}\otimes(\mathcal{V}_{\mathfrak{a}_{2}}^{\left(2\right)}\otimes\mathcal{V}_{\mathfrak{a}_{3}}^{\left(3\right)})$.
Under $\varphi_{(12)3}$ (resp. $\varphi_{1(23)}$), the action of $\mathcal{D}^{\omega}(G)$ defined by Eq.~\eqref{eq:pi_D} is given by $\left(\varDelta\otimes\text{id}\right)\circ\varDelta$
(resp. $\left(\text{id}\otimes\varDelta\right)\circ\varDelta$). Moreover,
the basis transformation is presented by the action of $\phi$ on
$\bigotimes_{n=1}^{3}\mathcal{V}_{\mathfrak{a}_{n}}^{\left(n\right)}$. 

The above discussion can be generalized to any finite number of excitations.
For example, the topological degrees of freedom associated with four topological 
charges $\mathfrak{a}_1$, $\mathfrak{a}_2$, $\mathfrak{a}_3$ and $\mathfrak{a}_4$
can be expressed in any one of the forms $((\mathcal{V}_{\mathfrak{a}_{1}}^{\left(1\right)}\otimes\mathcal{V}_{\mathfrak{a}_{2}}^{\left(2\right)})\otimes\mathcal{V}_{\mathfrak{a}_{3}}^{\left(3\right)})\otimes\mathcal{V}_{\mathfrak{a}_{4}}^{\left(4\right)}$,
$(\mathcal{V}_{\mathfrak{a}_{1}}^{\left(1\right)}\otimes\mathcal{V}_{\mathfrak{a}_{2}}^{\left(2\right)})\otimes(\mathcal{V}_{\mathfrak{a}_{3}}^{\left(3\right)}\otimes\mathcal{V}_{\mathfrak{a}_{4}}^{\left(4\right)})$
and $\mathcal{V}_{\mathfrak{a}_{1}}^{\left(1\right)}\otimes(\mathcal{V}_{\mathfrak{a}_{2}}^{\left(2\right)}\otimes(\mathcal{V}_{\mathfrak{a}_{3}}^{\left(3\right)}\otimes\mathcal{V}_{\mathfrak{a}_{4}}^{\left(4\right)}))$.

\subsubsection{Braiding and universal $R$-matrix}

\begin{figure}
\includegraphics[width=0.6\columnwidth]{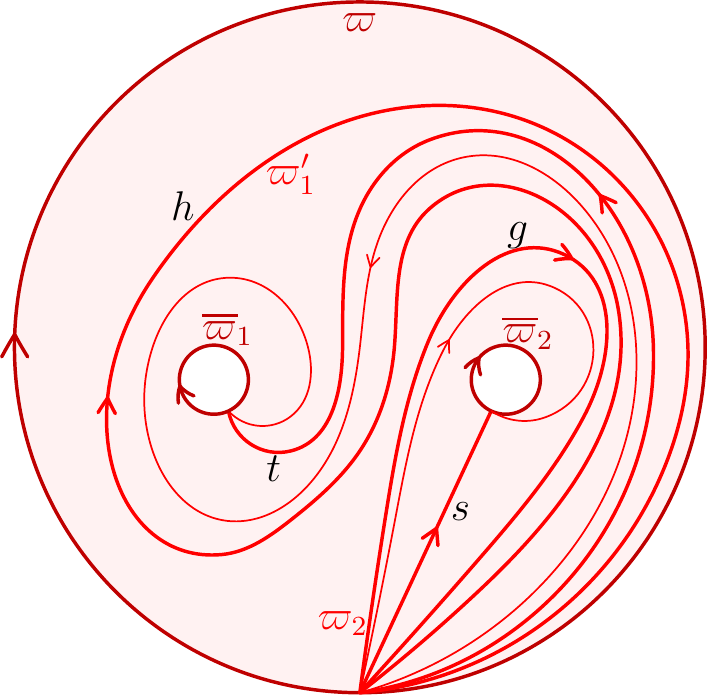}

\caption{Graphic representation of $\mathcal{R}\left|\mathtt{D}_{g}^{s}\otimes\mathtt{D}_{h}^{t}\right\rangle $. }

\label{fig:D2h_braiding}
\end{figure}

Let us define an operator $\mathcal{R}$ to decribe the braiding of
any two anyons. Graphically, $\mathcal{R}\left|\mathtt{D}_{g}^{s}\otimes\mathtt{D}_{h}^{t}\right\rangle $
is presented by Fig.~\ref{fig:D2h_braiding}. Explicitly, in the
original basis $\left\{ \left|\mathtt{D}_{g}^{s}\otimes\mathtt{D}_{h}^{t}\right\rangle \right\} _{g,h,s,t\in G}$
labeled by the colorings of the triangulated two-hole disk shown in
Fig.~\ref{fig:D2h}, we have 
\begin{equation}
\mathcal{R}\left|\mathtt{D}_{g}^{s}\otimes\mathtt{D}_{h}^{t}\right\rangle =\left|D^{g}\mathtt{D}_{h}^{t}\otimes\mathtt{D}_{g}^{s}\right\rangle ,
\end{equation}
where $D^{g}\coloneqq\sum_{f\in G}D_{f}^{g}$ acts on $\mathtt{D}_{h}^{t}$ as $D^{g}\mathtt{D}_{h}^{t}=\mathtt{D}_{ghg^{-1}}^{g}\mathtt{D}_{h}^{t}$,
describing the change
of $\mathtt{D}_{h}^{t}$ as it moves along loop $\varpi_{2}$. 

The universal $R$-matrix of $\mathcal{D}^{\omega}\left(G\right)$
is 
\begin{equation}
R=\sum_{g\in G}D_{g}^{e}\otimes D^{g}.
\end{equation}
In terms of $R$, the braiding operator can be express as
\begin{equation}
\mathcal{R}=\wp R,\label{eq:braiding_Op}
\end{equation}
where $\wp$ permutes the two factors of each basis vector $\left|\mathtt{D}_{g}^{s}\otimes\mathtt{D}_{h}^{t}\right\rangle $ (\emph{i.e.}, $\wp \left|\mathtt{D}_{g}^{s}\otimes\mathtt{D}_{h}^{t}\right\rangle = \left|\mathtt{D}_{h}^{t}\otimes\mathtt{D}_{g}^{s}\right\rangle$).
Under the action of local operators near $\overline{\varpi}_{1}$ and $\overline{\varpi}_{2}$, the Hilbert
space reduces into sectors labeled by particles types of the two
anyons. 

To summarize, the quantum double algebra $\mathcal{D}^{\omega}\left(G\right)$
is a quasi-Hopf algebra and its representations form a unitary modular tensor category---a special type of braided tensor
category---describing the behaviors of anyons that appear in the
lattice models of twisted gauge theories. Explicit examples can be
found in Appendix~\ref{subsec:ExamplesBTC}.

\subsection{Measuring invariants associated with topological charges\label{subsec:MeasuringTopSpin}}

To conclude the discussion of the lattice models based on twisted
gauge theories, we now explain how to define and detect some key
properties of topological charges by simple and universal measurements.

\subsubsection{Quantum dimension}

\begin{figure}
\includegraphics[width=.95\columnwidth]{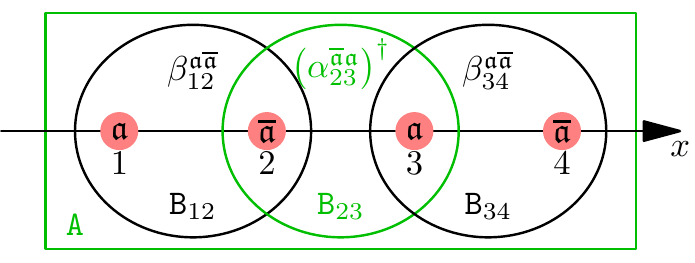}

\caption{Creating four quasiparticles of topological charges $\mathfrak{a}$,
$\overline{\mathfrak{a}}$, $\mathfrak{a}$, $\overline{\mathfrak{a}}$
at spots $1,2,3,4$ (red online) respectively in two different ways $\beta_{12}^{\mathfrak{a}\overline{\mathfrak{a}}}\beta_{34}^{\mathfrak{a}\overline{\mathfrak{a}}}$
and $(\alpha_{23}^{\overline{\mathfrak{a}}\mathfrak{a}})^{\dagger}\beta_{14}^{\mathfrak{a}\overline{\mathfrak{a}}}$.
The operator $\beta_{12}^{\mathfrak{a}\overline{\mathfrak{a}}}$ (resp.
$\beta_{34}^{\mathfrak{a}\overline{\mathfrak{a}}}$, $(\alpha_{23}^{\overline{\mathfrak{a}}\mathfrak{a}})^{\dagger}$,
$\beta_{14}^{\mathfrak{a}\overline{\mathfrak{a}}}$) supported within oval $\mathtt{B}_{12}$ (resp. oval $\mathtt{B}_{34}$, oval $\mathtt{B}_{23}$,
rectangle $\mathtt{A}$) creates a pair of quasiparticles at spots
$1,2$ (resp. spots $3,4$, spots $2,3$, spots $1,4$) carrying the labeled topological
charges.}

\label{fig:fusion4spots}
\end{figure}

To define and measure the quantum dimension associated with a topological
charge $\mathfrak{a}$, we consider two different processes creating
four quasiparticles of topological charges $\mathfrak{a}$, $\overline{\mathfrak{a}}$,
$\mathfrak{a}$, $\overline{\mathfrak{a}}$ at spots $1,2,3,4$ as shown in Fig.~\ref{fig:fusion4spots}. Let $\mathtt{B}_{12}$,
$\mathtt{B}_{23}$ and $\mathtt{B}_{34}$ be the three oval regions
containing spots $1,2$, spots $2,3$ and spots $3,4$ respectively.
Pick an operator $\beta_{12}^{\mathfrak{a}\overline{\mathfrak{a}}}$
(resp. $\beta_{34}^{\mathfrak{a}\overline{\mathfrak{a}}}$, $\beta_{14}^{\mathfrak{a}\overline{\mathfrak{a}}}$)
supported in oval $\mathtt{B}_{12}$ (resp. oval $\mathtt{B}_{34}$,
rectangle $\mathtt{A}$) that creates a pair of quasiparticles with
topological charges $\mathfrak{a},\overline{\mathfrak{a}}$ at spots
$1,2$ (resp. spots $3,4$, spots $1,4$). Moreover, pick an operator
$\alpha_{23}^{\overline{\mathfrak{a}}\mathfrak{a}}$ supported in oval $\mathtt{B}_{23}$ to annihilate a pair of anyons with topological
charges $\overline{\mathfrak{a}}$, $\mathfrak{a}$ at spots $2$,
$3$. The choice of these operators can be fixed up to some phase
factors by the normalization 
\begin{equation}
\alpha_{23}^{\overline{\mathfrak{a}}\mathfrak{a}}\left(\alpha_{23}^{\overline{\mathfrak{a}}\mathfrak{a}}\right)^{\dagger}=\beta^{\dagger}\beta=1,\forall\beta=\beta_{12}^{\mathfrak{a}\overline{\mathfrak{a}}},\beta_{34}^{\mathfrak{a}\overline{\mathfrak{a}}},\beta_{14}^{\mathfrak{a}\overline{\mathfrak{a}}}
\label{eq:ab_normalization}
\end{equation}
on the vacuum $\left|\Omega\right\rangle $. Then $\alpha_{23}^{\overline{\mathfrak{a}}\mathfrak{a}}\beta_{12}^{\mathfrak{a}\overline{\mathfrak{a}}}\beta_{34}^{\mathfrak{a}\overline{\mathfrak{a}}}\left|\Omega\right\rangle $
and $\beta_{14}^{\mathfrak{a}\overline{\mathfrak{a}}}\left|\Omega\right\rangle $
are the same state and hence there is $u_{\mathfrak{a}}\in\mathbb{C}$
such that
\begin{equation}
\alpha_{23}^{\overline{\mathfrak{a}}\mathfrak{a}}\beta_{12}^{\mathfrak{a}\overline{\mathfrak{a}}}\beta_{34}^{\mathfrak{a}\overline{\mathfrak{a}}}\left|\Omega\right\rangle =u_{\mathfrak{a}}\beta_{14}^{\mathfrak{a}\overline{\mathfrak{a}}}\left|\Omega\right\rangle .\label{eq:quantumdim}
\end{equation}
Let $d_{\mathfrak{a}}\coloneqq\frac{1}{\left|u_{\mathfrak{a}}\right|}$;
it is called the \emph{quantum dimension} associated with the topological
charge $\mathfrak{a}$. 

In other words, the overlap between $\left(\alpha_{23}^{\overline{\mathfrak{a}}\mathfrak{a}}\right)^{\dagger}\beta_{14}^{\mathfrak{a}\overline{\mathfrak{a}}}\left|\Omega\right\rangle $
and $\beta_{12}^{\mathfrak{a}\overline{\mathfrak{a}}}\beta_{34}^{\mathfrak{a}\overline{\mathfrak{a}}}\left|\Omega\right\rangle $
is $u_{\mathfrak{a}}$. Let $\mathfrak{q}_{23}$ be the total topological charge
of quasiparticles at spots $2,3$. Then in a basis labeled by $\mathfrak{q}_{23}$,
the only component of $\beta_{12}^{\mathfrak{a}\overline{\mathfrak{a}}}\beta_{34}^{\mathfrak{a}\overline{\mathfrak{a}}}\left|\Omega\right\rangle $
with $\mathfrak{q}_{23}$ trivial is $u_{\mathfrak{a}}\left(\alpha_{23}^{\overline{\mathfrak{a}}\mathfrak{a}}\right)^{\dagger}\beta_{14}^{\mathfrak{a}\overline{\mathfrak{a}}}\left|\Omega\right\rangle $.
Therefore, the quantum dimension $d_{\mathfrak{a}}$ can be defined and measured
by topological charge projectors. Since topological charges can be
detected by braiding, there exists a projector $P_{\mathfrak{q}}^{\mathtt{R}}$
supported near $\partial\mathtt{R}$ and commuting with the Hamiltonian requires that the total topological charge inside a finite region $\mathtt{R}$ is $\mathfrak{q}$. If
$\left|\Psi\right\rangle $ is a state with four excited spots as in
Fig.~\ref{fig:fusion4spots} satisfying 
$P_{\mathfrak{0}}^{\mathtt{A}}
=P_{\mathfrak{a}}^{\mathtt{B}_{1}}
=P_{\overline{\mathfrak{a}}}^{\mathtt{B}_{2}}
=P_{\mathfrak{a}}^{\mathtt{B}_{3}}
=P_{\overline{\mathfrak{a}}}^{\mathtt{B}_{4}}
=P_{\mathfrak{0}}^{\mathtt{B}_{12}}
=P_{\mathfrak{0}}^{\mathtt{B}_{34}}=1$,
then $d_{\mathfrak{a}}$ can also be defined and measured by 
\begin{equation}
\frac{1}{d_{\mathfrak{a}}}\coloneqq\frac{\left\langle \Psi\right|P_{\mathfrak{0}}^{\mathtt{B}_{23}}\left|\Psi\right\rangle }{\left\langle \Psi|\Psi\right\rangle },\label{eq:qd_projector}
\end{equation}
where $\mathfrak{0}$ denotes the trivial
topological charge and $\mathtt{B}_j$ is any oval region containing only spot $j$, for $j=1,2,3,4$.

Now let us compute $d_{\mathfrak{a}}$ for $\mathfrak{a}\in\mathfrak{Q}$
in a model of twisted gauge theory. Pick a representation $\mathcal{V}_{\mathfrak{a}}=\left(\rho_{\mathfrak{a}},V_{\mathfrak{a}}\right)$
for $\mathfrak{a}$. Let $\alpha_{\mathfrak{a}}:\mathcal{V}_{\mathfrak{a}}^{*}\otimes\mathcal{V}_{\mathfrak{a}}\rightarrow\mathbb{C}$
and $\beta_{\mathfrak{a}}:\mathbb{C}\rightarrow\mathcal{V}_{\mathfrak{a}}\otimes\mathcal{V}_{\mathfrak{a}}^{*}$
be the two intertwiners defined by Eqs.~\eqref{eq:ev_alpha} and
\eqref{eq:coev_beta}. Using the antipode $\left(S,\alpha,\beta\right)$
given in Eqs.~(\ref{eq:S_DG}-\ref{eq:S_beta_DG}), we have $\alpha_{\mathfrak{a}}\alpha_{\mathfrak{a}}^{\dagger}=\beta_{\mathfrak{a}}^{\dagger}\beta_{\mathfrak{a}}=\dim\mathcal{V}_{\mathfrak{a}}$. 
Thus, $\alpha_{23}^{\overline{\mathfrak{a}}\mathfrak{a}}\beta_{12}^{\mathfrak{a}\overline{\mathfrak{a}}}\beta_{34}^{\mathfrak{a}\overline{\mathfrak{a}}}$
acts on the vacuum as 
\begin{multline}
\mathbb{C}\xrightarrow{\frac{\beta_{\mathfrak{a}}}{\sqrt{\dim\mathcal{V}_{\mathfrak{a}}}}\otimes\frac{\beta_{\mathfrak{a}}}{\sqrt{\dim\mathcal{V}_{\mathfrak{a}}}}}\left(\mathcal{V}_{\mathfrak{a}}\otimes\mathcal{V}_{\mathfrak{a}}^{*}\right)\otimes\left(\mathcal{V}_{\mathfrak{a}}\otimes\mathcal{V}_{\mathfrak{a}}^{*}\right)=\\
\left(\left(\mathcal{V}_{\mathfrak{a}}\otimes\mathcal{V}_{\mathfrak{a}}^{*}\right)\otimes\mathcal{V}_{\mathfrak{a}}\right)\otimes\mathcal{V}_{\mathfrak{a}}^{*}\xrightarrow{\phi\otimes\text{id}_{\mathfrak{a}}}\left(\mathcal{V}_{\mathfrak{a}}\otimes\left(\mathcal{V}_{\mathfrak{a}}^{*}\otimes\mathcal{V}_{\mathfrak{a}}\right)\right)\otimes\mathcal{V}_{\mathfrak{a}}^{*}\\
\xrightarrow{\text{id}_{\mathfrak{a}}\otimes\frac{\beta_{\mathfrak{a}}}{\sqrt{\dim\mathcal{V}_{\mathfrak{a}}}}\otimes\text{id}_{\mathfrak{a}}}\mathcal{V}_{\mathfrak{a}}\otimes\mathcal{V}_{\mathfrak{a}}^{*}\label{eq:bab}
\end{multline}
up to a phase factor, where the equality in the first line is obtained by noticing that the state in $\mathcal{V}_{\mathfrak{a}}\otimes\mathcal{V}_{\mathfrak{a}}^{*}$ created from vacuum has trivial total topological charge.
It gets simplified to 
\begin{equation}
\mathbb{C}\xrightarrow{\left(\dim\mathcal{V}_{\mathfrak{a}}\right)^{-\frac{3}{2}}\beta_{\mathfrak{a}}}\mathcal{V}_{\mathfrak{a}}\otimes\mathcal{V}_{\mathfrak{a}}^{*},
\end{equation}
by the fact that the composition in Eq.~\eqref{eq:zigzag1} equals
identity. Therefore, $\alpha_{23}^{\overline{\mathfrak{a}}\mathfrak{a}}\beta_{12}^{\mathfrak{a}\overline{\mathfrak{a}}}\beta_{34}^{\mathfrak{a}\overline{\mathfrak{a}}}\left|\Omega\right\rangle =\left(\dim\mathcal{V}_{\mathfrak{a}}\right)^{-1}\beta_{14}^{\mathfrak{a}\overline{\mathfrak{a}}}\left|\Omega\right\rangle $
up to a phase factor and hence
\begin{equation}
d_{\mathfrak{a}}=\dim\mathcal{V}_{\mathfrak{a}}.
\end{equation}
Roughly, the quantum dimension $d_{\mathfrak{a}}$ tells how strongly $\mathfrak{a}$ and
$\overline{\mathfrak{a}}$ are entangled when they are
restricted to a trivial total topological charge.

The diagrammatic presentation used in tensor categories provides a useful tool in
describing the splitting, fusion, and braiding processes
of anyons~\cite{KITAEV20062,barkeshli}. Let \begin{align} 
\begin{tikzpicture}[baseline={([yshift=-0.2ex]current bounding box.center)}]
\draw[->-=0.8, line width=1pt, cap=round] (0,0) sin (-0.5,.5) cos (-1,0)node[xshift=-3bp, yshift=3bp]{$\mathfrak{a}$}; 
\end{tikzpicture}
\coloneqq \alpha_{\mathfrak{a}}, & &
\begin{tikzpicture}[baseline={([yshift=-0.8ex]current bounding box.center)}]
\draw[->-=0.8, line width=1pt, cap=round] (-1,0)node[xshift=-3bp, yshift=-3bp]{$\mathfrak{a}$} sin (-0.5,-0.5) cos (0,0);  
\end{tikzpicture}
\coloneqq \alpha_{\mathfrak{a}}^{\dagger},
\label{eq:graphic_annihilation}\\
\begin{tikzpicture}[baseline={([yshift=-0.8ex]current bounding box.center)}]
\draw[->-=0.8, line width=1pt, cap=round] (1,0) sin (0.5,-0.5) cos (0,0)node[xshift=-3bp, yshift=-3bp]{$\mathfrak{a}$};  
\end{tikzpicture}
\coloneqq \beta_{\mathfrak{a}}, & &
\begin{tikzpicture}[baseline={([yshift=-0.2ex]current bounding box.center)}]
\draw[->-=0.8, line width=1pt, cap=round] (0,0)node[xshift=-3bp, yshift=3bp]{$\mathfrak{a}$} sin (0.5,0.5) cos (1,0);  
\end{tikzpicture}
\coloneqq \beta_{\mathfrak{a}}^{\dagger},
\end{align}The pair of linear maps $\alpha_{\mathfrak{a}}$ and $\beta_{\mathfrak{a}}$
are picked such that the compositions in Eq.~\eqref{eq:zigzag1}
and \eqref{eq:zigzag2} equal identities, which are graphically presented
as \begin{align} 
\begin{tikzpicture}[baseline={([yshift=0ex]current bounding box.center)}]
\draw[->-=0.5, line width=1pt, cap=round] (0,-1) -- (0,0) sin (-0.25,0.5) cos (-0.5,0) sin (-0.75,-0.5) cos (-1,0)
-- (-1,1)node[xshift=4bp, yshift=-3bp]{$\mathfrak{a}$};  
\draw[->-=0.5, line width=1pt, cap=round] (0,-1) -- (0,0);
\draw[->-=0.5, line width=1pt, cap=round] (-1,0) -- (-1,1);
\end{tikzpicture}
=
\begin{tikzpicture}[baseline={([yshift=0ex]current bounding box.center)}]
\draw[->-=0.5, line width=1pt, cap=round] (0,-1) -- (0,1)node[xshift=4bp, yshift=-3bp]{$\mathfrak{a}$};
\end{tikzpicture}, \qquad
\begin{tikzpicture}[baseline={([yshift=0ex]current bounding box.center)}]
\draw[->-=0.5, line width=1pt, cap=round] (0,1)node[xshift=-4bp, yshift=-3bp]{$\mathfrak{a}$} -- (0,0) sin (-0.25,-0.5) cos (-0.5,0) sin (-0.75,0.5) cos (-1,0) -- (-1,-1); 
\draw[->-=0.5, line width=1pt, cap=round] (0,1) -- (0,0);
\draw[->-=0.5, line width=1pt, cap=round] (-1,0) -- (-1,-1);
\end{tikzpicture}
= 
\begin{tikzpicture}[baseline={([yshift=0ex]current bounding box.center)}]
\draw[->-=0.5, line width=1pt, cap=round] (0,1)node[xshift=4bp, yshift=-3bp]{$\mathfrak{a}$} -- (0,-1);
\end{tikzpicture}.
\label{eq:zigzag}
\end{align} Here a vertical line with label $\mathfrak{a}$ and an upward (resp.
downward) arrow is interpreted as the identity operator on the
topological charge $\mathfrak{a}$ (resp. $\overline{\mathfrak{a}}$). Convenient normalizations compatible with Eq.~\eqref{eq:zigzag} can be picked as
\begin{align} 
\begin{tikzpicture}[baseline={([yshift=0ex]current bounding box.center)}]
\draw[-{stealth'[flex=0.75]}, line width=1pt] (0,0)  arc (-180:-30:.6) ; 
\draw[line width=1pt, cap=round] (0,0)node[xshift=4bp, yshift=-3bp]{$\mathfrak{a}$}  arc (-180:180:.6) ; 
\end{tikzpicture}
= \alpha_{\mathfrak{a}}\alpha_{\mathfrak{a}}^{\dagger}=d_{\mathfrak{a}}, & & 
\begin{tikzpicture}[baseline={([yshift=0ex]current bounding box.center)}]
\draw[line width=1pt, cap=round] (0,0)node[xshift=4bp, yshift=-3bp]{$\mathfrak{a}$} arc (180:-180:.6); 
\draw[-{stealth'[flex=0.75]}, line width=1pt, cap=round] (1.2,0) arc (0:-150:.6);
\end{tikzpicture}
= \beta_{\mathfrak{a}}^{\dagger}\beta_{\mathfrak{a}}=d_{\mathfrak{a}}.
\label{eq:normalization2}
\end{align}
Notice that $\alpha_{\overline{\mathfrak{a}}}^{\dagger},\beta_{\mathfrak{a}}\in V_{\mathfrak{0}}^{\mathfrak{a}\overline{\mathfrak{a}}}$ and hence they just differ by a phase factor (\emph{i.e.}, $\beta_{\mathfrak{a}}=\varkappa_{\mathfrak{a}}\alpha_{\overline{\mathfrak{a}}}^{\dagger}$), where $\mathfrak{0}$ denotes the trivial topological charge. 
Further, if $\overline{\mathfrak{a}}=\mathfrak{a}$, then  $\alpha_{\overline{\mathfrak{a}}}=\alpha_{\mathfrak{a}}$
is already fixed by the choice of $\beta_{\mathfrak{a}}$ via Eq.~\eqref{eq:zigzag}.
In this case, $\varkappa_{\mathfrak{a}}$ is well-defined. It takes values
$\pm1$ and is called the \emph{Frobenius-Schur indicator}~\cite{KITAEV20062}.

For $\overline{\mathfrak{a}}=\mathfrak{a}$, we can measure $\varkappa_{\mathfrak{a}}$
by 
\begin{equation}
\left(\beta_{23}^{\mathfrak{a}\mathfrak{a}}\right)^{\dagger}\beta_{12}^{\mathfrak{a}\mathfrak{a}}\beta_{34}^{\mathfrak{a}\mathfrak{a}}\left|\Omega\right\rangle =\frac{\varkappa_{\mathfrak{a}}}{d_{\mathfrak{a}}}\beta_{14}^{\mathfrak{a}\mathfrak{a}}\left|\Omega\right\rangle .\label{eq:varkappa}
\end{equation}
in the setting of Fig.~\ref{fig:fusion4spots}, where $\left(\beta_{23}^{\mathfrak{a}\mathfrak{a}}\right)^{\dagger}$
is supported in oval $\mathtt{B}_{23}$ and annihilates a pair of anyons
both of topological charge $\mathfrak{a}$ at spots $2,3$. The operators
$\beta_{12}^{\mathfrak{a}\mathfrak{a}},\beta_{23}^{\mathfrak{a}\mathfrak{a}},\beta_{34}^{\mathfrak{a}\mathfrak{a}}$
can be compared with $\beta_{14}^{\mathfrak{a}\mathfrak{a}}$ by hopping
operators. For $i=1,2,3$, let $\mathcal{O}_{i}^{\mathfrak{a}}$ be
an operators supported on the oval $\mathtt{B}_{i\left(i+1\right)}$that
moves a quasiparticle of topological charge $\mathfrak{a}$ from spot
$i$ to spot $i+1$. We can require $\beta_{12}^{\mathfrak{a}\mathfrak{a}}=\left(\mathcal{O}_{3}^{\mathfrak{a}}\mathcal{O}_{2}^{\mathfrak{a}}\right)^{\dagger}\beta_{14}^{\mathfrak{a}\mathfrak{a}}$,
$\beta_{34}^{\mathfrak{a}\mathfrak{a}}=\mathcal{O}_{2}^{\mathfrak{a}}\mathcal{O}_{1}^{\mathfrak{a}}\beta_{14}^{\mathfrak{a}\mathfrak{a}}$
and $\beta_{23}^{\mathfrak{a}\mathfrak{a}}=\mathcal{O}_{1}^{\mathfrak{a}}\left(\mathcal{O}_{3}^{\mathfrak{a}}\right)^{\dagger}\beta_{14}^{\mathfrak{a}\mathfrak{a}}$
on $\left|\Omega\right\rangle $. Then it is easy to see that the
measurement of $\varkappa_{a}$ via Eq.~\eqref{eq:varkappa} is well-defined,
\emph{i.e.}, independent of the remaining freedom of adding phase factors
to $\mathcal{O}_{1}^{\mathfrak{a}}$, $\mathcal{O}_{2}^{\mathfrak{a}}$, $\mathcal{O}_{3}^{\mathfrak{a}}$
and $\beta_{14}^{\mathfrak{a}\mathfrak{a}}$.

\subsubsection{Braiding statistics}

\begin{figure}
\includegraphics[width=0.7\columnwidth]{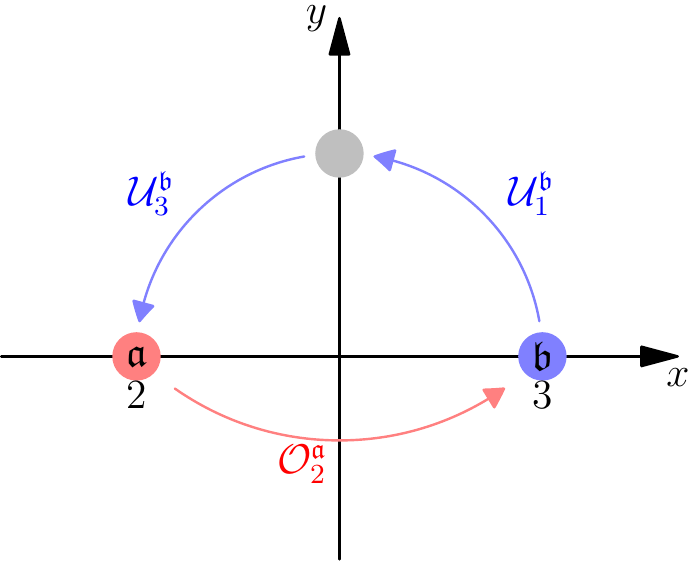}

\caption{Two anyons of topological charge $\mathfrak{a}$ and $\mathfrak{b}$ at spots $2$ and $3$ are
braided by $\mathcal{U}_{3}^{\mathfrak{b}}\mathcal{O}_{2}^{\mathfrak{a}}\mathcal{U}_{1}^{\mathfrak{b}}$,
where $\mathcal{U}_{1}^{\mathfrak{q}}$, $\mathcal{O}_{2}^{\mathfrak{q}}$ and $\mathcal{U}_{3}^{\mathfrak{q}}$
are the hopping operators (for topological charge $\mathfrak{q}$) indicated
by and supported near the three arrows respectively.}

\label{fig:braiding2D}
\end{figure}

In Fig.~\ref{fig:braiding2D}, the hopping processes of a single
quasiparticle of topological charge $\mathfrak{q}\in\mathfrak{Q}$
between the three spots (red, blue and grey online) can be made by
operators $\mathcal{U}_{1}^{\mathfrak{q}}$, $\mathcal{O}_{2}^{\mathfrak{q}}$, 
$\mathcal{U}_{3}^{\mathfrak{q}}$ supported near the corresponding arrows. 
To resolve the phase factor ambiguity, we require $\mathcal{U}_{3}^{\mathfrak{q}}\mathcal{U}_{1}^{\mathfrak{q}}\mathcal{O}_{2}^{\mathfrak{q}}=1$ in moving a single quasiparticle starting at spot $2$. 
Then 
\begin{gather}
\mathcal{U}_{3}^{\mathfrak{b}}\mathcal{O}_{2}^{\mathfrak{a}}\mathcal{U}_{1}^{\mathfrak{b}}:\mathcal{V}_{\mathfrak{a}}^{\left(2\right)}\otimes\mathcal{V}_{\mathfrak{b}}^{\left(3\right)}\xrightarrow{\mathcal{R}}\mathcal{V}_{\mathfrak{b}}^{\left(3\right)}\otimes\mathcal{V}_{\mathfrak{a}}^{\left(2\right)},\label{eq:Raa}\\
\mathcal{U}_{3}^{\mathfrak{a}}\mathcal{O}_{2}^{\mathfrak{b}}\mathcal{U}_{1}^{\mathfrak{a}}\mathcal{U}_{3}^{\mathfrak{b}}\mathcal{O}_{2}^{\mathfrak{a}}\mathcal{U}_{1}^{\mathfrak{b}}:\mathcal{V}_{\mathfrak{a}}^{\left(2\right)}\otimes\mathcal{V}_{\mathfrak{b}}^{\left(3\right)}\xrightarrow{\mathcal{R}^{2}}\mathcal{V}_{\mathfrak{a}}^{\left(2\right)}\otimes\mathcal{V}_{\mathfrak{b}}^{\left(3\right)}
\label{eq:R2}
\end{gather}
braid topological charges $\mathfrak{a},\mathfrak{b}\in\mathfrak{Q}$ initially at spots $2$ and $3$, 
where $\mathcal{V}_{\mathfrak{a}}^{\left(2\right)}$ and $\mathcal{V}_{\mathfrak{b}}^{\left(3\right)}$
are the corresponding representations. 
When $\mathfrak{a}=\mathfrak{b}$, the action
of $\mathcal{R}$, denoted $\mathcal{R}^{\mathfrak{aa}}$, is precisely captured
by Eq.~\eqref{eq:Raa}. When $\mathfrak{a}\neq \mathfrak{b}$, we cannot fixed the phase
factor of $\mathcal{U}_{1}^{\mathfrak{b}}\mathcal{O}_{2}^{\mathfrak{a}}\mathcal{U}_{3}^{\mathfrak{b}}$. Only $\mathcal{R}^{2}$, called the \emph{monodromy operator},
is well-defined via Eq.~\eqref{eq:R2}.

Graphically, $\mathcal{R}^{\mathfrak{b}\mathfrak{a}}:\mathcal{V}_{\mathfrak{a}}\otimes\mathcal{W}_{\mathfrak{b}}\rightarrow\mathcal{W}_{\mathfrak{b}}\otimes\mathcal{V}_{\mathfrak{a}}$
is presented as 
\begin{equation}
\begin{tikzpicture}[scale=0.8, baseline={([yshift=-.4ex]current bounding box.center)}]
\begin{knot}[ 
clip width =5,  
flip crossing=2 ] 
\strand[line width=1pt] (0,0) -- (1,1)node[xshift=4bp, yshift=-4bp]{$\mathfrak{a}$}; 
\strand[->-=0.9, line width=1pt] (1,0) -- (0,1)node[xshift=-4bp, yshift=-4bp]{$\mathfrak{b}$};  
\end{knot} 
\draw[->-=0.9, line width=1pt] (0,0) -- (1,1);
\end{tikzpicture}
\coloneqq \mathcal{R}^{\mathfrak{b}\mathfrak{a}},
\end{equation}where $\mathcal{V}_{\mathfrak{a}}$ and $\mathcal{W}_{\mathfrak{b}}$
are irreducible representations for $\mathfrak{a},\mathfrak{b}\in\mathfrak{Q}$
respectively. 

In general, $\mathcal{R}^{\mathfrak{a}\mathfrak{a}}$ is a matrix
even when restricted to a definite total topological charge $\mathfrak{c}$,
because $N_{\mathfrak{a}\mathfrak{a}}^{\mathfrak{c}}$ may be greater
than 1. To get a simple scalar out of $\mathcal{R}^{\mathfrak{a}\mathfrak{a}}$,
let us consider a process beginning and ending with the vacuum as
follows. First, we create four anyons at the spots positioned as in
Fig.~\ref{fig:fusion4spots}; a pair of anyons with topological charges
$\overline{\mathfrak{a}},\mathfrak{a}$ (resp. $\mathfrak{a},\overline{\mathfrak{a}}$)
are created at spots $1,2$ (resp. $3,4$) with an operator $(\alpha_{12}^{\overline{\mathfrak{a}}\mathfrak{a}})^{\dagger}$
(resp. $\beta_{34}^{\mathfrak{a}\overline{\mathfrak{a}}}$) supported
in oval $\mathtt{B}_{12}$ (resp. $\mathtt{B}_{34}$). We
normalize these operators by 
\begin{equation}
\alpha_{12}^{\overline{\mathfrak{a}}\mathfrak{a}}\left(\alpha_{12}^{\overline{\mathfrak{a}}\mathfrak{a}}\right)^{\dagger}=\left(\beta_{34}^{\mathfrak{a}\overline{\mathfrak{a}}}\right)^{\dagger}\beta_{34}^{\mathfrak{a}\overline{\mathfrak{a}}}=1\label{eq:normalization1}
\end{equation}
on the vacuum. Second, the anyons both with topological charge $\mathfrak{a}$
at spots $2,3$ are braided by $\mathcal{U}_{3}^{\mathfrak{a}}\mathcal{O}_{2}^{\mathfrak{a}}\mathcal{U}_{1}^{\mathfrak{a}}$
as in Fig.~\ref{fig:braiding2D}. Finally, $\alpha_{12}^{\overline{\mathfrak{a}}\mathfrak{a}}\left(\beta_{34}^{\mathfrak{a}\overline{\mathfrak{a}}}\right)^{\dagger}$
annihilates the four anyons. Then 
the \emph{topological spin} $\theta_{\mathfrak{a}}$ associated with
$\mathfrak{a}\in\mathfrak{Q}$ is the phase factor of 
\begin{equation}
\frac{\theta_{\mathfrak{a}}}{d_{\mathfrak{a}}}\coloneqq\left\langle \Omega\right|\alpha_{12}^{\overline{\mathfrak{a}}\mathfrak{a}}\left(\beta_{34}^{\mathfrak{a}\overline{\mathfrak{a}}}\right)^{\dagger}\mathcal{U}_{1}^{\mathfrak{a}}\mathcal{O}_{2}^{\mathfrak{a}}\mathcal{U}_{3}^{\mathfrak{a}}\left(\alpha_{12}^{\overline{\mathfrak{a}}\mathfrak{a}}\right)^{\dagger}\beta_{34}^{\mathfrak{a}\overline{\mathfrak{a}}}\left|\Omega\right\rangle ,\label{eq:measuretopspin}
\end{equation}
whose amplitude specifies the quantum dimension $d_{\mathfrak{a}}$
as well. Graphically, this equation is presented as
\begin{equation} 
\frac{\theta_{\mathfrak{a}}}{d_{\mathfrak{a}}}=\frac{1}{d_{\mathfrak{a}}^{2}} \, 
\begin{tikzpicture}[baseline={([yshift=-0.2ex]current bounding box.center)}]
\begin{knot}[clip width =4, consider self intersections=true, 
]
\strand[line width=1pt, cap=round] (0,0) to[out=90,in=0] (-0.5,.5) to[out=-180, in=0] (-1.5,-0.5) to[out=180,in=-90] (-2,0);
\strand[line width=1pt, cap=round] (-2,0) to[out=90,in=180] (-1.5,0.5) to[out=0, in=180] (-0.5,-0.5) to[out=0,in=-90] (0,0);
\end{knot} 
\draw[-<-=0.65, line width=1pt, cap=round] (-0.5,.5) to[out=-180, in=0] (-1.5,-0.5)node[xshift=0bp, yshift=6bp]{$\mathfrak{a}$};
\end{tikzpicture} \; ,
\label{eq:topspin_graph}
\end{equation}
where $\frac{1}{d_{\mathfrak{a}}^{2}}$ on the right hand side comes
from the different normalization conventions set by Eqs.~\eqref{eq:normalization2}
and \eqref{eq:normalization1}. For the models of twisted gauge theories,
\begin{equation}
\theta_{\mathfrak{a}}=\frac{1}{d_{\mathfrak{a}}}\text{tr}\left(\wp R,\mathcal{V}_{\mathfrak{a}}\otimes\mathcal{V}_{\mathfrak{a}}\right),
\end{equation}
where $\wp:\mathcal{V}_{\mathfrak{a}}\otimes\mathcal{V}_{\mathfrak{a}}\rightarrow\mathcal{V}_{\mathfrak{a}}\otimes\mathcal{V}_{\mathfrak{a}};\,v^{\left(1\right)}\otimes v^{\left(2\right)}\mapsto v^{\left(2\right)}\otimes v^{\left(1\right)}$
and $\text{tr}\left(\wp R,\mathcal{V}_{\mathfrak{a}}\otimes\mathcal{V}_{\mathfrak{a}}\right)$
is the trace of $\wp R$ over $\mathcal{V}_{\mathfrak{a}}\otimes\mathcal{V}_{\mathfrak{a}}$.

For $\mathfrak{a}\neq\mathfrak{b}$, the monodromy operator $\mathcal{R}^{2}=\mathcal{R}^{\mathfrak{a}\mathfrak{b}}\mathcal{R}^{\mathfrak{b}\mathfrak{a}}$
is diagonalized in the basis with definite total topological charge.
Explicitly,
\begin{equation}
\mathcal{R}^{2}=\mathcal{R}_{\mathfrak{c}}^{\mathfrak{a}\mathfrak{b}}\mathcal{R}_{\mathfrak{c}}^{\mathfrak{b}\mathfrak{a}}=\frac{\theta_{\mathfrak{c}}}{\theta_{\mathfrak{a}}\theta_{\mathfrak{b}}}\text{id}_{V_{\mathfrak{c}}^{\mathfrak{a}\mathfrak{b}}},
\end{equation}
on the sector with definite total topological charge $\mathfrak{c}$~\cite{KITAEV20062}. 

Analogously to the discussion of topological spin, we are interested in the following
process. First, four anyons with topological charges $\mathfrak{a}$,
$\overline{\mathfrak{a}}$, $\mathfrak{b}$, $\overline{\mathfrak{b}}$
are created at spots $1,2,3,4$ positioned as in Fig.~\ref{fig:fusion4spots}
by operators $\beta_{12}^{\mathfrak{a}\overline{\mathfrak{a}}}\beta_{34}^{\mathfrak{b}\overline{\mathfrak{b}}}$,
where $\beta_{12}^{\mathfrak{a}\overline{\mathfrak{a}}}$, $\beta_{34}^{\mathfrak{b}\overline{\mathfrak{b}}}$
are supported on ovals $\mathtt{B}_{12}$, $\mathtt{B}_{34}$
respectively and normalized in a similar way as in Eq.~\eqref{eq:normalization1}.
Second, a monodromy operator braiding $\overline{\mathfrak{a}}$ and $\mathfrak{b}$ is realized as in Eq.~\eqref{eq:R2}.
Finally, the four anyons are annihilated by $(\beta_{12}^{\mathfrak{a}\overline{\mathfrak{a}}}\beta_{34}^{\mathfrak{b}\overline{\mathfrak{b}}})^{\dagger}$.
The expectation value of the whole process on the vacuum $\left|\Omega\right\rangle $
is
\begin{equation}
\mathsf{S}_{\mathfrak{a}\mathfrak{b}}\coloneqq
\frac{1}{d_{\mathfrak{a}}d_{\mathfrak{b}}}
\begin{tikzpicture}[baseline={([yshift=-0.2ex]current bounding box.center)}]
\begin{knot}[ 
clip width =5,  
flip crossing=2 ] 
\strand[line width=1pt] (0.4,0) circle[radius=0.6]; 
\strand[line width=1pt] (-.4,0) circle[radius=0.6]; 
\end{knot} 
\draw[{stealth'[flex=0.75]}-, line width=1pt](-1,0) arc (180:190:0.6);
\draw[{stealth'[flex=0.75]}-, line width=1pt](-0.2,0) arc (180:190:0.6);
\draw (-0.4,0)node[xshift=0bp, yshift=0bp]{$\mathfrak{b}$};
\draw (-1.2,0)node[xshift=0bp, yshift=0bp]{$\mathfrak{a}$};
\end{tikzpicture} \,
=\sum_{\mathfrak{c}\in\mathfrak{Q}}N_{\overline{\mathfrak{a}}\mathfrak{b}}^{\mathfrak{c}}\frac{\theta_{\mathfrak{c}}}{\theta_{\mathfrak{a}}\theta_{\mathfrak{b}}}\frac{d_{\mathfrak{c}}}{d_{\mathfrak{a}}d_{\mathfrak{b}}},
\label{eq:S_graph}
\end{equation}
where the factor $\frac{1}{d_{\mathfrak{a}}d_{\mathfrak{b}}}$ is
from the normalization difference between $(\beta_{12}^{\mathfrak{a}\overline{\mathfrak{a}}})^{\dagger}\beta_{12}^{\mathfrak{a}\overline{\mathfrak{a}}}=(\beta_{34}^{\mathfrak{b}\overline{\mathfrak{b}}})^{\dagger}\beta_{34}^{\mathfrak{b}\overline{\mathfrak{b}}}=1$
on $\left|\Omega\right\rangle $ and Eq.~\eqref{eq:normalization2}.
In the literature~\cite{KITAEV20062,barkeshli}, $\mathsf{S}_{\mathfrak{a}\mathfrak{b}}$
are often rescaled to 
$\mathcal{S}_{\mathfrak{ab}}=\frac{d_{\mathfrak{a}}d_{\mathfrak{b}}}{\mathcal{D}} \mathsf{S}_{\mathfrak{ab}}$ 
and put into a matrix form $\mathcal{S}=\left(\mathcal{S}_{\mathfrak{ab}}\right)_{\mathfrak{a},\mathfrak{b}\in\mathfrak{Q}}$,
called the \emph{topological $S$-matrix}, which is closely related to a modular transformation of torus \cite{dijkgraaf1990,hu13twisted}. Here 
\begin{equation}
\mathcal{D}\coloneqq\sqrt{\sum_{\mathfrak{a}\in\mathfrak{Q}}d_{\mathfrak{a}}^{2}}
\end{equation}
is called the total quantum dimension. For the models of twisted gauge
theories,
\begin{equation}
\mathsf{S}_{\mathfrak{a}\mathfrak{b}}^{*}=\mathsf{S}_{\overline{\mathfrak{a}}\mathfrak{b}}=\frac{1}{d_{\mathfrak{a}}d_{\mathfrak{b}}}\text{tr}\left(\left(\wp R\right)^{2},\mathcal{V}_{\mathfrak{a}}\otimes\mathcal{V}_{\mathfrak{b}}\right),\label{eq:topS}
\end{equation}
where $\wp:\mathcal{V}\otimes\mathcal{W}\rightarrow\mathcal{W}\otimes\mathcal{V};\,v\otimes w\mapsto w\otimes v$
is the permutation operator acting on the tensor product of any two vectors spaces
and $\text{tr}\left(\left(\wp R\right)^{2},\mathcal{V}_{\mathfrak{a}}\otimes\mathcal{V}_{\mathfrak{b}}\right)$
is the trace of $\left(\wp R\right)^{2}$ over $\mathcal{V}_{\mathfrak{a}}\otimes\mathcal{V}_{\mathfrak{b}}$.

\section{Twisted fracton models}
\label{sec:twistedfracton}

We now introduce generalizations of two paradigmatic three-dimensional gapped fracton phases: the X-cube model~\cite{vijay16xcube} and the checkerboard model~\cite{fracton1,vijay16xcube}. Instead of reviewing the $\mathbb{Z}_2$ variants of these models, which have been intensely studied recently, we will here introduce general versions of these models based on a finite Abelian group $G$, whose identity element is denoted $0$. We note that in contrast with the original formulation of the X-cube model, where spins were defined to live on links of a cubic lattice, here we formulate this model on the dual lattice, where spins live on faces of the cubic lattice. 

\subsection{Twisted X-cube models}\label{subsec:X-cube_twisted_def}

\begin{figure}
\includegraphics[width=0.8\columnwidth]{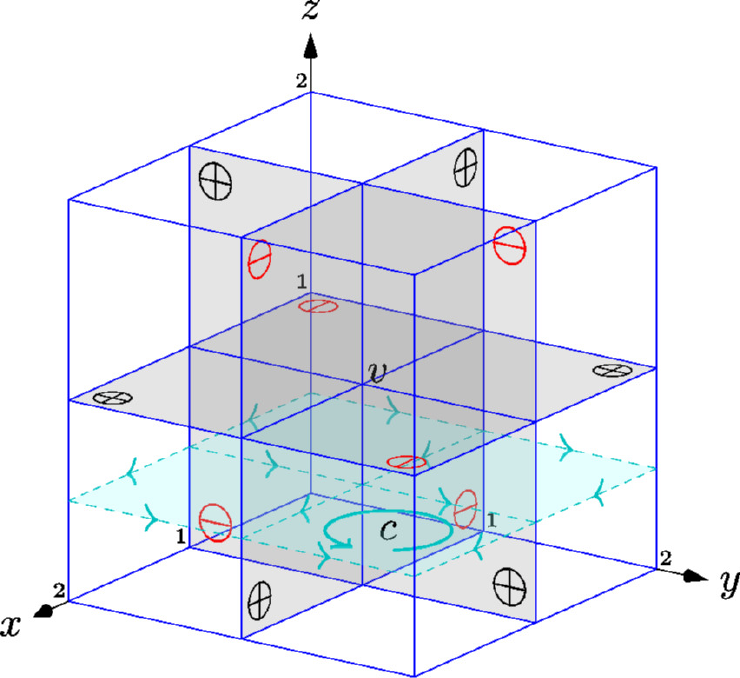}

\caption{The cubic lattice $\Lambda$ for X-cube models. Spins labeled by an
Abelian group $G$ lie on faces. The generalized gauge transformation
$A_{v}^{g}$ flips spins by $g$ (resp. $-g$) on the faces marked
$\oplus$ (resp. $\ominus$). The cross section $\Sigma_{1}^{z}$
is the intersection of $\Lambda$ with the plane $z=\frac{1}{2}$
(cyan online). The arrowed arc (cyan online) indicates the $z$-flux
of the associated cube. }

\label{fig:XC-1}
\end{figure}

Given a simple cubic lattice $\Lambda$, we pick the coordinates such
that its vertices are in $\mathbb{Z}^{3}$ as shown in Fig.~\ref{fig:XC-1}.
Each edge (resp. face, cube) is labeled by the coordinates of its
center. Let $\Lambda^{0}$ (resp. $\Lambda^{1}$, $\Lambda^{2}$,
$\Lambda^{3}$) be the label set for vertices (resp. edges, faces,
cubes), whose elements are usually denoted as $v$ (resp. $\ell$,
$p$, $c$). Then 
\begin{equation}
\Lambda^{1}=\Lambda_{x}^{1}\cup\Lambda_{y}^{1}\cup\Lambda_{z}^{1},
\end{equation}
where $\Lambda_{x}^{1}=\Lambda^{0}+\left(\frac{1}{2},0,0\right)$,
$\Lambda_{y}^{1}=\Lambda^{0}+\left(0,\frac{1}{2},0\right)$ and $\Lambda_{z}^{1}=\Lambda^{0}+\left(0,0,\frac{1}{2}\right)$
are the sets of $x$-, $y$- and $z$-edges (\emph{i.e.}, edges lying
in the $x$-, $y$- and $z$-direction) respectively. Similarly, 
\begin{equation}
\Lambda^{2}=\Lambda_{xy}^{2}\cup\Lambda_{yz}^{2}\cup\Lambda_{zx}^{2},
\end{equation}
where $\Lambda_{xy}^{2}=\Lambda^{0}+\left(\frac{1}{2},\frac{1}{2},0\right)$,
$\Lambda_{yz}^{1}=\Lambda^{0}+\left(0,\frac{1}{2},\frac{1}{2}\right)$
and $\Lambda_{zx}^{1}=\Lambda^{0}+\left(\frac{1}{2},0,\frac{1}{2}\right)$
are the sets of $xy$-, $yz$- and $zx$-faces (\emph{i.e.}, faces
perpendicular to the $z$-, $x$- and $y$-direction) respectively. In
addition, 
\begin{equation}
\Lambda^{3}=\Lambda^{0}+\frac{1}{2}\left(1,1,1\right).
\end{equation}
In the following, we would like to consider a cubic lattice on a three-dimensional torus $\mathtt{T}^{3}$,
obtained by identifying $\left(x,y,z\right)\sim\left(x+L_{x},y,z\right)\sim\left(x,y+L_{y},z\right)\sim\left(x,y,z+L_{z}\right)$ 
with $L_{x},L_{y},L_{z}\in \mathbb{Z}$ describing the system size. Such a lattice has vertices
$\Lambda^{0}=\mathbb{Z}_{L_{x}}\times\mathbb{Z}_{L_{y}}\times\mathbb{Z}_{L_{z}}$ and the infinite case can be viewed as its thermodynamic limit.

Given any region $\Gamma$ of $\Lambda$, let $\Lambda^{n}\left(\Gamma\right)$ for $n=0,1,2,3$ be the label sets of the vertices, edges, faces, cubes contained in $\Gamma$ respectively. 
Similarly, we may define sets $\Lambda_{x}^{1}\left(\Gamma\right)$,
$\Lambda_{xy}^{2}\left(\Gamma\right)$ and etc. 
For instance, $c\in \Lambda^{3}(\Gamma)$ means $c\subseteq\Gamma$ with $c\in\Lambda^{3}$ viewed as the region occupied by the cube $c$ (boundary included). In particular, $c\in \Lambda^{3}(\Gamma)$ implies $\Lambda^{n}\left(c\right)\subseteq\Lambda^{n}\left(\Gamma\right),\forall n=0,1,2,3$.  
For any cube $c$, $\Lambda^{3}(c)=\{c\}$ and $\Lambda^{2}(c)$ 
(resp. $\Lambda^{1}(c)$, $\Lambda^{0}(c)$) is the set of the 6 faces (resp. 12 edges, 8 vertices) of $c$.

\subsubsection{X-cube model based on a finite Abelian group}

Let $G$ be a finite Abelian group, with $0$ denoting its identity
element. A local Hilbert space (also called a \emph{spin} for short)
spanned by an orthonoraml basis $\left\{ \left|p,g\right\rangle \right\} _{g\in G}$ is
assigned to each face $p\in\Lambda^{2}$. Then the Hilbert space
associated with any region $\Gamma$ of $\Lambda$, denoted ${\cal H}\left(\Lambda^{2}\left(\Gamma\right),G\right)$,
is spanned by 
\begin{equation}
\left|\vartheta\right\rangle \coloneqq\bigotimes_{p\in\Lambda^{2}\left(\Gamma\right)}\left|p,\vartheta\left(p\right)\right\rangle ,
\end{equation}
with $\vartheta\in G^{\Lambda^{2}\left(\Gamma\right)}$, where $G^{\Lambda^{2}\left(\Gamma\right)}\coloneqq\text{Fun}\left(\Lambda^{2}\left(\Gamma\right),G\right)$
is the set of functions from $\Lambda^{2}\left(\Gamma\right)$  to $G$. Each element of 
$G^{\Lambda^{2}\left(\Gamma\right)}$ specifies a \emph{spin configuration} on $\Gamma$. On the whole lattice, the total Hilbert space
is $\mathcal{H}\left(\Lambda^{2},G\right)$. 

For each vertex $v$, we define a function $\kappa_{v}$ from $\Lambda^{2}$
to $\mathbb{Z}$, which maps $p=\left(p^{x},p^{y},p^{z}\right)\in\Lambda^{2}$
to 
\begin{equation}
\kappa_{v}\left(p\right)\coloneqq\sum_{s\in\Lambda^{0}\left(p\right)}\left(-1\right)^{s^{x}+s^{y}+s^{z}-p^{x}-p^{y}-p^{z}}\delta_{s,v}.
\end{equation}
Graphically, $\kappa_{v}$ is presented in Fig.~\ref{fig:XC-1};
its value is $+1$ (resp. $-1$) on faces marked $\oplus$ (resp. $\ominus$)
and zero on all faces not adjacent to $v$.  

In an untwisted X-cube model, $\forall g\in G$, a \emph{(generalized)
gauge transformation operator} $A_{v}^{g}$ associated with each vertex $v$ can
be defined as
\begin{equation}
A_{v}^{g}\coloneqq\sum_{\vartheta\in G^{\Lambda^{2}}}\left|\vartheta+\kappa_{v}g\right\rangle \left\langle \vartheta\right|.\label{eq:X_cube_A}
\end{equation}
Clearly, it is supported on the twelve faces adjacent to $v$. 

If $G=\mathbb{Z}_{2}=\left\{ 0,1\right\} $, then $A_{v}^{1}$ is the product of the Pauli operators $\sigma^{x}$ on the twelve
faces. As $v$ labels a cube of the dual lattice and the twelve faces
corresponds to the edges of this cube, $A_{v}^{1}$ is an X-cube
operator in the dual lattice. Thus, the original X-cube model~\cite{vijay16xcube} is a special case of the family of models we are
constructing here. 

In addition, supported on each cube $c\in\Lambda^{3}$, we have \emph{(generalized)
flux projectors} 
\begin{align}
B_{c}^{x} & \coloneqq\sum_{\vartheta\in G^{\Lambda^{2}\left(c\right)}}\delta_{\partial_{y}\vartheta\left(c\right)-\partial_{z}\vartheta\left(c\right),0}\left|\vartheta\right\rangle \left\langle \vartheta\right|,\\
B_{c}^{y} & \coloneqq\sum_{\vartheta\in G^{\Lambda^{2}\left(c\right)}}\delta_{\partial_{z}\vartheta\left(c\right)-\partial_{x}\vartheta\left(c\right),0}\left|\vartheta\right\rangle \left\langle \vartheta\right|,\\
B_{c}^{z} & \coloneqq\sum_{\vartheta\in G^{\Lambda^{2}\left(c\right)}}\delta_{\partial_{x}\vartheta\left(c\right)-\partial_{y}\vartheta\left(c\right),0}\left|\vartheta\right\rangle \left\langle \vartheta\right|,\\
B_{c} & \coloneqq B_{c}^{x}B_{c}^{y}B_{c}^{z},
\end{align}
where $\partial_{x}\vartheta\left(c\right)\coloneqq\vartheta\left(c+\left(\frac{1}{2},0,0\right)\right)-\vartheta\left(c-\left(\frac{1}{2},0,0\right)\right)$
and $\partial_{y}\vartheta$, $\partial_{z}\vartheta$ are defined
analogously. As in Fig.~\ref{fig:XC-1}, the $zx$- and $yz$-faces of cube $c$ can be thought as 
edges of the square with an arrowed arc; thus,  $B_{c}^{z}$ can be understood as a projector
requiring the $z$-flux of cube $c$ to be trivial. 

It is straightforward to check that $\forall v,v_{0},v_{1}\in \Lambda^{0}\left(\Lambda\right)$,
$\forall c,c_{0},c_{1}\in \Lambda^{3}\left(\Lambda\right)$, $\forall g,h\in G$,
$\forall\mu,\nu\in\left\{ x,y,z\right\} $, 
\begin{gather}
A_{v}^{g}A_{v}^{h}=A_{v}^{g+h},\;\left(A^{g}\right)^{\dagger}=A^{-g},\;\left(B_{c}^{\mu}\right)^{\dagger}=B_{c}^{\mu},\\
\left[A_{v_{0}}^{g},A_{v_{1}}^{h}\right]=\left[A_{v}^{g},B_{c}^{\mu}\right]=\left[B_{c_{0}}^{\mu},B_{c_{1}}^{\nu}\right]=0,
\end{gather}
Thus, we have mutually commuting Hermitian operators 
\begin{equation}
A_{v}\coloneqq\frac{1}{\left|G\right|}\sum_{g\in G}A_{v}^{g},
\end{equation}
associated with vertices, which also commute with flux projectors.
Finally, we arrive at the Hamiltonian of the X-cube model, which is
\begin{equation}
H=-\sum_{v\in \Lambda^{0}}A_{v}-\sum_{c\in \Lambda^{3}}B_{c},
\end{equation}
with ground states specified by $A_{v}=B_{c}^{x}=B_{c}^{y}=B_{c}^{z}=1$.
As we will compute, the ground state degeneracy of the untwisted
X-cube model on a lattice $\Lambda$ with underlying manifold $\mathtt{T}^{3}$ and vertices $\Lambda^{0}=\mathbb{Z}_{L_{x}}\times\mathbb{Z}_{L_{y}}\times\mathbb{Z}_{L_{z}}$
is 
\begin{equation}
GSD\left(\Lambda\right)=\left|G\right|^{2\left(L_{x}+L_{y}+L_{z}\right)-3}.\label{eq:GSD_Xcube}
\end{equation}
As $\log GSD(\Lambda)$ is negligible compared to $L_{x}L_{y}L_{z}$ in the the thermodynamic limit, the model is gapped.

\subsubsection{X-cube models twisted by 3-cocycles}

The X-cube model based on an Abelian group $G$ can be twisted by
3-cocycles slice by slice. Let $\Sigma_{i}^{x}$, $\Sigma_{j}^{y}$
and $\Sigma_{k}^{z}$ denote the intersection of $\Lambda$ with the
plane 
\begin{align}
x & =i-\frac{1}{2},\;\forall i\in\mathbb{Z}_{L_{x}},\\
y & =j-\frac{1}{2},\;\forall j\in\mathbb{Z}_{L_{y}},\\
z & =k-\frac{1}{2},\;\forall k\in\mathbb{Z}_{L_{z}},
\end{align}
respectively. For example, the cross section $\Sigma_{1}^{z}$ is
shown in Fig.~\ref{fig:XC-1}. For each $\ell\in\Lambda^{1}$, let
$\Sigma\left(\ell\right)$ be one of these cross sections which is perpendicular
to $\ell$ and contains the center of $\ell$. 

Each cross section is a square lattice, whose vertices (resp. edges,
plaquettes) correspond to the edges (resp. faces, cubes) of $\Lambda$
intersected by the plane. Let $E\left(\Sigma_{i}^{x}\right)$, $E\left(\Sigma_{j}^{y}\right)$
and $E\left(\Sigma_{k}^{z}\right)$ be the set of edges with orientation
chosen as in Fig.~\ref{fig:twisted_QD}. By restriction, each $\vartheta\in G^{\Lambda^{2}}$
gives $\vartheta_{i}^{x}\in G^{E\left(\Sigma_{i}^{x}\right)}$, $\vartheta_{j}^{y}\in G^{E\left(\Sigma_{j}^{y}\right)}$
and $\vartheta_{k}^{z}\in G^{E\left(\Sigma_{k}^{z}\right)}$. We notice
that $B_{c}^{x}$, $B_{c}^{y}$ and $B_{c}^{z}$ are then the flux
projectors for these square lattices. A complete triangulation of
each cross section is made as in Fig.~\ref{fig:twisted_QD}. 

Let $\omega$ be an assignment that assigns 3-cocycles $\omega_{i}^{x}$,
$\omega_{j}^{y}$, $\omega_{k}^{z}\in Z^{3}\left(G,U\left(1\right)\right)$
to the slices $\Sigma_{i}^{x}$, $\Sigma_{j}^{y}$, $\Sigma_{k}^{z}$
respectively.  For any region $\Gamma$ of $\Lambda$, let 
\begin{equation}
G_{B}^{\Lambda^{2}\left(\Gamma\right)}\coloneqq\left\{ \vartheta\in G^{\Lambda^{2}\left(\Gamma\right)}\;|\;B_{c}\left|\vartheta\right\rangle =\left|\vartheta\right\rangle ,\forall c\in\Lambda^{3}\left(\Gamma\right)\right\} ,
\end{equation}
whose elements are called \emph{locally flat} spin
configurations on $\Gamma$.
For each vertex $v$ and $g\in G$, we define an operator 
\begin{equation}
P_{v}^{g}\coloneqq\sum_{\vartheta\in G_{B}^{\Lambda^{2}\left(\Lambda\left[v\right]\right)}}\left|\vartheta\right\rangle \omega\left[\Lambda,v;\vartheta,g\right]\left\langle \vartheta-\kappa_{v}g\right|\label{eq:Xcube_P}
\end{equation}
supported on $\Lambda\left[v\right]$ (\emph{i.e.}, the region made of cubes adjacent to $v$ inside $\Lambda$), with 
\begin{equation}
\omega\left[\Lambda,v;\vartheta,g\right]\coloneqq\prod_{\mu=x,y,z}\frac{\omega\left[\Sigma_{v-}^{\mu},\overline{t_{\mu}}v;\vartheta,g\right]}{\omega\left[\Sigma_{v+}^{x},t_{\mu}v;\vartheta-\kappa_{v}g,g\right]},
\end{equation}
where $t_{\mu}v=v+(\frac{1}{2},0,0), v+(0,\frac{1}{2},0), v+(0,0,\frac{1}{2})$ and $\overline{t_{\mu}}v=v-(\frac{1}{2},0,0), v-(0,\frac{1}{2},0), v-(0,0,\frac{1}{2})$ for $\mu=x,y,z$ respectively. In addition,
$\Sigma_{v+}^{\mu}$ (resp. $\Sigma_{v-}^{\mu}$) is the cross section perpendicular to the $\mu$-direction and containing $t_{\mu}v$ (resp. $\overline{t_{\mu}}v$) as a vertex. 
Take $v=(1,1,1)$ in Fig.~\ref{fig:XC-1} for instance, $\Sigma_{v-}^z=\Sigma_{1}^z$ (cyan online).
Moreover, $\vartheta\in G_{B}^{\Lambda^{2}\left(\Lambda\left[v\right]\right)}$  (resp. $\vartheta-\kappa_{v}g\in G_{B}^{\Lambda^{2}\left(\Lambda\left[v\right]\right)}$) makes a local coloring of $\Sigma_{v-}^{\mu}$ (resp. $\Sigma_{v+}^{\mu}$) near $\overline{t_{\mu}}v$  (resp. $t_{\mu}v$). Then $\omega\left[\Sigma_{v-}^{\mu},t_{x}v;\vartheta,g\right]$ and  $\omega\left[\Sigma_{v+}^{\mu},\overline{t_{\mu}}v;\vartheta-\kappa_{v}g,g\right]$
denote the phase factors specified by Eq.~\eqref{eq:w-1} as in the quantum double model on
$\Sigma_{v\pm}^{\mu}$ twisted by the 3-cocycle assigned to
$\Sigma_{v\pm}^{\mu}$ by $\omega$.

It is straightforward to show that 
\begin{align}
\left(P_{v}^{g}\right)^{\dagger} & =P_{v}^{-g},\\
P_{v}^{g}P_{v}^{h} & =P_{v}^{g+h},\\
\left[P_{v_{0}}^{g},P_{v_{1}}^{h}\right] & =0,
\end{align}
$\forall v,v_{0},v_{1}\in\Lambda^{0}$,
$\forall g,h\in G$.
Thus, we have mutually commuting Hermitian local
projectors 
\begin{equation}
P_{v}\coloneqq\frac{1}{\left|G\right|}\sum_{g\in G}P_{v}^{g},
\end{equation}
labeled by vertives. 
If all 3-cocycles are trivial, then $P_v$ reduces to $A_{v}\prod_{c \ni v}B_{c}$, where $c \ni v$ means that $c$ is adjacent to $v$.

The Hamiltonian of the twisted X-cube model on $\Lambda$ is 
\begin{equation}
H=-\sum_{v\in \Lambda^{0}}P_{v},
\end{equation}
with ground states specified by $P_{v}=1$ for all vertices.

\subsection{Twisted checkerboard models}

\begin{figure}
\noindent\begin{minipage}[t]{1\columnwidth}%
\includegraphics[width=0.8\columnwidth]{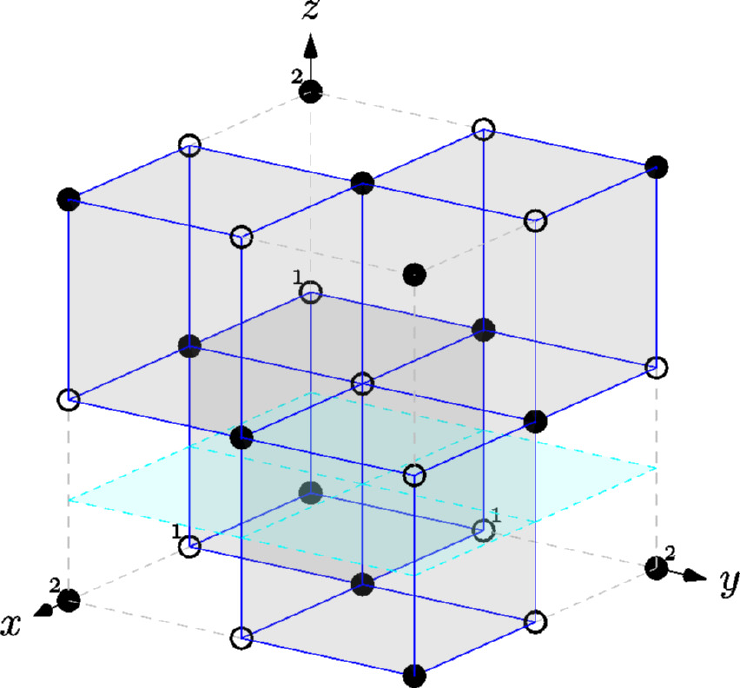}

(a) Three-dimensional checkerboard $\Lambda$ %
\end{minipage}\bigskip{}

\noindent\begin{minipage}[t]{1\columnwidth}%
\includegraphics[width=0.7\columnwidth]{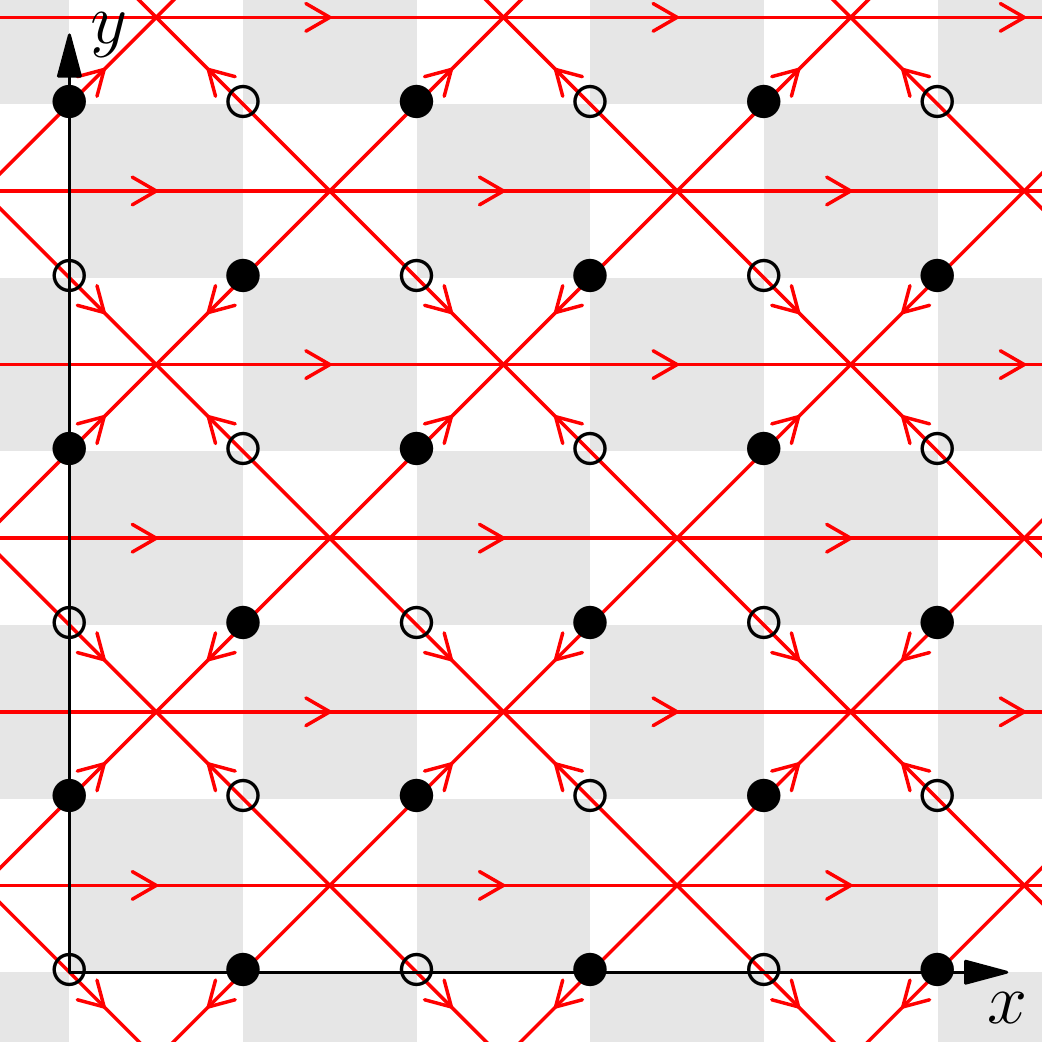}

(b) Triangulation of $\Sigma_{k}^{z}$ with $k$ odd.%
\end{minipage}\bigskip{}

\noindent\begin{minipage}[t]{1\columnwidth}%
\includegraphics[width=0.7\columnwidth]{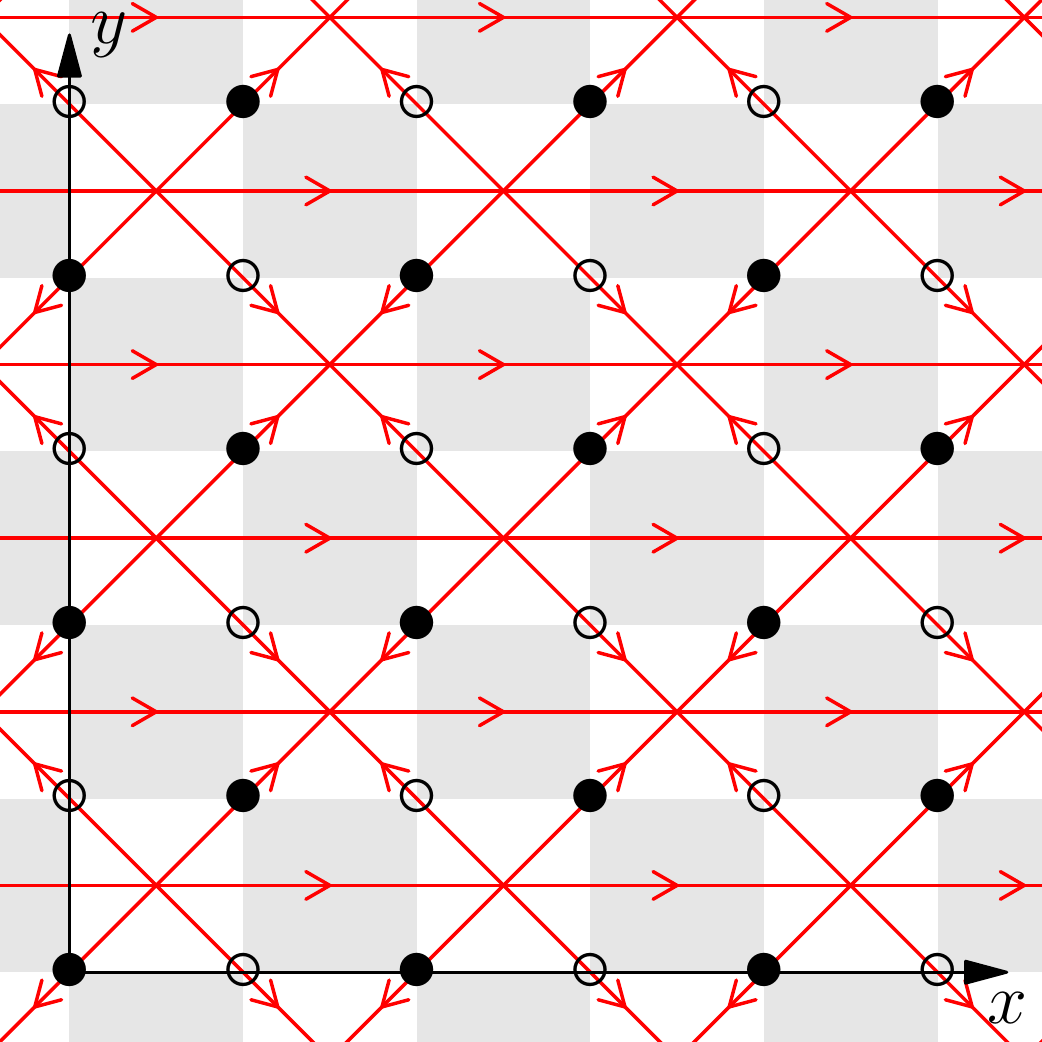}

(c) Triangulation of $\Sigma_{k}^{z}$ with $k$ even.%
\end{minipage}

\caption{(a) The three-dimensional checkerboard $\Lambda$ with vertices marked
as either $\bullet$ or $\circ$. The cross section $\Sigma_{k}^{z}$
is the intersection of $\Lambda$ with the plane $z=k-\frac{1}{2}$,
such as $\Sigma_{1}^{z}$ (cyan online). (b, c) A branched triangulation
(red online) of $\Sigma_{k}^{z}$ with the positions of vertices of
$\Lambda$ on the plane $z=k$ included.
Triangulations of $\Sigma_{i}^{x}$ and $\Sigma_{j}^{y}$ are obtained by permuting $x,y,z$ cyclically.
}

\label{fig:checker3d}
\end{figure}

A three-dimensional checkerboard $\Lambda$, as shown in Fig.~\ref{fig:checker3d}(a),
is obtained by coloring half of the cubes grey in a cubic lattice. Let
$\Lambda_{\bullet}^{3}$ (resp. $\Lambda_{\circ}^{3}$) be the set
of grey (resp. blank) cubes. Let $\Lambda^{0}$ be the set of
vertices. We also divide $\Lambda^{0}$ into two groups $\Lambda_{\bullet}^{0}$
and $\Lambda_{\circ}^{0}$, marked $\bullet$ and $\circ$ respectively
in Fig.~\ref{fig:checker3d}(a). In the chosen coordinates, 
\begin{align}
\Lambda_{\bullet}^{0} & =\left\{ \left(i,j,k\right)\in\Lambda^{0}\;|\;i+j+k\text{ is even}\right\} ,\\
\Lambda_{\circ}^{0} & =\left\{ \left(i,j,k\right)\in\Lambda^{0}\;|\;i+j+k\text{ is odd}\right\} .
\end{align}
Notice that all the grey (resp. uncolored) cubes are centered at $\Lambda_{\bullet}^{0}+\frac{1}{2}\left(1,1,1\right)$
(resp. $\Lambda_{\circ}^{0}+\frac{1}{2}\left(1,1,1\right)$). 

For an infinite system, $\Lambda^{0}=\mathbb{Z}^{3}$. In the following discussion, however, we prefer to identify
$\left(x,y,z\right)\sim\left(x+L_{x},y,z\right)\sim\left(x,y+L_{y},z\right)\sim\left(x,y,z+L_{z}\right)$ and consider the resulting checkerboard on $\mathtt{T}^{3}$,
where $L_{x}$, $L_{y}$ and $L_{y}$ need to be even integers in order to be compatible
with the checker pattern. Such a lattice has vertices 
$\Lambda^{0}=\mathbb{Z}_{L_{x}}\times\mathbb{Z}_{L_{y}}\times\mathbb{Z}_{L_{z}}$ and the infinite case can be viewed as its thermodynamic limit.

Since the checkerboard is just the cubic lattice with a checker pattern,
we can use notations introduced for the cubic lattice with or without decoration. For instance, $\Lambda^{3}\left(\Gamma\right)$  stands for the set of cubes inside region $\Gamma$ and its subset of grey cubes is denoted by
$\Lambda_{\bullet}^{3}(\Gamma)\coloneqq\Lambda_{\bullet}^{3}\cap \Lambda^{3}(\Gamma)$. 
In addition, $\Sigma_{i}^{x}$
still denotes the intersection of $\Lambda$ with the plane $x=i-\frac{1}{2}$,
but now it is not only a square lattice but also a two-dimensional checkerboard. 

\subsubsection{Checkerboard model based on a finite Abelian group}

Let $G$ be a finite Abelian group $G$, with $0$ denoting its identity
element. A local Hilbert space (also called a \emph{spin} for short)
spanned by an orthonormal basis $\left\{ \left|v,g\right\rangle \right\} _{g\in G}$ is
assigned to each vertex $v\in\Lambda^{0}$. Thus, the Hilbert space
associated to any region $\Gamma$ of $\Lambda$, denoted $\mathcal{H}\left(\Lambda^{0}\left(\Gamma\right),G\right)$,
is spanned by
\begin{equation}
\left|\vartheta\right\rangle \coloneqq\bigotimes_{v\in\Lambda^{0}\left(\Gamma\right)}\left|v,\vartheta\left(v\right)\right\rangle 
\end{equation}
with $\vartheta\in G^{\Lambda^{0}\left(\Gamma\right)}$, where $G^{\Lambda^{0}\left(\Gamma\right)}\coloneqq\text{Fun}\left(\Lambda^{0}\left(\Gamma\right),G\right)$
is the set of functions from $\Lambda^{0}\left(\Gamma\right)$ (\emph{i.e.},
vertices in $\Gamma$) to $G$. Each element of 
$G^{\Lambda^{0}\left(\Gamma\right)}$ specifies a \emph{spin configuration} on $\Gamma$.
On the whole lattice, the total Hilbert space
is $\mathcal{H}\left(\Lambda^{0},G\right)$.

For each grey cube $c$, let $\boldsymbol{1}_{c}:\Lambda^{0}\rightarrow\left\{ 0,1\right\} $
be the indicator function of $\Lambda^{0}\left(c\right)$, which has the
value 1 on each vertex of $c$ and 0 on any vertex not in c. For $g\in G$,
we define a \emph{(generalized) gauge transformation operator}
\begin{equation}
A_{c}^{g}\coloneqq\sum_{\vartheta\in G^{\Lambda^{0}}}\left|\vartheta+\boldsymbol{1}_{c}g\right\rangle \left\langle \vartheta\right|
\end{equation}
to flip spins on all the vertices of $c$. Clearly, it is supported
on $c$. In addition, let $\left(-1\right)^{v}:\Lambda^{0}\rightarrow\left\{ 1,-1\right\} $
be the function which has the value $1$ on each $v\in\Lambda_{\circ}^{0}$
and $-1$ on each $v\in\Lambda_{\bullet}^{0}$. For each grey cube
$c\in \Lambda_{\bullet}^{3}$, we have a \emph{(generalized) flux projector}
\begin{equation}
B_{c}\coloneqq\sum_{\vartheta\in G^{\Lambda^{0}}}\delta_{\sum_{v\in\Lambda^{0}\left(c\right)}\left(-1\right)^{v}\vartheta\left(v\right),0}\left|\vartheta\right\rangle \left\langle \vartheta\right|
\end{equation}
supported on $c$.

It is straightforward to check that 
\begin{align}
\left(A_{c}^{g}\right)^{\dagger}  =A_{c}^{-g},\;A_{c}^{g}A_{c}^{h}=A_{c}^{g+h},
\;\left(B_{c}\right)^{\dagger}=B_{c},\\
\left[A_{c_{0}}^{g},A_{c_{1}}^{h}\right]  =\left[A_{c_{0}}^{g},B_{c_{1}}\right]=\left[B_{c_{0}},B_{c_{1}}\right]=0,
\end{align}
$\forall c,c_{0},c_{1}\in\Lambda_{\bullet}^{3}$, $\forall g,h\in G$.
Thus, we have mutually commuting Hermitian operators 
\begin{equation}
A_{c}\coloneqq\frac{1}{\left|G\right|}\sum_{g\in G}A_{c}^{g},
\end{equation}
associated with grey cubes, which also commute with flux projectors
on grey cubes. The Hamiltonian of the checkerboard model
is then given by
\begin{equation}
H=-\sum_{c\in\Lambda_{\bullet}^{3}}\left(A_{c}+B_{c}\right),
\end{equation}
whose ground states are specified by $A_{c}=B_{c}=1$. As we will
compute later, the ground state degeneracy of this model
on a checkerboard $\Lambda$ with underlying manifold $\mathtt{T}^{3}$ and vertices $\Lambda^{0}=\mathbb{Z}_{L_{x}}\times\mathbb{Z}_{L_{y}}\times\mathbb{Z}_{L_{z}}$ is 
\begin{equation}
GSD\left(\Lambda\right)=\left|G\right|^{2\left(L_{x}+L_{y}+L_{z}\right)-6},\label{eq:GSD_checkerboard}
\end{equation}
As $\log GSD(\Lambda)$ is negligible compared to $L_{x}L_{y}L_{z}$ in the the thermodynamic limit, the model is gapped. 

For $G=\mathbb{Z}_{2}$, the model reduces to the original checkerboard
model defined by Vijay, Haah, and Fu~\cite{vijay16xcube}.

\subsubsection{Checkerboard models twisted by 3-cocycles}

To relate the checkerboard with a lattice model of gauge theory, let
us look at one cross section $\Sigma_{k}^{z}$ first and triangulate
it as in Fig.~\ref{fig:checker3d}(b) or (c). Let $\Delta^{n}\left(\Sigma_{k}^{z}\right)$
be the set of $n$-simplices in this triangulation. In addition, we
denote the set of edges with $\bullet$ or $\circ$ mark by $E\left(\Sigma_{k}^{z}\right)$,
which gives a new square lattice structure of the two-dimensional
checkerboard. Let $\Lambda_{\bullet}^{3}\left[\Sigma_{k}^{z}\right]$
(resp. $\Lambda_{\circ}^{3}\left[\Sigma_{k}^{z}\right]$) be the set
of grey (resp. blank) cubes intersecting with $\Sigma_{k}^{z}$,
which obviously labels the plaquettes (resp. vertices) of the new
square lattice structure of $\Sigma_{k}^{z}$. Similarly, let $\Lambda^{1}\left[\Sigma_{k}^{z}\right]$
be the set of edges in $\Lambda$ that intersect with $\Sigma_{k}^{z}$;
each $\ell\in\Lambda^{1}\left[\Sigma_{k}^{z}\right]$ is assumed oriented
toward the positive direction of $z$. Then there is a one-to-one
correspondence between $E\left(\Sigma_{k}^{z}\right)$ and $\Lambda^{1}\left[\Sigma_{k}^{z}\right]$;
hence we will use them interchangeably and simply write 
$E\left(\Sigma_{k}^{z}\right)=\Lambda^{1}\left[\Sigma_{k}^{z}\right]$.

Given $\vartheta\in G^{\Lambda^{0}}$, if we color $\ell=\left[v_{0}v_{1}\right]\in\Lambda^{1}\left[\Sigma_{k}^{z}\right]$
by 
\begin{equation}
\partial\vartheta\left(\ell\right)\coloneqq\vartheta\left(v_{1}\right)-\vartheta\left(v_{0}\right).\label{eq:d_vartheta}
\end{equation}
then, for $\Sigma_{k}^{z}$, we notice that $B_{c}$ at $c\in\Lambda_{\bullet}^{3}\left[\Sigma_{k}^{z}\right]$
works as a flux operator and that $A_{t_{z}c}^{g}$ (resp. $A_{\overline{t_{z}}c}^{g}$) with
$c\in\Lambda_{\circ}^{3}\left[\Sigma_{k}^{z}\right]$ works as a gauge
transformation operator, where $t_{z}c=c+\boldsymbol{z}$ (resp. $\overline{t_{z}}c=c-\boldsymbol{z}$)
is the grey cube above (resp. below) $c\in\Lambda_{\circ}^{3}\left[\Sigma_{k}^{z}\right]$.
Based on this observation, we construct a twisted version of $A_{c}^{g}$,
denoted $P_{c}^{g}$, below.

To prepare for the definition,  for any region $\Gamma$ of $\Lambda$, let
\begin{equation}
G_{B}^{\Lambda^{0}\left(\Gamma\right)}\coloneqq\left\{ \vartheta\in G^{\Lambda^{0}\left(\Gamma\right)}\;|\;B_{c}\left|\vartheta\right\rangle =\left|\vartheta\right\rangle ,\forall c\in\Lambda_{\bullet}^{3}\left(\Gamma\right)\right\} ,
\end{equation}
whose elements are called \emph{locally flat} spin
configurations on $\Gamma$. Moreover, for
each cube $c\in\Lambda^{3}$, let $\Lambda\left[c\right]$ be
the region made of the cubes whose intersection with $c$ is not
empty. In addition, translating cube $c$ by a unit in the positive (resp. negative) $\mu$-direction gives a cube denoted $t_{\mu}c$ (resp. $\overline{t_{\mu}}c$) of color different from $c$, for $\mu=x,y,z$. 

For grey cube $c$ centered at $(c^x,c^y,c^z)+\frac{1}{2}(1,1,1)$, let
\begin{equation}
P_{c}^{g}\coloneqq\sum_{\vartheta\in G_{B}^{\Lambda^{0}\left(\Lambda\left[c\right]\right)}}\left|\vartheta\right\rangle \omega\left[\Lambda,c;\vartheta,g\right]\left\langle \vartheta-\boldsymbol{1}_{c}g\right|,
\end{equation}
$\forall g\in G$, 
supported on $\Lambda\left[c\right]$, with 
\begin{multline}
\omega\left[\Lambda,c;\vartheta,g\right]\coloneqq\frac{\omega\left[\Sigma_{c-}^{x},\overline{t_{x}}c;\partial\vartheta,\left(-1\right)^{c^{z}}g\right]}{\omega\left[\Sigma_{c+}^{x},t_{x}c;\partial\left(\vartheta-\boldsymbol{1}_{c}g\right),\left(-1\right)^{c^{z}}g\right]}\cdot\\
\frac{\omega\left[\Sigma_{c-}^{y},\overline{t_{y}}c;\partial\vartheta,\left(-1\right)^{c^{x}}g\right]}{\omega\left[\Sigma_{c+}^{y},t_{y}c;\partial\left(\vartheta-\boldsymbol{1}_{c}g\right),\left(-1\right)^{c^{x}}g\right]}\cdot\\
\frac{\omega\left[\Sigma_{c-}^{z},\overline{t_{z}}c;\partial\vartheta,\left(-1\right)^{c^{y}}g\right]}{\omega\left[\Sigma_{c+}^{z},t_{z}c;\partial\left(\vartheta-\boldsymbol{1}_{c}g\right),\left(-1\right)^{c^{y}}g\right]},
\end{multline}
where $\Sigma_{c+}^{\mu}$ (resp. $\Sigma_{c-}^{\mu}$) is the cross section of $\Lambda$ passing the center
of $t_{\mu}c$ (resp. $\overline{t_{\mu}}c$) and perpendicular to the $\mu$-direction. 
The phase factors
$\omega\left[\Sigma_{c-}^{\mu},\overline{t_{\mu}}c;\partial\vartheta,(-1)^{c^\nu}g\right]$ and $\omega\left[\Sigma_{c+}^{\mu},t_{\mu}c;\partial\left(\vartheta-\boldsymbol{1}_{c}g\right),(-1)^{c^\nu}g\right]$ are defined by Eq.~\eqref{eq:w-1}; they are the phase factors appearing in the  twisted gauge transformation $(-1)^{c^{\nu}}g$ on $\Sigma_{c\pm}^{\mu}$ with triangulation in Fig.~\ref{fig:checker3d}(b) or (c).

To unpack this definition, let us look at a special case 
that $\omega$ is only nontrivial on the cross section $\Sigma_{1}^{z}$ shown
in Fig.~\ref{fig:checker3d}(a). Let us consider $P_{c}^{g}$ on
the cube centered at $c=\left(\frac{3}{2},\frac{1}{2},\frac{3}{2}\right)$
and compute $\omega\left[\Lambda,c;\vartheta,g\right]$. 
If the system size in the $z$-direction $L_{z}$ is larger than 2, then  $\omega\left[\Lambda,c;\vartheta,g\right]=
\omega\left[\Sigma_{c-}^{z},\overline{t_{z}}c;\partial\vartheta,g\right]$  due to the twisting of $\Sigma_{c-}^{z}=\Sigma_{1}^{x}$.
If $L_{z}=2$, then 
$c$ and $\Sigma_{1}^{z}$ share two faces. 
Thus, $t_{z}c=\overline{t_{z}}c$, 
$\Sigma_{c+}^{z}=\Sigma_{c-}^{z}=\Sigma_{1}^{x}$,  
$\omega\left[\Sigma_{c+}^{\mu},t_{\mu}c;\partial\left(\vartheta-\boldsymbol{1}_{c}g\right),g\right] =
\omega\left[\Sigma_{c-}^{\mu},\overline{t_{\mu}}c;\partial\vartheta,g\right] $ and hence $\omega\left[\Lambda,c;\vartheta,g\right]=1$.

The operators satisfy
\begin{align}
\left(P_{c}^{g}\right)^{\dagger} & =P_{c}^{-g},\\
P_{c}^{g}P_{c}^{h} & =P_{c}^{g+h},\\
\left[P_{c_{0}}^{g},P_{c_{1}}^{h}\right] & =0,
\end{align}
$\forall g,h\in G$, $\forall c,c_{0},c_{1}\in \Lambda_{\bullet}^{3}$. Therefore, we have mutually commuting Hermitian local projectors
\begin{equation}
P_{c}\coloneqq\frac{1}{\left|G\right|}\sum_{g\in G}P_{c}^{g},
\end{equation}
The Hamiltonian of the twisted checkerboard model on $\Lambda$ is then defined as
\begin{equation}
H=-\sum_{c\in \Lambda_{\bullet}^{3}} P_{c},
\end{equation}
with ground states specified by $P_{c}=1$ for all grey cubes. 

\section{Ground state degeneracy on $\mathtt{T}^{3}$: twisted X-cube models}
\label{sec:GSDXcube}

In this section, we consider a cubic lattice $\Lambda$ embedded on
a three-torus (\emph{i.e.}, three-dimensional torus) $\mathtt{T}^{3}$, whose vertex set is
$\Lambda^{0}=\mathbb{Z}_{L_{x}}\times\mathbb{Z}_{L_{y}}\times\mathbb{Z}_{L_{z}}$
with $L_{x},L_{y},L_{z}\in \mathbb{Z}$. A general method
is developed here to compute the ground state degeneracy, denoted
$GSD\left(\Lambda\right)$, of a twisted X-cube model on $\Lambda$.
In other words, we are going to determine $GSD\left(\Lambda\right)$
of a twisted X-cube model of system size $L_{x}\times L_{y}\times L_{z}$
with the periodic boundary condition identifying $\left(x+L_{x},y,z\right)\sim\left(x,y+L_{y},z\right)\sim\left(x,y,z+L_{z}\right)\sim\left(x,y,z\right)$. 

In particular, explicit computations will be given for examples based
on groups $\mathbb{Z}_{2}$ and $\mathbb{Z}_{2}^{3}\equiv\mathbb{Z}_{2}\times\mathbb{Z}_{2}\times\mathbb{Z}_{2}$.

\subsection{Generic setting}

The ground state Hilbert subspace is the image of the projector
\begin{equation}
P\left(\Lambda\right)\coloneqq\prod_{v\in\Lambda^{0}}P_{v}.
\end{equation}
Hence the ground state degeneracy $GSD\left(\Lambda\right)$ equals
the trace of $P\left(\Lambda\right)$. Explicitly,
\begin{equation}
\text{tr}P\left(\Lambda\right)=\frac{1}{\left|G^{\Lambda^{0}}\right|}\sum_{\vartheta\in G_{B}^{\Lambda^{2}}}\sum_{\eta\in G^{\Lambda^{0}}}\left\langle \vartheta\right|\prod_{v}P_{v}^{\eta\left(v\right)}\left|\vartheta\right\rangle ,\label{eq:trP_xcube}
\end{equation}
where $\left\langle \vartheta\right|\prod_{v}P_{v}^{\eta\left(v\right)}\left|\vartheta\right\rangle $
is nonzero if and only if $\vartheta\in G_{B}^{\Lambda^{2}}$ and
$\prod_{v}A_{v}^{\eta\left(v\right)}\left|\vartheta\right\rangle =\left|\vartheta\right\rangle .$
Let 
\begin{equation}
G_{A}^{\Lambda^{0}}\coloneqq\left\{ \eta\in G^{\Lambda^{0}}\;|\;\prod_{v\in\Lambda^{0}}A_{v}^{\eta\left(v\right)}\left|\vartheta\right\rangle =\left|\vartheta\right\rangle \right\}  .
\end{equation}
We notice that $G_{A}^{\Lambda^{0}}$ is independent of $\vartheta$
and that each $\eta\in G_{A}^{\Lambda^{0}}$ can be specified by its
values on three axes, or equivalently
\begin{align}
\eta_{o} & \coloneqq\eta\left(0,0,0\right),\\
\partial\eta_{i}^{x} & \coloneqq\eta\left(i,0,0\right)-\eta\left(i-1,0,0\right),\forall i\in\mathbb{Z}_{L_{x}},\\
\partial\eta_{j}^{y} & \coloneqq\eta\left(0,j,0\right)-\eta\left(0,j-1,0\right),\forall j\in\mathbb{Z}_{L_{y}},\\
\partial\eta_{k}^{z} & \coloneqq\eta\left(0,0,k\right)-\eta\left(0,0,k-1\right),\forall k\in\mathbb{Z}_{L_{z}},
\end{align}
subject to 
\begin{equation}
\sum_{n}\partial\eta_{n}^{\lambda}=0,\forall\lambda=x,y,z.
\end{equation}
Therefore, 
\begin{equation}
\left|G_{A}^{\Lambda^{0}}\right|=\left|G\right|^{L_{x}+L_{y}+L_{z}-2}.
\end{equation}

Any $\vartheta\in G_{B}^{\Lambda^{2}}$ assigns group elements to
the two non-contractible loops of $\Sigma_{k}^{z}$ in the $x$ and
$y$ directions respectively as
\begin{align}
\vartheta_{k}^{z}\left\langle x\right\rangle  & \coloneqq\sum_{i\in\mathbb{Z}_{L_{x}}}\vartheta\left(i-\frac{1}{2},0,k-\frac{1}{2}\right),\\
\vartheta_{k}^{z}\left\langle y\right\rangle  & \coloneqq\sum_{j\in\mathbb{Z}_{L_{y}}}\vartheta\left(0,j-\frac{1}{2},k-\frac{1}{2}\right),
\end{align}
$\forall k\in\mathbb{Z}_{L_{z}}$. Similarly, for $\Sigma_{i}^{x}$
and $\Sigma_{j}^{y}$, we have 
\begin{align}
\vartheta_{i}^{x}\left\langle y\right\rangle  & \coloneqq\sum_{j\in\mathbb{Z}_{L_{y}}}\vartheta\left(i-\frac{1}{2},j-\frac{1}{2},0\right),\\
\vartheta_{i}^{x}\left\langle z\right\rangle  & \coloneqq\sum_{k\in\mathbb{Z}_{L_{z}}}\vartheta\left(i-\frac{1}{2},0,k-\frac{1}{2}\right),\\
\vartheta_{j}^{y}\left\langle z\right\rangle  & \coloneqq\sum_{k\in\mathbb{Z}_{L_{z}}}\vartheta\left(0,j-\frac{1}{2},k-\frac{1}{2}\right),\\
\vartheta_{j}^{y}\left\langle x\right\rangle  & \coloneqq\sum_{i\in\mathbb{Z}_{L_{x}}}\vartheta\left(i-\frac{1}{2},j-\frac{1}{2},0\right),
\end{align}
$\forall i\in\mathbb{Z}_{L_{x}}, \forall j\in\mathbb{Z}_{L_{y}}$.
Clearly, they are subject to
\begin{align}
\vartheta^{xy} & = \sum_{i\in\mathbb{Z}_{L_{x}}}\vartheta_{i}^{x}\left\langle y\right\rangle =\sum_{j\in\mathbb{Z}_{L_{y}}}\vartheta_{j}^{y}\left\langle x\right\rangle ,\label{eq:theta_xy}\\
\vartheta^{yz} & = \sum_{j\in\mathbb{Z}_{L_{y}}}\vartheta_{j}^{y}\left\langle z\right\rangle =\sum_{k\in\mathbb{Z}_{L_{z}}}\vartheta_{k}^{z}\left\langle y\right\rangle ,\label{eq:theta_yz}\\
\vartheta^{zx} & = \sum_{k\in\mathbb{Z}_{L_{z}}}\vartheta_{k}^{z}\left\langle x\right\rangle =\sum_{i\in\mathbb{Z}_{L_{x}}}\vartheta_{i}^{x}\left\langle z\right\rangle ,\label{eq:theta_zx}
\end{align}
where $\vartheta^{xy}$ (resp. $\vartheta^{yz}$, $\vartheta^{zx}$) denotes the sum of $\vartheta(p)$ over faces lying in the plane $z=0$ (resp. $x=0$, $y=0$).

Thus, there are $\left|G\right|^{2\left(L_{x}+L_{y}+L_{z}\right)-3}$
choices of $\left\{ \vartheta_{n}^{\lambda}\left\langle \mu\right\rangle \right\} $
(\emph{i.e.}, the group elements assigned to non-contractible loops
of $\Sigma_{n}^{\lambda}$ for all possible $\lambda$, $n$). With $\left\{ \vartheta_{n}^{\lambda}\left\langle \mu\right\rangle \right\} $
fixed, we can pick: (1) $\vartheta\left(i-\frac{1}{2},j-\frac{1}{2},k\right)$
for $1\leq i<L_{x}$, $1\leq j<L_{y}$, $\forall k \in \mathbb{Z}_{L_{z}}$; (2) $\vartheta\left(i-\frac{1}{2},0,k-\frac{1}{2}\right)$
for $1\leq i<L_{x}$, $1\leq k<L_{z}$; (3) $\vartheta\left(0,j-\frac{1}{2},k-\frac{1}{2}\right)$
for $1\leq j<L_{y}$, $1\leq k<L_{z}$. In total, there are 
\begin{multline}
\left|G\right|^{\left(L_{x}-1\right)\left(L_{y}-1\right)L_{z}+\left(L_{x}-1\right)\left(L_{z}-1\right)+\left(L_{y}-1\right)\left(L_{z}-1\right)}\\
=\left|G\right|^{L_{x}L_{y}L_{z}-L_{x}-L_{y}-L_{z}+2}
\end{multline}
different choices of $\vartheta\in G_{B}^{\Lambda^{2}}$ corresponding to the
same $\left\{ \vartheta_{n}^{\lambda}\left\langle \mu\right\rangle \right\} $.
Therefore,
\begin{align}
\left|G_{B}^{\Lambda^{2}}\right| & =\left|G\right|^{2\left(L_{x}+L_{y}+L_{z}\right)-3}\times\left|G\right|^{L_{x}L_{y}L_{z}-L_{x}-L_{y}-L_{z}+2}\nonumber \\
 & =\left|G\right|^{L_{x}L_{y}L_{z}+L_{x}+L_{y}+L_{z}-1}.
\end{align}

\subsubsection{Untwisted X-cube models}

If the model is untwisted (\emph{i.e.}, $\omega\equiv1$), then Eq.~\eqref{eq:trP_xcube}
reduces to
\begin{align}
GSD\left(\Lambda\right) & =\text{tr}P\left(\Lambda\right)=\frac{\left|G_{A}^{\Lambda^{0}}\right|\left|G_{B}^{\Lambda^{2}}\right|}{\left|G^{\Lambda^{0}}\right|}\nonumber \\
 & =\left|G\right|^{2\left(L_{x}+L_{y}+L_{z}\right)-3}.\label{eq:GSD_G}
\end{align}
This ground state degeneracy was already mentioned in Eq.~\eqref{eq:GSD_Xcube}
as we introduced the model.

\subsubsection{Twisted X-cube models}

In a twisted X-cube model, each $\eta\in G_{A}^{\Lambda^{0}}$ makes
a gauge transformation labeled by $\partial\eta_{n}^{\lambda}$ uniformly
to each vertex of $\Sigma_{n}^{\lambda}$ for $\lambda=x,y,z$. Therefore,
\begin{align}
 & \left\langle \vartheta\right|\prod_{v}P_{v}^{\eta\left(v\right)}\left|\vartheta\right\rangle \nonumber \\
= & \prod_{i\in\mathbb{Z}_{L_{x}}}\omega_{i}^{x}\left[\mathtt{T^{3}};\vartheta_{i}^{x}\left\langle y\right\rangle ,\vartheta_{i}^{x}\left\langle z\right\rangle ,\partial\eta_{i}^{x}\right]\cdot\nonumber \\
 & \prod_{i\in\mathbb{Z}_{L_{y}}}\omega_{j}^{y}\left[\mathtt{T^{3}};\vartheta_{j}^{y}\left\langle z\right\rangle ,\vartheta_{j}^{y}\left\langle x\right\rangle ,\partial\eta_{j}^{y}\right]\cdot\nonumber \\
 & \prod_{i\in\mathbb{Z}_{L_{z}}}\omega_{k}^{z}\left[\mathtt{T^{3}};\vartheta_{k}^{z}\left\langle x\right\rangle ,\vartheta_{k}^{z}\left\langle y\right\rangle ,\partial\eta_{k}^{z}\right].\label{eq:L}
\end{align}
We notice that $\left\langle \vartheta\right|\prod_{v}P_{v}^{\eta\left(v\right)}\left|\vartheta\right\rangle $
is a one-dimensional representation of $\eta\in G_{A}^{\Lambda^{0}}$.
So $\sum_{\eta\in G_{A}^{\Lambda^{0}}}\left\langle \vartheta\right|\prod_{v}P_{v}^{\eta\left(v\right)}\left|\vartheta\right\rangle =0$
unless the representation is trivial. 

Let $\varTheta$ be the set of all possible choices of $\left\{ \vartheta_{n}^{\mu}\left\langle \nu\right\rangle \right\} $
making $\left\langle \vartheta\right|\prod_{v}P_{v}^{\eta\left(v\right)}\left|\vartheta\right\rangle $
a trivial representation of $G_{A}^{\Lambda^{0}}$. Since there are
$\left|G\right|^{L_{x}L_{y}L_{z}-L_{x}-L_{y}-L_{z}+2}$ choices of
$\vartheta\in G_{A}^{\Lambda^{0}}$ for each selected $\left\{ \vartheta_{n}^{\mu}\left\langle \nu\right\rangle \right\} $,
explicit computation shows that the ground state degeneracy on $\Lambda$
embedded in $\mathtt{T}^{3}$ is
\begin{multline}
GSD\left(\Lambda\right)=\text{tr}P\left(\Lambda\right)=\\
\frac{\left|G_{A}^{\Lambda^{0}}\right|\left|G\right|^{L_{x}L_{y}L_{z}-L_{x}-L_{y}-L_{z}+2}\left|\varTheta\right|}{\left|G^{\Lambda^{0}}\right|}=\left|\varTheta\right|.
\end{multline}
So we can get $GSD\left(\Lambda\right)$ by counting the cardinality
of $\varTheta$. By definition, $\left|\varTheta\right|\leq2\left(L_{x}+L_{y}+L_{z}\right)-3$.
So the ground state degeneracy of a twisted model is always less or
equal to that of its untwisted version. 

Technically, the triviality of $\left\langle \vartheta\right|\prod_{v}P_{v}^{\eta\left(v\right)}\left|\vartheta\right\rangle $
as a representation of $G_{A}^{\Lambda^{0}}$ is equivalent to 
\begin{equation}
\prod_{n\in\mathbb{Z}_{L_{\lambda}}}\omega_{n}^{\lambda}\left[\mathtt{T^{3}};\vartheta_{n}^{\lambda}\left\langle \mu\right\rangle ,\vartheta_{n}^{\lambda}\left\langle \nu\right\rangle ,\eta_{n}-\eta_{n-1}\right]=1,
\end{equation}
$\forall\left(\lambda,\mu,\nu\right)=\left(x,y,z\right),\left(y,z,x\right),\left(z,x,y\right)$,
$\forall\eta\in G^{L_{\lambda}}$. As 
\begin{align}
 & \prod_{n\in\mathbb{Z}_{L_{\lambda}}}\omega_{n}^{\lambda}\left[\mathtt{T^{3}};\vartheta_{n}^{\lambda}\left\langle \mu\right\rangle ,\vartheta_{n}^{\lambda}\left\langle \nu\right\rangle ,\eta_{n}-\eta_{n-1}\right]\nonumber \\
= & \prod_{n\in\mathbb{Z}_{L_{\lambda}}}\frac{\omega_{n}^{\lambda}\left[\mathtt{T^{3}};\vartheta_{n}^{\lambda}\left\langle \mu\right\rangle ,\vartheta_{n}^{\lambda}\left\langle \nu\right\rangle ,\eta_{n}\right]}{\omega_{n}^{\lambda}\left[\mathtt{T^{3}};\vartheta_{n}^{\lambda}\left\langle \mu\right\rangle ,\vartheta_{n}^{\lambda}\left\langle \nu\right\rangle ,\eta_{n-1}\right]}\nonumber \\
= & \prod_{n\in\mathbb{Z}_{L_{\lambda}}}\frac{\omega_{n}^{\lambda}\left[\mathtt{T^{3}};\vartheta_{n}^{\lambda}\left\langle \mu\right\rangle ,\vartheta_{n}^{\lambda}\left\langle \nu\right\rangle ,\eta_{n}\right]}{\omega_{n+1}^{\lambda}\left[\mathtt{T^{3}};\vartheta_{n+1}^{\lambda}\left\langle \mu\right\rangle ,\vartheta_{n+1}^{\lambda}\left\langle \nu\right\rangle ,\eta_{n}\right]},
\end{align}
the condition is further equivalent to that $\exists\gamma^{\lambda}\in\widehat{G}$,
\begin{equation}
\omega_{n}^{\lambda}\left[\mathtt{T^{3}};\vartheta_{n}^{\lambda}\left\langle \mu\right\rangle ,\vartheta_{n}^{\lambda}\left\langle \nu\right\rangle ,-\right]=\gamma^{\lambda},\forall n\in\mathbb{Z}_{L_{\lambda}},\label{eq:omega_tau}
\end{equation}
for $\lambda=x,y,z$ separately, where $\widehat{G}$ is the character
group of $G$ and $\omega_{n}^{\lambda}\left[\mathtt{T^{3}};\vartheta_{n}^{\lambda}\left\langle \mu\right\rangle ,\vartheta_{n}^{\lambda}\left\langle \nu\right\rangle ,-\right]$
is viewed as a one-dimensional representation of $G$ with $-$ denoting
a place holder for a group element. 

To take the constraints given by Eqs.~(\ref{eq:theta_xy}-\ref{eq:theta_zx})
into consideration, let 
\begin{multline}
\varTheta_{g,h,\gamma}^{\lambda}\coloneqq\left\{ \left(a,b\right)\in G^{L_{\lambda}}\times G^{L_{\lambda}}\right|\sum_{n}a_{n}=g,\sum_{n}b_{n}=h,\\
\left.\omega_{n}^{\lambda}\left[\mathtt{T^{3}};a_{n},b_{n},-\right]=\gamma,\forall n\in\mathbb{Z}_{L_{\lambda}}\right\} ,\label{eq:Theta_tau}
\end{multline}
for $g,h\in G$, $\gamma\in\widehat{G}$ and $\lambda=x,y,z$. In addition,
we write
\begin{equation}
\varTheta_{g,h}^{\lambda}\coloneqq\bigcup_{\gamma\in\widehat{G}}\varTheta_{g,h,\gamma}^{\lambda}.
\end{equation}
Then it is clear that 
\begin{equation}
\varTheta=\bigcup_{f,g,h\in G}\varTheta_{f,g}^{x}\times\varTheta_{g,h}^{y}\times\varTheta_{h,f}^{z}.
\end{equation}
Therefore, the cardinalities of these sets satisfy
\begin{gather}
\left|\varTheta_{g,h}^{\lambda}\right|=\sum_{\gamma\in\widehat{G}}\left|\varTheta_{g,h,\gamma}^{\lambda}\right|,\label{eq:varTheta_gh}\\
\left|\varTheta\right|=\sum_{f,g,h\in G}\left|\varTheta_{f,g}^{x}\right|\left|\varTheta_{g,h}^{y}\right|\left|\varTheta_{h,f}^{z}\right|.\label{eq:card_varTheta}
\end{gather}
Below, we will explain how to use Eq.~\eqref{eq:card_varTheta} to
count $\left|\varTheta\right|$ in the example based on $G=\mathbb{Z}_{2}^{3}$
with $\omega\left(f,g,h\right)=e^{i\pi f^{\left(1\right)}g^{\left(2\right)}h^{\left(3\right)}}$.

\subsection{Example: $G=\mathbb{Z}_{2}$ }

As discussed in Sec.~\ref{subsec:DW_Z2}, we always have 
\begin{equation}
\omega\left[\mathtt{T}^{3};f,g,h\right]=\frac{\omega_{h}\left(f,g\right)}{\omega_{h}\left(g,f\right)}=1,
\end{equation}
$\forall f,g,h\in G$. Therefore, the ground state degeneracy remains
unchanged from Eq.~\eqref{eq:GSD_G}, \emph{i.e.},
\begin{equation}
GSD\left(\Lambda\right)=2^{2\left(L_{x}+L_{y}+L_{z}\right)-3},
\end{equation}
no matter how we twist the model.

\subsection{Example: $G=\mathbb{Z}_{2}^{3}$ with $\omega\left(f,g,h\right)=e^{i\pi f^{\left(1\right)}g^{\left(2\right)}h^{\left(3\right)}}$}

As seen in Sec.~\ref{subsec:DW_Zm3}, $\forall f,g,h\in G=\mathbb{Z}_{2}\times\mathbb{Z}_{2}\times\mathbb{Z}_{2}$,
\begin{equation}
\omega\left[\mathtt{T}^{3};f,g,h\right]=e^{i\pi\left(f\times g\right)\cdot h}.
\end{equation}
We identify $\widehat{G}\cong G$; in particular, $\omega\left[\mathtt{T}^{3};f,g,-\right]\in\widehat{G}$ is identified with $f\times g\in G$.

To express $GSD\left(\Lambda\right)=\left|\varTheta\right|$ in the
form of Eq.~\eqref{eq:card_varTheta}, let us illustrate the calculation
of $\left|\varTheta_{g,h,\gamma}^{\lambda}\right|$, $\forall g,h,\gamma\in G$,
$\forall\lambda=x,y,z$. To be concrete, we would like to take $\lambda=z$
as an example. There are many ways to twist the model with $\omega$.
Let us discuss case by case.

\subsubsection{Some simple cases}

\paragraph{Case 1: none of $\Sigma_{k}^{z}$ are twisted.\\}

Clearly, $\varTheta_{g,h,\gamma}^{z}=\emptyset$ unless $\gamma=0\equiv\left(0,0,0\right)$.
For $\gamma=0$, there are $\left|G\right|^{L_{z}-1}$ ways to
pick $\left\{ \vartheta_{k}^{z}\left\langle x\right\rangle \right\} _{k\in\mathbb{Z}_{L_{z}}}$
subject to $\sum_{k}\vartheta_{k}^{z}\left\langle x\right\rangle =g$
and similarly $\left|G\right|^{L_{z}-1}$ ways to pick $\left\{ \vartheta_{k}^{z}\left\langle y\right\rangle \right\} _{k\in\mathbb{Z}_{L_{z}}}$
subject to $\sum_{k}\vartheta_{k}^{z}\left\langle x\right\rangle =h$.
Thus, in total
\begin{align}
\left|\varTheta_{g,h,\gamma}^{z}\right| & =\left|G\right|^{2L_{z}-2}\delta_{\gamma,0}=8^{2L_{z}-2}\delta_{\gamma,0}.\\
\left|\varTheta_{g,h}^{z}\right| & =\sum_{\gamma}\left|\varTheta_{g,h,\gamma}^{z}\right|=8^{2L_{z}-2}.
\end{align}

\paragraph{Case 2: $\Sigma_{k}^{z}$ partially twisted by $\omega$.\\}

Suppose that $\Sigma_{k}^{z}$ is twisted by $\omega$ for $k\in Z$ with $Z$ some proper subset of  $\mathbb{Z}_{L_{z}}$ (\emph{i.e.}, 
$Z\subsetneq\mathbb{Z}_{L_{z}}$).
For convenience of later discussions, let $\left\llbracket g,h,\gamma\right\rrbracket _{L}$
be the cardinality of 
\begin{multline}
\left[g,h,\gamma\right]_{L}\coloneqq\left\{ \left(a,b\right)\in G^{L}\times G^{L}\right|\\
\sum_{n}a_{n}=g,\sum_{n}b_{n}=h,\left.a_{n}\times b_{n}=\gamma,\forall n\right\} ,\label{eq:Theta_tau-1}
\end{multline}
for $g,h,\gamma\in G$ and $L$ a non-negative integer, where $a_{n}$
and $b_{n}$ are the components of $a$ and $b$ respectively. Then
$\left[g,h,\gamma\right]_{\left|Z\right|}$ labels the choices of
$\vartheta_{k}^{z}\left\langle x\right\rangle $ and $\vartheta_{k}^{z}\left\langle y\right\rangle $
for $k\in Z$, summed to $g$ and $h$ respectively. Each cross section
$\Sigma_{k}^{z}$ with $k\in\mathbb{Z}_{L_{z}}\backslash Z$ remains untwisted. Thus, $\varTheta_{g,h,\gamma}^{z}=\emptyset$ unless $\gamma=0\equiv\left(0,0,0\right)$. In detail,
\begin{equation}
\left|\varTheta_{g,h,\gamma}^{z}\right|=\delta_{\gamma,0}\sum_{g_{1},h_{1}\in G}\left\llbracket g_{1},h_{1},0\right\rrbracket _{\left|Z\right|}\left|G\right|^{2\left(L_{z}-\left|Z\right|-1\right)},\label{eq:varTheta_partial}
\end{equation}
where $\left|G\right|^{2\left(L_{z}-\left|Z\right|-1\right)}$ is
the number of ways to pick $\vartheta_{k}^{z}\left\langle x\right\rangle $
and $\vartheta_{k}^{z}\left\langle y\right\rangle $ for $k\in\mathbb{Z}_{L_{z}}\backslash Z$,
summed to $g-g_{1}$ and $h-h_{1}$ respectively. Further, we notice that
\begin{align}
 & \sum_{g,h\in G}\left\llbracket g,h,\gamma\right\rrbracket _{L}=\sum_{a,b\in G^{L}}\prod_{n=1}^{L}\delta_{a_{n}\times b_{n},\gamma}\nonumber \\
= & \left(\sum_{a_{1},b_{1}\in G}\delta_{a_{1}\times b_{1},\gamma}\right)^{L}=\begin{cases}
22^{L}, & \gamma=\left(0,0,0\right),\\
6^{L}, & \gamma\neq\left(0,0,0\right).
\end{cases}\label{eq:gh_gamma}
\end{align}
Therefore, Eq.~\eqref{eq:varTheta_partial} gives 
\begin{align}
\left|\varTheta_{g,h,\gamma}^{z}\right| & =22^{\left|Z\right|}\times8^{2\left(L_{z}-\left|Z\right|-1\right)}\delta_{\gamma,0},\\
\left|\varTheta_{g,h}^{z}\right| & =\sum_{\gamma}\left|\varTheta_{g,h,\gamma}^{z}\right|=22^{\left|Z\right|}\times8^{2\left(L_{z}-\left|Z\right|-1\right)}.\label{eq:varTheta_Z}
\end{align}

We notice that Eq.~\eqref{eq:varTheta_Z} does not depend on $g,h$
at all. We thus have a simple expression for the ground state degeneracy
if the model is twisted partially in all three directions. Explicitly,
if $\Sigma_{i}^{x}$ (resp. $\Sigma_{j}^{y}$, $\Sigma_{k}^{z}$)
is twisted for $i\in X\subsetneq\mathbb{Z}_{L_{x}}$ (resp. $j\in Y\subsetneq\mathbb{Z}_{L_{y}}$,
$k\in Z\subsetneq\mathbb{Z}_{L_{z}}$), the ground state degeneracy
for a system of size $L_{x}\times L_{y}\times L_{z}$ embedded on
$\mathtt{T}^{3}$, given by Eq.~\eqref{eq:card_varTheta}, gets simplified
to 
\begin{multline}
GSD\left(\Lambda\right)=\left|\varTheta\right|=\\
22^{\left|X\right|+\left|Y\right|+\left|Z\right|}\cdot8^{2\left(L_{x}+L_{y}+L_{z}-\left|X\right|-\left|Y\right|-\left|Z\right|\right)-3},
\label{eq:GSD_Xcube_partial}
\end{multline}
where $\left|X\right|$, $\left|Y\right|$ and $\left|Z\right|$ are the numbers of cross sections $\Sigma_{n}^{\lambda}$ twisted by $\omega$ in the three directions respectively.

\paragraph{Case 3: $\Sigma_{k}^{z}$ twisted by $\omega$ for each $k\in\mathbb{Z}_{L_{z}}$.\\}

By comparing definitions, we have
\begin{equation}
\left|\varTheta_{g,h,\gamma}^{z}\right|=\left\llbracket g,h,\gamma\right\rrbracket _{L_{z}}.
\end{equation}
Suppose that the model is partially twisted in the $x$ and $y$ directions.
Then $\left|\varTheta_{g,h}^{x}\right|$ and 
$\left|\varTheta_{g,h}^{y}\right|$ are given by the analogues of Eq.~\eqref{eq:varTheta_Z}. 
Together with Eqs.~\eqref{eq:varTheta_gh}, \eqref{eq:card_varTheta} and \eqref{eq:gh_gamma}, we get
\begin{multline}
GSD\left(\Lambda\right)=\left|\varTheta\right|=22^{\left|X\right|+\left|Y\right|}\cdot\\
8^{2\left(L_{x}+L_{y}-\left|X\right|-\left|Y\right|\right)-3}\cdot\left(22^{L_{z}}+7\times6^{L_{z}}\right).
\end{multline}
If both $X$ and $Y$ are empty, then the model is translation-invariant
and its ground state degeneracy is
\begin{equation}
GSD\left(\Lambda\right)=8^{2\left(L_{x}+L_{y}\right)-3}\cdot\left(22^{L_{z}}+7\times6^{L_{z}}\right)\label{eq:GSD_Xcube_Z}
\end{equation}
on a system of size $L_{x}\times L_{y}\times L_{z}$ embedded in $\mathtt{T}^{3}$.

\subsubsection{Computation of $\left\llbracket g,h,\gamma\right\rrbracket _{L}$}

If the model is fully twisted in more than one direction, its ground state
degeneracy is much more complicated. To find an efficient algorithm
to compute $\left\llbracket g,h,\gamma\right\rrbracket _{L}$, we
first notice that
\begin{align}
\left\llbracket g,h,\gamma\right\rrbracket _{1} & =\delta\left(g\times h,\gamma\right),\\
\left\llbracket g,h,\gamma\right\rrbracket _{L+L'} & =\sum_{p,q\in G}\left\llbracket p,q,\gamma\right\rrbracket _{L}\left\llbracket g-p,h-q,\gamma\right\rrbracket _{L'},\label{eq:recursion}
\end{align}
which follow from definitions. 

To organize the data about $\left\llbracket g,h,\gamma\right\rrbracket _{L}$,
let's consider the group ring $\mathbb{Z}G^{2}\coloneqq\mathbb{Z}\left[G\times G\right]$,
which admits a polynomial representation
\begin{equation}
\mathbb{Z}G^{2}\simeq\frac{\mathbb{Z}\left[s_{1},s_{2},s_{3},t_{1},t_{2},t_{3}\right]}{\left\langle \left\{ s_{j}^{2}-1,t_{j}^{2}-1\right\} _{j=1,2,3}\right\rangle }.
\end{equation}
We write $s^{a}\coloneqq s_{1}^{a^{\left(1\right)}}s_{2}^{a^{\left(2\right)}}s_{3}^{a^{\left(3\right)}}$,
$t^{a}\coloneqq t_{1}^{a^{\left(1\right)}}t_{2}^{a^{\left(2\right)}}t_{3}^{a^{\left(3\right)}}$ for short,
$\forall a=\left(a^{\left(1\right)},a^{\left(2\right)},a^{\left(3\right)}\right)\in G=\mathbb{Z}_{2}^{3}$
and construct a polynomial
\begin{equation}
\rho_{\gamma}\left(s_{1},s_{2},s_{3},t_{1},t_{2},t_{3}\right)\coloneqq\sum_{g,h\in G}\left\llbracket g,h,\gamma\right\rrbracket _{1}s^{g}t^{h}
\end{equation}
for each $\gamma\in G$. Because of Eq.~\eqref{eq:recursion}, $\rho_{\gamma}^{L}$
as in $\mathbb{Z}G^{2}$ (\emph{i.e.}, the $L^{\text{th}}$ power
of $\rho_{\theta}$ modulo $\left\{ s_{j}^{2}-1,t_{j}^{2}-1\right\} _{j=1,2,3}$)
is
\begin{multline}
\rho_{\gamma}^{L}=\text{\ensuremath{\sum}}_{g,h\in G}\left\llbracket g,h,\gamma\right\rrbracket _{L}s^{g}t^{h}\\
\mod\left\{ s_{j}^{2}-1,t_{j}^{2}-1\right\} _{j=1,2,3}.
\end{multline}
Thus, $\left\llbracket g,h,\gamma\right\rrbracket _{L}$ can be expressed
as a linear combination of $\left(\rho_{\gamma}\left(s,t\right)\right)^{L}$
with $s,t\in Z_{2}^{3}$, where $Z_{2}=\left\{ \pm1\right\} $. Explicitly,
\begin{equation}
\left\llbracket g,h,\gamma\right\rrbracket _{L}=\frac{1}{2^{6}}\sum_{s,t\in Z_{2}^{3}}s^{-g}t^{-h}\left(\rho_{\gamma}\left(s,t\right)\right)^{L},\label{eq:gh_gamma-1}
\end{equation}
where $\rho_{\gamma}\left(s,t\right)$ stands for the value of $\rho_{\gamma}$
at $s,t\in Z_{2}^{3}$. 

For example, $\rho_{0}\left(1,1,1,1,1,1\right)=\sum_{g,h\in G}\left\llbracket g,h,0\right\rrbracket _{1}=22$
and $\rho_{0}\left(1,1,1,1,1,-1\right)=\sum_{g,h\in G}\left\llbracket g,h,0\right\rrbracket _{1}\left(-1\right)^{g^{\left(3\right)}}=6$.
Then Eq.~\eqref{eq:gh_gamma-1} gives
\begin{equation}
\left\llbracket 0,0,0\right\rrbracket _{L}=2^{L-6}\left[42\times\left(-1\right)^{L}+21\times3^{L}+11^{L}\right],
\end{equation}
where each $0$ is short for $\left(0,0,0\right)\in G=\mathbb{Z}_{2}^{3}$.
\begin{widetext}
For convenience of later discussions, we write 
\begin{equation}
\left\llbracket g,h\right\rrbracket _{L}\coloneqq\sum_{\gamma\in G}\left\llbracket g,h,\gamma\right\rrbracket _{L}.
\end{equation}
By direct computation using Eq.~\eqref{eq:gh_gamma-1}, we get 
\begin{equation}
\left\llbracket g,h\right\rrbracket _{L}=\begin{cases}
2^{L-6}\cdot\left(11^{L}+49\cdot3^{L}+294\cdot\left(-1\right)^{L}+168\right), & \text{if }g=h=0,\\
2^{L-6}\cdot\left(11^{L}+17\cdot3^{L}+6\cdot\left(-1\right)^{L}-24\right), & \text{if }g\times h=0\text{ but }\left(g,h\right)\neq\left(0,0\right),\\
2^{L-6}\cdot\left(11^{L}+3^{L}-10\cdot\left(-1\right)^{L}+8\right), & \text{if }g\times h\neq0,
\end{cases}\label{eq:gh_L}
\end{equation}
where $0$ is short for the identity element $\left(0,0,0\right)$
of $G=\mathbb{Z}_{2}\times\mathbb{Z}_{2}\times\mathbb{Z}_{2}$.
\end{widetext}

\subsubsection{Translation-invariant cases}

The untwisted X-cube model has translation symmetries $\left(x,y,z\right)\rightarrow\left(x+1,y,z\right)$,
$\left(x,y,z\right)\rightarrow\left(x,y+1,z\right)$ and $\left(x,y,z\right)\rightarrow\left(x,y,z+1\right)$.
To keep the translational symmetries of the X-cube model, for each
direction $\lambda=x,y,z$, we either twist all $\Sigma_{n}^{\lambda}$
by $\omega\left(f,g,h\right)=e^{i\pi f^{\left(1\right)}g^{\left(2\right)}h^{\left(3\right)}}$ or twist none of them. We have seen the ground state
degeneracy Eq.~\eqref{eq:GSD_Xcube_Z} if the model is fully twisted
by $\omega$ in one direction. Now with Eqs.~\eqref{eq:varTheta_gh},
\eqref{eq:card_varTheta} and \eqref{eq:gh_L}, it is straightforward
to compute the ground state degeneracy $GSD\left(\Lambda\right)$
if the X-cube model is twisted by $\omega$ in two and three directions.
Let us summarize the results below.

\begin{widetext}
If we twist $\Sigma_{i}^{x},\forall i\in\mathbb{Z}_{L_{x}}$ and $\Sigma_{j}^{y}, \forall j\in\mathbb{Z}_{L_{y}}$ but none of $\Sigma_{k}^{z}$,
then the ground state degeneracy is 
\begin{multline}
GSD\left(\Lambda\right)=\left|G\right|^{2L_{z}-2}\sum_{f,g,h\in G}\left\llbracket f,g\right\rrbracket _{L_{x}}\left\llbracket g,h\right\rrbracket _{L_{y}}\\
=2^{L_{x}+L_{y}+6L_{z}-9}\Big[252\cdot\left(-1\right)^{L_{x}+L_{y}}+77\times3^{L_{x}+L_{y}}+11^{L_{x}+L_{y}}+84\cdot\left(-1\right)^{L_{x}}\cdot3^{L_{y}}+84\cdot\left(-1\right)^{L_{y}}\cdot3^{L_{x}}\\
+7\times3^{L_{x}}\times11^{L_{y}}+7\times3^{L_{y}}\times11^{L_{x}}\Big].\label{eq:GSD_Xcube_2}
\end{multline}
The result for twisting any other two directions, like $y$ and $z$,
can be obtained by permuting $x,y,z$. 

If the model is twisted in all three directions, then the expression
for its ground state degeneracy becomes
\begin{align}
GSD\left(\Lambda\right) & =\left|\varTheta\right|=\sum_{f,g,h\in G}\left\llbracket f,g\right\rrbracket _{L_{x}}\left\llbracket g,h\right\rrbracket _{L_{y}}\left\llbracket h,f\right\rrbracket _{L_{z}}\nonumber \\
 & =2^{L_{x}+L_{y}+L_{z}-9}\cdot\left\{ 11^{L_{x}+L_{y}+L_{z}}+1155\times3^{L_{x}+L_{y}+L_{z}}+49728\cdot\left(-1\right)^{L_{x}+L_{y}+L_{z}}\right.\nonumber \\
 & +9156\left[3^{L_{x}}\cdot(-1)^{L_{y}+L_{z}}+3^{L_{y}}\cdot(-1)^{L_{z}+L_{x}}+3^{L_{z}}\cdot(-1)^{L_{x}+L_{y}}\right]\nonumber \\
 & +2520\left[(-1)^{L_{x}}\cdot3^{L_{y}+L_{z}}+(-1)^{L_{y}}\cdot3^{L_{z}+L_{x}}+(-1)^{L_{z}}\cdot3^{L_{x}+L_{y}}\right]\nonumber \\
 & +252\left[11^{L_{x}}\cdot(-1)^{L_{y}+L_{z}}+11^{L_{y}}\cdot(-1)^{L_{z}+L_{x}}+11^{L_{z}}\cdot(-1)^{L_{x}+L_{y}}\right]\nonumber \\
 & +84\left[11^{L_{x}}\cdot3^{L_{y}}\cdot(-1)^{L_{z}}+11^{L_{y}}\cdot3^{L_{z}}\cdot(-1)^{L_{x}}+11^{L_{z}}\cdot3^{L_{x}}\cdot(-1)^{L_{y}}\right.\nonumber \\
 & \phantom{+84}\left.+11^{L_{y}}\cdot3^{L_{x}}\cdot(-1)^{L_{z}}+11^{L_{z}}\cdot3^{L_{y}}\cdot(-1)^{L_{x}}+11^{L_{x}}\cdot3^{L_{z}}\cdot(-1)^{L_{y}}\right]\nonumber \\
 & +77\left[11^{L_{x}}\cdot3^{L_{y}+L_{z}}+11^{L_{y}}\cdot3^{L_{z}+L_{x}}+11^{L_{z}}\cdot3^{L_{x}+L_{y}}\right]\nonumber \\
 & +7\left[11^{L_{x}+L_{y}}\cdot3^{L_{z}}+11^{L_{y}+L_{z}}\cdot3^{L_{x}}+11^{L_{z}+L_{x}}\cdot3^{L_{y}}\right]\nonumber \\
 & +27552\left[\left(-1\right)^{L_{x}+L_{y}}+\left(-1\right)^{L_{y}+L_{z}}+\left(-1\right)^{L_{z}+L_{x}}\right]\nonumber \\
 & +3360\left[3^{L_{x}}\cdot(-1)^{L_{y}}+3^{L_{y}}\cdot(-1)^{L_{z}}+3^{L_{z}}\cdot(-1)^{L_{x}}+3^{L_{y}}\cdot(-1)^{L_{x}}+3^{L_{z}}\cdot(-1)^{L_{y}}+3^{L_{x}}\cdot(-1)^{L_{z}}\right]\nonumber \\
 & +672\left[3^{L_{x}+L_{y}}+3^{L_{y}+L_{z}}+3^{L_{z}+L_{z}}\right]+17472\left[\left(-1\right)^{L_{x}}+\left(-1\right)^{L_{y}}+\left(-1\right)^{L_{z}}\right]\nonumber \\
 & \left.+1344\left[3^{L_{x}}+3^{L_{y}}+3^{L_{z}}\right]+13440\right\} .\label{eq:GSD_Xcube_3}
\end{align}
\end{widetext}

Both Eqs.~\eqref{eq:GSD_Xcube_2} and Eq.~\eqref{eq:GSD_Xcube_3}
are much more complicated than we originally expected. In order to double check the
validity of Eqs.~\eqref{eq:GSD_Xcube_Z}, \eqref{eq:GSD_Xcube_2},
and \eqref{eq:GSD_Xcube_3}, we can plug in $L_{x}=L_{y}=L_{z}=1$
and find that all of them give $GSD\left(\Lambda\right)=\left|G\right|^{3}$.
This is what we would expect, as in this reduced case all operators
$A_{v}$, $B_{c}$ and $P_{v}^{g}$ become the identity operator by
definition. Thus, $GSD\left(\Lambda\right)$ is just the dimension
of the total Hilbert space for $L_{x}=L_{y}=L_{z}=1$, which is $\left|G\right|^{3}$
as there are three faces in total. Despite the complexity of the GSD in the twisted case, it could be calculated straightforwardly within our framework. As in the second-to-last paragraph of Sec.~\ref{subsec:Top_charge}, it can also be proved stable to local perturbations with using the results in Sec.~\ref{subsec:X-cube_particle}.
The key point here is that the GSD for twisted fracton models depends explicitly on the system size, thus revealing the dependence of these phases on the geometry of the system. 

\section{Ground state degeneracy on $\mathtt{T}^{3}$: twisted checkerboard models}
\label{sec:GSDchecker}

In this section, we consider a checkerboard $\Lambda$ embedded on
a three-torus (\emph{i.e.}, three-dimensional torus) $\mathtt{T}^{3}$, whose vertex set is
$\Lambda^{0}=\mathbb{Z}_{L_{x}}\times\mathbb{Z}_{L_{y}}\times\mathbb{Z}_{L_{z}}$
with $L_{x},L_{y},L_{z}$ even integers. A general method
is developed here to compute the ground state degeneracy, denoted
$GSD\left(\Lambda\right)$, of a twisted checkerboard model on $\Lambda$.
In other words, we are going to determine $GSD\left(\Lambda\right)$
of a twisted checkerboard model of system size $L_{x}\times L_{y}\times L_{z}$
with the periodic boundary condition identifying $\left(x+L_{x},y,z\right)\sim\left(x,y+L_{y},z\right)\sim\left(x,y,z+L_{z}\right)\sim\left(x,y,z\right)$. 

Below, let us first describe our calculation method in a generic
setting and later 
illustrate it by explicit examples based
on groups $\mathbb{Z}_{2}$ and $\mathbb{Z}_{2}^{3}\equiv\mathbb{Z}_{2}\times\mathbb{Z}_{2}\times\mathbb{Z}_{2}$.

\subsection{Generic setting}

As a reminder, spins labeled by group elements of $G$ are on vertices
and the projectors $P_{c}$ are associated with grey cubes $c\in\Lambda_{\bullet}^{3}$.
In this section, we will use a triple of integers $\left(c^{x},c^{y},c^{z}\right)$
to label the cube centered at $\left(c^{x},c^{y},c^{z}\right)+\frac{1}{2}\left(1,1,1\right)$.

The ground state Hilbert subspace is the image of the projector
\begin{equation}
P\left(\Lambda\right)\coloneqq\prod_{c\in\Lambda_{\bullet}^{3}}P_{c}.
\end{equation}
So the ground state degeneracy $GSD\left(\Lambda\right)$ equals the
trace of $P\left(\Lambda\right)$. Explicitly,
\begin{equation}
\text{tr}P\left(\Lambda\right)=\frac{1}{\left|G^{\Lambda^{0}}\right|}\sum_{\vartheta\in G_{B}^{\Lambda^{0}}}\sum_{\eta\in G^{\Lambda_{\bullet}^{3}}}\left\langle \vartheta\right|\prod_{c\in\Lambda_{\bullet}^{3}}P_{c}^{\eta\left(c\right)}\left|\vartheta\right\rangle ,\label{eq:trP_xcube-1}
\end{equation}
where $\left\langle \vartheta\right|\prod_{v}P_{v}^{\eta\left(v\right)}\left|\vartheta\right\rangle $
is nonzero if and only if $\vartheta\in G_{B}^{\Lambda^{0}}$ and
$\prod_{c\in\Lambda_{\bullet}^{3}}A_{c}^{\eta\left(c\right)}\left|\vartheta\right\rangle =\left|\vartheta\right\rangle .$
Let 
\begin{equation}
G_{A}^{\Lambda_{\bullet}^{3}}\coloneqq\left\{ \eta\in G^{\Lambda_{\bullet}^{3}}\;|\;\prod_{c\in\Lambda_{\bullet}^{3}}A_{c}^{\eta\left(c\right)}\left|\vartheta\right\rangle =\left|\vartheta\right\rangle \right\} .
\end{equation}
We notice that $G_{A}^{\Lambda_{\bullet}^{3}}$ is independent of
$\vartheta$ and that each $\eta\in G_{A}^{\Lambda_{\bullet}^{3}}$
can be specified by 
\begin{gather}
\eta_{1}\coloneqq\eta\left(0,1,1\right),\;\eta_{2}\coloneqq\eta\left(1,0,1\right),\;\eta_{3}\coloneqq\eta\left(1,1,0\right),\\
\partial\eta_{i}^{x}\coloneqq\eta\left(i,0,0\right)-\eta\left(i-2,0,0\right),\forall i\text{ even},\\
\partial\eta_{i}^{x}\coloneqq\eta\left(i,1,0\right)-\eta\left(i-2,1,0\right),\forall i\text{ odd},\\
\partial\eta_{j}^{y}\coloneqq\eta\left(0,j,0\right)-\eta\left(0,j-2,0\right),\forall j\text{ even},\\
\partial\eta_{j}^{y}\coloneqq\eta\left(0,j,1\right)-\eta\left(0,j-2,1\right),\forall j\text{ odd},\\
\partial\eta_{k}^{z}\coloneqq\eta\left(0,0,k\right)-\eta\left(0,0,k-2\right),\forall k\text{ even},\\
\partial\eta_{k}^{z}\coloneqq\eta\left(1,0,k\right)-\eta\left(1,0,k-2\right),\forall k\text{ odd},
\end{gather}
subject to the constraints
\begin{equation}
\sum_{n\text{ odd}}\partial\eta_{n}^{\mu}=\sum_{n\text{ even}}\partial\eta_{n}^{\mu}=0,\forall\mu=x,y,z.
\end{equation}
Therefore, 
\begin{equation}
\left|G_{A}^{\Lambda_{\bullet}^{3}}\right|=\left|G\right|^{L_{x}+L_{y}+L_{z}-3}.
\end{equation}

Each $\vartheta\in G_{B}^{\Lambda^{0}}$ assigns group elements to
the two non-contractible loops of $\Sigma_{k}^{z}$ in the $x$ and
$\left(-1\right)^{k}y$ directions respectively as
\begin{align}
\vartheta_{k}^{z}\left\langle x\right\rangle  & \coloneqq\left(-1\right)^{k+1}\sum_{i\in\mathbb{Z}_{L_{x}}}\left(-1\right)^{i}\partial_{z}\vartheta\left(i,0,k\right),\\
\vartheta_{k}^{z}\left\langle y\right\rangle  & \coloneqq\left(-1\right)^{k+1}\sum_{j\in\mathbb{Z}_{L_{y}}}\left(-1\right)^{j}\partial_{z}\vartheta\left(0,j,k\right),
\end{align}
$\forall k\in\mathbb{Z}_{L_{z}}$, where $\partial_{z}\vartheta\left(i,j,k\right)\coloneqq\vartheta\left(i,j,k\right)-\vartheta\left(i,j,k-1\right)$.
The branching structure is shown in Fig.~\ref{fig:checker3d}. Similarly,
the group elements along the non-contractible loops of $\Sigma_{i}^{x}$ in the $y$ and $(-1)^{i}z$ directions are
\begin{align}
\vartheta_{i}^{x}\left\langle y\right\rangle  & \coloneqq\left(-1\right)^{i+1}\sum_{j\in\mathbb{Z}_{L_{y}}}\left(-1\right)^{j}\partial_{x}\vartheta\left(i,j,0\right),\\
\vartheta_{i}^{x}\left\langle z\right\rangle  & \coloneqq\left(-1\right)^{i+1}\sum_{k\in\mathbb{Z}_{L_{z}}}\left(-1\right)^{k}\partial_{x}\vartheta\left(i,0,k\right),
\end{align}
$\forall i \in \mathbb{Z}_{L_x}$, 
where $\partial_{x}\vartheta\left(i,j,k\right)\coloneqq\vartheta\left(i,j,k\right)-\vartheta\left(i-1,j,k\right)$. The group elements along the non-contractible loops of $\Sigma_{j}^{y}$ in the $z$ and $(-1)^{j}x$ directions are
\begin{align}
\vartheta_{j}^{y}\left\langle z\right\rangle  & \coloneqq\left(-1\right)^{j+1}\sum_{k\in\mathbb{Z}_{L_{z}}}\left(-1\right)^{k}\partial_{y}\vartheta\left(0,j,k\right),\\
\vartheta_{j}^{y}\left\langle x\right\rangle  & \coloneqq\left(-1\right)^{j+1}\sum_{i\in\mathbb{Z}_{L_{x}}}\left(-1\right)^{i}\partial_{y}\vartheta\left(i,0,k\right),
\end{align}
$\forall j \in \mathbb{Z}_{L_y}$,
where  $\partial_{y}\vartheta\left(i,j,k\right)\coloneqq\vartheta\left(i,j,k\right)-\vartheta\left(i,j-1,k\right)$.
Clearly, they are subject to 
\begin{align}
\vartheta^{xy} & =
\sum_{i\text{ even}}\vartheta_{i}^{x}\left\langle y\right\rangle =\sum_{j\text{ even}}\vartheta_{j}^{y}\left\langle x\right\rangle ,\nonumber \\
& =\sum_{i\text{ odd}}\vartheta_{i}^{x}\left\langle y\right\rangle =\sum_{j\text{ odd}}\vartheta_{j}^{y}\left\langle x\right\rangle ,\label{eq:theta_xy-1}\\
\vartheta^{yz} & =
\sum_{j\text{ even}}\vartheta_{j}^{y}\left\langle z\right\rangle =\sum_{k\text{ even}}\vartheta_{k}^{z}\left\langle y\right\rangle ,\nonumber \\
& =\sum_{j\text{ odd}}\vartheta_{j}^{y}\left\langle z\right\rangle =\sum_{k\text{ odd}}\vartheta_{k}^{z}\left\langle y\right\rangle ,\label{eq:theta_yz-1}\\
\vartheta^{zx} & =
\sum_{k\text{ even}}\vartheta_{k}^{z}\left\langle x\right\rangle =\sum_{i\text{ even}}\vartheta_{i}^{x}\left\langle z\right\rangle ,\nonumber \\
& =\sum_{k\text{ odd}}\vartheta_{k}^{z}\left\langle x\right\rangle =\sum_{i\text{ odd}}\vartheta_{i}^{x}\left\langle z\right\rangle .\label{eq:theta_zx-1}
\end{align}
where $\vartheta^{xy}$ (resp. $\vartheta^{yz}$, $\vartheta^{zx}$) denotes the sum of $\left(-1\right)^{v}\vartheta\left(v\right)$ over vertices in the plane $z=0$ (resp. $x=0$, $y=0$).
So there are $\left|G\right|^{2\left(L_{x}+L_{y}+L_{z}\right)-9}$
choices of $\left\{ \vartheta_{n}^{\mu}\left\langle \nu\right\rangle \right\} $
(\emph{i.e.}, the group elements assigned to non-contractible loops
of $\Sigma_{n}^{\mu}$ for all possible $\mu$, $n$). 

There are $G^{L_{x}L_{y}L_{z}-\left(L_{x}-1\right)\left(L_{y}-1\right)\left(L_{z}-1\right)-2\left(L_{x}+L_{y}+L_{z}\right)+9}$
ways to color vertices on the planes $x=0$, $y=0$ and $z=0$ for
each chosen $\left\{ \vartheta_{n}^{\mu}\left\langle \nu\right\rangle \right\} $.
Further, the number of choices of $\partial_{z}\vartheta$ to complete
the coloring of $\Sigma_{k}^{z}$ for each $k=1,2,\cdots,L_{z}-2$
equals $\left|G\right|^{\frac{1}{2}\left(L_{x}-2\right)\left(L_{y}-2\right)}$,
where $\frac{1}{2}(L_x-2)(L_y-2)$ is the number of cubes in $\Lambda_{\circ}^{3}$ cut by
$\Sigma_{k}^{z}$ but not touching the planes $x=0$ and $y=0$. At this point
we have actually specified $\vartheta\in G_{B}^{\Lambda^{0}}$ already;
in total
\begin{multline}
\left|G_{B}^{\Lambda^{0}}\right|=\left|G\right|^{\frac{1}{2}\left(L_{x}-2\right)\left(L_{y}-2\right)\left(L_{z}-2\right)}\cdot\left|G\right|^{2\left(L_{x}+L_{y}+L_{z}\right)-9}\cdot\\
\left|G\right|^{L_{x}L_{y}L_{z}-\left(L_{x}-1\right)\left(L_{y}-1\right)\left(L_{z}-1\right)-2\left(L_{x}+L_{y}+L_{z}\right)+9},
\end{multline}
which simplifies to 
\begin{equation}
\left|G_{B}^{\Lambda^{0}}\right|=\left|G\right|^{\frac{1}{2}L_{x}L_{y}L_{z}+L_{x}+L_{y}+L_{z}-3}.
\end{equation}

\subsubsection{Untwisted checkerboard models}

If the model is untwisted (\emph{i.e.}, $\omega\equiv1$), then Eq.~\eqref{eq:trP_xcube-1}
reduces to
\begin{align}
GSD\left(\Lambda\right) & =\text{tr}P\left(\Lambda\right)=\frac{\left|G_{A}^{\Lambda_{\bullet}^{3}}\right|\left|G_{B}^{\Lambda^{0}}\right|}{\left|G^{\Lambda_{\bullet}^{3}}\right|}\nonumber \\
 & =\left|G\right|^{2\left(L_{x}+L_{y}+L_{z}\right)-6}.\label{eq:GSD_G-1}
\end{align}
This ground state degeneracy was already mentioned in Eq.~\eqref{eq:GSD_checkerboard}
as we introduced the model.

\subsubsection{Twisted checkerboard models}

In a twisted checkerboard model, each $\eta\in G_{A}^{\Lambda_{\bullet}^{3}}$
makes a gauge transformation labeled by $\partial\eta_{n}^{\lambda}$
uniformly to $\Sigma_{n}^{\lambda}$ for $\lambda=x,y,z$. Therefore,
\begin{align}
 & \left\langle \vartheta\right|\prod_{c\in\Lambda_{\bullet}^{3}}P_{c}^{\eta\left(c\right)}\left|\vartheta\right\rangle \nonumber \\
= & \prod_{i\in\mathbb{Z}_{L_{x}}}\omega_{i}^{x}\left[\mathtt{T^{3}};\vartheta_{i}^{x}\left\langle y\right\rangle ,\vartheta_{i}^{x}\left\langle z\right\rangle ,\partial\eta_{i}^{x}\right]\cdot\nonumber \\
 & \prod_{i\in\mathbb{Z}_{L_{y}}}\omega_{j}^{y}\left[\mathtt{T^{3}};\vartheta_{j}^{y}\left\langle z\right\rangle ,\vartheta_{j}^{y}\left\langle x\right\rangle ,\partial\eta_{j}^{y}\right]\cdot\nonumber \\
 & \prod_{i\in\mathbb{Z}_{L_{z}}}\omega_{k}^{z}\left[\mathtt{T^{3}};\vartheta_{k}^{z}\left\langle x\right\rangle ,\vartheta_{k}^{z}\left\langle y\right\rangle ,\partial\eta_{k}^{z}\right].\label{eq:L-2}
\end{align}
We can view $\left\langle \vartheta\right|\prod_{c\in\Lambda_{\bullet}^{3}}P_{c}^{\eta\left(c\right)}\left|\vartheta\right\rangle $
as a one-dimensional representation of $\eta\in G_{A}^{\Lambda_{\bullet}^{3}}$.
Therefore, $\sum_{\eta\in G_{A}^{\Lambda_{\bullet}^{3}}}\left\langle \vartheta\right|\prod_{c\in\Lambda_{\bullet}^{3}}P_{c}^{\eta\left(c\right)}\left|\vartheta\right\rangle =0$
unless the representation is trivial. 

Let $\varTheta$ collect all possible choices of $\left\{ \vartheta_{n}^{\lambda}\left\langle \mu\right\rangle \right\} $
making $\left\langle \vartheta\right|\prod_{c\in\Lambda_{\bullet}^{3}}P_{c}^{\eta\left(c\right)}\left|\vartheta\right\rangle $
the trivial representation of $G_{A}^{\Lambda_{\bullet}^{3}}$. As
there are $\left|G\right|^{L_{x}L_{y}L_{z}-\left(L_{x}-1\right)\left(L_{y}-1\right)\left(L_{z}-1\right)-2\left(L_{x}+L_{y}+L_{z}\right)+9}\cdot\left|G\right|^{\frac{1}{2}\left(L_{x}-2\right)\left(L_{y}-2\right)\left(L_{z}-2\right)}$
choices of $\vartheta\in G_{A}^{\Lambda_{\bullet}^{3}}$ for each
chosen $\left\{ \vartheta_{n}^{\mu}\left\langle \nu\right\rangle \right\} $,
explicit computation shows that the ground state degeneracy on $\Lambda$
with underlying space $\mathtt{T}^{3}$ is
\begin{multline}
GSD\left(\Lambda\right)=\text{tr}P\left(\Lambda\right)=\\
\left|G\right|^{L_{x}L_{y}L_{z}-\left(L_{x}-1\right)\left(L_{y}-1\right)\left(L_{z}-1\right)-2\left(L_{x}+L_{y}+L_{z}\right)+9}\cdot\\
\left|G\right|^{\frac{1}{2}\left(L_{x}-2\right)\left(L_{y}-2\right)\left(L_{z}-2\right)}\cdot\frac{\left|G_{A}^{\Lambda_{\bullet}^{3}}\right|\left|\varTheta\right|}{\left|G^{\Lambda_{\bullet}^{3}}\right|}=\left|G\right|^{3}\left|\varTheta\right|.\label{eq:GSD_checker}
\end{multline}
Therefore, we can get $GSD\left(\Lambda\right)$ by counting the cardinality
of $\varTheta$. By definition, $\left|\varTheta\right|\leq2\left(L_{x}+L_{y}+L_{z}\right)-9$.
So the ground state degeneracy of a twisted model is always less or
equal to that of its untwisted version. 

Technically, the triviality of $\left\langle \vartheta\right|\prod_{c\in\Lambda_{\bullet}^{3}}P_{c}^{\eta\left(c\right)}\left|\vartheta\right\rangle $
as a representation of $G_{A}^{\Lambda^{0}}$ is equivalent to requiring that
\begin{equation}
\prod_{n\in\mathbb{Z}_{L_{\lambda}}}\omega_{n}^{\lambda}\left[\mathtt{T^{3}};\vartheta_{n}^{\lambda}\left\langle \mu\right\rangle ,\vartheta_{n}^{\lambda}\left\langle \nu\right\rangle ,\eta_{n}-\eta_{n-2}\right]=1,
\end{equation}
$\forall\left(\lambda,\mu,\nu\right)=\left(x,y,z\right),\left(y,z,x\right),\left(z,x,y\right)$,
$\forall\eta\in G^{L_{\lambda}}$. Since
\begin{align}
 & \prod_{n\in\mathbb{Z}_{L_{\lambda}}}\omega_{n}^{\lambda}\left[\mathtt{T^{3}};\vartheta_{n}^{\lambda}\left\langle \mu\right\rangle ,\vartheta_{n}^{\lambda}\left\langle \nu\right\rangle ,\eta_{n}-\eta_{n-2}\right]\nonumber \\
= & \prod_{n\in\mathbb{Z}_{L_{\lambda}}}\frac{\omega_{n}^{\lambda}\left[\mathtt{T^{3}};\vartheta_{n}^{\lambda}\left\langle \mu\right\rangle ,\vartheta_{n}^{\lambda}\left\langle \nu\right\rangle ,\eta_{n}\right]}{\omega_{n}^{\lambda}\left[\mathtt{T^{3}};\vartheta_{n}^{\lambda}\left\langle \mu\right\rangle ,\vartheta_{n}^{\lambda}\left\langle \nu\right\rangle ,\eta_{n-2}\right]}\nonumber \\
= & \prod_{n\in\mathbb{Z}_{L_{\lambda}}}\frac{\omega_{n}^{\lambda}\left[\mathtt{T^{3}};\vartheta_{n}^{\lambda}\left\langle \mu\right\rangle ,\vartheta_{n}^{\lambda}\left\langle \nu\right\rangle ,\eta_{n}\right]}{\omega_{n+2}^{\lambda}\left[\mathtt{T^{3}};\vartheta_{n+2}^{\lambda}\left\langle \mu\right\rangle ,\vartheta_{n+2}^{\lambda}\left\langle \nu\right\rangle ,\eta_{n}\right]},
\end{align}
the condition is further equivalent to that $\exists\gamma_{0}^{\lambda},\gamma_{1}^{\lambda}\in\widehat{G}$,
\begin{equation}
\omega_{n}^{\lambda}\left[\mathtt{T^{3}};\vartheta_{n}^{\lambda}\left\langle \mu\right\rangle ,\vartheta_{n}^{\lambda}\left\langle \nu\right\rangle ,-\right]=\gamma_{n\,(\text{mod }2)}^{\lambda},\forall n\in\mathbb{Z}_{L_{\lambda}},\label{eq:omega_tau-1}
\end{equation}
for $\lambda=x,y,z$ separately, where $\omega_{n}^{\lambda}\left[\mathtt{T^{3}};\vartheta_{n}^{\lambda}\left\langle \mu\right\rangle ,\vartheta_{n}^{\lambda}\left\langle \nu\right\rangle ,-\right]$
is viewed as a one-dimensional representation of $G$ with $-$ denoting
a place holder for a group element and $\widehat{G}$ stands for the
character group of $G$. 

To take the constraints given by Eqs.~(\ref{eq:theta_xy-1}-\ref{eq:theta_zx-1})
into consideration, let 
\begin{multline}
\varTheta_{g,h,\gamma}^{\lambda,\kappa}\coloneqq\left\{ \left(a,b\right)\in G^{\frac{L_{\mu}}{2}}\times G^{\frac{L_{\mu}}{2}}\right|\sum_{n}a_{n}=g,\sum_{n}b_{n}=h,\\
\left.\omega_{n}^{\lambda}\left[\mathtt{T^{3}};a_{n},b_{n},-\right]=\gamma,\forall n\in2\mathbb{Z}_{L_{\lambda}}+\kappa\vphantom{G^{\frac{L_{\mu}}{2}}}\right\} ,\label{eq:Theta_tau-2}
\end{multline}
for $g,h\in G$, $\gamma\in\widehat{G}$, $\lambda=x,y,z$ and $\kappa=0,1$.
In addition, we write
\begin{align}
\varTheta_{g,h}^{\lambda,\kappa} & \coloneqq\bigcup_{\gamma\in\widehat{G}}\varTheta_{g,h,\gamma}^{\lambda,\kappa},
\end{align}
It is straightforward to see that 
\begin{equation}
\varTheta=\bigcup_{f,g,h\in G}\varTheta_{f,g}^{x,0}\times\varTheta_{f,g}^{x,1}\times\varTheta_{g,h}^{y,0}\times\varTheta_{g,h}^{y,1}\times\varTheta_{h,f}^{z,0}\times\varTheta_{h,f}^{z,1}.
\end{equation}
Therefore, the cardinalities of these sets satisfy
\begin{gather}
\left|\varTheta_{g,h}^{\lambda,\kappa}\right|=\sum_{\gamma\in\widehat{G}}\left|\varTheta_{g,h,\gamma}^{\lambda,\kappa}\right|,\label{eq:varTheta_gh-1}\\
\left|\varTheta\right|=\sum_{f,g,h\in G}\prod_{\kappa\in\mathbb{Z}_{2}}\left|\varTheta_{f,g}^{x,\kappa}\right|\left|\varTheta_{g,h}^{y,\kappa}\right|\left|\varTheta_{h,f}^{z,\kappa}\right|.\label{eq:card_varTheta-1}
\end{gather}
Below, we will explain how to use Eq.~\eqref{eq:card_varTheta-1}
to count $\left|\varTheta\right|$ in the example based on $G=\mathbb{Z}_{2}^{3}$
with $\omega\left(f,g,h\right)=e^{i\pi f^{\left(1\right)}g^{\left(2\right)}h^{\left(3\right)}}$. 

\subsection{Example: $G=\mathbb{Z}_{2}$ }

As discussed in Sec.~\ref{subsec:DW_Z2}, we always have 
\begin{equation}
\omega\left[\mathtt{T}^{3};f,g,h\right]=\frac{\omega_{h}\left(f,g\right)}{\omega_{h}\left(g,f\right)}=1,
\end{equation}
$\forall f,g,h\in G$. Therefore, $\varTheta$ includes all possible
choices of $\left\{ \vartheta_{n}^{\lambda}\left(\mu\right)\right\} $
and hence $\left|\varTheta\right|=\left|G\right|^{2\left(L_{x}+L_{y}+L_{z}\right)-9}$.
Then 
\begin{equation}
GSD\left(\Lambda\right)=\left|G\right|^{3}\left|\varTheta\right|=2^{2\left(L_{x}+L_{y}+L_{z}\right)-6},
\end{equation}
which remains unchanged, no matter how we twist the model.

\subsection{Example: $G=\mathbb{Z}_{2}^{3}$ with $\omega\left(f,g,h\right)=e^{i\pi f^{\left(1\right)}g^{\left(2\right)}h^{\left(3\right)}}$}

As seen in Sec.~\ref{subsec:DW_Zm3}, $\forall f,g,h\in G=\mathbb{Z}_{2}\times\mathbb{Z}_{2}\times\mathbb{Z}_{2}$,
\begin{equation}
\omega\left[\mathtt{T}^{3};f,g,h\right]=e^{i\pi\left(f\times g\right)\cdot h}.
\end{equation}
We identify $\widehat{G}\cong G$; in particular, $\omega\left[\mathtt{T}^{3};f,g,-\right]\in\widehat{G}$
is identified with $f\times g\in G$. 

First, let us illustrate the calculation of $\left|\varTheta_{g,h,\gamma}^{z,\kappa}\right|$,
$\forall g,h,\gamma\in G$, $\forall\kappa=0,1$ for some simple cases.
The computation of $\left|\varTheta_{g,h,\gamma}^{\lambda,\kappa}\right|$
for $\lambda=x,y$ is similar.

\subsubsection{Some simple cases}

\paragraph{Case 1: none of $\Sigma_{k}^{z}$ are twisted.\\}

Clearly, $\varTheta_{g,h,\gamma}^{z,\kappa}=\emptyset$ unless $\gamma=0\equiv\left(0,0,0\right)$.
For $\gamma=0$, there are $\left|G\right|^{\frac{1}{2}L_{z}-1}$
ways to pick $\left\{ \vartheta_{k}^{z}\left\langle x\right\rangle \right\} _{k\in2\mathbb{Z}_{L_{z}}+\kappa}$
subject to $\sum_{k\in2\mathbb{Z}_{L_{z}}+\kappa}\vartheta_{k}^{z}\left\langle x\right\rangle =g$
and similarly $\left|G\right|^{\frac{1}{2}L_{z}-1}$ ways to pick
$\left\{ \vartheta_{k}^{z}\left\langle y\right\rangle \right\} _{k\in2\mathbb{Z}_{L_{z}}+\kappa}$
subject to $\sum_{k\in2\mathbb{Z}_{L_{z}}+\kappa}\vartheta_{k}^{z}\left\langle x\right\rangle =h$.
In total, $\forall\kappa \in \{0,1\}$, $\forall g,h\in G$,
\begin{align}
\left|\varTheta_{g,h,\gamma}^{z,\kappa}\right| & =\left|G\right|^{L_{z}-2}\delta_{\gamma,0}=8^{L_{z}-2}\delta_{\gamma,0},\\
\left|\varTheta_{g,h}^{z,\kappa}\right| & =\sum_{\gamma}\left|\varTheta_{g,h,\gamma}^{z,\kappa}\right|=8^{L_{z}-2}.
\end{align}

\paragraph{Case 2: $\Sigma_{k}^{z}$ partially twisted by $\omega$.\\}

Suppose that $\Sigma_{k}^{z}$ is twisted by $\omega$ for $k\in Z_{\kappa}\subsetneq2\mathbb{Z}_{L_{z}}+\kappa$,
where $\kappa=0,1$. We would like to express $\left|\varTheta_{g,h,\gamma}^{z,\kappa}\right|$
in terms of $\left\llbracket g,h,\gamma\right\rrbracket _{L}$, the
cardinality of the set $\left[g,h,\gamma\right]_{L}$ defined by Eq.~\eqref{eq:Theta_tau-1}.
We notice that $\left[g_{1},h_{1},\gamma\right]_{\left|Z_{\kappa}\right|}$
labels the choices of $\vartheta_{k}^{z}\left\langle x\right\rangle $
and $\vartheta_{k}^{z}\left\langle y\right\rangle $ 
for $k\in Z_{\kappa}$, satisfying $\vartheta_{k}^{z}\left\langle x\right\rangle \times \vartheta_{k}^{z}\left\langle y\right\rangle = \gamma$ and 
summed to $g_{1}$ and $h_{1}$ respectively. The remaining untwisted $\Sigma_{k}^{z}$
with $k\in\left(2\mathbb{Z}_{L_{z}}+\kappa\right)\backslash Z_{\kappa}$
still requires $\gamma=0\equiv\left(0,0,0\right)$. Thus,
\begin{equation}
\left|\varTheta_{g,h,\gamma}^{z,\kappa}\right|=\delta_{\gamma,0}\sum_{g_{1},h_{1}\in G}\left\llbracket g_{1},h_{1},0\right\rrbracket _{\left|Z_{\kappa}\right|}\left|G\right|^{L_{z}-2\left|Z_{\kappa}\right|-2},\label{eq:varTheta_partial-1}
\end{equation}
where $\left|G\right|^{L_{z}-2\left|Z_{\kappa}\right|-2}$ is the
number of ways to pick $\vartheta_{k}^{z}\left\langle x\right\rangle $
and $\vartheta_{k}^{z}\left\langle y\right\rangle $ for $k\in\left(2\mathbb{Z}_{L_{z}}+\kappa\right)\backslash Z_{\kappa}$,
summed to $g-g_{1}$ and $h-h_{1}$ respectively. With Eq.~\eqref{eq:gh_gamma},
it gets simplified to 
\begin{align}
\left|\varTheta_{g,h,\gamma}^{z,\kappa}\right| & =22^{\left|Z_{\kappa}\right|}\times8^{L_{z}-2\left|Z_{\kappa}\right|-2}\delta_{\gamma,0},\\
\left|\varTheta_{g,h}^{z,\kappa}\right| & =\sum_{\gamma}\left|\varTheta_{g,h,\gamma}^{z,\kappa}\right|=22^{\left|Z_{\kappa}\right|}\times8^{L_{z}-2\left|Z_{\kappa}\right|-2}.\label{eq:varTheta_Z-1}
\end{align}

We notice that Eq.~\eqref{eq:varTheta_Z-1} does not depend on $g,h$
at all. Thus, if $\Sigma_{i}^{x}$ (resp. $\Sigma_{j}^{y}$, $\Sigma_{k}^{z}$)
is twisted for $i\in X_{\kappa}\subsetneq2\mathbb{Z}_{L_{x}}+\kappa$
(resp. $j\in Y_{\kappa}\subsetneq2\mathbb{Z}_{L_{y}}+\kappa$, $k\in Z_{\kappa}\subsetneq\mathbb{Z}_{L_{z}}+\kappa$),
then $GSD\left(\Lambda\right)$ for a system of size $L_{x}\times L_{y}\times L_{z}$
embedded on $\mathtt{T}^{3}$, given by Eqs.~\eqref{eq:GSD_checker}
and \eqref{eq:card_varTheta-1}, gets simplified to 
\begin{multline}
GSD\left(\Lambda\right)=\left|G\right|^{3}\left|\varTheta\right|=\\
22^{\left|X\right|+\left|Y\right|+\left|Z\right|}\cdot8^{2\left(L_{x}+L_{y}+L_{z}-\left|X\right|-\left|Y\right|-\left|Z\right|\right)-6},\label{eq:GSD_checker_partial}
\end{multline}
where $X\coloneqq X_{0}\cup X_{1}$, $Y\coloneqq Y_{0}\cup Y_{1}$,
$Z=Z_{0}\cup Z_{1}$. In particular, it reduces to Eq.~\eqref{eq:GSD_G-1}
as expected, if $X,Y,Z$ are all empty.

\paragraph{Case 3: $\Sigma_{k}^{z}$ twisted by $\omega$ for each $k\in2\mathbb{Z}_{L_{z}}$.\\}

By comparing definitions, we have
\begin{equation}
\left|\varTheta_{g,h,\gamma}^{z,0}\right|=\left\llbracket g,h,\gamma\right\rrbracket _{\frac{1}{2}L_{z}}.
\end{equation}
In addition, $\Sigma_{k}^{z}$ (resp. $\Sigma_{i}^{x}$ , $\Sigma_{j}^{y}$)
may be twisted by $\omega$ for $k\in Z_{1}\subsetneq2\mathbb{Z}_{L_{z}}+1$
(resp. $i\in X_{\kappa}\subsetneq2\mathbb{Z}_{L_{x}}+\kappa$, $j\in Y_{\kappa}\subsetneq2\mathbb{Z}_{L_{y}}+\kappa$
with $\kappa=0,1$) as well. Then $\left|\varTheta_{g,h}^{z,1}\right|$
(resp. $\left|\varTheta_{g,h}^{x,\kappa}\right|$ and $\left|\varTheta_{g,h}^{y,\kappa}\right|$)
is given by Eq.~\eqref{eq:varTheta_Z-1} (resp. its analogue for
the $x$ and $y$ direction). In total, Eqs.~\eqref{eq:gh_gamma}
and \eqref{eq:card_varTheta-1} give
\begin{multline}
\left|\varTheta\right|=8^{2\left(L_{x}+L_{y}-\left|X\right|-\left|Y\right|-\left|Z_{1}\right|\right)+L_{z}-9}\cdot22^{\left|X\right|+\left|Y\right|+\left|Z_{1}\right|}\\
\cdot\left(22^{\frac{1}{2}L_{z}}+7\times6^{\frac{1}{2}L_{z}}\right),
\end{multline}
where $X\coloneqq X_{0}\cup X_{1}$ and $Y=Y_{0}\cup Y_{1}$. Therefore,
the ground state degeneracy is 
\begin{multline}
GSD\left(\Lambda\right)=\left|G\right|^{3}\left|\varTheta\right|=8^{2\left(L_{x}+L_{y}-\left|X\right|-\left|Y\right|-\left|Z_{1}\right|\right)+L_{z}-6}\\
\cdot22^{\left|X\right|+\left|Y\right|+\left|Z_{1}\right|}\cdot\left(22^{\frac{1}{2}L_{z}}+7\times6^{\frac{1}{2}L_{z}}\right).
\end{multline}
If $\left|X\right|=\left|Y\right|=\left|Z_{1}\right|=0$, the model
is translation-invariant and its ground state degeneracy reduces to
\begin{equation}
GSD\left(\Lambda\right)=8^{2\left(L_{x}+L_{y}\right)+L_{z}-6}\left(22^{\frac{1}{2}L_{z}}+7\times6^{\frac{1}{2}L_{z}}\right)\label{eq:GSD_checker_Z0}
\end{equation} 
with system size $L_{x}\times L_{y}\times L_{z}$ embedded on $\mathtt{T}^{3}$.

\subsubsection{Translation-invariant cases}

The untwisted checkerboard model has the translation symmetries $\left(x,y,z\right)\rightarrow\left(x+2,y,z\right)$,
$\left(x,y,z\right)\rightarrow\left(x,y+2,z\right)$ and $\left(x,y,z\right)\rightarrow\left(x,y,z+2\right)$.
To keep these translation symmetries, we either twist all $\Sigma_{n}^{\lambda}$
for $n\in2\mathbb{Z}_{L_{\lambda}}+\kappa$ together or twist none of
them, where $\kappa=0,1$. With Eqs.~\eqref{eq:gh_gamma-1}, \eqref{eq:gh_L},
\eqref{eq:GSD_checker}, \eqref{eq:varTheta_gh-1}, and \eqref{eq:card_varTheta-1},
we can compute the ground state degeneracies $GSD\left(\Lambda\right)$
of each translation-invariant case. Let us list the results for some
examples below.

\paragraph{Case 1: half of $\Sigma_{k}^{z}$'s are twisted by $\omega$ (e.g., $\Sigma_{k}^{z}$ is twisted by $\omega$ for $k\in 2\mathbb{Z}_{L_z}$).\\}

The ground state degeneracy is given by Eq.~\eqref{eq:GSD_checker_Z0}.

\paragraph{Case 2: half of $\Sigma_{n}^{\lambda}$'s are twisted by $\omega$ in both the $x$
and $y$ directions (e.g., both $\Sigma_{i}^{x}$ and $\Sigma_{j}^{y}$
are twisted by $\omega$ for $i\in2\mathbb{Z}_{L_{x}}$ and $j\in2\mathbb{Z}_{L_{y}}$).\\}

The ground state degeneracy is 
\begin{multline}
GSD\left(\Lambda\right)=\left|G\right|^{3}\left|\varTheta\right|\\
=\left|G\right|^{3}\sum_{f,g,h\in G}\left\llbracket f,g\right\rrbracket _{\frac{1}{2}L_{x}}\left\llbracket g,h\right\rrbracket _{\frac{1}{2}L_{y}}\left|G\right|^{L_{x}+L_{y}+2L_{z}-8}\\
=2^{\frac{7}{2}L_{x}+\frac{7}{2}L_{y}+6L_{z}-18}\Big[252\cdot\left(-1\right)^{\frac{L_{x}+L_{y}}{2}}+77\times3^{\frac{L_{x}+L_{y}}{2}}\\
+11^{\frac{L_{x}+L_{y}}{2}}+84\cdot\left(-1\right)^{\frac{1}{2}L_{x}}\cdot3^{\frac{1}{2}L_{y}}+84\cdot\left(-1\right)^{\frac{1}{2}L_{y}}\cdot3^{\frac{1}{2}L_{x}}\\
+7\times3^{\frac{1}{2}L_{x}}\times11^{\frac{1}{2}L_{y}}+7\times3^{\frac{1}{2}L_{y}}\times11^{\frac{1}{2}L_{x}}\Big].
\end{multline}
The result for twisting by half any other two directions, like $y$
and $z$, can be obtained by permuting $x,y,z$.

\paragraph{Case 3: half of $\Sigma_{n}^{\lambda}$'s are twisted in all the three
directions.\\}

The ground state degeneracy in this case is 
\begin{align}
 & GSD\left(\Lambda\right)=\left|G\right|^{3}\left|\varTheta\right|\nonumber \\
= & \left|G\right|^{3}\sum_{f,g,h\in G}\left\llbracket f,g\right\rrbracket _{\frac{L_{x}}{2}}\left\llbracket g,h\right\rrbracket _{\frac{L_{y}}{2}}\left\llbracket h,f\right\rrbracket _{\frac{L_{z}}{2}}\left|G\right|^{L_{x}+L_{y}+L_{z}-6}\nonumber \\
= & 8^{L_{x}+L_{y}+L_{z}-3}\sum_{f,g,h\in G}\left\llbracket f,g\right\rrbracket _{\frac{L_{x}}{2}}\left\llbracket g,h\right\rrbracket _{\frac{L_{y}}{2}}\left\llbracket h,f\right\rrbracket _{\frac{L_{z}}{2}},
\label{eq:GSD_checker_half3}
\end{align}
where $\sum_{f,g,h\in G}\left\llbracket f,g\right\rrbracket _{\frac{1}{2}L_{x}}\left\llbracket g,h\right\rrbracket _{\frac{1}{2}L_{y}}\left\llbracket h,f\right\rrbracket _{\frac{1}{2}L_{z}}$can
either be calculated with Eq.~\eqref{eq:gh_L} directly or be expressed
by Eq.~\eqref{eq:GSD_Xcube_3} with $L_{x}$, $L_{y}$ and $L_{z}$
replaced by $\frac{1}{2}L_{x}$, $\frac{1}{2}L_{y}$ and $\frac{1}{2}L_{z}$
respectively.

\paragraph{Case 4: each $\Sigma_{k}^{z}$ is twisted by $\omega$ for $k\in\mathbb{Z}_{L_{z}}$.\\}

Here, the ground state degeneracy is give by
\begin{align}
 & GSD\left(\Lambda\right)=\left|G\right|^{3}\left|\varTheta\right|\nonumber \\
= & \left|G\right|^{3}\sum_{f,g,h\in G}\left\llbracket f,g\right\rrbracket _{\frac{L_{z}}{2}}^{2}\left|G\right|^{2L_{x}+2L_{y}-8}\nonumber \\
= & 8^{2L_{x}+2L_{y}-6}\cdot2^{L_{z}}\left(11^{L_{z}}+14\times33^{\frac{L_{z}}{2}}+133\times3^{L_{z}}\right.\nonumber \\
 & \left.+1344\cdot\left(-1\right)^{\frac{L_{z}}{2}}+504\cdot\left(-1\right)^{\frac{L_{z}}{2}}\cdot3^{\frac{L_{z}}{2}}+2100\right), 
 \label{eq:GSD_checker_z}
\end{align}
where $\left\llbracket f,g\right\rrbracket _{\frac{L_{z}}{2}}^{2}$
is the square of $\left\llbracket f,g\right\rrbracket _{\frac{L_{z}}{2}}$
specified by Eq.~\eqref{eq:gh_L}.

\paragraph{Case 5: both $\Sigma_{i}^{x}$ and $\Sigma_{j}^{y}$ twisted by $\omega$
for $i\in\mathbb{Z}_{L_{x}}$ and $j\in\mathbb{Z}_{L_{y}}$.\\}

The ground state degeneracy for this case is 
\begin{align}
 & GSD\left(\Lambda\right)=\left|G\right|^{3}\left|\varTheta\right|\nonumber \\
= & \left|G\right|^{3}\sum_{f,g,h\in G}\left\llbracket f,g\right\rrbracket _{\frac{L_{x}}{2}}^{2}\left\llbracket g,h\right\rrbracket _{\frac{L_{y}}{2}}^{2}\left|G\right|^{2L_{z}-4}\nonumber \\
= & 8^{2L_{z}-1}\sum_{f,g,h\in G}\left\llbracket f,g\right\rrbracket _{\frac{L_{x}}{2}}^{2}\left\llbracket g,h\right\rrbracket _{\frac{L_{y}}{2}}^{2},\label{eq:GSD_checker2}
\end{align}
where $\left\llbracket f,g\right\rrbracket _{\frac{L_{x}}{2}}$ and
$\left\llbracket g,h\right\rrbracket _{\frac{L_{y}}{2}}$ are given
by Eq.~\eqref{eq:gh_L}. Explicitly, $GSD\left(\Lambda\right)$ can
be expressed as a long polynomial in terms of  $2^{L_{\lambda}}$, $\left(-1\right)^{\frac{1}{2}L_{\lambda}}$,
$3^{\frac{1}{2}L_{\lambda}}$, $11^{\frac{1}{2}L_{\lambda}}$ with $\lambda=x,y$
and $2^{L_{z}}$. 

\paragraph{Case 6: all $\Sigma_{i}^{x}$, $\Sigma_{j}^{y}$ and $\Sigma_{k}^{z}$
twisted by $\omega$ for $i\in\mathbb{Z}_{L_{x}}$ $j\in\mathbb{Z}_{L_{y}}$
and $k\in\mathbb{Z}_{L_{z}}$.\\}

The ground state degeneracy is 
\begin{align}
 & GSD\left(\Lambda\right)=\left|G\right|^{3}\left|\varTheta\right|\nonumber \\
= & \left|G\right|^{3}\sum_{f,g,h\in G}\left\llbracket f,g\right\rrbracket _{\frac{L_{x}}{2}}^{2}\left\llbracket g,h\right\rrbracket _{\frac{L_{y}}{2}}^{2}\left\llbracket h,f\right\rrbracket _{\frac{L_{z}}{2}}^{2},\label{eq:GSD_checker3}
\end{align}
where $\left\llbracket f,g\right\rrbracket _{\frac{L_{x}}{2}}$, $\left\llbracket g,h\right\rrbracket _{\frac{L_{y}}{2}}$
and $\left\llbracket h,f\right\rrbracket _{\frac{L_{z}}{2}}$ are given
by Eq.~\eqref{eq:gh_L}. Explicitly, $GSD\left(\Lambda\right)$ can
be expressed as a long polynomial in terms of $2^{L_{\lambda}}$, $\left(-1\right)^{\frac{1}{2}L_{\lambda}}$,
$3^{\frac{1}{2}L_{\lambda}}$ and $11^{\frac{1}{2}L_{\lambda}}$ with $\lambda=x,y,z$.

To conclude, we note that our formalism allows us to explicitly calculate the GSD of each twisted checkerboard model, which is also stable to local perturbations by the argument in the second-to-last paragraph of Sec.~\ref{subsec:Top_charge} using the results in Sec.~\ref{subsec:checker_particle}.
Once again,
we emphasize that the dependence of the GSD on the system size clearly reflect the geometric nature of gapped three-dimensional fracton orders. 

\section{Quasiparticles in twisted X-cube models}
\label{sec:QPXcube}

We have seen that there is a lot of freedom in twisting the X-cube model
by 3-cocycles. Below, by an X-cube model based on an Abelian group
$G$, we refer to any of these twisted versions, including the original
untwisted one. We are going to develop a universal method for analyzing
the properties of quasiparticles in these models. Technically, by
a quasiparticle, we mean a finite excited region. Without loss of
generality, we can simply study excited cuboids.

To study all possible excited states of a cuboid $\mathtt{C}$, such as
the grey one of size $2\times2\times2$ in the center of Fig.~\ref{fig:XC-cage-1-1},
we remove all the requirements $P_{v}=1$ for $v\in\mathtt{C}$. In
addition, we would like that the other excitations are far away from
$\mathtt{C}$. So we pick a much larger cuboid $\mathtt{C}^{\prime}$
containing $\mathtt{C}$ deep inside, as shown in Fig.~\ref{fig:XC-cage-1-1},
and  study the Hilbert subspace selected by $P_{v}=1$ for $v\in\mathtt{C}^{\prime}-\mathtt{C}^{\circ}$,
where $\mathtt{C}^{\circ}$ is the interior of $\mathtt{C}$. 

Such a Hilbert subspace describes an isolated excited cuboid $\mathtt{C}$
and it may be decomposed into more than one irreducible sector according
to the actions of all local operators near $\mathtt{C}$, which leads
to the notation of \emph{particle type}. An excited spot (\emph{i.e.},
\emph{quasiparticle}) is called \emph{simple} if it is already projected
into a definite particle type, which cannot be changed locally. In
the following, we will work out the classification of particle types in the twisted fracton models.
It turns out that each particle type can be labeled by the $x$, $y$
and $z$ \emph{topological charges} subject to some constraints. Then
the fusion of topological charges can be described by the coproduct
of $\mathcal{D}^{\omega}\left(G\right)$. 

Further, we notice that a quasiparticle is mobile in the $x$ (resp.
$y$, $z$) direction if and only if its $x$ (resp. $y$, $z$) topological
charge is trivial. A quasiparticle is called a \emph{fracton} if it
is not a fusion result of mobile quasiparticles. Necessarily, a fracton has to be immobile; it has non-trivial topological charges in all three directions. In addition, we will
also describe some novel braiding processes of mobile quasiparticles with restricted mobilities in this section.

\subsection{Particle type and topological charges}\label{subsec:X-cube_particle}

\begin{figure}
\includegraphics[width=0.8\columnwidth]{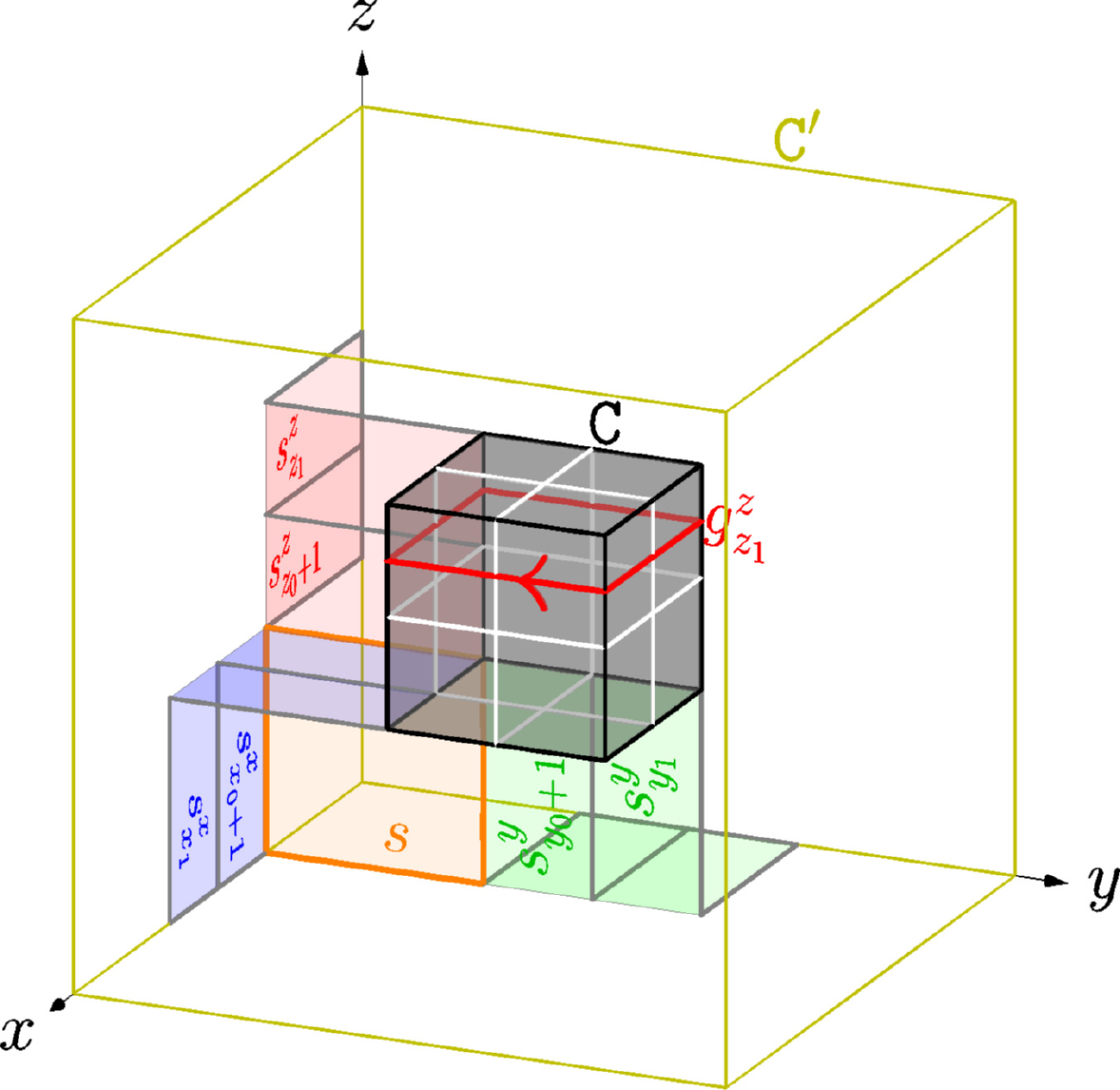}

\caption{An excited cuboid $\mathtt{C}=\left[x_{0},x_{1}\right]\times\left[y_{0},y_{1}\right]\times\left[z_{0},z_{1}\right]$
isolated from other excitations outside $\mathtt{C}'$ in an X-cube
model. The Hilbert space is spanned by the states
$\left|\chi,\overline{\chi};s,D_{\boldsymbol{g}}^{\boldsymbol{s}}\right\rangle $ with 
$s$, $\boldsymbol{s}=(s_{x_{0}+1}^{x},\cdots,s_{x_{1}}^{x},s_{y_{0}+1}^{y},\cdots,s_{y_{1}}^{y},s_{z_{0}+1}^{z},\cdots,s_{z_{1}}^{z})
$ specifying the sums of group elements on the faces in the corresponding membranes
(colored orange, blue, green and red online), $\boldsymbol{g}=\left(g_{x_{0}+1}^{x},\cdots,g_{x_{1}}^{x},g_{y_{0}+1}^{y},\cdots,g_{y_{1}}^{y},g_{z_{0}+1}^{z},\cdots,g_{z_{1}}^{z}\right)$
describing fluxes around $\partial\mathtt{C}$ and 
$\chi$ (resp. $\overline{\chi}$) coloring $\partial\mathtt{C}'$ (resp. $\partial\mathtt{C}$) compatible with fluxes $\boldsymbol{g}$.}

\label{fig:XC-cage-1-1}
\end{figure}

Let $\mathtt{C}=\left[x_{0},x_{1}\right]\times\left[y_{0},y_{1}\right]\times\left[z_{0},z_{1}\right]$
be a generic cuboid and $\mathtt{C}'=\left[x_{0}^{\prime},x_{1}^{\prime}\right]\times\left[y_{0}^{\prime},y_{1}^{\prime}\right]\times\left[z_{0}^{\prime},z_{1}^{\prime}\right]$
a much larger cube containing $\mathtt{C}$, as shown in Fig.~\ref{fig:XC-cage-1-1}.
Further, let $\mathtt{M}=\mathtt{C}'-\mathtt{C}^{\circ}$, where $\mathtt{X}^{\circ}$
denotes the interior of any topological space $\mathtt{X}$. Then
$\mathtt{M}$ is a three-dimensional manifold with boundary. We denote the set of cubes (resp.
faces, edges, vertices) inside $\mathtt{M}$ by $\Lambda^{3}\left(\mathtt{M}\right)$ (resp.
$\Lambda^{2}\left(\mathtt{M}\right)$, $\Lambda^{1}\left(\mathtt{M}\right)$,
$\Lambda^{0}\left(\mathtt{M}\right)$). Let $\mathcal{H}(\Lambda^{2}\left(\mathtt{M}\right), G)$
be the Hilbert space describing all the physical degrees of freedom
on $\mathtt{M}$. To classify generic excitations within $\mathtt{C}$,
we need to analyze the subspace of $\mathcal{H}(\Lambda^{2}\left(\mathtt{M}\right), G)$
selected by the projector 
\begin{equation}
P\left(\mathtt{M}\right)\coloneqq\prod_{v\in\Lambda^{0}\left(\mathtt{M}^{\circ}\right)}P_{v}.
\end{equation}
Let $\mathcal{H}_{0}\left(\mathtt{M}\right)$ denote this subspace, \emph{i.e.}, the image of $P\left(\mathtt{M}\right)$.

Let $\mathtt{M}_{i}^{x}$ with $i\in\left\{ x_{0}^{\prime}+1,x_{0}^{\prime}+2,\cdots,x_{1}^{\prime}\right\} $
(resp. $\mathtt{M}_{j}^{y}$ with $j\in\left\{ y_{0}^{\prime}+1,y_{0}^{\prime}+2,\cdots,y_{1}^{\prime}\right\} $,
$\mathtt{M}_{k}^{z}$ with $k\in\left\{ z_{0}^{\prime}+1,z_{0}^{\prime}+2,\cdots,z_{1}^{\prime}\right\} $)
be the intersection of $\mathtt{M}$ with the plane $x=i-\frac{1}{2}$
(resp. $y=j-\frac{1}{2}$, $z=k-\frac{1}{2}$), \emph{i.e.}, the region of $\Sigma_{i}^{x}$ (resp. $\Sigma_{j}^{y}$, $\Sigma_{k}^{z}$) inside $\mathtt{M}$.
As in Fig.~\ref{fig:tqd_annulus}(b),
we embed $\mathtt{M}_{i}^{x}$ (resp. $\mathtt{M}_{j}^{y}$, $\mathtt{M}_{k}^{z}$)
into a triangulated annulus $\overline{\mathtt{M}}_{i}^{x}$ (resp.
$\overline{\mathtt{M}}_{j}^{y}$, $\overline{\mathtt{M}}_{k}^{z}$).
If the plane does not cut $\mathtt{C}$, then $\overline{\mathtt{M}}_{i}^{x}$
(resp. $\overline{\mathtt{M}}_{j}^{y}$, $\overline{\mathtt{M}}_{k}^{z}$)
reduces to a topological sphere. We pick the base point of the outer/inner
boundary of $\overline{\mathtt{M}}_{i}^{x}$ (resp. $\overline{\mathtt{M}}_{j}^{y}$,
$\overline{\mathtt{M}}_{k}^{z}$) to be in the line $\left(y,z\right)=\left(y_{0}^{\prime},z_{0}^{\prime}\right)/\left(y_{0},z_{0}\right)$
(resp. $\left(z,x\right)=\left(z_{0}^{\prime},x_{0}^{\prime}\right)/\left(z_{0},x_{0}\right)$,
$\left(x,y\right)=\left(x_{0}^{\prime},y_{0}^{\prime}\right)/\left(x_{0},y_{0}\right)$). 

For convenience, we write $\mathtt{C}^{x}\coloneqq\{x_{0}+1,x_{0}+2,\cdots,x_{1}\}$,
$\mathtt{C}^{y}\coloneqq\{y_{0}+1,y_{0}+2,\cdots,y_{1}\}$ and $\mathtt{C}^{z}\coloneqq\{z_{0}+1,z_{0}+2,\cdots,z_{1}\}$.
Let $g_{i}^{x}$ (resp. $g_{j}^{y}$, $g_{k}^{z}$) be the group element
associated with the inner boundary of $\overline{\mathtt{M}}_{i}^{x}$
(resp. $\overline{\mathtt{M}}_{j}^{y}$, $\overline{\mathtt{M}}_{k}^{z}$).
For $i\notin\mathtt{C}^{x}$ (resp. $j\notin\mathtt{C}^{y}$, $k\notin\mathtt{C}^{z}$),
we write $g_{i}^{x}=0$ (resp. $g_{j}^{y}=0$, $g_{k}^{z}=0$) because
$\partial\overline{\mathtt{M}}_{i}^{x}=\emptyset$ (resp. $\partial\overline{\mathtt{M}}_{j}^{y}=\emptyset$,
$\partial\overline{\mathtt{M}}_{k}^{z}=\emptyset$). Hence, to describe
the fluxes, we need 
\begin{align}
\boldsymbol{g} & \coloneqq\left(\boldsymbol{g}^{x},\boldsymbol{g}^{y},\boldsymbol{g}^{z}\right)\in G^{\mathtt{C}^{x}}\times G^{\mathtt{C}^{y}}\times G^{\mathtt{C}^{z}},\label{eq:f}\\
\boldsymbol{g}^{x} & \coloneqq\left(g_{x_{0}+1}^{x},g_{x_{0}+2}^{x},\cdots,g_{x_{1}}^{x}\right)\in G^{\mathtt{C}^{x}},\label{eq:fx}\\
\boldsymbol{g}^{y} & \coloneqq\left(g_{y_{0}+1}^{y},g_{y_{0}+2}^{y},\cdots,g_{y_{1}}^{y}\right)\in G^{\mathtt{C}^{y}},\label{eq:fy}\\
\boldsymbol{g}^{z} & \coloneqq\left(g_{z_{0}+1}^{z},g_{z_{0}+2}^{z},\cdots,g_{z_{1}}^{z}\right)\in G^{\mathtt{C}^{z}}.\label{eq:fz}
\end{align}
Often, $\boldsymbol{g}$ (resp.
$\boldsymbol{g}^{x}$) is also written as $\left(g_{n}^{\mu}\right)_{n\in\mathtt{C}^{\mu}}^{\mu=x,y,z}$
(resp. $\left(g_{i}^{x}\right)_{i\in\mathtt{C}^{x}}$).
These data are subject to the constraint
\begin{equation}
\sum_{\mu=x,y,z}\sum_{n\in\mathtt{C}^{\mu}}g_{n}^{\mu}=0.\label{eq:valid_flux}
\end{equation}
We denote the set of all allowed values of $\boldsymbol{g}$ by $F\left(\mathtt{C}\right)$.
As a group, $F\left(\mathtt{C}\right)$ is isomorphic to $G^{x_{1}-x_{0}+y_{1}-y_{0}+z_{1}-z_{0}-1}$. 

Using the triangulations of $\overline{\mathtt{M}}_{i}^{x}$,
$\overline{\mathtt{M}}_{j}^{y}$ and $\overline{\mathtt{M}}_{k}^{z}$, 
we define a set of vectors forming an orthonormal basis of $\mathcal{H}_{0}(\mathtt{M})$ by 
\begin{equation}
\left|\chi,\overline{\chi};s,\mathtt{D}_{\boldsymbol{g}}^{\boldsymbol{s}}\right\rangle \coloneqq\sum_{\vartheta\in G_{B}^{\Lambda^{2}\left(\mathtt{M}\right)}\left(s,\chi,\overline{\chi}\right)}\frac{Z\left(\vartheta;\mathtt{D}_{\boldsymbol{g}}^{\boldsymbol{s}}\right)}{\left|G\right|^{\frac{1}{2}\left|\Lambda^{0}\left(\mathtt{M}^{\circ}\right)\right|}}\left|\vartheta\right\rangle \label{eq:xcube_exicitation}
\end{equation}
with $\boldsymbol{g}\in F(\mathtt{C})$, $s\in G$, 
$\boldsymbol{s}\in G^{\mathtt{C}^{x}}\times G^{\mathtt{C}^{y}}\times G^{\mathtt{C}^{z}}$ and 
$\chi \in G^{\partial\mathtt{C'}}$, $\overline{\chi} \in G^{\partial\mathtt{C}}$ compatible with $\boldsymbol{g}$.
In detail, $s$ specifies the sum of group elements on the faces
in the lower left square region (orange online) and $G_{B}^{\Lambda^{2}\left(\mathtt{M}\right)}\left(s,\chi,\overline{\chi}\right)$
denotes the set of $\vartheta\in G_{B}^{\Lambda^{2}\left(\mathtt{M}\right)}$
compatible with $s$ and coinciding with $\overline{\chi}$, $\chi$ on $\partial\mathtt{C}$,
$\partial\mathtt{C}'$. In addition, 
\begin{multline}
Z\left(\vartheta;\mathtt{D}_{\boldsymbol{g}}^{\boldsymbol{s}}\right)\coloneqq\prod_{i=x_{0}^{\prime}+1}^{x_{1}^{\prime}}Z_{i}^{x}\left(\vartheta;\mathtt{D}_{\boldsymbol{g}}^{\boldsymbol{s}}\right)\cdot\\
\prod_{j=y_{0}^{\prime}+1}^{y_{1}^{\prime}}Z_{j}^{y}\left(\vartheta;\mathtt{D}_{\boldsymbol{g}}^{\boldsymbol{s}}\right)\cdot\prod_{k=z_{0}^{\prime}+1}^{z_{1}^{\prime}}Z_{k}^{z}\left(\vartheta;\mathtt{D}_{\boldsymbol{g}}^{\boldsymbol{s}}\right),\label{eq:Z_xcube}
\end{multline}
where $Z_{n}^{\mu}(\vartheta;\mathtt{D}_{\boldsymbol{g}}^{\boldsymbol{s}})$
is the Dijkgraaf-Witten partition function of a ball with surface $-\overline{\mathtt{M}}_{n}^{\mu}$
for $n\notin\mathtt{C}^{\mu}$ or a solid torus with surface $(-\overline{\mathtt{M}}_{n}^{\mu})\cup_{\varpi\overline{\varpi}}\mathtt{D}_{g_{n}^{\mu}}^{s_{n}^{\mu}}$
(as in Fig.~\ref{fig:D_ht}(b)) for $n\in\mathtt{C}^{\mu}$ in the coloring specified by $\vartheta$.
The minus sign before $\overline{\mathtt{M}}_{n}^{\mu}$ means that the orientation
of $\overline{\mathtt{M}}_{n}^{\mu}$ is pointing toward the inside
of the solid according to the right hand rule. 

To manipulate the states within $\mathcal{H}_{0}\left(\mathtt{M}\right)$,
we can define a collection of operators 
$P_{v}^{g}$ for $v\in \Lambda^{0}\left(\partial\mathtt{C}\right)$ (resp.  $v\in \Lambda^{0}\left(\partial\mathtt{C'}\right)$), commuting with $P(\mathtt{M})$ and supported near $\partial\mathtt{C}$ (reps. $\partial\mathtt{C'}$),
by Eq.~\eqref{eq:Xcube_P} with $\Lambda$ replaced by
$\mathtt{M}$ and using the triangulations of $\overline{\mathtt{M}}_{i}^{x}$,
$\overline{\mathtt{M}}_{j}^{y}$, $\overline{\mathtt{M}}_{k}^{z}$.
Clearly, $\overline{\chi}$ and $\chi$ can be manipulated by $P_{v}^{g}$ for $v\in \Lambda^{0}\left(\partial\mathtt{C}\right)$ and  $v\in \Lambda^{0}\left(\partial\mathtt{C'}\right)$ respectively. Thus, they are local degrees of freedom and can be neglected in the discussion of particle types. The reduced Hilbert space, denoted by $\mathcal{H}_{*}(\overline{\mathtt{M}})$, is spanned by
$\left|s,\mathtt{D}_{\boldsymbol{g}}^{\boldsymbol{s}}\right\rangle $
with $\boldsymbol{g} \in F(\mathtt{C})$, $s\in G$ and $\boldsymbol{s}\in G^{\mathtt{C}^{x}}\times G^{\mathtt{C}^{y}}\times G^{\mathtt{C}^{z}}$.

As in the twisted quantum double models, we can define states
$\left|s,\mathtt{D}_{\boldsymbol{g}}^{\boldsymbol{s}}\mathtt{D}_{\boldsymbol{h}}^{\boldsymbol{t}}\right\rangle $
by replacing $\mathtt{D}_{\boldsymbol{g}}^{\boldsymbol{s}}$ by $\mathtt{D}_{\boldsymbol{g}}^{\boldsymbol{s}}\mathtt{D}_{\boldsymbol{h}}^{\boldsymbol{t}}$
in Eqs.~\eqref{eq:xcube_exicitation} and \eqref{eq:Z_xcube}. 
Analogously, we have
\begin{equation}
\left|s,\mathtt{D}_{\boldsymbol{g}}^{\boldsymbol{s}}\mathtt{D}_{\boldsymbol{h}}^{\boldsymbol{t}}\right\rangle =\delta_{\boldsymbol{g},\boldsymbol{h}}\prod_{\mu,n}\omega_{n,g_{n}^{\mu}}^{\mu}\left(s_{n}^{\mu},t_{n}^{\mu}\right)\left|s,\mathtt{D}_{\boldsymbol{g}}^{\boldsymbol{s}+\boldsymbol{t}}\right\rangle. 
\end{equation}
This motivates us to consider the algebra
\begin{align}
\mathcal{D}\left[\mathtt{C}\right] & \coloneqq\mathbb{C}G\otimes\mathcal{D}_{x}\left[\mathtt{C}\right]\otimes\mathcal{D}_{y}\left[\mathtt{C}\right]\otimes\mathcal{D}_{z}\left[\mathtt{C}\right]
\end{align}
with each factor $\mathcal{D}_{\mu}\left[\mathtt{C}\right]$ and its basis given by
\begin{align}
\mathcal{D}_{\mu}\left[\mathtt{C}\right]\coloneqq\bigotimes_{n\in\mathtt{C}^{\mu}}\mathcal{D}^{\omega_{n}^{\mu}}\left(G\right), &  & D_{\boldsymbol{g}^{\mu}}^{\boldsymbol{s}^{\mu}}\coloneqq\bigotimes_{n\in\mathtt{C}^{\mu}}D_{g_{n}^{\mu}}^{s_{n}^{\mu}},
\end{align}
$\forall \mu=x,y,z$.
For short, we write $D_{\boldsymbol{g}}^{\boldsymbol{s}}\coloneqq D_{\boldsymbol{g}^{x}}^{\boldsymbol{s}^{x}}\otimes D_{\boldsymbol{g}^{y}}^{\boldsymbol{s}^{y}}\otimes D_{\boldsymbol{g}^{z}}^{\boldsymbol{s}^{z}}$,
where $\boldsymbol{g}=\left(\boldsymbol{g}^{x},\boldsymbol{g}^{y},\boldsymbol{g}^{z}\right), 
\boldsymbol{s}=\left(\boldsymbol{s}^{x},\boldsymbol{s}^{y},\boldsymbol{s}^{z}\right) \in G^{\mathtt{C}^{x}}\times G^{\mathtt{C}^{y}}\times G^{\mathtt{C}^{z}}$. 

In addition, $\forall t\in G$, we have operators
\begin{align}
P_{z\geq k}^{t}&\coloneqq\prod_{v\in\Lambda^{0}\left(\partial\mathtt{C}^{\prime},k\leq z\leq z_{1}\right)}P_{v}^{t},\label{eq:Xcube_Zalpha-1}\\
\overline{P}_{z\geq k}^{t}&\coloneqq\prod_{v\in\Lambda^{0}\left(\partial\mathtt{C},z\geq k\right)}\left(P_{v}^{t}\right)^{\dagger},\label{eq:Xcube_Zbeta-1}
\end{align}
with $\Lambda^{0}\left(\partial\mathtt{C},z\geq k\right)\coloneqq\left\{ \left(x,y,z\right)\in\Lambda^{0}\left(\partial\mathtt{C}\right)|z\geq k\right\} $~and $\Lambda^{0}\left(\partial\mathtt{C}',k\leq z\leq z_{1}\right)\coloneqq\left\{ \left(x,y,z\right)\in\Lambda^{0}\left(\partial\mathtt{C}'\right)|k\leq z\leq z_{1}\right\}  $.
which do not change $\chi$, $\overline{\chi}$. They act on $\mathcal{H}_{*}\left(\overline{\mathtt{M}}\right)$
as
\begin{align}
P_{z\geq k}^{t}\left|s,\mathtt{D}_{\boldsymbol{g}}^{\boldsymbol{s}}\right\rangle  & =\begin{cases}
\left|t+s,\mathtt{D}_{\boldsymbol{g}}^{\boldsymbol{s}}\right\rangle , & k\leq z_{0},\vphantom{\frac{1}{\frac{2}{3}}}\\
\left|s,\mathtt{D}_{\boldsymbol{g}}^{t\delta_{k}^{z}}\mathtt{D}_{\boldsymbol{g}}^{\boldsymbol{s}}\right\rangle , & z_{0}<k\leq z_{1},\\
\left|s,\mathtt{D}_{\boldsymbol{g}}^{\boldsymbol{s}}\right\rangle , & k>z_{1},\vphantom{\frac{\frac{1}{2}}{2}}
\end{cases}\label{eq:Pzk}\\
\overline{P}_{z\geq k}^{t}\left|s,\mathtt{D}_{\boldsymbol{g}}^{\boldsymbol{s}}\right\rangle  & =\begin{cases}
\left|s+t,\mathtt{D}_{\boldsymbol{g}}^{\boldsymbol{s}}\right\rangle , & k\leq z_{0},\vphantom{\frac{1}{\frac{2}{3}}}\\
\left|s,\mathtt{D}_{\boldsymbol{g}}^{\boldsymbol{s}}\mathtt{D}_{\boldsymbol{g}}^{t\delta_{k}^{z}}\right\rangle , & z_{0}<k\leq z_{1},\\
\left|s,\mathtt{D}_{\boldsymbol{g}}^{\boldsymbol{s}}\right\rangle , & k>z_{1}.\vphantom{\frac{\frac{1}{2}}{2}}
\end{cases}
\end{align}
Similarly, replacing $z\geq k$ by $x\geq i$ and $y\geq j$, we have
operators $P_{x\geq i}^{h}$, $P_{y\geq j}^{h}$ supported near $\partial\mathtt{C}^{\prime}$
and $\overline{P}_{x\geq i}^{h}$, $\overline{P}_{y\geq j}^{h}$ supported
near $\partial\mathtt{C}$. Moreover, there is clearly a projector $T_{\boldsymbol{h}}$ (resp. $\overline{T}_{\boldsymbol{h}}$)
supported on $\partial\mathtt{C}^{\prime}$ (resp. $\partial\mathtt{C}$)
that acts as
\begin{equation}
\overline{T}_{\boldsymbol{h}}\left|s,\mathtt{D}_{\boldsymbol{g}}^{\boldsymbol{s}}\right\rangle
=
T_{\boldsymbol{h}}\left|s,\mathtt{D}_{\boldsymbol{g}}^{\boldsymbol{s}}\right\rangle =\delta_{\boldsymbol{h},\boldsymbol{g}}\left|s,\mathtt{D}_{\boldsymbol{g}}^{\boldsymbol{s}}\right\rangle .
\end{equation}

In terms of these operators, we can define a left action $\pi$ and a right action $\overline{\pi}$
of $\mathcal{D}\left(\mathtt{C}\right)$ on $\mathcal{H}_{0}\left(\overline{\mathtt{M}}\right)$ as 
\begin{align}
\pi\left(s\otimes D_{\boldsymbol{g}}^{\boldsymbol{s}}\right) & \coloneqq T_{\boldsymbol{g}}P_{z\geq z_{0}}^{s}\prod_{i\in\mathtt{C}^{x}}P_{x\geq i}^{s_{i}^{x}}\prod_{j\in\mathtt{C}^{y}}P_{y\geq j}^{s_{j}^{y}}\prod_{k\in\mathtt{C}^{z}}P_{z\geq k}^{s_{k}^{z}},\label{eq:DC}\\
\overline{\pi}\left(s\otimes D_{\boldsymbol{g}}^{\boldsymbol{s}}\right) & \coloneqq \overline{T}_{\boldsymbol{g}}\overline{P}_{z\geq z_{0}}^{s}\prod_{i\in\mathtt{C}^{x}}\overline{P}_{x\geq i}^{s_{i}^{x}}\prod_{j\in\mathtt{C}^{y}}\overline{P}_{y\geq j}^{s_{j}^{y}}\prod_{k\in\mathtt{C}^{z}}\overline{P}_{z\geq k}^{s_{k}^{z}},
\end{align}
$\forall s\in G, \forall \boldsymbol{s},\boldsymbol{g}\in
G^{\mathtt{C}^{x}}\times G^{\mathtt{C}^{y}}\times G^{\mathtt{C}^{z}}$. By construction, 
\begin{align}
\pi\left(s\otimes D_{\boldsymbol{g}}^{\boldsymbol{s}}\right)\left|t,\mathtt{D}_{\boldsymbol{h}}^{\boldsymbol{t}}\right\rangle  & =\left|s+t,\mathtt{D}_{\boldsymbol{g}}^{\boldsymbol{s}}\mathtt{D}_{\boldsymbol{h}}^{\boldsymbol{t}}\right\rangle ,\\
\overline{\pi}\left(s\otimes D_{\boldsymbol{g}}^{\boldsymbol{s}}\right)\left|t,\mathtt{D}_{\boldsymbol{h}}^{\boldsymbol{t}}\right\rangle  & =\left|t+s,\mathtt{D}_{\boldsymbol{h}}^{\boldsymbol{t}}\mathtt{D}_{\boldsymbol{g}}^{\boldsymbol{s}}\right\rangle .
\end{align}
Thus, $\mathcal{H}_{*}\left(\overline{\mathtt{M}}\right)$ is equivalent
to $\mathcal{A}\left[\mathtt{C}\right]$ as a $\mathcal{D}\left(\mathtt{C}\right)$-$\mathcal{D}\left(\mathtt{C}\right)$
bimodule by the obvious map
\begin{equation}
\mathcal{H}_{*}\left(\overline{\mathtt{M}}\right)
\xrightarrow{\sim}\mathcal{A}\left[\mathtt{C}\right]
:\,
\left|s,\mathtt{D}_{\boldsymbol{g}}^{\boldsymbol{s}}\right\rangle
\mapsto
s\otimes D_{\boldsymbol{g}}^{\boldsymbol{s}} ,\label{eq:isometry}
\end{equation}
where $\mathcal{A}\left[\mathtt{C}\right]$ is the subalgebra of
$\mathcal{D}\left[\mathtt{C}\right]$ spanned by $s\otimes D_{\boldsymbol{g}}^{\boldsymbol{s}}$
with $\boldsymbol{g}$ constrained by Eq.~\eqref{eq:valid_flux},
\emph{i.e.}, $\boldsymbol{g}\in F\left(\mathtt{C}\right)$. 

Since both $\mathbb{C}G$ and $\mathcal{D}^{\omega_{n}^{\mu}}\left(G\right)$ are semisimple, 
\begin{equation}
\rho=\bigoplus_{\left(q,\mathfrak{a}_{n}^{\mu}\right)_{n\in\mathtt{C}^{\mu}}^{\mu=x,y,z}}\varrho_{q}\otimes\bigotimes_{\mu,n}\rho_{\mathfrak{a}_{n}^{\mu}}
\end{equation} 
gives an isomorphism of algebras
\begin{equation}
\mathcal{D}\left[\mathtt{C}\right]\simeq\bigoplus_{\left(q,\mathfrak{a}_{n}^{\mu}\right)_{n\in\mathtt{C}^{\mu}}^{\mu=x,y,z}}\mathcal{L}\left(\mathcal{V}_{q}\right)\otimes\bigotimes_{\mu,n}\mathcal{L}\left(\mathcal{V}_{\mathfrak{a}_{n}^{\mu}}\right).\label{eq:DC_sectors}
\end{equation}
In detail, the character group $\widehat{G}$ of a group $G$ collects all its one-dimensional representations and 
$\mathcal{V}_{q}=(\varrho_{q},V_{q})$ is a representation corresponding to $q\in\widehat{G}$ acting on Hilbert space $V_{q}$. Moreover, $\mathfrak{a}_{n}^{\mu}$ labels equivalent classes
of irreducible representations of $\mathcal{D}^{\omega_{n}^{\mu}}\left(G\right)$ and 
$\mathcal{V}_{\mathfrak{a}_{n}^{\mu}}=\left(\rho_{\mathfrak{a}_{n}^{\mu}},V_{\mathfrak{a}_{n}^{\mu}}\right)$ is an explicit representation on a Hilbert space $V_{\mathfrak{a}_{n}^{\mu}}$ corresponding to $\mathfrak{a}_{n}^{\mu}$. Explicitly, 
$\mathfrak{a}_{n}^{\mu}$
is specified by a pair $\left(g_{n}^{\mu},\varrho_{n}^{\mu}\right)$ with $g_{n}^{\mu}\in G$ describing the flux and $\varrho_{n}^{\mu}$ an irreducible $\omega_{n,g_{n}^{\mu}}^{\mu}$-representation (up
to isomorphism) of $G$. Refer to Appendix~\ref{subsec:Reps_DG}
for details of these representations. 

\begin{figure}
	\includegraphics[width=0.8\columnwidth]{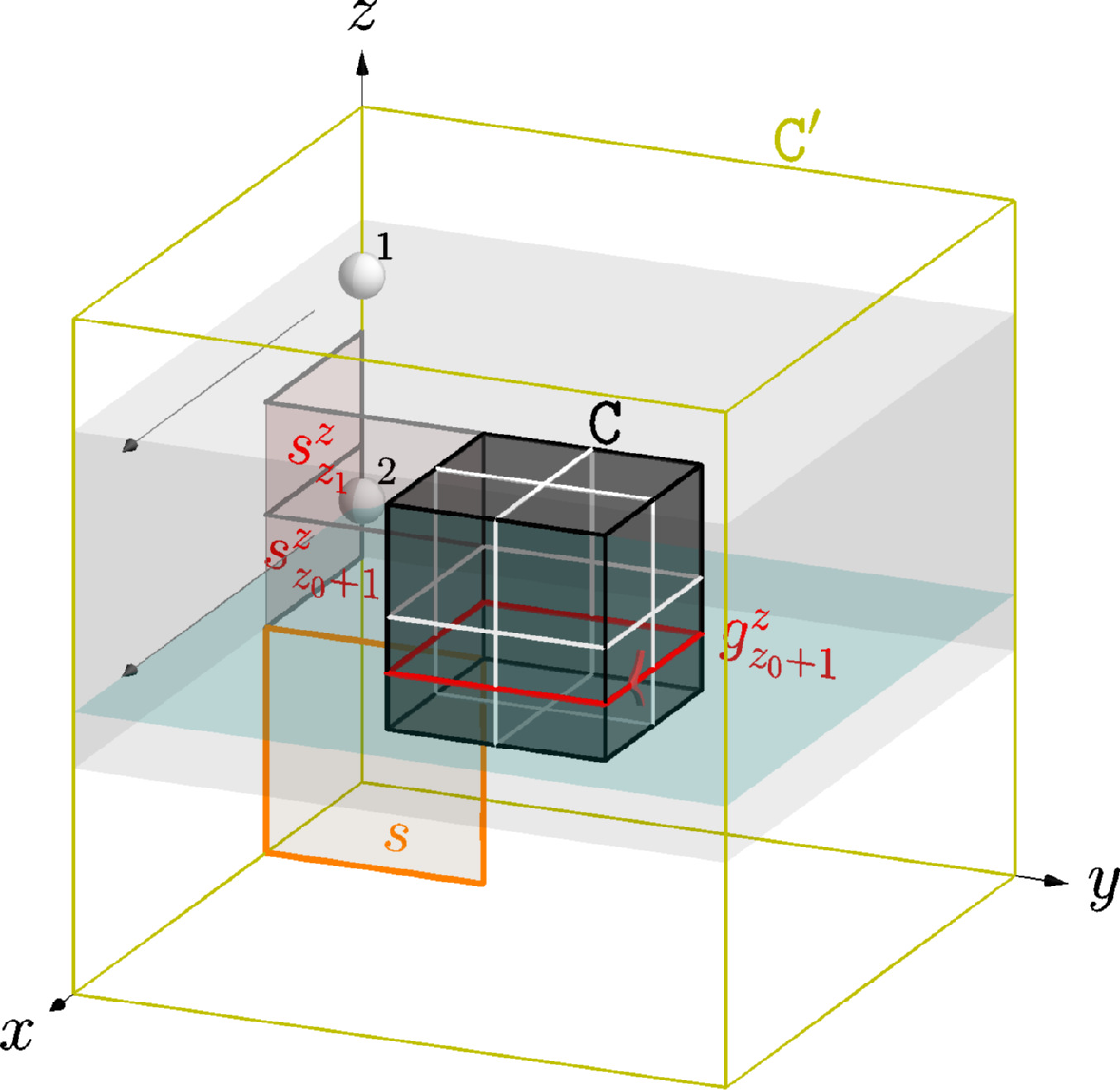}
	
	\caption{A $z$ topological charge $\mathfrak{a}_{k}^{z}$ of an excited cuboid
		$\mathtt{C}=\left[x_{0},x_{1}\right]\times\left[y_{0},y_{1}\right]\times\left[z_{0},z_{1}\right]$
		can be detected by braiding a pair of quasiparticles along $\partial\mathtt{C'}$ with operator supported near the grey region in $\partial\mathtt{C}'$.
		Quasiparticle $1$ is kept above $z=z_{1}$ and quasiparticle $2$ moves near the plane $z=k$ (cyan online).}
	
	\label{fig:topcharges}
\end{figure}

Denote the set of  $\boldsymbol{\mathfrak{a}}=(q,\mathfrak{a}_{n}^{\mu})_{n\in\mathtt{C}^{\mu}}^{\mu=x,y,z}=\left(q,g_{n}^{\mu},\varrho_{n}^{\mu}\right)_{n\in\mathtt{C}^{\mu}}^{\mu=x,y,z}$
with $\left(g_{n}^{\mu}\right)_{n\in\mathtt{C}^{\mu}}^{\mu=x,y,z}\in F\left(\mathtt{C}\right)$ by $\mathfrak{Q}\left[\mathtt{C}\right]$. 
Then the composition
\begin{align}
\mathcal{H}_{*}\left(\overline{\mathtt{M}}\right)\xrightarrow[\sim]{\left|s,\mathtt{D}_{\boldsymbol{g}}^{\boldsymbol{s}}\right\rangle \mapsto s\otimes D_{\boldsymbol{g}}^{\boldsymbol{s}}}\mathcal{A}\left[\mathtt{C}\right] & \xrightarrow[\sim]{\tilde{\rho}}\nonumber \\
\bigoplus_{\boldsymbol{\mathfrak{a}}\in\mathfrak{Q}\left[\mathtt{C}\right]}\mathcal{L}\left(\mathcal{V}_{q}\right)\otimes\bigotimes_{\mu,n}\mathcal{L}\left(\mathcal{V}_{\mathfrak{a}_{n}^{\mu}}\right) & =\bigoplus_{\boldsymbol{\mathfrak{a}}\in\mathfrak{Q}\left[\mathtt{C}\right]}\mathcal{V}_{\boldsymbol{\mathfrak{a}}}\otimes\mathcal{V}_{\boldsymbol{\mathfrak{a}}}^{*}
\end{align}
is an ismorphism of Hilbert spaces respecting both left and right actions of $\mathcal{D}\left[\mathtt{C}\right]$, where
\begin{gather}
\tilde{\rho} 
\coloneqq\bigoplus_{\boldsymbol{\mathfrak{a}}\in\mathfrak{Q}\left[\mathtt{C}\right]}\frac{1}{\sqrt{\left|G\right|}}\varrho_{q}\otimes\bigotimes_{\mu,n}\sqrt{\frac{\dim_{\mathbb{C}}\mathcal{V}_{\mathfrak{a}_{n}^{\mu}}}{\left|G\right|}}\rho_{\mathfrak{a}_{n}^{\mu}},\label{eq:rho_xcube}\\
\mathcal{V}_{\boldsymbol{\mathfrak{a}}} 
\coloneqq\mathcal{V}_{q}\otimes\bigotimes_{\mu,n}\mathcal{V}_{\mathfrak{a}_{n}^{\mu}}.\label{eq:V_1}
\end{gather}
The normalization for each sector in $\tilde{\rho}$ is picked such that the inner product structure is respected.
Clearly, $\mathfrak{Q}\left[\mathtt{C}\right]$ labels particle types of the excited cuboid $\mathtt{C}$ and 
$\mathcal{V}_{\boldsymbol{\mathfrak{a}}}$ (resp. $\mathcal{V}_{\boldsymbol{\mathfrak{a}}}^{*}$)
describes the degrees of freedom near $\partial\mathtt{C'}$ (resp. $\partial\mathtt{C}$).
Physically, $\mathfrak{a}_{k}^{z}=(g_{k}^{z},\varrho_{k}^{z})$ can be detected by braiding a pair of quasiparticles
in the $x$ and $y$ directions via operator supported near grey region in $\partial\mathtt{C'}$ as in Fig.~\ref{fig:topcharges}. Thus, $\mathfrak{a}_{k}^{z}$ is called a $z$ topological charge.
Actually, $q$ can also be viewed as a $z$ topological charge, since  $\mathfrak{a}_{k}^{z}$ reduces to $(0,\varrho_{q})$ when quasiparticle $2$ is lowered below $z=z_0$.
Similarly, $\mathfrak{a}_{i}^{x}$ (resp. $\mathfrak{a}_{j}^{y}$) can be detected by braiding processes in the $y,z$ (resp. $z,x$) directions and is called a $x$ (resp. $y$) topological charge. Also, $q$ can be viewed as an $x$ and a $y$ topological charge.

Distinct from conventional topological orders, the number of allowed particle types of a finite excited region $\mathtt{C}$ in a fracton model increases/decreases as the size of $\mathtt{C}$ grows/shrinks. If a quasiparticle can be localized in a smaller cuboid $\mathtt{C}_{a}=\left[a_{0}^{x},a_{1}^{x}\right]\times\left[a_{0}^{y},a_{1}^{y}\right]\times\left[a_{0}^{z},a_{1}^{z}\right] \subset \mathtt{C}$, then its particle type $\boldsymbol{\mathfrak{a}}=(q,\mathfrak{a}_{n}^{\mu})_{n\in\mathtt{C}^{\mu}}^{\mu=x,y,z} \in \mathfrak{Q}\left[\mathtt{C}\right]$ satisfies
\begin{equation}
\mathfrak{a}_{n}^{\mu}=\begin{cases}
\left(0,\varrho_{q}\right), & n\leq a_{0}^{\mu},\\
\mathfrak{0}, & n>a_{1}^{\mu},
\end{cases}
\end{equation}
$\forall \mu=x,y,z$, where $\mathfrak{0}$ denotes the trivial representation
(\emph{i.e.}, the counit) of any $\mathcal{D}^{\omega_{n}^{\mu}}(G)$.
In other words, $\mathfrak{Q}\left[\mathtt{C}_{a}\right]$ can be viewed as a subset of $\mathfrak{Q}\left[\mathtt{C}\right]$; each 
$\mathcal{V}_{\mathfrak{a}}$ for $\mathfrak{a}\in \mathfrak{Q}\left[\mathtt{C}_{a}\right]$ carries an irreducible representation of $\mathcal{D}\left[\mathtt{C}\right]$ for any cuboid $\mathtt{C}$ containing $\mathtt{C}_{a}$.

\subsection{Fusion of quasiparticles}

Suppose that there are two spatially separated excited cuboids $\mathtt{C}_{a}$
and $\mathtt{C}_{b}$ containing deep inside a much larger cuboid
$\mathtt{C}'$. Let $\mathtt{M}\coloneqq\mathtt{C}^{\prime}-\mathtt{C}_{a}^{\circ}-\mathtt{C}_{b}^{\circ}$.
The discussion in the above subsection can be repeated here for the
two-hole manifold $\mathtt{M}$. With the coloring on $\partial\mathtt{C}'$
and the local degrees of freedom near $\mathtt{C}_{a}$ and $\mathtt{C}_{b}$
fixed, we are left with Hilbert spaces $\mathcal{V}\left[\boldsymbol{\mathfrak{a}},\boldsymbol{\mathfrak{b}}\right]$
labeled by $\boldsymbol{\mathfrak{a}}\in\mathfrak{Q}\left[\mathtt{C}_{a}\right]$
and $\boldsymbol{\mathfrak{b}}\in\mathfrak{Q}\left[\mathtt{C}_{b}\right]$. Using
two copies of Eq.~\eqref{eq:rho_xcube}, we have 
\begin{equation}
\mathcal{V}\left[\boldsymbol{\mathfrak{a}},\boldsymbol{\mathfrak{b}}\right]\simeq\mathcal{V}_{\boldsymbol{\mathfrak{a}}}\otimes\mathcal{V}_{\boldsymbol{\mathfrak{b}}},
\end{equation}
where $\mathcal{V}_{\boldsymbol{\mathfrak{a}}}$ and $\mathcal{V}_{\boldsymbol{\mathfrak{b}}}$
are defined by Eq.~\eqref{eq:V_1}.

All these states can be viewed as an excited cuboid $\mathtt{C}$,
where $\mathtt{C}$ is cuboid containing both $\mathtt{C}_{a}$ and
$\mathtt{C}_{b}$ inside $\mathtt{C}'$. With operators
supported on $\mathtt{C}$, the Hilbert space $\mathcal{V}_{\boldsymbol{\mathfrak{a}}}\otimes\mathcal{V}_{\boldsymbol{\mathfrak{b}}}$
may be further reduced. To determine the total charge of $\mathtt{C}$,
we study the action of $\mathcal{D}\left[\mathtt{C}\right]$ on via
$\pi$ defined in Eq.~\eqref{eq:DC}.  Analogous to Sec.~\ref{subsec:Fusions-and-braidings},
it is specified by the coproduct 
\begin{equation}
\varDelta\coloneqq\varDelta_{o}\otimes\bigotimes_{\mu=x,y,z}\bigotimes_{n\in\mathtt{C}^{\mu}}\varDelta_{n}^{\mu},
\end{equation}
where $\varDelta_{o}:\mathbb{C}G\rightarrow\mathbb{C}G\otimes\mathbb{C}G,\,g\mapsto g\otimes g$
is the default coproduct of $\mathbb{C}G$. The vector space of intertwiners
between the representations $\mathcal{V}_{\boldsymbol{\mathfrak{c}}}$ and $\mathcal{V}_{\boldsymbol{\mathfrak{a}}}\otimes\mathcal{V}_{\boldsymbol{\mathfrak{b}}}$
of $\mathcal{D}\left(\mathtt{C}\right)$
\begin{equation}
V_{\boldsymbol{\mathfrak{c}}}^{\boldsymbol{\mathfrak{a}}\boldsymbol{\mathfrak{b}}}\coloneqq\text{Hom}\left(\mathcal{V}_{\boldsymbol{\mathfrak{c}}},\mathcal{V}_{\boldsymbol{\mathfrak{a}}}\otimes\mathcal{V}_{\boldsymbol{\mathfrak{b}}}\right)
\end{equation}
encodes the ways of fusing $\boldsymbol{\mathfrak{a}}$ and $\boldsymbol{\mathfrak{b}}$ into
$\boldsymbol{\mathfrak{c}}\in\mathfrak{Q}\left[\mathtt{C}\right]$. In particular,
$N_{\boldsymbol{\mathfrak{a}}\boldsymbol{\mathfrak{b}}}^{\boldsymbol{\mathfrak{c}}}\coloneqq\dim_{\mathbb{C}}V_{\boldsymbol{\mathfrak{c}}}^{\boldsymbol{\mathfrak{a}}\boldsymbol{\mathfrak{b}}}$
is the corresponding fusion rule. It is possible to fuse $\boldsymbol{\mathfrak{a}}$
and $\boldsymbol{\mathfrak{b}}$ into $\boldsymbol{\mathfrak{c}}$ if and only if $N_{\boldsymbol{\mathfrak{a}}\boldsymbol{\mathfrak{b}}}^{\boldsymbol{\mathfrak{c}}}\geq1$. Moreover,  $N_{\boldsymbol{\mathfrak{a}}\boldsymbol{\mathfrak{b}}}^{\boldsymbol{\mathfrak{c}}}\geq1$ implies $q_{c}=q_{a}+q_{b}$, where $\boldsymbol{\mathfrak{a}}=(q_{a},\mathfrak{a}_{n}^{\mu})_{n\in \mathtt{C}_{a}^{\mu}}^{\mu}$, 
$\boldsymbol{\mathfrak{b}}=(q_{b},\mathfrak{b}_{n}^{\mu})_{n\in \mathtt{C}_{b}^{\mu}}^{\mu}$ and
$\boldsymbol{\mathfrak{c}}=(q_{c},\mathfrak{c}_{n}^{\mu})_{n\in \mathtt{C}^{\mu}}^{\mu}$.

Similarly to the discussion in Sec.~\ref{subsec:Fusions-and-braidings}, in order to describe
three or more excitations, we need to be careful with their associations. 

\subsection{Mobility of quasiparticles}
\label{subsec:Mobility-Xcube}

Now let us think about moving a quasiparticle from one cuboid $\mathtt{C}_{a}=\left[a_{0}^{x},a_{1}^{x}\right]\times\left[a_{0}^{y},a_{1}^{y}\right]\times\left[a_{0}^{z},a_{1}^{z}\right]$
to another $\mathtt{C}_{b}=\left[b_{0}^{x},b_{1}^{x}\right]\times\left[b_{0}^{y},b_{1}^{y}\right]\times\left[b_{0}^{z},b_{1}^{z}\right]$.
The movement can be made by a local operator if and only if the initial
and final states have the same particle type $\left(q,g_{n}^{\mu},\varrho_{n}^{\mu}\right)_{n\in\mathtt{C}^{\mu}}^{\mu=x,y,z}\in\mathfrak{Q}\left(\mathtt{C}\right)$
as an excited cuboid $\mathtt{C}$, where $\mathtt{C}$ is a larger
cuboid containing both $\mathtt{C}_{a}$ and $\mathtt{C}_{b}$. Because
$\mathtt{C}_{a}\cap\mathtt{C}_{b}=\emptyset$, we have $\left[a_{0}^{\mu},a_{1}^{\mu}\right]\cap\left[b_{0}^{\mu},b_{1}^{\mu}\right]=\emptyset$
for at least one of $\mu=x,y,z$, in which case we say that the quasiparticle
is mobile in the $\mu$ direction. 

For instance, suppose $a_{0}^{z}>b_{1}^{z}$. Then the position
of $\mathtt{C}_{a}$ implies that $P_{z\geq a_{0}^{z}}^{t}$ acts
as $\varrho_{q}\left(t\right)$, while the position of $\mathtt{C}_{b}$
implies that $P_{z\geq a_{0}^{z}}^{t}$ acts trivially. Hence $q\in\widehat{G}$
has to be trivial. Obviously, it follows that $q$ is trivial if the
excited cuboid $\mathtt{C}$ of type $\left(q,g_{n}^{\mu},\varrho_{n}^{\mu}\right)_{n\in\mathtt{C}^{\mu}}^{\mu=x,y,z}\in\mathfrak{Q}\left[\mathtt{C}\right]$
is a fusion result of mobile quasiparticles. In fact, it is not hard
to see that the converse is true as well. Therefore, an excitation
is a \emph{fracton} (\emph{i.e.}, a finite excited region that is
not a fusion result of mobile quasiparticles) if and only if $q$
is not trivial. 

In fact, the mobility of an excited cuboid $\mathtt{C}_{a}$ in the
$z$ direction implies that $\mathfrak{a}_{k}^{z}$ is trivial for
all $k\in\mathtt{C}^{z}$ as well. To see this, we notice that the
operators $\pi(\mathcal{D}_{x}\left[\mathtt{C}_{a}\right])$ in Eq.~\eqref{eq:DC}
are supported near $\partial\mathtt{C}'\cap\left\{ \left(x,y,z\right)|z_{0}\leq z\leq z_{1}\right\} $,
the excitation can be moved away along the $z$ direction without
touching the support region of $\pi(\mathcal{D}_{x}\left[\mathtt{C}_{a}\right])$
and hence $z$ topological charges $\mathfrak{a}_{k}^{z}$ are conserved.
Thus, if $\mathfrak{a}_{k}^{\mu}$ is nontrivial, then it is not possible
to move the excitation away along the $z$ direction. In general,
all $\mu$ topological charges must be trivial in order for a quasiparticle
to be mobile in the $\mu$ direction. In addition, if a quasiparticle
is mobile in two directions, then only topological charges in the
third direction can be nontrivial. This is an important result of our work, since it relates the \emph{mobility} of quasiparticles to their \emph{topological charges}.

\subsection{Braiding of mobile quasiparticles}
\label{subsec:Braiding_general}

If an excited spot 
is mobile in
the $\mu$ direction (resp. in both the $\mu$ and $\nu$ directions),
we call it a $\mu$-particle (resp. $\mu\nu$-particle). 

\subsubsection{Braiding of 2d mobile quasiparticles}

For braiding of 2d mobile quasiparticles (\emph{i.e.}, excitations
mobile in two dimensions), the discussion in Sec.~\ref{subsec:MeasuringTopSpin}
can be repeated. For example, the result of the measurement described by Eq.~\eqref{eq:measuretopspin} involving an exchange of two identical $xy$-particles with $z$
topological charges $\{\mathfrak{a}_{k}^{z}\}_{k\in\mathtt{C}^{z}}$ is
\begin{equation}
\prod_{k\in\mathtt{C}^{z}}\frac{\theta_{\mathfrak{a}_{k}^{z}}}{\dim_{\mathbb{C}}\mathcal{V}_{\mathfrak{a}_{k}^{z}}}=\prod_{k\in\mathtt{C}^{z}}\frac{\text{tr}\left(\wp R_{k}^{z},\mathcal{V}_{\mathfrak{a}_{k}^{z}}\otimes\mathcal{V}_{\mathfrak{a}_{k}^{z}}\right)}{\left(\dim_{\mathbb{C}}\mathcal{V}_{\mathfrak{a}_{k}^{z}}\right)^{2}},
\end{equation}
where $R_{k}^{z}$ is the universal $R$-matrix for $\mathcal{D}^{\omega_{k}^{z}}\left(G\right)$
and $\theta_{\mathfrak{a}_{k}^{z}}$ is the topological spin associated
with the representation $\mathcal{V}_{\mathfrak{a}_{k}^{z}}$ defined
in Eq.~\eqref{eq:topspin_irrep}. The quantum dimension and topological
spin of the $xy$-particle are 
\begin{align}
d_{\boldsymbol{\mathfrak{a}}}^{z} & =\prod_{k\in\mathtt{C}^{z}}\dim_{\mathbb{C}}\mathcal{V}_{\mathfrak{a}_{k}^{z}},\label{eq:d_z}\\
\theta_{\boldsymbol{\mathfrak{a}}}^{z} & =\prod_{k\in\mathtt{C}^{z}}\theta_{\mathfrak{a}_{k}^{z}}=\prod_{k\in\mathtt{C}^{z}}\frac{\text{tr}\left(\wp R_{k}^{z},\mathcal{V}_{\mathfrak{a}_{k}^{z}}\otimes\mathcal{V}_{\mathfrak{a}_{k}^{z}}\right)}{\dim_{\mathbb{C}}\mathcal{V}_{\mathfrak{a}_{k}^{z}}}.\label{eq:theta_z}
\end{align}
The results for $yz$-particles and $zx$-particles are analogous.

In general, the topological charges of a quasiparticle can be detected
by braiding 2d particles around it. We may measure the quantum dimension $d_{\mathfrak{a}_{n}^{\mu}}$
associated with each $\mu$ topological charge $\mathfrak{a}_{n}^{\mu}$ through Eq.~\eqref{eq:quantumdim}. Further,
this leads to a notion of quantum dimension of any particle type $\boldsymbol{\mathfrak{a}}=\left(q,\mathfrak{a}_{n}^{\mu}\right)=\left(q,g_{n}^{\mu},\varrho_{n}^{\mu}\right)$,
defined by
\begin{equation}
d_{\boldsymbol{\mathfrak{a}}}\coloneqq\prod_{\mu,n}d_{\mathfrak{a}_{n}^{\mu}}.
\end{equation}
For twisted fracton models based on an Abelian group, the quantum dimension of $\mathfrak{a}_{n}^{\mu}=\left(g_{n}^{\mu},\varrho_{n}^{\mu}\right)$
equals the degree (\emph{i.e.}, the dimension) of the representation
$\varrho_{n}^{\mu}$. 

Crucially, the quantum dimension of fracton $(q,\mathfrak{0})$ (\emph{i.e.}, with $\mathfrak{a}_{n}^{\mu}=\mathfrak{0},\forall n\in\mathtt{C}^{\mu},\forall\mu=x,y,z$ but $q\neq0$) is one, where $\mathfrak{0}$ denote trivial topological charge and $0$ the identity element of the character group $\widehat{G}\simeq G$. Thus, every fracton $(q,\mathfrak{a}_{n}^{\mu})_{n\in \mathtt{C}^{\mu}}^{\mu}$ is a fusion result of a fracton with quantum dimension 1 and some mobile particles; explicitly, 
$(q,\mathfrak{a}_{n}^{\mu})_{n\in \mathtt{C}^{\mu}}^{\mu}=(q,\mathfrak{0})\times\prod_{\mu,n} (0,\mathfrak{a}_{n}^{\mu})$. 
Therefore, there is no inextricably non-Abelian fracton  in any twisted X-cube model. 

\subsubsection{Full braiding of 1d mobile quasiparticles}

\begin{figure}
	\includegraphics[width=0.7\columnwidth]{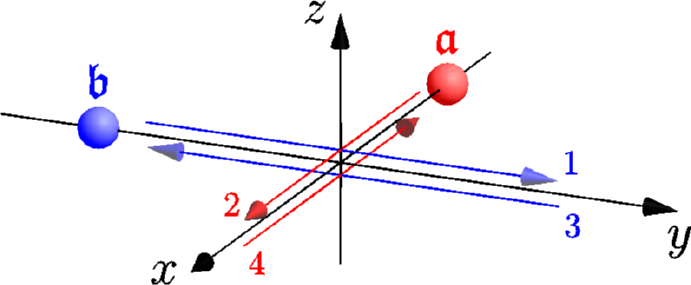}
	
	\caption{
		Arrows $1,3$ (resp. $2,4$) indicate that the movements of quasiparticle $\boldsymbol{\mathfrak{b}}$
		(resp. $\boldsymbol{\mathfrak{a}}$) made by operators $\mathcal{O}_{y}^{\boldsymbol{\mathfrak{b}}}$, 
		$(\mathcal{O}_{y}^{\boldsymbol{\mathfrak{b}}})^{\dagger}$ 
		(resp. $\mathcal{O}_{x}^{\boldsymbol{\mathfrak{a}}}$,
		$(\mathcal{O}_{x}^{\boldsymbol{\mathfrak{a}}})^{\dagger}$)
		supported near the $y$ (resp. $x$) axis. 
		A full braiding of the $z$ topological
		charges can be realized by $\left(\mathcal{O}_{x}^{\boldsymbol{\mathfrak{a}}}\right)^{\dagger}\left(\mathcal{O}_{y}^{\boldsymbol{\mathfrak{b}}}\right)^{\dagger}\mathcal{O}_{x}^{\boldsymbol{\mathfrak{a}}}\mathcal{O}_{y}^{\boldsymbol{\mathfrak{b}}}$.}
	
	\label{fig:cb_m}
\end{figure}

Given two quasiparticles of types $\boldsymbol{\mathfrak{a}}$ and $\boldsymbol{\mathfrak{b}}$
mobile along two different directions (\emph{e.g.}, the $x$ and $y$
directions) respectively, a full braiding of them can be easily made,
as depicted in Fig.~\ref{fig:cb_m}. Let $\mathcal{O}_{x}^{\boldsymbol{\mathfrak{a}}}$
be an operator supported near the $x$-axis that moves the $x$-particle
in the way indicated by arrow $1$ pointing towards the positive $x$ direction.
This operator is normalized such that $\left(\mathcal{O}_{x}^{\boldsymbol{\mathfrak{a}}}\right)^{\dagger}\mathcal{O}_{x}^{\boldsymbol{\mathfrak{a}}}=1$
on the $x$-particle. Similarly, we have operators $\mathcal{O}_{y}^{\boldsymbol{\mathfrak{b}}}$
and $\left(\mathcal{O}_{y}^{\boldsymbol{\mathfrak{b}}}\right)^{\dagger}$ supported
near the $y$-axis that move the $y$-particle forth and back as indicated by arrows $1,3$
in Fig.~\ref{fig:cb_m}; they are normalized by $\left(\mathcal{O}_{y}^{\boldsymbol{\mathfrak{b}}}\right)^{\dagger}\mathcal{O}_{y}^{\boldsymbol{\mathfrak{b}}}=1$
on the initial state of the $y$-particle. Then $\left(\mathcal{O}_{x}^{\boldsymbol{\mathfrak{a}}}\right)^{\dagger}\left(\mathcal{O}_{y}^{\boldsymbol{\mathfrak{b}}}\right)^{\dagger}\mathcal{O}_{x}^{\boldsymbol{\mathfrak{a}}}\mathcal{O}_{y}^{\boldsymbol{\mathfrak{b}}}$
describes a full braiding of the $z$ topological charges of $\boldsymbol{\mathfrak{a}}$ and $\boldsymbol{\mathfrak{b}}$. If the two quasiparticles carry $z$ topological charges
$\mathfrak{a}_{k}^{z}$, $\mathfrak{b}_{k}^{z}$ separately and a
definite total $z$ topological charge $\mathfrak{c}_{k}^{z}$ together,
then the full braiding acts as a scalar
\begin{equation}
\left(\mathcal{O}_{x}^{\boldsymbol{\mathfrak{a}}}\right)^{\dagger}\left(\mathcal{O}_{y}^{\boldsymbol{\mathfrak{b}}}\right)^{\dagger}\mathcal{O}_{x}^{\boldsymbol{\mathfrak{a}}}\mathcal{O}_{y}^{\boldsymbol{\mathfrak{b}}} =
\prod_{k}\frac{\theta_{\mathfrak{c}_{k}^{z}}}{\theta_{\mathfrak{a}_{k}^{z}}\theta_{\mathfrak{b}_{k}^{z}}},
\end{equation}
where $\theta_{\mathfrak{a}_{k}^{z}},\theta_{\mathfrak{b}_{k}^{z}}$
and $\theta_{\mathfrak{c}_{k}^{z}}$ are the topological spins associated
with the representation $\mathcal{V}_{\mathfrak{a}_{k}^{z}}$, $\mathcal{V}_{\mathfrak{b}_{k}^{z}}$,
and $\mathcal{V}_{\mathfrak{c}_{k}^{z}}$ respectively, defined in
Eq.~\eqref{eq:topspin_irrep}.

\begin{figure}
\includegraphics[width=0.7\columnwidth]{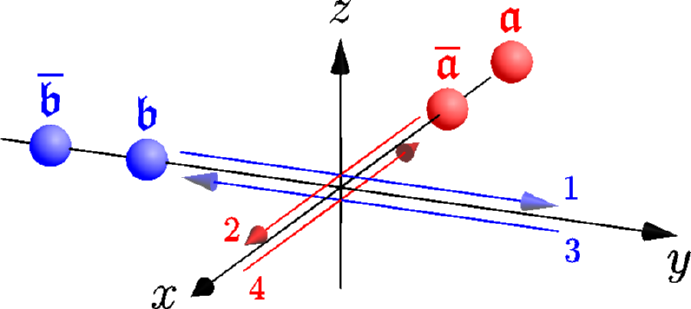}

\caption{An $S$-matrix measurement $\mathsf{S}_{\boldsymbol{\mathfrak{a}\mathfrak{b}}}^{z}$
is associated with the process made of three steps: (1) creating a
pair of $x$-particles $\boldsymbol{\mathfrak{a}},\overline{\boldsymbol{\mathfrak{a}}}$ and
a pair of $y$-particles $\boldsymbol{\mathfrak{b}},\overline{\boldsymbol{\mathfrak{b}}}$ from
vacuum; (2) a full braiding of $\overline{\boldsymbol{\mathfrak{a}}}$ and $\boldsymbol{\mathfrak{b}}$, \emph{i.e.}, moving them according to arrows $1,2,3,4$ in order;
(3) annihilating the pairs $\boldsymbol{\mathfrak{a}},\overline{\boldsymbol{\mathfrak{a}}}$
and $\boldsymbol{\mathfrak{b}},\overline{\boldsymbol{\mathfrak{b}}}$ back to vacuum. }

\label{fig:cb_S}
\end{figure}

Similarly, we can make the $S$-matrix measurements. For example,
$\mathsf{S}_{\boldsymbol{\mathfrak{a}\mathfrak{b}}}^{z}$ is the expectation value
(on the vacuum) of the process shown in Fig.~\ref{fig:cb_S}, in
the normalization that $\mathcal{O}^{\dagger}\mathcal{O}=1$ for any
step $\mathcal{O}$ on its initial state. The result is
\begin{equation}
\mathsf{S}_{\boldsymbol{\mathfrak{a}\mathfrak{b}}}^{z}=\prod_{k}\mathsf{S}_{\mathfrak{a}_{k}^{z}\mathfrak{b}_{k}^{z}},
\end{equation}
where $\mathsf{S}_{\mathfrak{a}_{k}^{z}\mathfrak{b}_{k}^{z}}$ can
be computed by Eq.~\eqref{eq:topS} on representations $\mathcal{V}_{\mathfrak{a}_{k}^{z}}$
and $\mathcal{V}_{\mathfrak{b}_{k}^{z}}$ of $\mathcal{D}^{\omega_{k}^{z}}\left(G\right)$.
Analogously, we have $\mathsf{S}_{\boldsymbol{\mathfrak{a}\mathfrak{b}}}^{x}$ (resp.
$\mathsf{S}_{\boldsymbol{\mathfrak{a}\mathfrak{b}}}^{y}$) for braidings in the
$yz$ directions (resp. $zx$ directions). 

\subsubsection{Half braiding of 1d mobile quasiparticles}

\begin{figure}
\includegraphics[width=0.7\columnwidth]{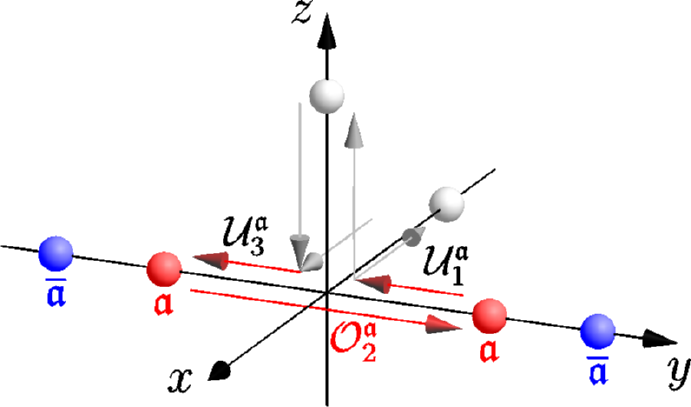}

\caption{Two $y$-particles both of type $\boldsymbol{\mathfrak{a}}$ are braided by $\mathcal{U}_{3}^{\boldsymbol{\mathfrak{a}}}\mathcal{O}_{2}^{\boldsymbol{\mathfrak{a}}}\mathcal{U}_{1}^{\boldsymbol{\mathfrak{a}}}$,
where $\mathcal{U}_{1}^{\boldsymbol{\mathfrak{a}}}$ splits the $y$-particle of
type  $\boldsymbol{\mathfrak{a}}$ on the right into an $x$-particle and a $z$-particle,
$\mathcal{U}_{3}^{\boldsymbol{\mathfrak{a}}}$ fuses the $x$-particle and the
$z$-particle into the $y$-particle of type $\boldsymbol{\mathfrak{a}}$ on the
left and $\mathcal{O}_{2}^{\boldsymbol{\mathfrak{a}}}$ is hopping operator for
the $y$-particle of type $\boldsymbol{\mathfrak{a}}$. In addition, $\mathcal{U}_{1}^{\boldsymbol{\mathfrak{a}}}$,
$\mathcal{U}_{3}^{\boldsymbol{\mathfrak{a}}}$ and $\mathcal{O}_{2}^{\boldsymbol{\mathfrak{a}}}$ are supported
near the corresponding arrows respectively.}

\label{fig:cb_y}
\end{figure}

It is also possible to make a half braiding in order to exchange two 1d mobile
particles. For example, two $y$-particles, both of type $\boldsymbol{\mathfrak{a}}$,
can be braided by $\mathcal{U}_{3}^{\boldsymbol{\mathfrak{a}}}\mathcal{O}_{2}^{\boldsymbol{\mathfrak{a}}}\mathcal{U}_{1}^{\boldsymbol{\mathfrak{a}}}$
as illustrated in Fig.~\ref{fig:cb_y}. Naturally, we require $\mathcal{U}_{3}^{\boldsymbol{\mathfrak{a}}}\mathcal{U}_{1}^{\boldsymbol{\mathfrak{a}}}\mathcal{O}_{2}^{\boldsymbol{\mathfrak{a}}}=1$
on a single $y$-particle of type $\boldsymbol{\mathfrak{a}}$ on the left. All
fracton models considered in this paper allow splitting a 1d mobile quasiparticle
into two 1d mobile quasiparticles in the other two directions (e.g. a particle mobile along the $x$ direction can split into one mobile along $y$ and another mobile along $z$). Thus,
we can make a topological spin measurement described by the expression
on the right hand side of Eq.~\eqref{eq:measuretopspin}. The result
for the situation shown in Fig.~\ref{fig:cb_y} is $\frac{\theta_{\boldsymbol{\mathfrak{a}}}^{x}}{d_{\boldsymbol{\mathfrak{a}}}^{x}}\cdot\frac{\theta_{\boldsymbol{\mathfrak{a}}}^{z}}{d_{\boldsymbol{\mathfrak{a}}}^{z}}$,
where $d_{\boldsymbol{\mathfrak{a}}}^{z}$, $\theta_{\boldsymbol{\mathfrak{a}}}^{z}$, $d_{\boldsymbol{\mathfrak{a}}}^{x}$,
and $\theta_{\boldsymbol{\mathfrak{a}}}^{x}$ are computed by Eqs.~\eqref{eq:d_z}, \eqref{eq:theta_z} and their analogues. In particular, the quantum dimension of a $y$-particle is $d_{\boldsymbol{\mathfrak{a}}}^{x}d_{\boldsymbol{\mathfrak{a}}}^{z}$, which can also be simply defined in the same way as in Eq.~\eqref{eq:quantumdim}.

\subsection{Examples}

For concreteness, we now consider examples of twisted X-cube models which host quasiparticles exhibiting novel and interesting behaviors. In particular, we will study models wherein the one-dimensional particles carry either semionic or non-Abelian statistics. But there is no inextricably non-Abelian fracton in these models; each fracton can split into a fracton of quantum dimension one and several mobile particles. 

\subsubsection{$G=\mathbb{Z}_{2}$: 1d mobile semions}

Let us start with the simplest nontrivial group $G=\mathbb{Z}_{2}=\left\{ 0,1\right\} $.
It is well-known~\cite{chen2013} that $H^{3}\left(\mathbb{Z}_{2},U\left(1\right)\right)=\mathbb{Z}_{2}$
and its nontrivial element is presented by the 3-cocyle 
\begin{equation}
\omega\left(f,g,h\right)=\begin{cases}
-1, & f=g=h=1,\\
1, & \text{otherwise}.
\end{cases}\label{eq:cocycleZ2}
\end{equation}
The group structure of $H^{3}\left(\mathbb{Z}_{2},U\left(1\right)\right)$
is given by $\left[\omega\right]+\left[\omega\right]=\left[\omega^{2}\right]=0$,
where $\left[\omega\right]$ and $\left[\omega^{2}\right]$ denotes
the elements of the cohomology group presented by $\omega$ and $\omega^{2}\left(f,g,h\right)\coloneqq\left(\omega\left(f,g,h\right)\right)^{2}$
respectively. Obviously, $\omega^{2}\equiv1$ presents the identity
element $0$ of $H^{3}\left(\mathbb{Z}_{2},U\left(1\right)\right)$. 

Along each cross section $\Sigma_{n}^{\mu}$ of the lattice $\Lambda$,
the model can be either untwisted or twisted. The pure charges (\emph{i.e.},
the flux is trivial everywhere) behave in the same way, no matter
whether the model is twisted or not. Thus, we are more interested in excitations
that violate $B_{c}=1$ below.

To compare the untwisted case (\emph{i.e.}, the original X-cube model~\cite{vijay16xcube}) with the fully twisted case, we may consider
the braiding of an $x$-particle of type $\boldsymbol{\mathfrak{a}}$ and a $y$-particle
of type $\boldsymbol{\mathfrak{b}}$. In either case, we can identify their $z$
topological charges by requiring that they fuse into a $z$-particle.
In other words, $\mathfrak{a}_{k}^{z}\times\mathfrak{b}_{k}^{z}=\mathfrak{0}$
implies that $\mathfrak{a}_{k}^{z}=\mathfrak{b}_{k}^{z}$, where $\mathfrak{0}$
denote the trivial topological charge. Then we may braid them as in
Fig.~\ref{fig:cb_S} and use this as a diagnostic of the effect of twisting on $\mathsf{S}_{\boldsymbol{\mathfrak{a}\mathfrak{b}}}^{z}$.
When $\mathfrak{a}_{k}^{z}\times\mathfrak{b}_{k}^{z}=\mathfrak{0}$, 
\begin{equation}
\mathsf{S}_{\mathfrak{a}_{k}^{z}\mathfrak{b}_{k}^{z}}=\frac{\theta_{\mathfrak{0}}}{\theta_{\mathfrak{a}_{k}^{z}}\theta_{\mathfrak{b}_{k}^{z}}}=\theta_{\mathfrak{a}_{k}^{z}}^{-2}.
\end{equation}
In the untwisted case, $\theta_{\mathfrak{a}_{k}^{z}}^{-2}$ is always $1$ and hence $\mathsf{S}_{\boldsymbol{\mathfrak{a}\mathfrak{b}}}^{z}=1$
if $\mathfrak{a}_{k}^{z}\times\mathfrak{b}_{k}^{z}=\mathfrak{0},\forall k$. 

In the twisted case, $\theta_{\mathfrak{a}_{k}^{z}}$ may be $\pm i$.
Explicitly, in the notation used in Sec.~\ref{subsec:semion}, $\theta_{\left(1,0\right)}=i$
and $\theta_{\left(1,1\right)}=-i$. Further, we may imagine an $x$-particle
of type $\boldsymbol{\mathfrak{a}}$ centered at $\left(x+\frac{1}{2},\frac{1}{2},\frac{1}{2}\right)$
whose only nontrivial topological charges are $\mathfrak{a}_{1}^{z}=\mathfrak{a}_{1}^{y}=\left(1,0\right)$
and a $y$-particle of type $\boldsymbol{\mathfrak{b}}$ at $\left(\frac{1}{2},y+\frac{1}{2},\frac{1}{2}\right)$
whose only nontrivial topological charges are $\mathfrak{b}_{1}^{z}=\mathfrak{b}_{1}^{x}=\left(1,0\right)$.
The braiding is made on the plane $z=\frac{1}{2}$. Then we have $\mathsf{S}_{\boldsymbol{\mathfrak{a}\mathfrak{b}}}^{z}=-1$
even when $\mathfrak{a}_{k}^{z}\times\mathfrak{b}_{k}^{z}=\mathfrak{0}$, $\forall k$.
This behavior demonstrates the effect of twisting along the plane $z=\frac{1}{2}$, thereby revealing the existence of excitations with semionic mutual statistics, which are restricted to move along one-dimensional sub-manifolds. 

We note that this twisted X-cube model, based on $G=\mathbb{Z}_{2}$, can also be realized by coupling interpenetrating layers of doubled semion string-net models~\cite{han}.

\subsubsection{$G=\mathbb{Z}_{2}\times\mathbb{Z}_{2}\times\mathbb{Z}_{2}$: non-Abelian 1d mobile quasiparticles}

An example of a twisted X-cube model with non-Abelian one-dimensional particles can
be constructed based on the group $G=\mathbb{Z}_{2}\times\mathbb{Z}_{2}\times\mathbb{Z}_{2}$
with the 3-cocycle
\begin{equation}
\omega\left(f,g,h\right)=e^{i\pi\left(f^{\left(1\right)}g^{\left(2\right)}h^{\left(3\right)}\right)},\label{eq:w_z2_3}
\end{equation}
where $f=\left(f^{\left(1\right)},f^{\left(2\right)},f^{\left(3\right)}\right),g=\left(g^{\left(1\right)},g^{\left(2\right)},g^{\left(3\right)}\right),h=\left(h^{\left(1\right)},h^{\left(2\right)},h^{\left(3\right)}\right)\in G$.
We also write the elements of $G$ simply as $000$, $100$,
$110$ and so on for short. As examples, we have $\omega\left(100,010,001\right)=-1$
and $\omega\left(100,001,010\right)=1$ in these notations. 

We notice that a quasiparticle with nontrivial fluxes $g_{1}^{y}=g_{1}^{z}=100$
is an $x$-particle. Moreover, it is not a 2d mobile particle
or a fusion result of 2d mobile particles, because the fluxes of any
2d particle satisfy the constraint
\begin{equation}
\sum_{i}g_{i}^{x}=\sum_{j}g_{j}^{y}=\sum_{k}g_{k}^{z}=0
\end{equation}
which easily follows from Eq.~\eqref{eq:valid_flux}. Further, if
either $\Sigma_{1}^{y}$ or $\Sigma_{1}^{z}$ is twisted by the 3-cocycle
in Eq.~\eqref{eq:w_z2_3}, then either $\varrho_{1}^{y}$ or $\varrho_{1}^{z}$
has to be two-dimensional, as shown in Table~\ref{table:irrep}. Thus,
this $x$-particle has quantum dimension greater than 1, clearly reflecting its non-Abelian character. 
The braiding properties of such non-Abelian 1d particles can be computed following the methods described in Sec.~\ref{subsec:Braiding_general}. Details of a similar calculation will be given later for the twisted
checkerboard based on the same group and the same 3-cocycle. 

However, this $x$-particle is still \emph{not inextricably non-Abelian} if the model is only \emph{partially} twisted. Suppose that the non-Abelian behavior comes from the twisting of $\Sigma_{1}^{z}$ and that there is a nearby parallel plane, say $\Sigma_{2}^{z}$, which remains untwisted. Then the $x$-particle can be split into an Abelian $x$-particle with fluxes $g_{1}^{y}=g_{2}^{z}=100$ and a non-Abelian $xy$-particle with fluxes $g_{1}^{z}=g_{2}^{z}=100$, implying that it is not inextricably non-Abelian according to the definition in Sec.~\ref{subsec:intro_sum}. Contrarily, if the model is fully twisted in at least one direction, then such a splitting is no longer possible and hence the $x$-particle becomes inextricably non-Abelian.
In this case, we call the corresponding fracton phase \emph{non-Abelian}; it is clearly distinct from an Abelian fracton phase with some layers of conventional non-Abelian topological states inserted.
This dramatic change between the fully and partially twisted cases is also reflected in their GSD on $\mathtt{T}^{3}$, which is explicitly given by Eqs.~\eqref{eq:GSD_Xcube_Z}, \eqref{eq:GSD_Xcube_2}, \eqref{eq:GSD_Xcube_3} for three fully twisted cases and Eq.~\eqref{eq:GSD_Xcube_partial} for a generic partially twisted case.
Both the presence of inextricably non-Abelian 1d mobile quasiparticles and their exotic GSD establishes that these non-Abelian fracton phases are a completely new type of quantum states.

Moreover, we emphasize that \emph{no} twisted X-cube model, defined in Sec.~\ref{subsec:X-cube_twisted_def}, hosts inextricably non-Abelian fractons. To see this, we notice that a quasiparticle $\left(q,\mathfrak{a}_{n}^{\mu}\right)$
is a fracton if and only if $q\neq0$. However, such excitations can always be viewed as a fusion result of a fracton $\left(q,\mathfrak{0}\right)$ of quantum dimension
1---thus, an Abelian fracton---and some mobile quasiparticles. In other words, there is no inextricably non-Abelian fracton in twisted X-cube models. Thus, in order to find a model with inextricably non-Abelian fractons, we look to the twisted checkerboard models, which we consider next. 

\section{Quasiparticles in twisted checkerboard models}
\label{sec:QPchecker}

We now study quasiparticles in the twisted checkerboard models, proceeding analogously to the previous section. Here, the particle types can also be labeled
by their $x$, $y$, and $z$ topological charges, subject to certain constraints. After systematically analyzing the mobility, fusion and braiding of quasiparticles in terms of their topological charges, we will then study specific examples to elucidate the plethora of novel phenomena which may occur in the twisted checkerboard models.

\subsection{Particle type and topological charges}\label{subsec:checker_particle}

\begin{figure}
\noindent\begin{minipage}[t]{1\columnwidth}%
\includegraphics[width=0.7\columnwidth]{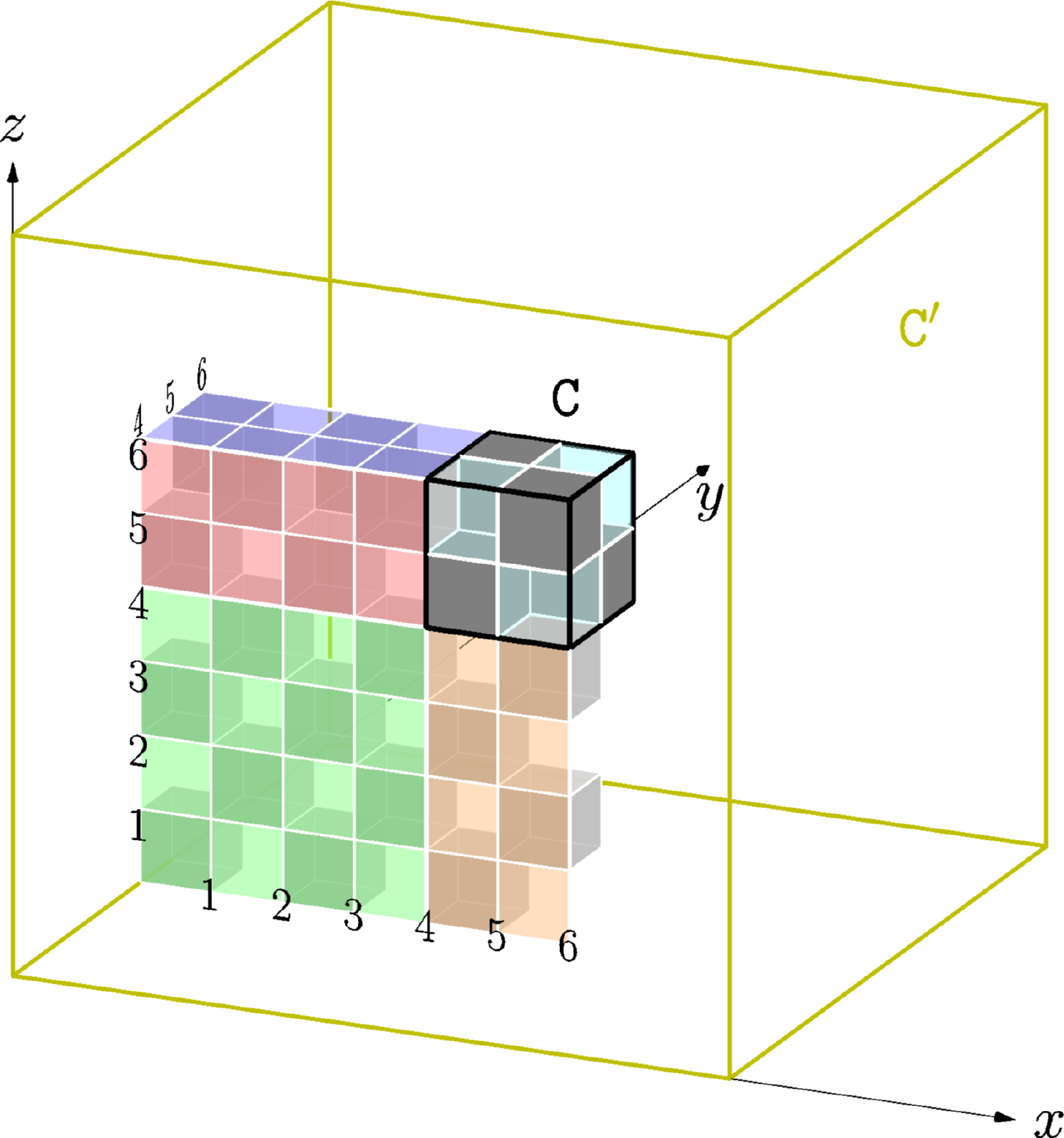}

\vspace{-0.4cm}

(a) %
\end{minipage}\bigskip{}

\noindent\begin{minipage}[t]{1\columnwidth}%
\includegraphics[width=0.7\columnwidth]{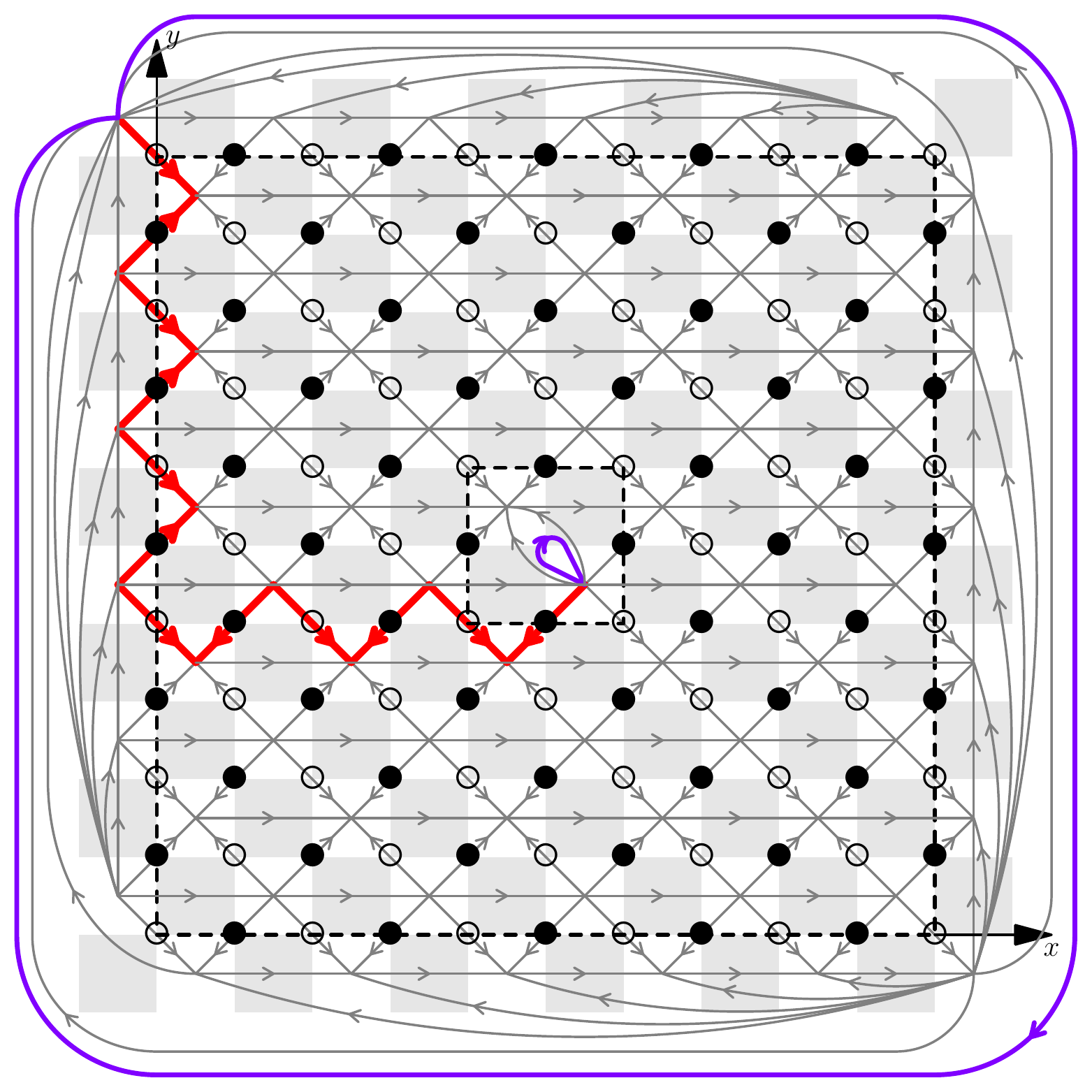}

(b) $\overline{\mathtt{M}}_{k}^{z}$ with $k=5$%
\end{minipage}\bigskip{}

\noindent\begin{minipage}[t]{1\columnwidth}%
\includegraphics[width=0.7\columnwidth]{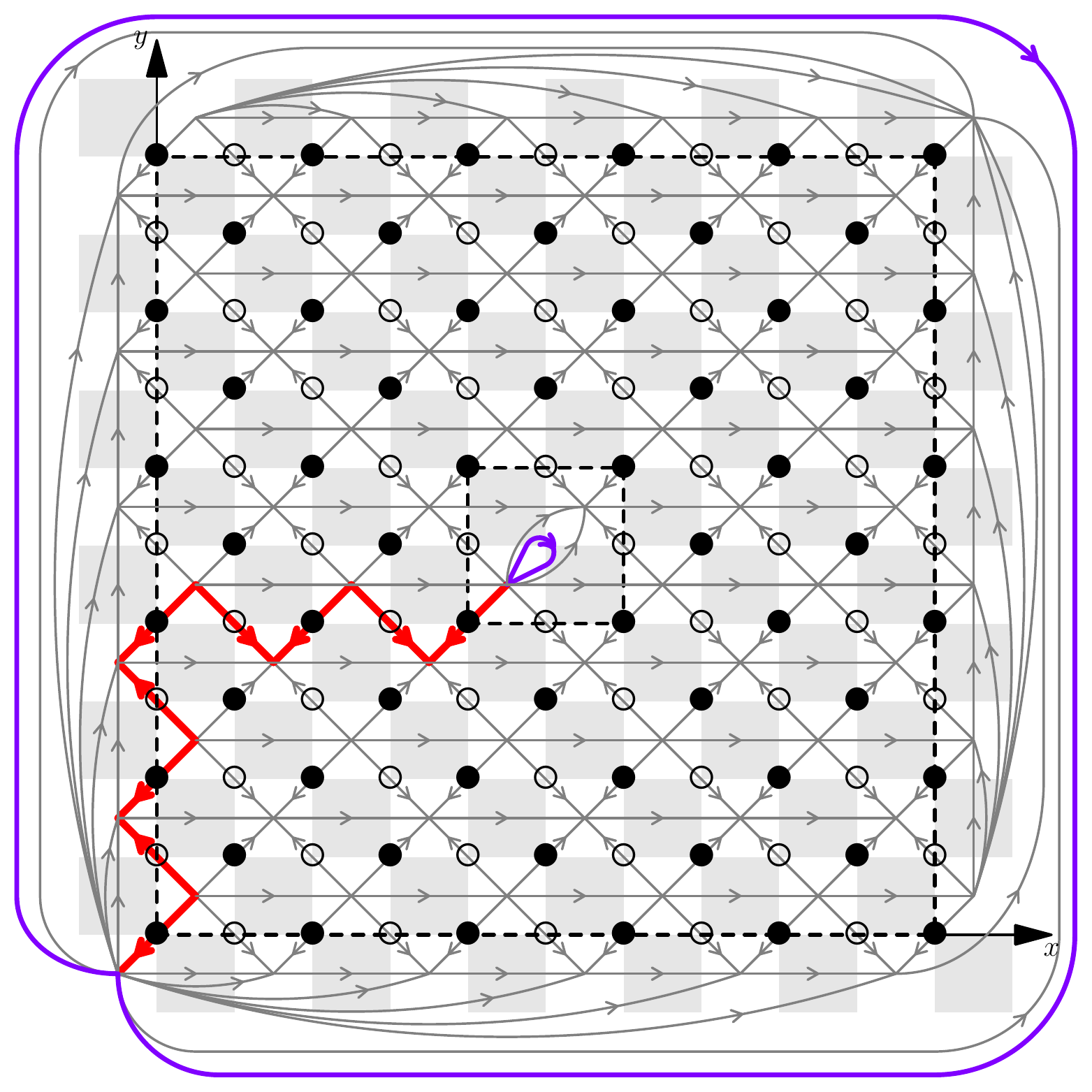}

(c) $\overline{\mathtt{M}}_{k}^{z}$ with $k=6$%
\end{minipage}

\caption{(a) An excited cuboid $\mathtt{C}=\left[4,6\right]\times\left[4,6\right]\times\left[4,6\right]$
isolated from other excitations outside $\mathtt{C}'$ in a checkerboard
model. The membranes (orange, blue, green and red online) between
$\partial\mathtt{C}$ and $\partial\mathtt{C}'$ indicates the bulk
degrees of freedom used for defining $s_{i}^{x}$, $s_{j}^{y}$ and
$s_{k}^{z}$ in the main text. (b, c) A triangulation of cross section $\overline{\mathtt{M}}_{k}^{z}$.
Some edges are thickened to highlight the outer and inner boundaries
(purple online) of $\overline{\mathtt{M}}_{k}^{z}$ and a path (red
online) in between. }

\label{fig:cb_M0}
\end{figure}

Any excited spot (\emph{i.e.}, quasiparticle) can be enclosed in a
finite cuboid $\mathtt{C}=\left[x_{0},x_{1}\right]\times\left[y_{0},y_{1}\right]\times\left[z_{0},z_{1}\right]$.
Let $\mathtt{C}'=\left[x_{0}^{\prime},x_{1}^{\prime}\right]\times\left[y_{0}^{\prime},y_{1}^{\prime}\right]\times\left[z_{0}^{\prime},z_{1}^{\prime}\right]$
be a much larger cuboid containing $\mathtt{C}$. Without loss of generality,
$x_{0},x_{1},y_{0},y_{1},z_{0},z_{1},x_{0}',x_{1}',y_{0}',y_{1}',z_{0}',z_{1}'$
are picked to be even integers. In the following, we use Fig.~\ref{fig:cb_M0}(a)
for illustration, where $\mathtt{C}=\left[4,6\right]\times\left[4,6\right]\times\left[4,6\right]$,
$\mathtt{C}'=\left[0,10\right]\times\left[0,10\right]\times\left[0,10\right]$
and the coordinates are chosen such that the grey unit cubes are centered
at $\left(x,y,z\right)+\frac{1}{2}\left(1,1,1\right)$ with $x,y,z\in\mathbb{Z}$
and $x+y+z$ even. 

For convenience, we write $\mathtt{C}^{x}\coloneqq\left\{ x_{0},x_{0}+1,\cdots,x_{1}\right\} $,
$\mathtt{C}^{y}\coloneqq\left\{ y_{0},y_{0}+1,\cdots,y_{1}\right\} $
and $\mathtt{C}^{z}\coloneqq\left\{ z_{0},z_{0}+1,\cdots,z_{1}\right\} $.
Let $\mathtt{M}=\mathtt{C}'-\mathtt{C}^{\circ}$, where $\mathtt{X}^{\circ}$ denotes the interior
of any topological space $\mathtt{X}$. Then $\mathtt{M}$ is
a three-dimensional manifold with boundary.
Let $\mathcal{H}\left(\Lambda^{0}(\mathtt{M}),G\right)$ be the
Hilbert space describing all the physical degrees of freedom on $\mathtt{M}$ and $G_{B}^{\Lambda^{0}\left(\mathtt{M}\right)}\coloneqq\left\{ \vartheta\in G^{V\left(\mathtt{M}\right)}|B_{c}\left|\vartheta\right\rangle =\left|\vartheta\right\rangle ,\forall c\in\Lambda_{\bullet}^{3}\left(\mathtt{M}\right)\right\}  $.
To classify excitations inside $\mathtt{C}$, we analyze the
subspace $\mathcal{H}_{0}\left(\mathtt{M}\right)$ selected out of $\mathcal{H}\left(\Lambda^{0}(\mathtt{M}),G\right)$  by the
projector 
\begin{equation}
P\left(\mathtt{M}\right)\coloneqq\prod_{c\in\Lambda_{\bullet}^{3}\left(\mathtt{M}^{\circ}\right)}P_{c},
\end{equation}
where $c$ labels grey cubes in the interior of $\mathtt{M}$. 

The Hilbert space $\mathcal{H}\left(\Lambda^{0}(\mathtt{M}),G\right)$ 
has an orthonormal basis
$\left\{ \left|\vartheta\right\rangle |\vartheta\in G^{\Lambda^{0}\left(\mathtt{M}\right)}\right\} $,
where $\Lambda^{0}\left(\mathtt{M}\right)$ is the set of vertices
in $\mathtt{M}$ and $G^{\Lambda^{0}\left(\mathtt{M}\right)}$ is
the set of functions from $\Lambda^{0}\left(\mathtt{M}\right)$ to
$G$. Let  $\mathtt{M}_{i}^{x}$, $\mathtt{M}_{j}^{y}$ and $\mathtt{M}_{k}^{z}$
be the intersection of $\mathtt{M}$ with the plane 
\begin{align}
x & =i-\frac{1}{2},\;\forall i=x_{0}^{\prime}+1,x_{0}^{\prime}+2,\cdots,x_{1}^{\prime};\\
y & =j-\frac{1}{2},\;\forall j=y_{0}^{\prime}+1,y_{0}^{\prime}+2,\cdots,y_{1}^{\prime};\\
z & =k-\frac{1}{2},\;\forall k=z_{0}^{\prime}+1,z_{0}^{\prime}+2,\cdots,z_{1}^{\prime}; 
\end{align}
(\emph{i.e.}, the region of $\Sigma_{i}^{x}$, $\Sigma_{j}^{y}$ and $\Sigma_{k}^{z}$ inside $\mathtt{M}$) respectively. Each of them is either a disk or an annulus as a topological
space and a region of two-dimensional checkerboard. Respectively, 
we can embed it into either a triangulated sphere or a triangulated annulus with two loops as boundary, denoted $\overline{\mathtt{M}}_{n}^{\mu}$
for $\mu=x,y,z$. Let $\Delta^{m}\left(\overline{\mathtt{M}}_{n}^{\mu}\right)$
be the set of $m$-simplices in this triangulation.

Examples of $\overline{\mathtt{M}}_{k}^{z}$ 
are given in Fig.~\ref{fig:cb_M0}(b) and (c). Marks $\bullet$ and
$\circ$ are added to show the positions of vertices of $\mathtt{M}$
on the plane $z=k$ and the value of $\left(-1\right)^{v}$ on these
vertices. Let $E\left(\mathtt{M}_{k}^{z}\right)$ be the subset of
$\Delta^{1}\left(\overline{\mathtt{M}}_{k}^{z}\right)$ containing
the edges with a $\bullet$ or $\circ$ mark; it has a one-to-one
correspondence with $\Lambda^{1}\left[\mathtt{M}_{k}^{z}\right]$ (\emph{i.e.},
the set of edges of $\mathtt{M}$ intersecting with the plane $z=k-\frac{1}{2}$).
Given $\vartheta\in G_{B}^{\Lambda^{0}\left(\mathtt{M}\right)}$,
we color $\Lambda^{1}\left[\mathtt{M}_{k}^{z}\right]$ and hence $E\left(\mathtt{M}_{k}^{z}\right)$
by $\partial\vartheta$ as in Eq.~\eqref{eq:d_vartheta}, which extends
uniquely to a coloring of $\overline{\mathtt{M}}_{k}^{z}$. The triangulation
and coloring for $\overline{\mathtt{M}}_{i}^{x}$ and $\overline{\mathtt{M}}_{j}^{y}$
are obtained analogously. After picking $\overline{\mathtt{M}}_{i}^{x}$
(resp. $\overline{\mathtt{M}}_{j}^{y}$, $\overline{\mathtt{M}}_{k}^{z}$)
for each cross section $\mathtt{M}_{i}^{x}$ (resp. $\mathtt{M}_{j}^{y}$,
$\mathtt{M}_{k}^{z}$), we can define $P_{c}^{g}$ for $c$ touching
$\partial\mathtt{M}$.

To give a basis of $\mathcal{H}_{0}\left(\mathtt{M}\right)$, we pick
paths (going inwards) connecting base points of the outer and inner
boundaries of each annulus $\overline{\mathtt{M}}_{i}^{x}$, $\overline{\mathtt{M}}_{j}^{y}$
and $\overline{\mathtt{M}}_{k}^{z}$. Let $s_{i}^{x}$ for $x_{0}<i\leq x_{1}$,
$s_{j}^{y}$ for $y_{0}<j\leq y_{1}$ and $s_{k}^{z}$ for $z_{0}<i\leq z_{1}$
be the group elements associated with these paths. Examples of such
paths are shown by the zigzag sequences of thick (red online) edges
in Fig.~\ref{fig:cb_M0}(b) and (c); correspondingly, 
\begin{align}
s_{k}^{z} & =\sum_{\left(x,y\right)\in\text{path}}\left(-1\right)^{\left(x,y,k\right)}\vartheta_{k}^{z}\left(x,y\right)\\
& = \sum_{\left(x,y\right)\in\text{path}}\left(-1\right)^{\left(x,y,k\right)}\left[\vartheta\left(x,y,k\right)-\vartheta\left(x,y,k-1\right)\right],
\end{align}
where $\left(x,y\right)\in\mathbb{Z}\times\mathbb{Z}$ labels vertices
of $\mathtt{M}_{k}^{z}$.

In Fig.~\ref{fig:cb_M0}(a), a choice of such paths for all annuli
$\overline{\mathtt{M}}_{i}^{x}$ (resp. $\overline{\mathtt{M}}_{j}^{y}$,
$\overline{\mathtt{M}}_{k}^{z}$) is illustrated as a rectangular
ribbon, colored orange (resp. blue, red) online, connecting $\partial\mathtt{C}$
and $\partial\mathtt{C'}$. The ribbon shows the positions of spins
(inside $\mathtt{M}^{\circ}$) in terms of which $s_{i}^{x}$ (resp.
$s_{j}^{y}$, $s_{k}^{z}$) can be expressed; for cleanness, we do
not draw the part of these paths on $\partial\mathtt{M}$. In addition,
we need 
\begin{align}
s_{x_0}^{x} & =\sum_{x_{0}^{\prime}<i\leq x_{0}}\sum_{z_{0}^{\prime}<k<z_{0}}\left(-1\right)^{\left(i,y_{0},k\right)}\vartheta\left(i,y_{0},k\right),\\
s_{y_0}^{y} &  =-\sum_{x_{0}^{\prime}<i<x_{0}}\sum_{z_{0}^{\prime}<k<z_{0}}\left(-1\right)^{\left(i,y_{0},k\right)}\vartheta\left(i,y_{0},k\right),\label{eq:g2-1}\\
s_{z_0}^{z} & =\sum_{x_{0}^{\prime}<i<x_{0}}\sum_{z_{0}^{\prime}<k\leq z_{0}}\left(-1\right)^{\left(i,y_{0},k\right)}\vartheta\left(i,y_{0},k\right),\label{eq:g3-1}
\end{align}
in terms of spins on the square (green online) sheet in Fig.~\ref{fig:cb_M0}(a).
Collectively, we write 
\begin{align}
\boldsymbol{s} & \coloneqq\left(\boldsymbol{s}^{x},\boldsymbol{s}^{y},\boldsymbol{s}^{z}\right)\in G^{\mathtt{C}^{x}}\times G^{\mathtt{C}^{y}}\times G^{\mathtt{C}^{z}},\\
\boldsymbol{s}^{x} & \coloneqq\left(s_{x_{0}}^{x},s_{x_{0}+1}^{x},\cdots,s_{x_{1}}^{x}\right)\in G^{\mathtt{C}^{x}},\\
\boldsymbol{s}^{y} & \coloneqq\left(s_{y_{0}}^{y},s_{y_{0}+1}^{y},\cdots,s_{y_{1}}^{y}\right)\in G^{\mathtt{C}^{y}},\\
\boldsymbol{s}^{z} & \coloneqq\left(s_{z_{0}}^{z},s_{z_{0}+1}^{z},\cdots,s_{z_{1}}^{z}\right)\in G^{\mathtt{C}^{z}}.
\end{align}

Moreover, we need to label fluxes as well. Let $g_{i}^{x}$ (resp.
$g_{j}^{y}$, $g_{k}^{z}$) be the group element associated with either the inner or the outer boundary of $\overline{\mathtt{M}}_{i}^{x}$ (resp. $\overline{\mathtt{M}}_{j}^{y}$,
$\overline{\mathtt{M}}_{k}^{z}$). By assumption, $B_{c}=1$ for $c$ outside
$\mathtt{C}=\left[x_{0},x_{1}\right]\times\left[y_{0},y_{1}\right]\times\left[z_{0},z_{1}\right]$,
so we write $g_{i}^{x}=0$ (resp. $g_{j}^{y}=0$, $g_{k}^{z}=0$)
unless $x_{0}<i\leq x_{1}$ (resp. $y_{0}<j\leq y_{1}$, $z_{0}<k\leq z_{1}$).
The data needed for describing fluxes are collected as
\begin{align}
\boldsymbol{g} & \coloneqq\left(\boldsymbol{g}^{x},\boldsymbol{g}^{y},\boldsymbol{g}^{z}\right)\in G^{\mathtt{C}^{x}}\times G^{\mathtt{C}^{y}}\times G^{\mathtt{C}^{z}},\label{eq:f_checker}\\
\boldsymbol{g}^{x} & \coloneqq\left(0,g_{x_{0}+1}^{x},\cdots,g_{x_{1}}^{x}\right)\in G^{\mathtt{C}^{x}},\label{eq:f_checker1}\\
\boldsymbol{g}^{y} & \coloneqq\left(0,g_{y_{0}+1}^{y},\cdots,g_{y_{1}}^{y}\right)\in G^{\mathtt{C}^{y}},\label{eq:f_checker2}\\
\boldsymbol{g}^{z} & \coloneqq\left(0,g_{z_{0}+1}^{z},\cdots,g_{z_{1}}^{z}\right)\in G^{\mathtt{C}^{z}}.\label{eq:f_checker3}
\end{align}
Not all of them are independent; they are subject to the constraints
\begin{align}
\sum_{k\text{ even}}g_{k}^{z} & =\sum_{i\text{ odd}}g_{i}^{x}-\sum_{j\text{ odd}}g_{j}^{y},\label{eq:cb_f1}\\
\sum_{i\text{ even}}g_{i}^{x} & =\sum_{j\text{ odd}}g_{j}^{y}-\sum_{k\text{ odd}}g_{k}^{z},\label{eq:cb_f2}\\
\sum_{j\text{ even}}g_{j}^{y} & =\sum_{k\text{ odd}}g_{k}^{z}-\sum_{i\text{ odd}}g_{i}^{x}.\label{eq:cb_f3}
\end{align}
Let $F\left[\mathtt{C}\right]$ be the set of $\boldsymbol{g}$ in
the form of Eqs.~(\ref{eq:f_checker}-\ref{eq:f_checker3})  satisfying
Eqs.~(\ref{eq:cb_f1}-\ref{eq:cb_f3}). As a group, $F\left(\mathtt{C}\right)$
is isomorphic to $G^{x_{1}-x_{0}+y_{1}-y_{0}+z_{1}-z_{0}-3}$. 

Let $G_{B}^{\Lambda^{0}\left(\mathtt{M}\right)}\left(\boldsymbol{s},\chi,\overline{\chi}\right)$
be the set of $\vartheta\in G_{B}^{\Lambda^{0}\left(\mathtt{M}\right)}$
compatible with $\boldsymbol{s}$ and coinciding with $\chi\in G^{\Lambda^{0}\left(\partial\mathtt{C'}\right)}$
and $\overline{\chi}\in G^{\Lambda^{0}\left(\partial\mathtt{C}\right)}$ on
$\partial\mathtt{M}=\partial\mathtt{C}\cup\partial\mathtt{C}'$. Now
we have enough notations to give the basis vectors of $\mathcal{H}_{0}\left(\mathtt{M}\right)$;
they are defined by
\begin{equation}
\left|\chi,\overline{\chi};\mathtt{D}_{\boldsymbol{g}}^{\boldsymbol{s}}\right\rangle \coloneqq
\sum_{\vartheta\in G_{B}^{\Lambda^{0}\left(\mathtt{M}\right)}\left(\boldsymbol{s},\chi,\overline{\chi}\right)}Z\left(\vartheta;\mathtt{D}_{\boldsymbol{g}}^{\boldsymbol{s}}\right)\left|\vartheta\right\rangle 
\end{equation}
with $\boldsymbol{s}\in G^{\mathtt{C}^{x}}\times G^{\mathtt{C}^{y}}\times G^{\mathtt{C}^{z}}$,
$\boldsymbol{g}\in F\left(\mathtt{C}\right)$, $\chi\in G^{\Lambda^{0}\left(\partial\mathtt{C'}\right)}$
and $\overline{\chi}\in G^{\Lambda^{0}\left(\partial\mathtt{C}\right)}$. To
complete the definition, the phase factor $Z\left(\vartheta;\mathtt{D}_{\boldsymbol{g}}^{\boldsymbol{s}}\right)$
is given by
\begin{multline}
Z\left(\vartheta;\mathtt{D}_{\boldsymbol{g}}^{\boldsymbol{s}}\right)\coloneqq\prod_{i=x_{0}^{\prime}+1}^{x_{1}^{\prime}}Z_{i}^{x}\left(\vartheta;\mathtt{D}_{\boldsymbol{g}}^{\boldsymbol{s}}\right)\cdot\\
\prod_{j=y_{0}^{\prime}+1}^{y_{1}^{\prime}}Z_{j}^{y}\left(\vartheta;\mathtt{D}_{\boldsymbol{g}}^{\boldsymbol{s}}\right)\cdot\prod_{k=z_{0}^{\prime}+1}^{z_{1}^{\prime}}Z_{k}^{z}\left(\vartheta;\mathtt{D}_{\boldsymbol{g}}^{\boldsymbol{s}}\right),\label{eq:Z_cb}
\end{multline}
where $Z_{i}^{x}\left(\vartheta;\mathtt{D}_{\boldsymbol{g}}^{\boldsymbol{s}}\right)$
is the Dijkgraaf-Witten partition function of a ball with surface colored by
$\vartheta_{i}^{x}$ if $\overline{\mathtt{M}}_{i}^{x}$ is a sphere
or a solid torus with surface colored by $\left(\vartheta_{i}^{x};\mathtt{D}_{g_{i}^{x}}^{s_{i}^{x}}\right)$
if $\overline{\mathtt{M}}_{i}^{x}$ is a annulus. Clearly, the vector $\left|\mathtt{D}_{\boldsymbol{g}}^{\boldsymbol{s}};\chi,\overline{\chi}\right\rangle \neq 0$ if and only if $\chi\in G^{\Lambda^{0}\left(\partial\mathtt{C}'\right)}$
and $\overline{\chi}\in G^{\Lambda^{0}\left(\partial\mathtt{C}\right)}$ are
compatible with $\boldsymbol{g}\in F\left(\mathtt{C}\right)$.
Analogously,  $Z_{j}^{y}\left(\vartheta;\mathtt{D}_{\boldsymbol{g}}^{\boldsymbol{s}}\right)$
and $Z_{k}^{z}\left(\vartheta;\mathtt{D}_{\boldsymbol{g}}^{\boldsymbol{s}}\right)$ are defined.

For the purposes of finding all local operators, let
\begin{equation}
\left|\chi,\overline{\chi};\mathtt{D}_{\boldsymbol{g}}^{\boldsymbol{s}}\mathtt{D}_{\boldsymbol{g}'}^{\boldsymbol{s}'}\right\rangle \coloneqq
\sum_{\vartheta\in G_{B}^{\Lambda^{0}\left(\mathtt{M}\right)}\left(\boldsymbol{s}+\boldsymbol{s}',\chi,\overline{\chi}\right)}Z\left(\vartheta;\mathtt{D}_{\boldsymbol{g}}^{\boldsymbol{s}}\mathtt{D}_{\boldsymbol{g}'}^{\boldsymbol{s}'}\right)\left|\vartheta\right\rangle ,
\end{equation}
where $Z\left(\vartheta;\mathtt{D}_{\boldsymbol{g}}^{\boldsymbol{s}}\mathtt{D}_{\boldsymbol{g}'}^{\boldsymbol{s}'}\right)$
is given by Eq.~\eqref{eq:Z_cb} with $\mathtt{D}_{\boldsymbol{g}}^{\boldsymbol{s}}$
replaced by $\mathtt{D}_{\boldsymbol{g}}^{\boldsymbol{s}}\mathtt{D}_{\boldsymbol{g}'}^{\boldsymbol{s}'}$.
Clearly, $\left|\chi,\overline{\chi};\mathtt{D}_{\boldsymbol{g}}^{\boldsymbol{s}}\mathtt{D}_{\boldsymbol{g}'}^{\boldsymbol{s}'}\right\rangle =0$
if $\boldsymbol{g}\neq\boldsymbol{g}'$. Moreover, $\chi$ (resp. $\overline{\chi}$)
can be changed by $P_{v}^{g}$ on $\partial\mathtt{C}'$ (resp. $\partial\mathtt{C}$) and hence describes degrees of freedom near $\partial\mathtt{C}'$ (resp. $\partial\mathtt{C}$).
To classify particles, we can keep
$\chi,\overline{\chi}$ fixed and consider the subspace $\mathcal{H}_{*}\left(\overline{\mathtt{M}}\right)$
spanned by $\left|\mathtt{D}_{\boldsymbol{g}}^{\boldsymbol{s}};\chi,\overline{\chi}\right\rangle $
for $\boldsymbol{g}\in F\left(\mathtt{C}\right)$ and $\boldsymbol{s}\in G^{\mathtt{C}^{x}}\times G^{\mathtt{C}^{y}}\times G^{\mathtt{C}^{z}}$.
Since $\chi,\overline{\chi}$ are fixed, we omit them in notation and simply write $\left|\mathtt{D}_{\boldsymbol{g}}^{\boldsymbol{s}}\right\rangle $. 

Next, let us construct a set
of operators supported near either $\partial\mathtt{C}$ or $\partial\mathtt{C}'$ to distinguish states in $\mathcal{H}_{*}\left(\overline{\mathtt{M}}\right)$.
For $t\in G$, we have gauge
transformation operators given by
\begin{align}
Z_{k}^{t} & \coloneqq\prod_{c\sim\left(\partial\mathtt{C}',z=k+\frac{1}{2}\right)}P_{c}^{\left(-1\right)^{c_{y}}t},\\
\overline{Z}_{k}^{t} & \coloneqq\prod_{c\sim\left(\partial\mathtt{C},z=k+\frac{1}{2}\right)}\left(P_{c}^{\left(-1\right)^{c_{y}}t}\right)^{\dagger},
\end{align}
where $c\sim\left(\partial\mathtt{C}',z=k+\frac{1}{2}\right)$
(resp. $c\sim\left(\partial\mathtt{C},z=k+\frac{1}{2}\right)$)
means that the cube $c$ is cut by plane $z=k+\frac{1}{2}$ and touches
$\partial\mathtt{C}^{\prime}$ (resp. $\partial\mathtt{C}$). They
act as the identity on $\mathcal{H}_{*}\left(\overline{\mathtt{M}}\right)$ unless
$z_{0}-1\leq k\leq z_{1}$. In addition, for 
$h\in G$, let $T_{h}\left[\mathtt{M}_{k}^{z}\right]$ (resp. $\overline{T}_{h}\left[\mathtt{M}_{k}^{z}\right]$) be the
projector supported on $\partial\mathtt{C}'$ (resp. $\partial\mathtt{C}$) requiring that the group element associated with the outer
(resp. inner) boundary of $\overline{\mathtt{M}}_{k}^{z}$ is $h$. 

Then the two sets of operators 
\begin{align}
D_{k,h}^{z,t} & \coloneqq T_{h}\left[\mathtt{M}_{k}^{z}\right]\prod_{l\in k+2\mathbb{N}}Z_{l}^{t},\label{eq:DZ}\\
\overline{D}_{k,h}^{z,t} & \coloneqq \overline{T}_{h}\left[\mathtt{M}_{k}^{z}\right]\prod_{l\in k+2\mathbb{N}}\overline{Z}_{l}^{t}
\end{align}
commute with $P\left(\mathtt{M}\right)$, keep $\chi,\overline{\chi}$ fixed and hence act on $\mathcal{H}_{*}\left(\overline{\mathtt{M}}\right)$, where $\mathbb{N}$ denotes the set of non-negative integers.
For $z_{0}<k\leq z_{1}$,
\begin{align}
D_{k,h}^{z,t}\left|\mathtt{D}_{\boldsymbol{g}}^{\boldsymbol{s}}\right\rangle  & =\delta_{h,g_{k}^{z}}\left|\mathtt{D}_{\boldsymbol{g}}^{t\delta_{k}^{z}}\mathtt{D}_{\boldsymbol{g}}^{\boldsymbol{s}}\right\rangle ,\label{eq:pi_1}\\
\overline{D}_{k,h}^{z,t}\left|\mathtt{D}_{\boldsymbol{g}}^{\boldsymbol{s}}\right\rangle  & =\delta_{h,g_{k}^{z}}\left|\mathtt{D}_{\boldsymbol{g}}^{\boldsymbol{s}}\mathtt{D}_{\boldsymbol{g}}^{t\delta_{k}^{z}}\right\rangle ,\label{eq:pi_2}
\end{align}
where the component of $h\delta_{k}^{z}\in G^{\mathtt{C}^{x}}\times G^{\mathtt{C}^{y}}\times G^{\mathtt{C}^{z}}$
corresponding to $k\in\mathtt{C}^{z}$ is $h$ and the other components
are zero. For $k>z_{1}$ or $k\leq z_{0}$, these operators act as
\begin{align}
D_{k,h}^{z,t}\left|\mathtt{D}_{\boldsymbol{g}}^{\boldsymbol{s}}\right\rangle  & =\overline{D}_{k,h}^{z,t}\left|\mathtt{D}_{\boldsymbol{g}}^{\boldsymbol{s}}\right\rangle \nonumber \\
& =\begin{cases}
\delta_{h,0}\left|\mathtt{D}_{\boldsymbol{g}}^{\boldsymbol{s}}\right\rangle , & k>z_{1},\\
\delta_{h,0}\left|\mathtt{D}_{\boldsymbol{g}}^{t\delta_{z_{0}}^{z}}\mathtt{D}_{\boldsymbol{g}}^{\boldsymbol{s}}\right\rangle , & k\leq z_{0}\text{ even},\\
\delta_{h,0}\left|\mathtt{D}_{\boldsymbol{g}}^{t\delta_{x_{0}}^{x}-t\delta_{y_{0}}^{y}}\mathtt{D}_{\boldsymbol{g}}^{\boldsymbol{s}}\right\rangle , & k\leq z_{0}\text{ odd}.
\end{cases}\label{eq:pi_chargez}
\end{align}
Analogously, we can define operators $D_{i,n}^{x,h}$, $\overline{D}_{i,n}^{x,h}$,
$D_{j,n}^{y,h}$ and $\overline{D}_{j,n}^{y,h}$. Their actions on $\mathcal{H}_{*}\left(\overline{\mathtt{M}}\right)$
are obtained by permuting $x,y,z$ cyclically in Eqs.~(\ref{eq:pi_1}-\ref{eq:pi_chargez}).
In particular, for $i>x_1,i\leq x_{0}$ and 
$j>y_1,j\leq y_{0}$, we have 
\begin{align}
D_{i,h}^{x,t}\left|\mathtt{D}_{\boldsymbol{g}}^{\boldsymbol{s}}\right\rangle  & =\overline{D}_{i,h}^{x,t}\left|\mathtt{D}_{\boldsymbol{g}}^{\boldsymbol{s}}\right\rangle \nonumber \\
& =\begin{cases}
\delta_{h,0}\left|\mathtt{D}_{\boldsymbol{g}}^{\boldsymbol{s}}\right\rangle , & i>x_{1}\\
\delta_{h,0}\left|\mathtt{D}_{\boldsymbol{g}}^{t\delta_{x_{0}}^{x}}\mathtt{D}_{\boldsymbol{g}}^{\boldsymbol{s}}\right\rangle , & i\leq x_{0}\text{ even},\\
\delta_{h,0}\left|\mathtt{D}_{\boldsymbol{g}}^{t\delta_{y_{0}}^{y}-t\delta_{z_{0}}^{z}}\mathtt{D}_{\boldsymbol{g}}^{\boldsymbol{s}}\right\rangle , & i\leq x_{0}\text{ odd},
\end{cases}\label{eq:pi_chargex}\\
D_{j,h}^{y,t}\left|\mathtt{D}_{\boldsymbol{g}}^{\boldsymbol{s}}\right\rangle  & =\overline{D}_{j,h}^{y,t}\left|\mathtt{D}_{\boldsymbol{g}}^{\boldsymbol{s}}\right\rangle \nonumber \\
& =\begin{cases}
\delta_{h,0}\left|\mathtt{D}_{\boldsymbol{g}}^{\boldsymbol{s}}\right\rangle , & j>y_{1}\\
\delta_{h,0}\left|\mathtt{D}_{\boldsymbol{g}}^{t\delta_{y_{0}}^{y}}\mathtt{D}_{\boldsymbol{g}}^{\boldsymbol{s}}\right\rangle , & j\leq y_{0}\text{ even},\\
\delta_{h,0}\left|\mathtt{D}_{\boldsymbol{g}}^{t\delta_{z_{0}}^{z}-t\delta_{x_{0}}^{x}}\mathtt{D}_{\boldsymbol{g}}^{\boldsymbol{s}}\right\rangle , & j\leq y_{0}\text{ odd}.
\end{cases}\label{eq:pi_chargey}
\end{align}

To get these operators organized,
we consider the algebra $\mathcal{D}\left[\mathtt{C}\right]\coloneqq\mathcal{D}_{x}\left[\mathtt{C}\right]\otimes\mathcal{D}_{y}\left[\mathtt{C}\right]\otimes\mathcal{D}_{z}\left[\mathtt{C}\right]$ with each factor $\mathcal{D}_{\mu}\left[\mathtt{C}\right]$ and its basis given by
\begin{align}
\mathcal{D}_{\mu}\left[\mathtt{C}\right]\coloneqq\bigotimes_{n\in\mathtt{C}^{\mu}}\mathcal{D}^{\omega_{n}^{\mu}}\left(G\right), &  & D_{\boldsymbol{h}^{\mu}}^{\boldsymbol{t}^{\mu}}\coloneqq\bigotimes_{i\in\mathtt{C}^{\mu}}D_{h_{n}^{\mu}}^{t_{n}^{\mu}},
\end{align}
$\forall \mu=x,y,z$. We write $D_{\boldsymbol{h}}^{\boldsymbol{t}}\coloneqq D_{\boldsymbol{h}^{x}}^{\boldsymbol{t}^{x}}\otimes D_{\boldsymbol{h}^{y}}^{\boldsymbol{t}^{y}}\otimes D_{\boldsymbol{h}^{z}}^{\boldsymbol{t}^{z}}$
for short, where $\boldsymbol{h}=\left(\boldsymbol{h}^{x},\boldsymbol{h}^{y},\boldsymbol{h}^{z}\right), \boldsymbol{t}=\left(\boldsymbol{t}^{x},\boldsymbol{t}^{y},\boldsymbol{t}^{z}\right)
\in G^{\mathtt{C}^{x}}\times G^{\mathtt{C}^{y}}\times G^{\mathtt{C}^{z}}
$.
The left and right actions of ${\cal D}\left(\mathtt{C}\right)$ on
$\mathcal{H}_{*}\left(\overline{\mathtt{M}}\right)$ are 
\begin{align}
\pi\left(D_{\boldsymbol{h}}^{\boldsymbol{t}}\right) & =\prod_{i=x_{0}}^{x_{1}}D_{i,h_{i}^{x}}^{x,t_{i}^{x}}\cdot\prod_{j=y_{0}}^{y_{1}}D_{j,h_{j}^{y}}^{y,t_{j}^{y}}\cdot\prod_{k=z_{0}}^{z_{1}}D_{k,h_{k}^{z}}^{z,t_{k}^{z}},\\
\overline{\pi}\left(D_{\boldsymbol{h}}^{\boldsymbol{t}}\right) & =\prod_{i=x_{0}}^{x_{1}}\overline{D}_{i,h_{i}^{x}}^{x,t_{i}^{x}}\cdot\prod_{j=y_{0}}^{y_{1}}\overline{D}_{j,h_{j}^{y}}^{y,t_{j}^{y}}\cdot\prod_{k=z_{0}}^{z_{1}}\overline{D}_{k,h_{k}^{z}}^{z,t_{k}^{z}},
\end{align}
supported near $\partial\mathtt{C}'$ and $\partial\mathtt{C}$ respectively. More precisely, 
$\pi\left(D_{\boldsymbol{h}}^{\boldsymbol{t}}\right)$
can be realized by operators near the region on $\partial\mathtt{C}'$ between planes $z=z_0$ and $z=z_1$, since $Z_{l}^{t}$ in Eq.~\eqref{eq:DZ} acts trivially unless $z_{0}-1\leq l\leq z_{1}$.
By construction,
\begin{align}
\pi\left(D_{\boldsymbol{h}}^{\boldsymbol{t}}\right)\left|\mathtt{D}_{\boldsymbol{g}}^{\boldsymbol{s}}\right\rangle  & =\left|\mathtt{D}_{\boldsymbol{h}}^{\boldsymbol{t}}\mathtt{D}_{\boldsymbol{g}}^{\boldsymbol{s}}\right\rangle ,\\
\overline{\pi}\left(D_{\boldsymbol{h}}^{\boldsymbol{t}}\right)\left|\mathtt{D}_{\boldsymbol{g}}^{\boldsymbol{s}}\right\rangle  & =\left|\mathtt{D}_{\boldsymbol{g}}^{\boldsymbol{s}}\mathtt{D}_{\boldsymbol{h}}^{\boldsymbol{t}}\right\rangle .
\end{align} In particular, they are zero
if $\boldsymbol{h}\notin F\left(\mathtt{C}\right)$. 

Let $\mathcal{A}\left[\mathtt{C}\right]$ be the subalgebra of $\mathcal{D}\left[\mathtt{C}\right]$
spanned by $D_{\boldsymbol{h}}^{\boldsymbol{t}}$ with $\boldsymbol{h}\in F\left(\mathtt{C}\right)$ and $\boldsymbol{t}
\in G^{\mathtt{C}^{x}}\times G^{\mathtt{C}^{y}}\times G^{\mathtt{C}^{z}}
$.
Then
\begin{equation}
\mathcal{H}_{*}\left(\overline{\mathtt{M}}\right)\xrightarrow{\sim}\mathcal{A}\left[\mathtt{C}\right]:\,\left|\mathtt{D}_{\boldsymbol{g}}^{\boldsymbol{s}}\right\rangle \mapsto D_{\boldsymbol{g}}^{\boldsymbol{s}}
\label{eq:bimodule_iso}
\end{equation}
is an isomorphism between 
$\mathcal{H}_{*}\left(\mathtt{M}\right)$ and 
$\mathcal{A}\left[\mathtt{C}\right]$ as a $\mathcal{D}\left[\mathtt{C}\right]$-$\mathcal{D}\left[\mathtt{C}\right]$-bimodule
(\emph{i.e.}, a vector space carrying both left and right actions
of $\mathcal{D}\left[\mathtt{C}\right]$). 


Let $\mathfrak{Q}_{n}^{\mu}$ be the isomorphism classes of irreducible
representations of $\mathcal{D}^{\omega_{n}^{\mu}}\left(G\right)$.
We call each element of $\mathfrak{Q}_{n}^{\mu}$ a \emph{$\mu$ topological
charge}. Pick an irreducible representation $\mathcal{V}_{\mathfrak{a}_{n}^{\mu}}=\left(\rho_{\mathfrak{a}_{n}^{\mu}},V_{\mathfrak{a}_{n}^{\mu}}\right)$
on a Hilbert space with $\dagger$ respected for each $\mathfrak{a}_{n}^{\mu}\in\mathfrak{Q}_{n}^{\mu}$.
Then 
\begin{equation}
\rho=\bigoplus_{\left(\mathfrak{a}_{n}^{\mu}\right)_{n\in\mathtt{C}^{\mu}}^{\mu=x,y,z}}\bigotimes_{\mu,n}\rho_{\mathfrak{a}_{n}^{\mu}}
\end{equation}
gives an isomorphism of algebras 
\begin{equation}
\mathcal{D}\left[\mathtt{C}\right]\simeq\bigoplus_{\left(\mathfrak{a}_{n}^{\mu}\right)_{n\in\mathtt{C}^{\mu}}^{\mu=x,y,z}}\bigotimes_{\mu,n}\mathcal{L}\left(\mathcal{V}_{\mathfrak{a}_{n}^{\mu}}\right).
\label{eq:DC_sectors-1}
\end{equation}
Explicitly, a $\mu$ topological charge $\mathfrak{a}_{n}^{\mu}\in\mathfrak{Q}_{n}^{\mu}$
is labeled by a pair $\left(g_{n}^{\mu},\varrho_{n}^{\mu}\right)$,
where $g_{n}^{\mu}\in G$ describes the flux and $\varrho_{n}^{\mu}$
is an irreducible $\omega_{n,g_{n}^{\mu}}^{\mu}$-representation (up
to isomorphism) of $G$. Refer to Appendix~\ref{subsec:Reps_DG}
for the details of the labels.

Let $\mathfrak{Q}\left[\mathtt{C}\right]$ be the set of $\boldsymbol{\mathfrak{a}}=\left(\mathfrak{a}_{n}^{\mu}\right)_{n\in\mathtt{C}^{\mu}}^{\mu=x,y,z}=\left(g_{n}^{\mu},\varrho_{n}^{\mu}\right)_{n\in\mathtt{C}^{\mu}}^{\mu=x,y,z}$ with $\left(g_{n}^{\mu}\right)_{n\in\mathtt{C}^{\mu}}^{\mu=x,y,z}\in F\left(\mathtt{C}\right)$. Then the composition 
\begin{align}
\mathcal{H}_{*}\left(\overline{\mathtt{M}}\right)\xrightarrow[\sim]{\left|\mathtt{D}_{\boldsymbol{g}}^{\boldsymbol{s}}\right\rangle \mapsto  D_{\boldsymbol{g}}^{\boldsymbol{s}}}\mathcal{A}\left[\mathtt{C}\right] & \xrightarrow[\sim]{\tilde{\rho}}\nonumber \\
\bigoplus_{\boldsymbol{\mathfrak{a}}\in\mathfrak{Q}\left[\mathtt{C}\right]}\bigotimes_{\mu,n}\mathcal{L}\left(\mathcal{V}_{\mathfrak{a}_{n}^{\mu}}\right) & =\bigoplus_{\boldsymbol{\mathfrak{a}}\in\mathfrak{Q}\left[\mathtt{C}\right]}\mathcal{V}_{\boldsymbol{\mathfrak{a}}}\otimes\mathcal{V}_{\boldsymbol{\mathfrak{a}}}^{*}
\end{align}
is an ismorphism of Hilbert spaces respecting both the left and right actions of $\mathcal{D}\left[\mathtt{C}\right]$, where 
$\mathcal{V}_{\boldsymbol{\mathfrak{a}}}\coloneqq\bigotimes_{\mu,n}\mathcal{V}_{\mathfrak{a}_{n}^{\mu}}$ and
\begin{equation}
\tilde{\rho}\coloneqq\bigoplus_{\boldsymbol{\mathfrak{a}}\in\mathfrak{Q}\left[\mathtt{C}\right]}\bigotimes_{\mu,n}\sqrt{\frac{\dim_{\mathbb{C}}\mathcal{V}_{\mathfrak{a}_{n}^{\mu}}}{\left|G\right|}}\rho_{\mathfrak{a}_{n}^{\mu}}.\label{eq:rho_xcube-1}
\end{equation}
The normalization for each sector in $\tilde{\rho}$ is picked such that the inner product structure is respected.
Clearly, $\mathfrak{Q}\left[\mathtt{C}\right]$ labels particle types of the excited cuboid $\mathtt{C}$ and 
$\mathcal{V}_{\boldsymbol{\mathfrak{a}}}$ (resp. $\mathcal{V}_{\boldsymbol{\mathfrak{a}}}^{*}$)
describes the degrees of freedom near $\partial\mathtt{C'}$ (resp. $\partial\mathtt{C}$).
Physically, $\mathfrak{a}_{k}^{z}=(g_{k}^{z},\varrho_{k}^{z})$ can be detected by braiding a pair of quasiparticles
in the $x$ and $y$ directions via operator supported near grey region in $\partial\mathtt{C'}$ as in Fig.~\ref{fig:topcharges}. Thus, $\mathfrak{a}_{k}^{z}$ is called a $z$ topological charge. Similarly, $\mathfrak{a}_{i}^{x}$ (resp. $\mathfrak{a}_{j}^{y}$) is called a $x$ (resp. $y$) topological charge. 
Since $g_{x_0}^x=g_{y_0}^y=g_{z_0}^z=0$, we have 
\begin{equation}
\begin{array}{ccc}
\mathfrak{a}_{x_{0}}^{x}=\left(0,\varrho_{q^{x}}\right), & \mathfrak{a}_{y_{0}}^{y}=\left(0,\varrho_{q^{y}}\right), & \mathfrak{a}_{z_{0}}^{z}=\left(0,\varrho_{q^{z}}\right),\end{array} \label{eq:a3q}
\end{equation}
where $q^{x},q^{y},q^{z}\in G\simeq \widehat{G}$ with $G$ identified with its character group $\widehat{G}$ and $\varrho_{q^{x}},\varrho_{q^{y}},\varrho_{q^{z}}$ denoting the corresponding representations.

Distinct from conventional topological orders, the number of allowed particle types of a finite excited region $\mathtt{C}$ in a fracton model increases/decreases as the size of $\mathtt{C}$ grows/shrinks. If a quasiparticle can be localized in a smaller cuboid $\mathtt{C}_{a}=\left[a_{0}^{x},a_{1}^{x}\right]\times\left[a_{0}^{y},a_{1}^{y}\right]\times\left[a_{0}^{z},a_{1}^{z}\right] \subset \mathtt{C}$, then  the analogues of 
Eqs.~(\ref{eq:pi_chargez}-\ref{eq:pi_chargey})
for this smaller excitation imply that
its particle type $\boldsymbol{\mathfrak{a}}=(\mathfrak{a}_{n}^{\mu})_{n\in\mathtt{C}^{\mu}}^{\mu=x,y,z} \in \mathfrak{Q}\left[\mathtt{C}\right]$ satisfies
\begin{equation}
\mathfrak{a}_{k}^{z}=\begin{cases}
\mathfrak{0}, & k>a_{1}^{z},\\
\left(0,\varrho_{q^{z}}\right), & k\leq a_{0}^{z}\text{ even},\\
\left(0,\varrho_{q^{x}-q^{y}}\right), & k\leq a_{0}^{z}\text{ odd}
\end{cases} \label{eq:az}
\end{equation}
and the counterparts obtained by permuting $x,y,z$ cyclically, where $\mathfrak{0}$ denotes the trivial representation
(\emph{i.e.}, the counit) of any $\mathcal{D}^{\omega_{n}^{\mu}}(G)$.
In other words, $\mathfrak{Q}\left[\mathtt{C}_{a}\right]$ can be viewed as a subset of $\mathfrak{Q}\left[\mathtt{C}\right]$; each 
$\mathcal{V}_{\boldsymbol{\mathfrak{a}}}$ for $\boldsymbol{\mathfrak{a}}\in \mathfrak{Q}\left[\mathtt{C}_{a}\right]$ carries an irreducible representation of $\mathcal{D}\left[\mathtt{C}\right]$ for any cuboid $\mathtt{C}$ containing $\mathtt{C}_{a}$.

\subsection{Fusion of quasiparticles}
\label{subsec:Fusion-cb}

Suppose that there are two spatially separated excited cuboids $\mathtt{C}_{a}$
and $\mathtt{C}_{b}$ contained deep inside a much larger cuboid
$\mathtt{C}'$. Let $\mathtt{M}\coloneqq\mathtt{C}^{\prime}-\mathtt{C}_{a}^{\circ}-\mathtt{C}_{b}^{\circ}$.
The discussion in the above subsection can be repeated here for the
two-hole manifold $\mathtt{M}$. With the coloring on $\partial\mathtt{C}'$
and local degrees of freedom near $\mathtt{C}_{a}$ and $\mathtt{C}_{b}$
fixed separately, we are left with Hilbert spaces $\mathcal{V}\left[\boldsymbol{\mathfrak{a}},\boldsymbol{\mathfrak{b}}\right]$
labeled by $\boldsymbol{\mathfrak{a}}\in\mathfrak{Q}\left[\mathtt{C}_{a}\right]$
and $\boldsymbol{\mathfrak{b}}\in\mathfrak{Q}\left[\mathtt{C}_{b}\right]$. Using
two copies of Eq.~\eqref{eq:rho_xcube-1}, we have 
\begin{equation}
\mathcal{V}\left[\boldsymbol{\mathfrak{a}},\boldsymbol{\mathfrak{b}}\right]\simeq\mathcal{V}_{\boldsymbol{\mathfrak{a}}}\otimes\mathcal{V}_{\boldsymbol{\mathfrak{b}}},
\end{equation}
where $\mathcal{V}_{\boldsymbol{\mathfrak{a}}}\coloneqq\bigotimes_{\mu,n}\mathcal{V}_{\mathfrak{a}_{n}^{\mu}}$ and $\mathcal{V}_{\boldsymbol{\mathfrak{b}}}\coloneqq\bigotimes_{\mu,n}\mathcal{V}_{\mathfrak{b}_{n}^{\mu}}$.

The action of $\mathcal{D}\left[\mathtt{C}\right]$ on $\mathcal{V}_{\boldsymbol{\mathfrak{a}}}\otimes\mathcal{V}_{\boldsymbol{\mathfrak{b}}}$
via $\pi$ defined in Eq.~\eqref{eq:DC} is specified by the coproduct
\begin{equation}
\varDelta\coloneqq\bigotimes_{\mu=x,y,z}\bigotimes_{n\in\mathtt{C}^{\mu}}\varDelta_{n}^{\mu}.
\end{equation}
The linear space of intertwiners between the representations $\mathcal{V}_{\boldsymbol{\mathfrak{c}}}$
and $\mathcal{V}_{\boldsymbol{\mathfrak{a}}}\otimes\mathcal{V}_{\boldsymbol{\mathfrak{b}}}$
of $\mathcal{D}\left[\mathtt{C}\right]$
\begin{equation}
V_{\boldsymbol{\mathfrak{c}}}^{\boldsymbol{\mathfrak{a}\mathfrak{b}}}\coloneqq\text{Hom}\left(\mathcal{V}_{\boldsymbol{\mathfrak{c}}},\mathcal{V}_{\boldsymbol{\mathfrak{a}}}\otimes\mathcal{V}_{\boldsymbol{\mathfrak{b}}}\right)
\end{equation}
encodes the ways of fusing $\boldsymbol{\mathfrak{a}}$ and $\boldsymbol{\mathfrak{b}}$ into
$\boldsymbol{\mathfrak{c}}\in\mathfrak{Q}\left[\mathtt{C}\right]$. In particular,
$N_{\boldsymbol{\mathfrak{a}\mathfrak{b}}}^{\boldsymbol{\mathfrak{c}}}\coloneqq\dim_{\mathbb{C}}V_{\boldsymbol{\mathfrak{c}}}^{\boldsymbol{\mathfrak{a}\mathfrak{b}}}$
is the corresponding fusion rule. It is possible to fuse $\boldsymbol{\mathfrak{a}}$
and $\boldsymbol{\mathfrak{b}}$ into $\boldsymbol{\mathfrak{c}}$ if and only if $N_{\boldsymbol{\mathfrak{a}\mathfrak{b}}}^{\boldsymbol{\mathfrak{c}}}\geq1$.
Moreover,
$N_{\boldsymbol{\mathfrak{a}\mathfrak{b}}}^{\boldsymbol{\mathfrak{c}}}\geq1$ implies $q_{c}^{\mu}=q_{a}^{\mu}+q_{b}^{\mu},\forall \mu=x,y,z$, where
$q_{a}^{\mu}, q_{b}^{\mu}, q_{c}^{\mu}\in G\simeq \widehat{G}$ are specified by $\mathfrak{a}_{\mu_0}^{\mu}$, $\mathfrak{b}_{\mu_0}^{\mu}$, $\mathfrak{c}_{\mu_0}^{\mu}$ as in Eq.~\eqref{eq:a3q}.

\subsection{Mobility of quasiparticles}

Given an excitation of type $\boldsymbol{\mathfrak{a}}\in\mathfrak{Q}\left[\mathtt{C}_{a}\right]$ inside a cuboid $\mathtt{C}_{a}=\left[a_{0}^{x},a_{1}^{x}\right]\times\left[a_{0}^{y},a_{1}^{y}\right]\times\left[a_{0}^{z},a_{1}^{z}\right]$,
it follows from the same argument in Sec.~\ref{subsec:Mobility-Xcube}
that $\boldsymbol{\mathfrak{a}}$ is mobile in the $z$ direction if and only if
all its $z$ topological charges $\mathfrak{a}_{k}^{z}=\left(g_{\mu}^{z},\varrho_{k}^{z}\right)$
are trivial. With Eqs.~(\ref{eq:cb_f1}-\ref{eq:cb_f3}), $g_{k}^{z}=0,\forall k$
implies that
\begin{align}
\sum_{i\text{ odd}}g_{i}^{x}+\sum_{j\text{ odd}}g_{j}^{y}+\sum_{k\text{ odd}}g_{k}^{z} & \in2G,\label{eq:flux_odd}
\end{align}
where $2G\coloneqq\left\{ 2g|g\in G\right\} $. 

The form of $\mathfrak{a}_{k}^{z}$ for $k\leq a_{0}^{z}$ is given
by Eq.~\eqref{eq:az}. Then $\mathfrak{a}_{k}^{z}=\mathfrak{0},\forall k\leq a_{0}^{z}$
implies that $q^{z}=0$, $q^{z}=q^{y}$ and hence
\begin{equation}
q^{x}+q^{y}+q^{z}\in2\widehat{G},\label{eq:charge_even}
\end{equation}
where
$q^{\mu}\in G\simeq \widehat{G}$ is specified by $\mathfrak{a}_{\mu_0}^{\mu}$ as in Eq.~\eqref{eq:a3q}.
Clearly, the conditions~\eqref{eq:flux_odd} and \eqref{eq:charge_even}
hold as well if the excitation is movable in the $x$ or $y$ direction. In fact, both hold if and only if the excitation
is a fusion result of movable quasiparticles.
Thus, an excitation
is a fracton (\emph{i.e.}, \emph{not} a fusion result of mobile quasiparticles) if and only if either 
$\sum_{i\text{ odd}}g_{i}^{x}+\sum_{j\text{ odd}}g_{j}^{y}+\sum_{k\text{ odd}}g_{k}^{z}\notin2G$
or $q^{x}+q^{y}+q^{z}\notin2\widehat{G}$.  
Similarly to the excitations in the twisted X-cube models, we thus see that the mobility of quasiparticles is determined by their topological charges, thereby allowing us to utilize familiar concepts from the study of topological order to reveal the intriguing phenomenology of fracton order.

\subsection{Braiding of mobile quasiparticles}

The general discussion of braidings in Sec.~\ref{subsec:Braiding_general}
applies here as well. What changes are the constraints on topological charges
$\boldsymbol{\mathfrak{a}}=\left(\mathfrak{a}_{n}^{\mu}\right)_{n\in\mathtt{C}^{\mu}}^{\mu=x,y,z}$.
Rather than analyzing the implications of these modified constraints abstractly, we study the physical consequences directly through examples below. 

\subsection{Examples}

We now illustrate through examples how fracton excitations in the twisted checkerboard models can exhibit semionic or non-Abelian braiding statistics.
Importantly, distinct from the twisted X-cube models, it is possible to construct a twisted checkerboard model with inextricably non-Abelian fractons that is not a fusion result of immobile excitations of quantum dimension 1 and mobile quasiparticles. 
\subsubsection{$G=\mathbb{Z}_{2}$ (untwisted)}

For $G=\mathbb{Z}_{2}=\left\{ 0,1\right\} $, it is known~\cite{chen2013}
that the third cohomology group $H^{3}\left(G,U\left(1\right)\right)=\mathbb{Z}_{2}$,
whose nontrivial element is presented by the 3-cocycle in Eq.~\eqref{eq:cocycleZ2}.
Each layer of cubes in the checkerboard model can be either untwisted
or twisted. The pure charges (\emph{i.e.}, quasiparticles without nontrivial flux)
behave in the same way, no matter whether the model is twisted or
not. Thus, we are more interested in excitations that violate $B_{c}=1$
below.

\begin{figure}
\noindent\begin{minipage}[t]{1\columnwidth}%
\includegraphics[width=0.8\columnwidth]{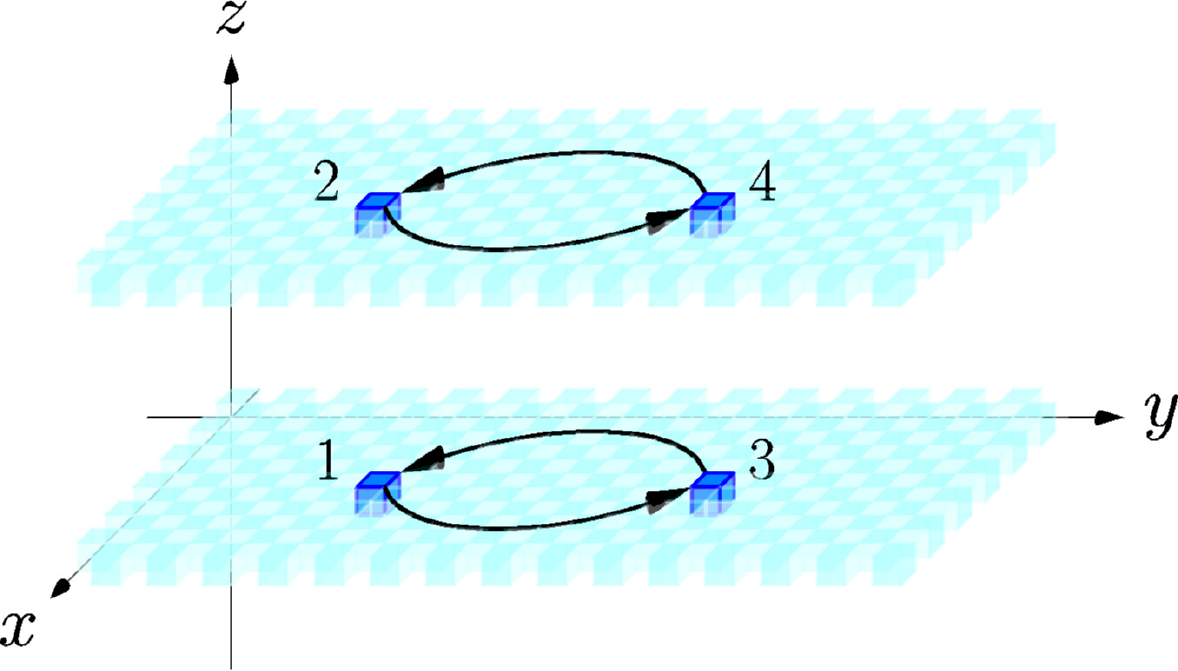}

(a) %
\end{minipage}\medskip{}

\noindent\begin{minipage}[t]{1\columnwidth}%
\includegraphics[width=0.8\columnwidth]{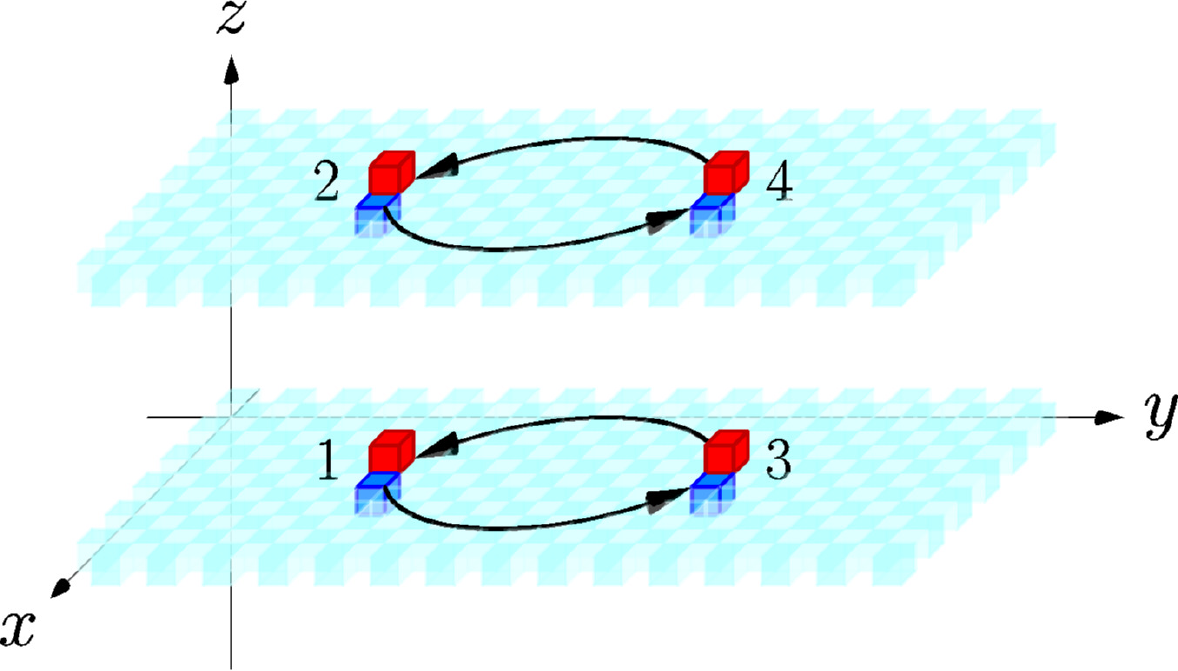}

(b)%
\end{minipage}

\medskip{}

\noindent\begin{minipage}[t]{1\columnwidth}%
\includegraphics[width=0.8\columnwidth]{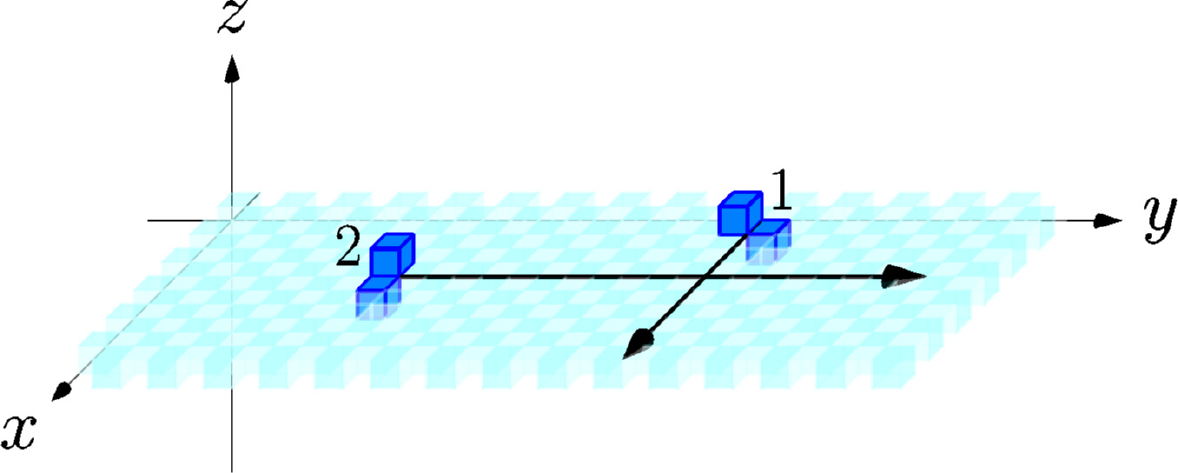}

(c)%
\end{minipage}

\caption{Braidings of particles in twisted or untwisted checkerboard models
with $G=\mathbb{Z}_{2}=\{0,1\}$. (a) Each labeled cube (blue online) in the
illustrated checkerboard layers (cyan online) carries a nontrivial
flux, \emph{i.e.}, the sum of group elements on its vertices
equals $1\neq0$. In the untwisted model, $A_{c}=1$ can be kept on all
cubes. If a labeled cube is in a twisted layer $\Sigma_{k}^{z}$,
it carries the projective representation $\varrho_{k}^{z}\left(s\right)=i^{s}$
and the trivial representation $\varrho_{k'}^{z}\left(s\right)=1$
for $k'\protect\neq k$, where $s\in G$. (b) Each cube (red
online) on top of the drawn layers (cyan online) indicates a
violation of $A_{c}=1$ in the untwisted model. (c) Quasiparticle $1$
(resp. $2$) is movable in the $x$-direction (resp. $y$-direction).}

\label{fig:CB-1}
\end{figure}

First, let us explain the braiding process of two $xy$-particles
in the original (untwisted) checkerboard model in order for the readers
to get familiar with our notations. We consider the four fractons
shown as the cubes (blue online) labeled as $1,2,3,4$ in Fig.~\ref{fig:CB-1}(a),
where only $B_{c}=1$ is violated. In the chosen coordinates, the four cubes are 
centered at $\frac{1}{2}\left(1,1,1\right)+\{ \left(6,8,0\right),\left(6,8,10\right),\left(6,20,0\right),\left(6,20,10\right)\} $
respectively. 

To study the braidings in the $x$ and $y$ directions, we group them into
two pairs $\mathtt{C}_{12}$ (containing cubes $1,2$) and $\mathtt{C}_{34}$
(containing cubes $3,4$). Both $\mathtt{C}_{12}$ and $\mathtt{C}_{34}$
are $xy$-particles; they have trivial fluxes in both the $x$ and $y$ directions. 
Explicitly, the particle type of $\mathtt{C}_{12}$ is specified by its fluxes in each cross section $\mathtt{M}_{k}^{z}$ 
\begin{equation}
g_{k}^{z}=\begin{cases}
1, & k=1\text{ or }11,\\
0, & \text{otherwise}.
\end{cases}\label{eq:C12_flux}
\end{equation}
In fact, Eq.~\eqref{eq:C12_flux} also holds for $\mathtt{C}_{34}$
for the configuration shown in Fig.~\ref{fig:CB-1}(a); it is easy
to see that we can move $\mathtt{C}_{12}$ to $\mathtt{C}_{34}$ and
that they are of the the same particle type. Also, $\mathtt{C}_{12}$
and $\mathtt{C}_{34}$ can fuse into a completely trivial particle.
Without violation of $A_{c}=1$, these are pure fluxes. Hence the braiding
operator $\mathcal{R}$, exchanging $\mathtt{C}_{12}$ and $\mathtt{C}_{34}$
illustrated by the arrows in Fig.~\ref{fig:CB-1}(a), acts trivially
(\emph{i.e.}, $\mathcal{R}=1$).

If $A_{c}=1$ is also violated on the extra cubes (red
online) shown in Fig.~\ref{fig:CB-1}(b), then both $\mathtt{C}_{12}$
and $\mathtt{C}_{34}$ have charges as well. 
Explicitly, the relevant representation of $G=\mathbb{Z}_{2}$, carried
by both $\mathtt{C}_{12}$ and $\mathtt{C}_{34}$, is specified by
\begin{equation}
\varrho_{k}^{z}\left(1\right)=\begin{cases}
-1, & 1<k\leq11\text{ and }k\text{ odd},\\
1, & \text{otherwise}.
\end{cases}
\end{equation}
Therefore, the braiding operator acts as 
\begin{equation}
\mathcal{R}=\bigotimes_{k}\varrho_{k}^{z}\left(g_{k}^{z}\right)=-1.
\end{equation}
Hence, $\mathtt{C}_{12}$ and $\mathtt{C}_{34}$ shown in Fig.~\ref{fig:CB-1}(b)
behave like fermions as for braiding in the $x$ and $y$ directions.

\subsubsection{$G=\mathbb{Z}_{2}$ ($\Sigma_{k}^{z}$ twisted for $k>z_{0}$)}
\label{subsec:brading_half}

Suppose $1<z_0<11$. In particular, the upper layer $\Sigma_{11}^{z}$ (resp. the lower layer $\Sigma_{1}^{z}$) drawn in Fig.~\ref{fig:CB-1}(a)
is twisted (untwisted). Still, we can pair fractons $1,2$ (resp. $3,4$) 
into an $xy$-particle $\mathtt{C}_{12}$ (resp. $\mathtt{C}_{34}$).
We also assume that $\mathtt{C}_{12}$ and $\mathtt{C}_{34}$ are
of the same particle type with their flux configuration given by
Eq.~\eqref{eq:C12_flux}. 

Since the relevant quantum double algebra is twisted on $\mathtt{M}_{11}^{z}$ 
satisfying 
\begin{equation}
D_{11,1}^{z,s}D_{11,1}^{z,t}=\omega_{1}\left(s,t\right)D_{11,1}^{z,s+t},\;\forall s,t\in G,
\end{equation}
an allowed collection of representations carried by both $\mathtt{C}_{12}$
and $\mathtt{C}_{34}$ can be specified by
\begin{equation}
\varrho_{k}^{z}\left(1\right)=\begin{cases}
i, & k=11,\\
1, & \text{otherwise}.
\end{cases}
\end{equation}
The braiding operator that exchanges $\mathtt{C}_{12}$ and $\mathtt{C}_{34}$
as illustrated in Fig.~\ref{fig:CB-1}(a) is 
\begin{equation}
\mathcal{R}=\bigotimes_{k}\varrho_{k}^{z}\left(g_{k}^{z}\right)=\varrho_{11}^{z}\left(1\right)=i,\label{eq:semion}
\end{equation}
which shows a semionic behavior. However, the semionic behavior cannot
appear in an untwisted model, where $\varrho_{k}^{z}\left(g_{k}^{z}\right)$
is always $\pm1$. This clearly shows that there is no continuous
path of gapped local Hamiltonians connecting the untwisted model and
a partially twisted model, in which $\mathtt{M}_{k}^{z}$ is twisted
for $k\geq z_{0}$ and untwisted for $k<z_{0}$. 

\subsubsection{$G=\mathbb{Z}_{2}$ ($\Sigma_{k}^{z}$ twisted for $k$ odd)}

There are many different ways of twisting the checkerboard model. Another simple case
is to twist all $\mathtt{M}_{k}^{z}$ with $k$ odd but to leave all
$\Sigma_{k}^{z}$ with $k$ even untwisted, which we called the half
twisted model for short. In this case, there are no semionic 2d mobile
particles. However, we could instead consider 1d mobile particles,
as shown in Fig.~\ref{fig:CB-1}(c). The nontrivial fluxes and representations
carried by quasiparticle $1$ are 
\begin{align}
g_{1}^{z} & =g_{2}^{z}=1,\\
g_{n}^{y} & =g_{n+1}^{y}=1,\\
\varrho_{1}^{z}\left(s\right) & =i^{s},\forall s\in\mathbb{Z}_{2},
\end{align}
while quasiparticle $2$ carries
\begin{align}
h_{1}^{z} & =h_{2}^{z}=1,\\
h_{m}^{x} & =h_{m+1}^{x}=1,\\
\varsigma_{1}^{z}\left(t\right) & =i^{t},\forall t\in\mathbb{Z}_{2},
\end{align}
with some $m,n\in \mathbb{Z}$. Together, they fuse into a $z$-particle. 

Let $\mathcal{O}_{x}$ (resp. $\mathcal{O}_{y}$) be the operator moving
quasiparticle $1$ (resp. $2$) along the $x$ (resp. $y$) direction.
They (\emph{i.e.}, $\mathcal{O}_{x}$ and $\mathcal{O}_{y}$) are supported near the corresponding arrows in Fig.~\ref{fig:CB-1}(c). Then  $\mathcal{O}_{x}\mathcal{O}_{y}$ and $\mathcal{O}_{y}\mathcal{O}_{x}$
differ by a full braiding of the identical $z$-topological
charges of the two quasiparticles, which equals
\begin{equation}
\mathcal{R}^{2}=\bigotimes_{k}\left[\varrho_{k}^{z}\left(h_{k}^{z}\right)\otimes\varsigma_{k}^{z}\left(g_{k}^{z}\right)\right]=-1.
\end{equation}

On the other hand, if the model is untwisted, $\mathcal{R}^{2}=1$
as long as the two one-dimensional particles $1,2$ fuse to a particle
mobile in the third direction. This is because the fusion condition
implies $g_{k}^{z} =h_{k}^{z}\in G$  and 
$\varrho_{k}^{z}=\varsigma_{k}^{z}$ as one-dimensional representations of $G$, $\forall k$. Thus, $\varrho_{k}^{z}\left(h_{k}^{z}\right)\otimes\varsigma_{k}^{z}\left(g_{k}^{z}\right)=\left[\varrho_{k}^{z}\left(g_{k}^{z}\right)\right]^{2}=1$ in the untwisted model.

In summary, in the untwisted model, if an $x$-particle and a $y$-particle
fuse into a $z$-particle, then the corresponding hopping operators
always commute
\begin{equation}
\mathcal{O}_{x}\mathcal{O}_{y}=\mathcal{O}_{y}\mathcal{O}_{x},
\end{equation}
while in the model with $\mathtt{M}_{k}^{z}$ twisted alternately,
we can have 
\begin{equation}
\mathcal{O}_{x}\mathcal{O}_{y}=-\mathcal{O}_{y}\mathcal{O}_{x}.
\end{equation}
This clearly distinguishes the half twisted model from the untwisted model. 

\subsubsection{$G=\mathbb{Z}_{2}$ (fully twisted)}

Suppose that $\Sigma_{k}^{z}$ is twisted for all $k\in\mathbb{Z}$.
Here, we have not found a characteristic braiding process which distinguishes
this model from the untwisted case.

However, we can still argue that there is no continuous path of gapped
local Hamiltonians connecting the untwisted model and the fully twisted
model. Let us give a proof by contradiction. Suppose there exists
a continuous change of gapped local Hamiltonians $\mathcal{H}\left(\tau\right)$
parameterized by $\tau\in\left[0,1\right]$ such that $\mathcal{H}\left(0\right)$
and $\mathcal{H}\left(1\right)$ are the untwisted and the fully twisted
checkerboard models respectively. Then there exists a local unitary
transformation $U^{\text{loc}}$, which can be described as a finite-depth
quantum circuit, such that $\mathcal{H}\left(1\right)=(U^{\text{loc}})^{\dagger}\mathcal{H}\left(0\right)U^{\text{loc}}$
~\cite{chen2010}. Let $U_{z>0}^{\text{loc}}$ be a local unitary operator
obtained by keeping only operators in $U^{\text{loc}}$ supported
on the region $z>0$. Then $(U_{z>0}^{\text{loc}})^{\dagger}\mathcal{H}\left(0\right)U_{z>0}^{\text{loc}}$
describes a model which is untwisted for $z\leq0$ and twisted for $z\geq L$,
where $L$ is a finite positive number characterizing the correlation
length. 

Moreover, let us consider the braiding process shown in Fig.~\ref{fig:CB-1}(a)
with fractons $1,3$ located in $z\leq0$ and fractons $2,4$ in $z\geq L$.
If the braiding is made on the ground state of $\mathcal{H}\left(0\right)$
(resp. $(U_{z>0}^{\text{loc}})^{\dagger}\mathcal{H}\left(0\right)U_{z>0}^{\text{loc}}$),
then the exchange of $\mathtt{C}_{12}$ and $\mathtt{C}_{34}$ cannot be semionic (resp. can be semionic). However, the \emph{local} unitary transformation
cannot change the braiding statistics, which are a non-local property of the topological charges. This leads to a contradiction, which proves the non-existence of a continuous path of gapped local Hamiltonians
connecting the untwisted model and the fully twisted model. It remains to be seen whether there exists a braiding process which clearly distinguishes these two cases. 

\subsubsection{$G=\mathbb{Z}_{2}\times\mathbb{Z}_{2}\times\mathbb{Z}_{2}$: non-Abelian fractons}

Finally, let us give an example of a model which provides an explicit realization of non-Abelian fractons, one of the central results of our work. It
is constructed with the group $G=\mathbb{Z}_{2}\times\mathbb{Z}_{2}\times\mathbb{Z}_{2}$
and the 3-cocycle
\begin{equation}
\omega\left(f,g,h\right)=e^{i\pi\left(f^{\left(1\right)}g^{\left(2\right)}h^{\left(3\right)}\right)}, \label{eq:cocycle_nonAb}
\end{equation}
where $f=\left(f^{\left(1\right)},f^{\left(2\right)},f^{\left(3\right)}\right),g=\left(g^{\left(1\right)},g^{\left(2\right)},g^{\left(3\right)}\right),h=\left(h^{\left(1\right)},h^{\left(2\right)},h^{\left(3\right)}\right)\in G$.
We will also interchangeably write the elements of $G$ simply as $000$, $100$,
$110$ and so on for short. As examples, we have $\omega\left(100,010,001\right)=-1$
and $\omega\left(100,001,010\right)=1$ in such notations. Now, to work out
an explicit example, we twist $\Sigma_{k}^{z}$ for all $k$. 

\begin{figure}
	\noindent\begin{minipage}[t]{1\columnwidth}%
		\includegraphics[width=0.8\columnwidth]{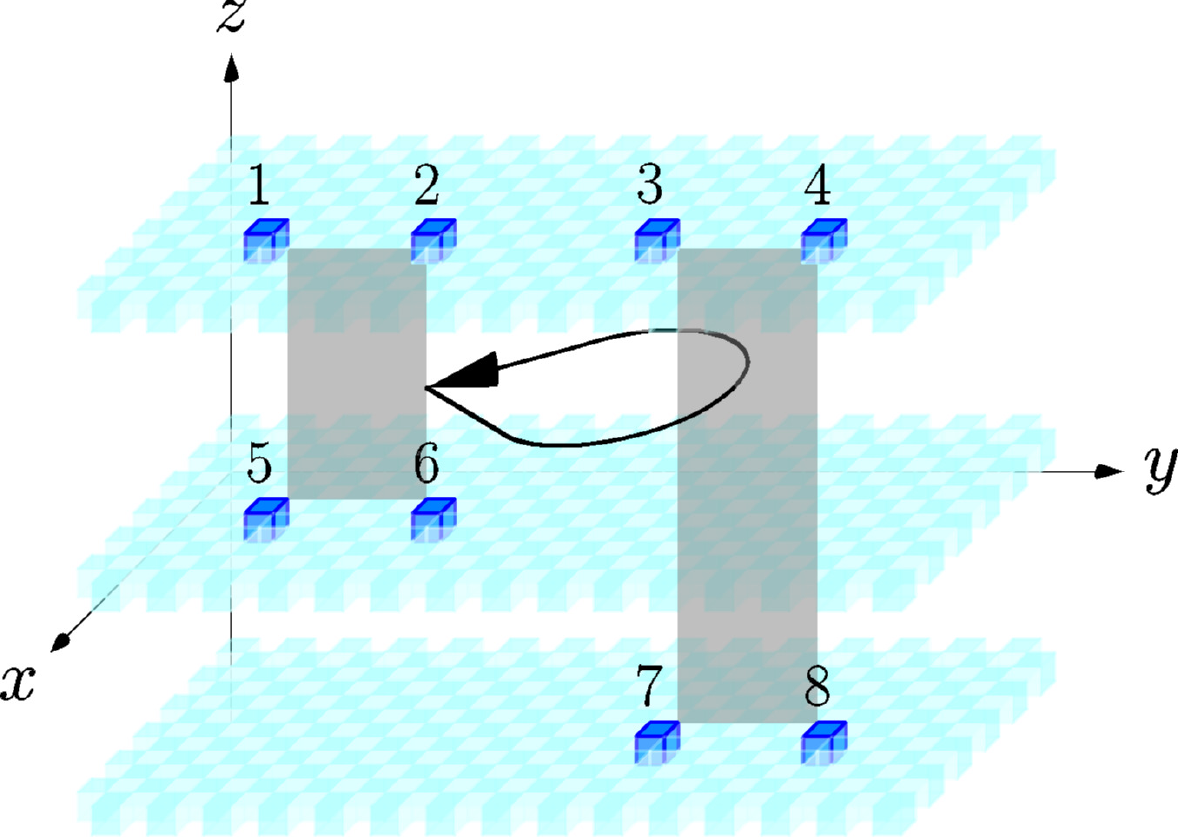}
		
		(a) %
	\end{minipage}\medskip{}
	
	\noindent\begin{minipage}[t]{1\columnwidth}%
		\includegraphics[width=0.8\columnwidth]{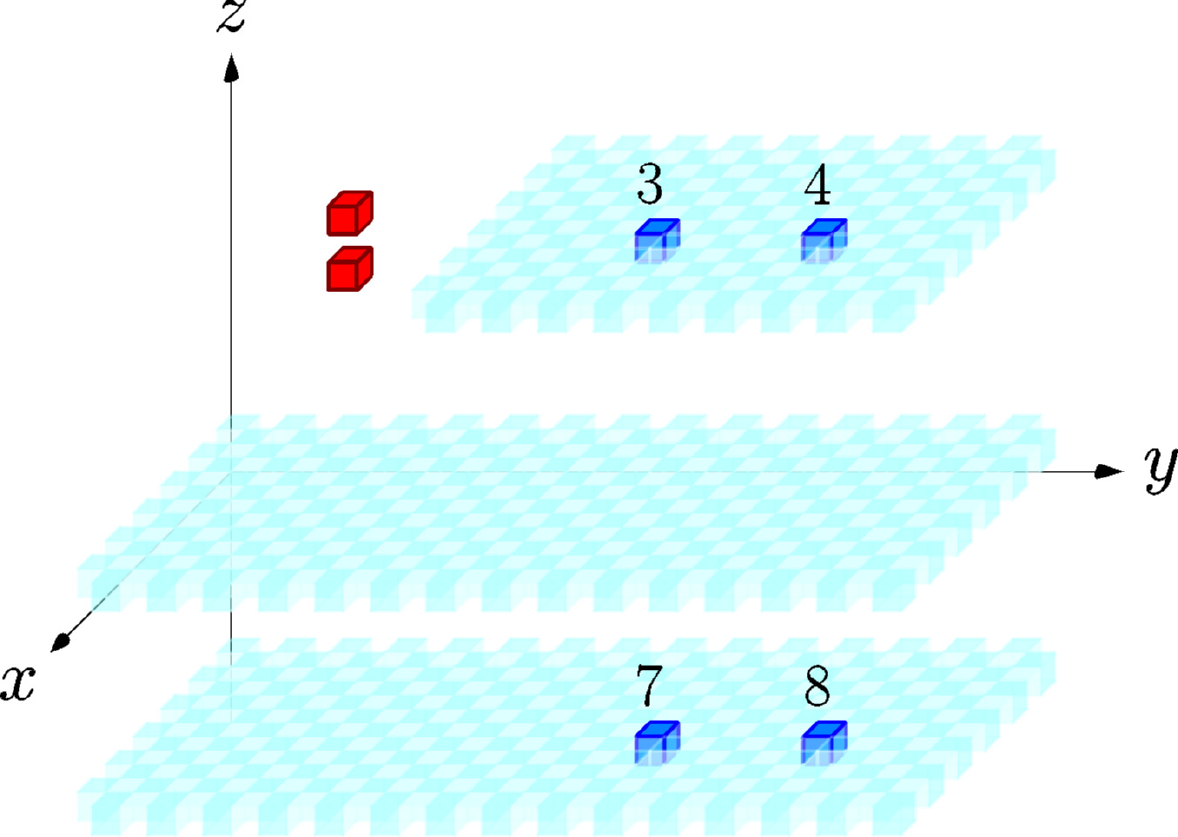}
		
		(b)%
	\end{minipage}
	
	\caption{(a) Two groups of fractons with nontrivial flux $1,2,5,6$ and $3,4,7,8$ are created from
		vacuum separately by an operator supported near the corresponding
		grey membrane. Fractons $2$ and $6$ are paired into a quasiparticle
		mobile in two dimensions and move around fracton $3$ along the path
		indicated by the arrow. (b) After the braiding, fractons $1$, $2$,
		$5$ and $6$ cannot fuse back into vacuum any more but left with
		a pair of pure charges (red online) just above and below the checkerboard layer
		(cyan online) where fractons $1$, $2$, $3$ and $4$ violate $B_{c}=1$,
		if the checkerboard model based on the group $G=\mathbb{Z}_{2}\times\mathbb{Z}_{2}\times\mathbb{Z}_{2}$
		is twisted by $\omega\left(f,g,h\right)=e^{i\pi\left(f^{\left(1\right)}g^{\left(2\right)}h^{\left(3\right)}\right)}$
		along this layer.}
	
	\label{fig:CB-2}
\end{figure}

Let us consider the eight fractons labeled as $1,2,\cdots,8$ in Fig.~\ref{fig:CB-2}(a),
divided into two groups. Each group is created by an operator supported
near the corresponding grey sheet. In the left (resp. right) group,
each fracton carries a flux $100$ (resp. $011$) and a projective
representation $\varrho_{100}$ (resp. $\varrho_{011}$) satisfying
\begin{align}
\varrho_{100}\left(g\right)\varrho_{100}\left(h\right) & =\omega_{100}\left(g,h\right)\varrho_{100}\left(gh\right),\\
\varrho_{011}\left(g\right)\varrho_{011}\left(h\right) & =\omega_{011}\left(g,h\right)\varrho_{011}\left(gh\right).
\end{align}
 for $g,h\in G$. A choice of $\varrho_{100}$ and $\varrho_{011}$
is given by
\begin{gather}
\varrho_{100}\left(100\right)=\sigma_{0},\,\varrho_{100}\left(010\right)=\sigma_{1},\,\varrho_{100}\left(001\right)=\sigma_{3},\\
\varrho_{011}\left(100\right)=\sigma_{3},\,\varrho_{011}\left(010\right)=\sigma_{1},\,\varrho_{011}\left(001\right)=\sigma_{1},
\end{gather}
where $\sigma_{0},\sigma_{1},\sigma_{2},\sigma_{3}$ denote the $2\times2$ identity matrix and the three Pauli matrices.
Together with fluxes, they correspond to $\rho_{100}^{+}$ and $\rho_{011}^{+}$
in Table~\ref{table:irrep} of Appendix~\ref{subsec:ExamplesBTC}.

In particular, each fracton with flux $100$ (resp. $011$) carries
a two-dimensional Hilbert space $\mathcal{V}_{100}$ (resp. $\mathcal{V}_{011}$),
whose basis is denoted as $\left\{ \left|100;\uparrow\right\rangle ,\left|100;\downarrow\right\rangle \right\} $
(resp. $\left\{ \left|011;\uparrow\right\rangle ,\left|011;\downarrow\right\rangle \right\} $).
When the flux is clear from context, we simply write $\left|\uparrow\right\rangle $,
$\left|\downarrow\right\rangle $ for short. We pair
fractons $1$ and $5$ into an $xy$-particle, denoted $\mathtt{C}_{15}$.
Similarly, we have $xy$-particles $\mathtt{C}_{26}$, $\mathtt{C}_{37}$,
$\mathtt{C}_{48}$. 

Since fractons $1,2,5,6$ are created together from the ground state,
their total topological charge is trivial and hence they are in the state $\left|\uparrow\uparrow+\downarrow\downarrow\right\rangle _{12}\otimes\left|\uparrow\uparrow+\downarrow\downarrow\right\rangle _{56}$.
For example, in the chosen coordinates, the fractons $1,2$ carry flux $100$ each
in $\Sigma_{11}^{z}$; direct computation
shows
\begin{align}
 & \varDelta\left(D_{11,h}^{z,g}\right)\left|\uparrow\uparrow+\downarrow\downarrow\right\rangle _{12}\nonumber \\
= & \delta_{h,000}\omega^{g}\left(100,100\right)\varrho_{100}\left(g\right)\otimes\varrho_{100}\left(g\right)\left|\uparrow\uparrow+\downarrow\downarrow\right\rangle _{12}\nonumber \\
= & \delta_{h,000}\left|\uparrow\uparrow+\downarrow\downarrow\right\rangle _{12}\label{eq:fracton12}
\end{align}
for $g=100,010,001$ and hence the $z$ topological charge of $\left|\uparrow\uparrow+\downarrow\downarrow\right\rangle _{12}$
is trivial. Similarly, the state of fractons $3,4,7,8$ with trivial
topological charge is $\left|\uparrow\downarrow+\downarrow\uparrow\right\rangle _{34}\otimes\left|\uparrow\downarrow+\downarrow\uparrow\right\rangle _{78}$. 

Next, let us consider the monodromy operator braiding $\mathtt{C}_{26}$
around $\mathtt{C}_{37}$ as shown in  Fig.~\ref{fig:CB-2}(a). It acts
trivially on $5,6,7,8$ and nontrivially on $1,2,3,4$ as
\begin{multline}
\sigma_{0}\otimes\varrho_{100}\left(011\right)\otimes\varrho_{011}\left(100\right)\otimes\sigma_{0}\left|\uparrow\uparrow+\downarrow\downarrow\right\rangle _{12}\left|\uparrow\downarrow+\downarrow\uparrow\right\rangle _{34}\\
=-\left|\uparrow\downarrow-\downarrow\uparrow\right\rangle _{12}\left|\uparrow\downarrow-\downarrow\uparrow\right\rangle _{34},
\end{multline}
where we have used 
\begin{equation}
\varrho_{100}\left(011\right)=\omega_{100}\left(010,001\right)\varrho_{100}\left(010\right)\varrho_{100}\left(001\right)=-\sigma_{1}\sigma_{3}
\end{equation}
and the associator between $\left(\mathcal{V}_{100}\otimes\mathcal{V}_{100}\right)\otimes\left(\mathcal{V}_{011}\otimes\mathcal{V}_{011}\right)$
and $\left(\mathcal{V}_{100}\otimes\left(\mathcal{V}_{100}\otimes\mathcal{V}_{011}\right)\right)\otimes\mathcal{V}_{011}$
equals the identity. 

As in Eq.~\eqref{eq:fracton12}, direct computation shows that $D_{11,h}^{z,g}$ acts as
\begin{align}
 & \varDelta\left(D_{11,h}^{z,g}\right)\left|\uparrow\downarrow-\downarrow\uparrow\right\rangle _{12}\nonumber \\
= & \delta_{h,000}\omega^{g}\left(100,100\right)\varrho_{100}\left(g\right)\otimes\varrho_{100}\left(g\right)\left|\uparrow\downarrow-\downarrow\uparrow\right\rangle _{12}\nonumber \\
= & \delta_{h,000}\left(-1\right)^{g^{\left(2\right)}+g^{\left(3\right)}}\left|\uparrow\downarrow-\downarrow\uparrow\right\rangle _{12},\label{eq:fracton12-1}\\
 & \varDelta\left(D_{11,h}^{z,g}\right)\left|\uparrow\downarrow-\downarrow\uparrow\right\rangle _{34}\nonumber \\
= & \delta_{h,000}\omega^{g}\left(011,011\right)\varrho_{011}\left(g\right)\otimes\varrho_{011}\left(g\right)\left|\uparrow\downarrow-\downarrow\uparrow\right\rangle _{34}\nonumber \\
= & \delta_{h,000}\left(-1\right)^{g^{\left(2\right)}+g^{\left(3\right)}}\left|\uparrow\downarrow-\downarrow\uparrow\right\rangle _{34}.\label{eq:fracton34}
\end{align}

This implies that the group of fractons $1,2,5,6$ cannot fuse back
into the vacuum any more after braiding. At best, we can annihilate all
fluxes, resulting in a pair of pure charges, as depicted in Fig.~\ref{fig:CB-2}(b).
This analysis holds for the group of fractons $3,4,7,8$ as well. Thus, the braiding process described here is a clear and unambiguous signature of fractons with a quantum dimension greater than 1.
A similar braiding was discussed in the context of anyons in twisted gauge theory  \cite{hep-th/9511195}.

Clearly, if the model is twisted fully by Eq.~\eqref{eq:cocycle_nonAb} in at least one direction (e.g. $\Sigma_{k}^{z}$ is twisted for all $k$), then the above fractons are all \emph{inextricably} non-Abelian, \emph{i.e.,} their quantum dimension cannot be reduced to 1 by adding or removing some mobile quasiparticles. However, if $\Sigma_{k}^{z}$ is alternately twisted between even and odd layers, then the above fractons can be made Abelian by adding or removing a 1d mobile quasiparticle and these are hence not inextricably non-Abelian fractons. But the added or removed 1d mobile quasiparticle is inextricably non-Abelian. Thus, such a model still displays a non-Abelian fracton phase. Finally, if $\Sigma_{k}^{z}$ is only partially twisted for both even and odd layers, then each fracton (resp. 1d mobile quasiparticle) can be viewed as a fusion result of an Abelian fracton (resp. 1d mobile quasiparticle) and non-Abelian anyons; thus, the fracton phase is not strictly non-Abelian. This crucial distinction between these three cases is also reflected in their GSD on $\mathtt{T}^{3}$, as given by Eq.~\eqref{eq:GSD_checker_z}, Eq.~\eqref{eq:GSD_checker_Z0}, and Eq.~\eqref{eq:GSD_checker_partial}.

\section{Conclusions and Outlook}
\label{cncls}

In this work, we have constructed a large class of novel three-dimensional quantum phases of matter exhibiting fracton order. Here, we shall summarize our main results and discuss some open questions which go beyond the scope of this paper but deserve further investigation. 

The key result of our work is the construction of ``twisted fracton models,'' which represent a general class of type-I fracton phases of matter, including those with inextricably non-Abelian fractons. In particular, we have constructed and studied the twisted versions of both the X-cube and checkerboard models, 
with spins---labeled by elements of a finite Abelian group $G$---on the faces (resp. vertices) of a cubic (resp. checkerboard) lattice for the X-cube (resp. checkerboard) model. For either case, the untwisted Hamiltonian consists of local (generalized) gauge transformations and local flux projections. Their twisted versions are obtained by adding to the gauge transformations an extra phase factor specified by 3-cocycles $\omega \in H^3(G,U(1))$ and locally flat spin configurations.


Both families of models are then carefully studied. We have made an exact computation of their ground state degeneracy (GSD) on the three-torus $\mathtt{T}^{3}$, which depends sub-extensively on the system size. In particular, our computation discovers for the first time the exotic GSD (e.g. Eqs.~\eqref{eq:GSD_Xcube_Z}, \eqref{eq:GSD_Xcube_2}, \eqref{eq:GSD_Xcube_3}, and \eqref{eq:GSD_checker3}) of non-Abelian fracton phases (\emph{i.e.}, fracton phases hosting either inextricably non-Abelian fractons or inextricably non-Abelian 1d mobile quasiparticles) on $\mathtt{T}^{3}$. 

In addition, we have systematically analyzed the braiding and fusion properties of quasiparticles in twisted fracton phases and, in the process of doing so, defined necessary notions such as topological charge, quantum dimension, and inextricably non-Abelian fractons and 1d mobile quasiparticles. Thus, our work also provides the first systematic route for describing the braiding and fusion of quasiparticles in type-I fracton phases, including those which are non-Abelian. As such, our work provides a general framework within which future studies of fracton order may be conducted. As an important intermediate step, we have also provided a detailed derivation of anyon properties of lattice models of twisted gauge theories in two spatial dimensions, which is then readily applicable to the twisted fracton models.

For possible future directions, we first notice that non-Abelian type-I fracton phases may also be constructed by coupling $d=2$ topological orders, a procedure followed in Refs.~\cite{nonabelian,cagenet}. For instance, Ref.~\cite{cagenet} constructs so-called ``cage-net'' fracton models, a distinct non-Abelian generalisation of the X-cube model, by coupling together layers of string-net models, which realizes inextricably non-Abelian 1d mobile quasiparticles (but not non-Abelian fractons). It remains an open question whether inextricable non-Abelian fractons (in particular, twisted checkerboard models) can be realized by a similar coupled layer approach. Similarly, the correspondence between the twisted X-cube models and the ``cage-net'' fracton models also remains unclear. In general, understanding the generic possibilities for type-I fracton phases and their constructions remains an interesting open question.

Further generalizing our results regarding the properties of twisted fracton models to generic type-I fracton phases constitutes an important open direction. A first step towards this goal would be to apply our systematic approach for describing quasiparticles to other fracton phases which lie beyond our construction here, such as the cage-net models, in order to understand the generic features of excitations. In addition, it is desirable to precisely derive the GSD of a non-Abelian twisted fracton model on $\mathtt{T}^{3}$, which depends exotically on the system size, 
in terms of its quasiparticle properties. This will likely be crucial in determining the GSD of a generic non-Abelian fracton phase on $\mathtt{T}^{3}$. 

Moreover, generalizations of the X-cube model on more generic lattices~\cite{slagle3} and on general three-dimensional manifolds~\cite{shirley} were proposed recently. Both these works found that the X-cube Hamiltonian may be defined on a lattice or manifold where the vertices locally resemble the vertex of a cubic lattice. In the language of Ref.~\cite{shirley}, this construction involves the notion of a ``singular compact total foliation'' of the spatial manifold, wherein the lattice may be understood as being constructed with transversely
intersecting stacks of parallel surfaces. In principle, there appears to be no obstruction to generalizing the twisted variants of the X-cube and checkerboard models to generic spatial manifolds; however, it will be an interesting challenge to understand the dependence of their GSD on both the global topology and the foliation. 





With a rich landscape of type-I fracton systems uncovered by our construction, a better classification scheme for fracton phases is more needed than ever. While we have focused on the braiding-related differences amongst twisted fracton phases here, we can see, for instance, that twisted X-cube models based on a given group $G$ share certain similarities, such as how the topological charges of a quasiparticle are constrained. Roughly speaking, these similarities reflect the 3d information inherent in these states while the 
differences treat the 2d features. Based on these ideas, we expect that an improved classification scheme would explicitly inform us how information at different levels is organized, which would allow for a more systematic study of fracton phases. Such investigations may lead to an instructive quantum field theoretical description capturing the universal properties of these phases.

As with most recent studies of fracton orders, we have focused here on type-I fracton phases, \emph{i.e.}, those where fractons are created at the corners of membrane operators and whose full spectrum contains additional quasiparticles with restricted mobility. It remains to be seen whether insights from this work can be extended to type-II fracton models, such as Haah's code~\cite{haah}, which host type-II fractons, \emph{i.e.}, those created by fractal operators. One possible route for realizing twisted versions of type-II fracton phases may be to study dual theories of the recently introduced ``fractal SPT'' states~\cite{kubica,devakulfractal}.
It will be especially interesting to see whether a multi-channel fusion rule is allowed for type-II fractons. 

Besides searching for, and studying mechanisms of, new fracton phases, it is also important to explore possible realizations and potential applications of twisted fracton models. This line of investigation leads to several interesting questions worthy of future studies, such as the possibility of making quantum simulations of twisted fracton phases in cold atomic systems and of using non-Abelian fracton phases for 
quantum information storage or topological quantum computation to achieve better resilience to noise and decoherence.

\begin{acknowledgments}
We are grateful to Danny Bulmash, Fiona Burnell, Meng Cheng, Guillaume Dauphinais, Trithep Devakul, Michael Hermele, Yuting Hu, Rahul Nandkishore, Wilbur Shirley, Kevin Slagle, Shivaji Sondhi, Oscar Viyuela,  Juven Wang, and Yizhi You for stimulating conversations and correspondence. H.S. and M.A.M.-D. acknowledge financial support from the Spanish MINECO grant FIS2015-67411, and the CAM research consortium QUITEMAD+,
Grant No. S2013/ICE-2801. The research of M.A.M.-D. has been supported
in part by the U.S. Army Research Office through Grant No. W911N F-14-1-0103. A.P. acknowledge support from the University of Colorado at Boulder. S.-J. H. is supported by the U.S. Department of Energy, Office of Science, Basic Energy Sciences (BES) under Award number DE- SC0014415.

\end{acknowledgments}

\appendix

\section{Group cohomology and Dijkgraaf-Witten weight}
\label{sec:GCDW}

\subsection{Definition of group cohomology}

Let $G$ be a finite group with its identity element denoted as $e$
and $U\left(1\right)\coloneqq\left\{ z\in\mathbb{C}\,|\;\left|z\right|=1\right\} $
be the Abelian group of phase factors. For each nonnegative integer
$n$, let $C^{n}\left(G,U\left(1\right)\right)$ be the set of functions
from $G^{n}$ (\emph{i.e.}, direct product of $n$ copies of $G$)
to $U\left(1\right)$. Also, for each $n$, there is a so-called coboundary
map 
\begin{align}
\delta:C^{n}\left(G,U\left(1\right)\right) & \rightarrow C^{n+1}\left(G,U\left(1\right)\right)\nonumber \\
\omega & \mapsto\delta\omega
\end{align}
given by
\begin{multline}
\delta\omega\left(g_{1},g_{2},\cdots,g_{n+1}\right)=\omega\left(g_{2},g_{3},\cdots,g_{n+1}\right)\cdot\\
\prod_{j=1}^{n}\omega\left(g_{1},\cdots,g_{j-1},g_{j}g_{j+1},g_{j+2},g_{j+3},\cdots,g_{n+1}\right)^{\left(-1\right)^{j}}\\
\cdot\omega\left(g_{1},g_{2},\cdots,g_{n}\right)^{\left(-1\right)^{n+1}}
\end{multline}
In addition, let 
\begin{equation}
Z^{n}\left(G,U\left(1\right)\right)\coloneqq\left\{ \omega\in C^{n}\left(G,U\left(1\right)\right)|\delta\omega=1\right\} ,
\end{equation}
whose elements are called $n$-cocycles, and $\delta\omega=1$ is
called the cocycle condition. We denote the image of $C^{n-1}\left(G,U\left(1\right)\right)$
under $\delta$ by 
\begin{equation}
B^{n}\left(G,U\left(1\right)\right)\coloneqq\delta C^{n-1}\left(G,U\left(1\right)\right),
\end{equation}
whose elements are called $n$-coboundaries. Induced by the Abelian
group structure of $U\left(1\right)$, all $C^{n}\left(G,U\left(1\right)\right)$,
$Z^{n}\left(G,U\left(1\right)\right)$ and $B^{n}\left(G,U\left(1\right)\right)$
can be viewed as Abelian groups with coboundary maps viewed as homomorphisms
of Abelian groups. 

It can be checked that applying $\delta$ twice always gives a trivial
map $\delta^{2}:C^{n-1}\left(G,U\left(1\right)\right)\rightarrow C^{n+1}\left(G,U\left(1\right)\right)$,
\emph{i.e.}, $\delta^{2}c=1,\forall c\in C^{n-1}\left(G,U\left(1\right)\right)$. Hence $B^{n}\left(G,U\left(1\right)\right)\subset Z^{n}\left(G,U\left(1\right)\right)$.
The quotient of them
\begin{equation}
H^{n}\left(G,U\left(1\right)\right)\coloneqq\frac{Z^{n}\left(G,U\left(1\right)\right)}{B^{n}\left(G,U\left(1\right)\right)}
\end{equation}
is called $n$-th cohomology group of $G$ with coefficients in $U\left(1\right)$,
whose elements can be labeled by the coset of $\omega\in Z^{n}\left(G,U\left(1\right)\right)$, \emph{i.e.},
\begin{equation}
\left[\omega\right]\coloneqq\omega\cdot B^{n}\left(G,U\left(1\right)\right).
\end{equation}
Dijkgraaf-Witten
topological quantum field theories constructed from $\omega,\omega'\in Z^{3}\left(G,U\left(1\right)\right)$
with $\left[\omega\right]=\left[\omega'\right]$ are equivalent. 

An $n$-cocycle $\omega\in Z^{n}\left(G,U\left(1\right)\right)$ is
called normalized if $\omega\left(g_{1},g_{2},\cdots,g_{n}\right)=1$
whenever any of $g_{1},\cdots,g_{n}$ is the identity element $e$
of $G$. It is a standard result that any element of an $n$-th cohomology
group can be presented by a normalized $n$-cocycle. For simplicity,
we always work with normalized cocycles without loss of generality. 

\subsection{Triangulated manifold}
\label{subsec:Triangulation}

Roughly, a triangulation of a topological space $\mathtt{X}$ is a decomposition
of $\mathtt{X}$ into simplices. A\emph{ $k$-simplex} is the $k$-dimensional
analogue of triangle; for lower dimensions, a $0$-simplex (resp. $1$-simplex,
$2$-simplex, $3$-simplex) is a point (resp. segment, triangle, tetrahedron).
In algebraic topology, such a decomposition is called a simplical
structure. A topological space with a simplical structure is called
a simplicial complex. Keeping only $k$-simplices with $k\leq m$ in a
simplicial complex $\mathtt{X}$ results in a subcomplex called the $m$-skeleton
of $\mathtt{X}$ and denoted $\mathtt{X}^{m}$. By definition, there is a sequence of
inclusions $\mathtt{X}^{0}\subset \mathtt{X}^{1}\subset \mathtt{X}^{2}\cdots$. Since $\mathtt{X}^{1}$
is a graph, terminology from graph theory is used; $0$-simplices
(resp. $1$-simplices) are usually called vertices (resp. edges). 

To work with topological quantum field theories of Dijkgraaf-Witten
type, we want to decompose manifolds into \emph{ordered simplices},
\emph{i.e.}, simplices whose vertices are ordered. Such a decomposition
is called an \emph{ordered simplicial structure}. It is equivalent
to a simplicial structure together with a branching structure. A \emph{branching
structure} is a choice of orientation of each edge in the simplicial
complex so that there is no triangle whose three edges form a closed
walk~\cite{chen2013}. A topological space with ordered simplicial
structure is called an ordered simplicial complex. 

Technically, the notion of a simplicial complex is too restrictive;
no vertices of a simplex can coincide. A slight generalization of an ordered simplicial complex, dropping this restriction, leads to the
notion of a $\Delta$-complex. The definition of a $\Delta$-complex structure
can be found in Ref.~\cite{Hatcher}. To ensure everything is well-defined,
we always work with finite $\Delta$-complexes,\emph{ i.e.}, those
with finite number of simplices. To summarize, in this paper, the
precise meaning of a \emph{triangulation} of a topological space is
a finite $\Delta$-complex structure on it; vertices of each simplex
are assumed ordered and allowed to coincide. In particular, triangles
can be singular, whose vertices may coincide.

\subsection{Dijkgraaf-Witten weight}
\label{subsec:DW}

Let us now consider a gauge field labeled by a finite group $G$ on an
$n$-dimensional triangulated oriented manifold $\mathtt{X}$ (probably with
boundaries $\partial \mathtt{X}\neq\emptyset$). Let $\Delta^{1}\left(\mathtt{X}\right)$
be the set of its $1$-simplices (\emph{i.e.}, ordered edges). A gauge field
configuration is specified by an assignment $\xi:\Delta^{1}\left(\mathtt{X}\right)\rightarrow G$.
It will be called a \emph{coloring} of $\mathtt{X}$, if it is \emph{locally flat},
\emph{i.e.}, $\xi\left(\left[v_{0}v_{1}\right]\right)\xi\left(\left[v_{1}v_{2}\right]\right)=\xi\left(\left[v_{0}v_{2}\right]\right)$
for any $2$-simplex (\emph{i.e.}, ordered triangle) $\left[v_{0}v_{1}v_{2}\right]$
in $\mathtt{X}$. The sets of all colorings of $\mathtt{X}$ and its boundary $\partial \mathtt{X}$
are denoted $\text{Col}\left(\mathtt{X};G\right)$ and $\text{Col}\left(\partial \mathtt{X};G\right)$ (or simply $\text{Col}\left(\mathtt{X}\right)$ and $\text{Col}\left(\partial \mathtt{X}\right)$)
respectively. Let $\zeta\in\text{Col}\left(\partial \mathtt{X}\right)$, then
we write $\text{Col}\left(\mathtt{X},\zeta\right)$ for the set of colorings
of $\mathtt{X}$ which coincide with $\zeta$ on $\partial \mathtt{X}$.

Given $\omega\in Z^{n}\left(G,U\left(1\right)\right)$, the \emph{Dijkgraaf-Witten
weight} 
\begin{equation}
\omega\left[\mathtt{X},\xi\right]\equiv\left\langle \omega,\xi_{\#}\mathtt{X}\right\rangle \coloneqq\prod_{\sigma}\left\langle \omega,\xi\sigma\right\rangle ^{\text{sgn}\left(\sigma\right)} \label{eq:DW_weight}
\end{equation}
is assigned to each $\xi\in\text{Col}\left(\mathtt{X}\right)$, where the product
is over all $n$-simplices $\sigma$ in $\mathtt{X}$. The sign $\text{sgn}\left(\sigma\right)=1$
(resp. $-1$) if the orientation of $\sigma=\left[\sigma_{0}\sigma_{1}\cdots\sigma_{n}\right]$
determined by the ordering of its vertices is the same as (resp. opposite
to) the orientation of $\mathtt{X}$. In addition, 
\begin{equation}
\left\langle \omega,\xi\sigma\right\rangle \coloneqq\omega\left(\xi\left(\left[\sigma_{0}\sigma_{1}\right]\right),\xi\left(\left[\sigma_{1}\sigma_{2}\right]\right),\cdots,\xi\left(\left[\sigma_{n-1}\sigma_{n}\right]\right)\right),
\end{equation}
which is often simply written as $\left[\sigma_{0}\sigma_{1}\cdots\sigma_{n}\right]$
to avoid heavy notations in concrete calculations, when $\omega$ and
$\xi$ are clear from context. 

Before proceeding further, let us further elucidate the dependence
of $\omega\left[\mathtt{X},\xi\right]$ on $\xi\in\text{Col}\left(\mathtt{X},\zeta\right)$.
To borrow terms from topology, $\xi\in\text{Col}\left(\mathtt{X}\right)$
can be viewed as a continuous map from $\mathtt{X}$ to the classifying space
$BG$ for the group and it maps $\mathtt{X}^{0}$ to a base point of $BG$.
To be concrete, we always refer to the standard $\Delta$-complex
realization of $BG$. In general, if $\xi,\xi'\in\text{Col}\left(\mathtt{X},\zeta\right)$
are homotopic to each other relative to $\partial \mathtt{X}$, then $\omega\left[\mathtt{X},\xi\right]=\omega\left[\mathtt{X},\xi'\right]$.

We notice that such a homotopy can be presented as a coloring on $\mathtt{X}\times \mathtt{I}$, where $\mathtt{I}=\left[0,1\right]$. The $\Delta$-complex structure of $\mathtt{X}\times \mathtt{I}$ is induced by that of $\mathtt{X}$ as follows:
if $\left[v_{0}v_{1}\cdots v_{k}\right]$ is a $k$-simplex of $\mathtt{X}$,
then $\left[v_{0}v_{1}\cdots v_{k}v'_{k}\right]$, $\left[v_{0}v_{1}\cdots v_{k-1}v_{k-1}'v_{k}'\right]$,
..., $\left[v_{0}v_{1}'\cdots v_{k-1}'v_{k}'\right]$ are $\left(k+1\right)$-simplices
of $\mathtt{X}\times \mathtt{I}$. Here  $\mathtt{X}\times\left\{ 0\right\} $
is identified with $\mathtt{X}$; we write $v$ (resp. $v'$) for vertices
in $\mathtt{X}\times\left\{ 0\right\} $ (resp. $\mathtt{X}\times\left\{ 1\right\} $).
A homotopy $\vartheta$ from $\xi$ to $\xi'$ relative to $\partial \mathtt{X}$
can be presented as a coloring of $\mathtt{X}\times \mathtt{I}$ such that $\mathtt{X}\times\left\{ 0\right\} $
(resp. $\mathtt{X}\times\left\{ 1\right\} $) is colored as $\xi$ (resp. $\xi'$)
and such that each $\left[vv'\right]$ in $\left(\partial \mathtt{X}\right)\times \mathtt{I}$
is colored by the identity element of $G$. If such a homotopy exists
for $\xi,\xi'\in\text{Col}\left(\mathtt{X}\right)$, we say $\xi,\xi'$ are
homotopic to each other relative to $\partial \mathtt{X}$. 

To see $\omega\left[\mathtt{X},\xi\right]=\omega\left[\mathtt{X},\xi'\right]$, let
$C_{n+1}\left(\mathtt{X}\times \mathtt{I}\right)$ be the group of $\left(n+1\right)$-chains
of $\mathtt{X}\times \mathtt{I}$, \emph{i.e.}, the free Abelian group generated by
$\Delta^{n+1}\left(\mathtt{X}\times \mathtt{I}\right)$. We also use $\mathtt{X}\times \mathtt{I}$ to
denote the orientation-dependent sum of all $\left(n+1\right)$-simplices
in $\mathtt{X}\times \mathtt{I}$; we write $\mathtt{X}\times \mathtt{I}=\sum_{\sigma}\text{sgn}\left(\sigma\right)\sigma\in C_{n+1}\left(\mathtt{X}\times \mathtt{I}\right)$.
Using the Abelian group homormophism $\vartheta_{\#}:C_{n+1}\left(\mathtt{X}\times \mathtt{I}\right)\rightarrow C_{n+1}\left(BG\right)$ induced by the homotopy $\vartheta: \mathtt{X}\times\mathtt{I}\rightarrow BG$,
 we have 
\begin{equation}
\left\langle \omega,\vartheta_{\#}\partial\left(\mathtt{X}\times \mathtt{I}\right)\right\rangle =\left\langle \delta\omega,\vartheta_{\#}\left(\mathtt{X}\times \mathtt{I}\right)\right\rangle =1.\label{eq:boundary_co}
\end{equation}
We notice that $\partial\left(\mathtt{X}\times \mathtt{I}\right)=\mathtt{X}\times\left\{ 1\right\} -\mathtt{X}\times\left\{ 0\right\} +\left(\partial \mathtt{X}\right)\times \mathtt{I}$
and that $\vartheta_{\#}$ coincides with group homomorphism $\xi_{\#}$ (resp. $\xi_{\#}^{\prime}$) induced by $\xi$ (resp. $\xi^{\prime}$)
on $\mathtt{X}\times\left\{ 0\right\} $ (resp. $\mathtt{X}\times\left\{ 1\right\} $).
In addition, $\omega$ gives the trivial phase factor 1 on all $n$-simplices
in $\left(\partial \mathtt{X}\right)\times \mathtt{I}$. Hence $\left\langle \omega,\vartheta_{\#}\partial\left(\mathtt{X}\times \mathtt{I}\right)\right\rangle =\left\langle \omega,\xi_{\#}\mathtt{X}\right\rangle /\left\langle \omega,\xi_{\#}^{\prime}\mathtt{X}\right\rangle $,
so we get $\left\langle \omega,\xi_{\#}\mathtt{X}\right\rangle =\left\langle \omega,\xi_{\#}^{\prime}\mathtt{X}\right\rangle $. 

Suppose that $\mathtt{X}$ and $\mathtt{X}'$ are the same manifold with probably different $\Delta$-complex structures that coincide on boundary. Let us consider an $\left(n+1\right)$-dimensional
$\Delta$-manifold $\mathtt{X}\times \mathtt{I}$ whose boundary is triangulated as
$\mathtt{X}$ (resp. $\mathtt{X}'$) on $\mathtt{X}\times\left\{ 0\right\} $ (resp. $\mathtt{X}\times\left\{ 1\right\} $)
and the induced $\Delta$-complex structure of $\left(\partial \mathtt{X}\right)\times \mathtt{I}$.\footnote{It is known that the $\Delta$-complex structure on the boundary of
manifold can be extend to the whole of the manifold.} Then a homotopy $\vartheta$ from $\xi\in\text{Col}\left(\mathtt{X}\right)$
to $\xi'\in\text{Col}\left(\mathtt{X}'\right)$ relative to $\partial \mathtt{X}$ can
be defined as a coloring of this $\Delta$-manifold $\mathtt{X}\times \mathtt{I}$ that
coincides with $\xi$ (resp. $\xi'$) on $\mathtt{X}\times\left\{ 0\right\} $
(resp. $\mathtt{X}\times\left\{ 1\right\} $) and colors each $\left[vv'\right]\in\left(\partial \mathtt{X}\right)\times \mathtt{I}$
by the identity element of $G$. Again, repeating the above argument using
Eq.~\eqref{eq:boundary_co}, we get $\left\langle \omega,\xi_{\#}\mathtt{X}\right\rangle =\left\langle \omega,\xi_{\#}^{\prime}\mathtt{X}\right\rangle $
if $\xi$ and $\xi'$ are homotopic to each other.

To determine when two colorings on $\mathtt{X}$ are homotopic, we pick a vertex $s\in \mathtt{X}$ as the base point
and a path $\left\langle s,v\right\rangle $ from $s$ to every vertex $v$ other than $s$, where $\mathtt{X}$ is assumed to be connected. 
Let $\pi_{1}\left(\mathtt{X},s\right)$ be the fundamental group of $\mathtt{X}$ based at $s$.
For any subset $W\subseteq \mathtt{X}^{0}\backslash\left\{ s\right\}$, we write
$\left\langle s,W\right\rangle \coloneqq\left\{ \left\langle s,v\right\rangle |v\in W\right\} $. 
Then $\text{Col}\left(\mathtt{X};G\right)$ is in one-to-one correspondence to
$\text{Hom}\left(\pi_{1}\left(\mathtt{X},s\right),G\right)\times G^{\left\langle s,\mathtt{X}^{0}\backslash\left\{ s\right\}\right\rangle }$, where
$\text{Hom}\left(\pi_{1}\left(\mathtt{X},s\right),G\right)$ is the set
of group homomorphisms from $\pi_{1}\left(\mathtt{X},s\right)$ to $G$.

Further, suppose that $\mathtt{X}$ and $\mathtt{X}'$ are the same manifold with probably different $\Delta$-complex structures that coincide on boundary. 
If $\partial \mathtt{X}\neq\emptyset$, the base point $s$ is picked in $\partial \mathtt{X}$.
Then
$\xi\in\text{Col}\left(\mathtt{X};G\right)$ and $\xi'\in\text{Col}\left(\mathtt{X}';G\right)$
are homotopic relative to $\partial \mathtt{X}$ if and only if they assign
the same group element to each path in $\pi_{1}\left(\mathtt{X},s\right)$
and $\left\langle s,\partial \mathtt{X}^{0}\backslash{s}\right\rangle $, where $\partial \mathtt{X}^{0}$ is the set of vertices in $\partial \mathtt{X}$. If $\partial \mathtt{X}=\partial \mathtt{X}'=\emptyset$, then
$\xi\in\text{Col}\left(\mathtt{X},G\right)$ and $\xi'\in\text{Col}\left(\mathtt{X}',G\right)$
are homotopic if and only if there exists $g\in G$ such that $\xi'\left(q\right)=g\xi\left(q\right)g^{-1}$
for any $q\in\pi_{1}\left(\mathtt{X},s\right)$. 

The summation of $\left\langle \omega,\xi_{\#}\mathtt{X}\right\rangle $ over
$\xi\in\text{Col}\left(\mathtt{X},\zeta\right)$ gives the \emph{Dijkgraaf-Witten
partition function }
\begin{equation}
Z_{\omega}\left(\mathtt{X},\zeta\right)\coloneqq\frac{1}{\left|G\right|^{\left|\mathtt{X}^{0}\backslash\partial \mathtt{X}^{0}\right|}}\sum_{\xi\in\text{Col}\left(\mathtt{X},\zeta\right)}\left\langle \omega,\xi_{\#}\mathtt{X}\right\rangle ,\label{eq:Z_DW}
\end{equation}
where $\left|\mathtt{X}^{0}\backslash\partial \mathtt{X}^{0}\right|$ is the number
of vertices of $\mathtt{X}$ not in $\partial \mathtt{X}$. From the discussion above,
it is clear that $Z_{\omega}\left(\mathtt{X},\zeta\right)$ does not depend
on how $\mathtt{X}\backslash\partial \mathtt{X}$ is triangulated.

\section{Algebra preliminaries for $\mathcal{D}^{\omega}\left(G\right)$}
\label{sec:AlgebraDG}

\subsection{Some definitions for quasi-bialgebras}

A \emph{quasi-bialgebra} $\left({\cal A},\varDelta,\varepsilon,\phi\right)$
is an algebra $\mathcal{A}$ over $\mathbb{C}$ equipped with algebra
homomorphisms $\varDelta:\mathcal{A}\rightarrow\mathcal{A}\otimes{\cal A}$,
$\varepsilon:\mathcal{A}\rightarrow\mathbb{C}$ and an invertible
element $\phi\in\mathcal{A}\otimes\mathcal{A}\otimes\mathcal{A}$
such that
\begin{gather}
\left(\text{id}\otimes\varDelta\right)\left(\varDelta\left(a\right)\right)=\phi\left(\varDelta\otimes\text{id}\right)\left(\varDelta\left(a\right)\right)\phi^{-1},\forall a\in\mathcal{A},\\
\left(\text{id}\otimes\text{id}\otimes\varDelta\right)\left(\phi\right)\left(\varDelta\otimes\text{id}\otimes\text{id}\right)\left(\phi\right)\nonumber \\
=\left(1\otimes\phi\right)\left(\text{id}\otimes\varDelta\otimes\text{id}\right)\left(\phi\right)\left(\phi\otimes1\right),\\
\left(\varepsilon\otimes\text{id}\right)\circ\varDelta=\text{id}=\left(\text{id}\otimes\varepsilon\right)\circ\varDelta,\\
\left(\text{id}\otimes\varepsilon\otimes\text{id}\right)\left(\phi\right)=1\otimes1,
\end{gather}
where $1$ denotes the identity element of $\mathcal{A}$. Respectively,
$\varDelta$, $\varepsilon$ and $\phi$ are called the \emph{copruduct},
the\emph{ counit}, and the \emph{Drinfeld associator}. A quasi-bialgebra
is a generalization of bialgebra; it relaxes the coassociativity condition. 

An \emph{antipode} on a quasi-bialgebra $\left({\cal A},\varDelta,\varepsilon,\phi\right)$
is a triple $\left(S,\alpha,\beta\right)$, where $S:\mathcal{A}\rightarrow\mathcal{A}$
is an algebra antihomomorphism and $\alpha,\beta\in\mathcal{A}$,
satisfying 
\begin{align}
\sum_{j}S\left(a_{j}^{\left(1\right)}\right)\alpha a_{j}^{\left(2\right)} & =\varepsilon\left(a\right)\alpha,\label{eq:antipode1}\\
\sum_{j}a_{j}^{\left(1\right)}\beta S\left(a_{j}^{\left(2\right)}\right) & =\varepsilon\left(a\right)\beta,\label{eq:antipode2}\\
\sum_{j}\phi_{j}^{\left(1\right)}\beta S\left(\phi_{j}^{\left(2\right)}\right)\alpha\phi_{j}^{\left(3\right)} & =1,\label{eq:antipode3}\\
\sum_{j}S\left(\bar{\phi}_{j}^{\left(1\right)}\right)\alpha\bar{\phi}_{j}^{\left(2\right)}\beta S\left(\bar{\phi}_{j}^{\left(3\right)}\right) & =1,\label{eq:antipode4}
\end{align}
for any $a\in\mathcal{A}$, where $\sum_{j}a_{j}^{\left(1\right)}\otimes a_{j}^{\left(2\right)}=\varDelta\left(a\right)$,
$\sum_{j}\phi_{j}^{\left(1\right)}\otimes\phi_{j}^{\left(2\right)}\otimes\phi_{j}^{\left(3\right)}=\phi$
and $\sum_{j}\bar{\phi}_{j}^{\left(1\right)}\otimes\bar{\phi}_{j}^{\left(2\right)}\otimes\bar{\phi}_{j}^{\left(3\right)}=\phi^{-1}$.
A \emph{quasi-Hopf} algebra $\left({\cal A},\varDelta,\varepsilon,\phi,S,\alpha,\beta\right)$
is a quasi-bialgebra with an antipode $\left(S,\alpha,\beta\right)$
such that $S$ is bijective.

A \emph{quasi-triangular} quasi-bialgebra $\left({\cal A},\varDelta,\varepsilon,\phi,R\right)$
is a quasi-bialgebra equipped with an invertible element $R\in\mathcal{A}\otimes\mathcal{A}$,
called the\emph{ universal $R$-matrix}, satisfying 
\begin{align}
\varDelta^{\text{op}}\left(a\right) & =R\varDelta\left(a\right)R^{-1},\\
\left(\varDelta\otimes\text{id}\right)\left(R\right) & =\phi_{312}R_{13}\phi_{132}^{-1}R_{23}\phi,\\
\left(\text{id}\otimes\varDelta\right)\left(R\right) & =\phi_{231}^{-1}R_{13}\phi_{213}R_{12}\phi^{-1},
\end{align}
where $\varDelta^{\text{op}}\coloneqq\wp\circ\varDelta$ with $\wp\left(a_{1}\otimes a_{2}\right)\coloneqq a_{2}\otimes a_{1}$
and $R_{ij}$ stands for $R$ acting non-trivially in the $i$-th
and $j$-th slot of $\mathcal{A}\otimes\mathcal{A}\otimes\mathcal{A}$.
In addition, if $\sigma$ denotes a permutation of $\left\{ 1,2,3\right\} $
and $\phi=\sum_{j}\phi_{j}^{\left(1\right)}\otimes\phi_{j}^{\left(2\right)}\otimes\phi_{j}^{\left(3\right)}$,
then $\phi_{\sigma\left(1\right)\sigma\left(2\right)\sigma\left(3\right)}\coloneqq\sum_{j}\phi_{j}^{\left(\sigma^{-1}\left(1\right)\right)}\otimes\phi_{j}^{\left(\sigma^{-1}\left(2\right)\right)}\otimes\phi_{j}^{\left(\sigma^{-1}\left(3\right)\right)}$. 

\subsection{Tensor product of quasi-bialgebras}
\label{subsec:Tensor-bialg}

Given two quasi-bialgebras $\left({\cal A}_{1},\varDelta_{1},\varepsilon_{1},\phi_{1}\right)$
and $\left({\cal A}_{2},\varDelta_{2},\varepsilon_{2},\phi_{2}\right)$,
their tensor product $\mathcal{A}\coloneqq\mathcal{A}_{1}\otimes\mathcal{A}_{2}$
is also a quasi-bialgebra equipped with the coproduct $\varDelta:\mathcal{A}\rightarrow\mathcal{A}\otimes\mathcal{A}$
given by the composition of the following two maps
\begin{multline}
\mathcal{A}=\mathcal{A}_{1}\otimes\mathcal{A}_{2}\xrightarrow{\varDelta_{1}\otimes\varDelta_{2}}\left(\mathcal{A}_{1}\otimes\mathcal{A}_{1}\right)\otimes\left(\mathcal{A}_{2}\otimes\mathcal{A}_{2}\right)\\
\rightarrow\left(\mathcal{A}_{1}\otimes\mathcal{A}_{2}\right)\otimes\left(\mathcal{A}_{1}\otimes\mathcal{A}_{2}\right)=\mathcal{A}\otimes\mathcal{A},
\end{multline}
where the second map swaps the middle two tensor factors $\left(\mathcal{A}_{1}\otimes\mathcal{A}_{2}\right)\otimes\left(\mathcal{A}_{1}\otimes\mathcal{A}_{2}\right)=\mathcal{A}_{1}\otimes\mathcal{A}_{2}\otimes\mathcal{A}_{1}\otimes\mathcal{A}_{2}$.
The counit $\varepsilon$ is 
\begin{equation}
\mathcal{A}_{1}\otimes\mathcal{A}_{2}\xrightarrow{\varepsilon_{1}\otimes\varepsilon_{2}}\mathbb{C}\otimes\mathbb{C}\rightarrow\mathbb{C},
\end{equation}
where the second map is the multiplication of $\mathbb{C}$. The Drinfeld
associator $\phi$ is also given by the tensor product of $\phi_{1}$
and $\phi_{2}$; more precisely, $\phi$ is the image of $\phi_{1}\otimes\phi_{2}$
under the map 
\begin{equation}
\left(\mathcal{A}_{1}^{\otimes3}\right)\otimes\left(\mathcal{A}_{2}^{\otimes3}\right)\rightarrow\left(\mathcal{A}_{1}\otimes\mathcal{A}_{2}\right)^{\otimes3}=\mathcal{A}\otimes\mathcal{A}\otimes\mathcal{A}
\end{equation}
swapping corresponding factors.

For notational compactness, we do not express the identification maps
$\mathcal{A}_{1}^{\otimes n}\otimes\mathcal{A}_{2}^{\otimes n}\cong\mathcal{A}^{\otimes n}$
and $\mathbb{C}\otimes\mathbb{C}\cong\mathbb{C}$ explicitly. Thus,
we can simply write $\varDelta=\varDelta_{1}\otimes\varDelta_{2}$,
$\varepsilon=\varepsilon_{1}\otimes\varepsilon_{2}$ and $\phi=\phi_{1}\otimes\phi_{2}$.
This convention of notation simplification will be used below.

If $\left({\cal A}_{1},\varDelta_{1},\varepsilon_{1},\phi_{1}\right)$
and $\left({\cal A}_{2},\varDelta_{2},\varepsilon_{2},\phi_{2}\right)$
have antipodes $\left(S_{1},\alpha_{1},\beta_{1}\right)$ and $\left(S_{2},\alpha_{2},\beta_{2}\right)$
respectively, then their tensor product $\left({\cal A},\varDelta,\varepsilon,\phi\right)$
is also a quai-Hopf algebra with antipode $\left(S_{1}\otimes S_{2},\alpha_{1}\otimes\alpha_{2},\beta_{1}\otimes\beta_{2}\right)$.

In addition, if $\left({\cal A}_{1},\varDelta_{1},\varepsilon_{1},\phi_{1}\right)$
and $\left({\cal A}_{2},\varDelta_{2},\varepsilon_{2},\phi_{2}\right)$
are quasi-triangular with universal matrices $R_{1}\in\mathcal{A}_{1}\otimes\mathcal{A}_{1}$
and $R_{2}\in\mathcal{A}_{2}\otimes\mathcal{A}_{2}$ respectively,
then their tensor product $\left({\cal A},\varDelta,\varepsilon,\phi\right)$
is also quasi-triangular with a universal matrix $R=R_{1}\otimes R_{2}$.

This discussion here generalizes to the tensor product of a finite
number of quasi-bialgebras.

\subsection{Representation category of quasi-bialgebra}

Below, all vector spaces are assumed to be finite-dimensional for simplicity.
A representation $\left(\rho,V\right)$ of $\mathcal{A}$ is a vector
space $V$ over $\mathbb{C}$ equipped with an algebra homomorphism
$\rho:\mathcal{A}\rightarrow\text{End}(V)\equiv\mathcal{L}(V)$, where $\text{End}(V)\equiv\mathcal{L}(V)$ is the algebra of all linear operators on $V$. A morphism $f:\left(\rho_{1},V_{1}\right)\rightarrow\left(\rho_{2},V_{2}\right)$
is a linear map that commutes with the action of $\mathcal{A}$, \emph{i.e.},
\begin{equation}
f\circ\rho_{1}\left(a\right)=\rho_{2}\left(a\right)\circ f,\forall a\in\mathcal{A}.
\end{equation}
Such a map is called an \emph{intertwiner} in representation theory. By the
representation category of $\mathcal{A}$, we mean the the category whose
objects are the representations of $\mathcal{A}$ and whose morphisms
are the intertwiners between them. As it fits in a more general setting on the categories
of modules, the representation category of $\mathcal{A}$ is denoted
by $\mathcal{A}$\textsf{-Mod}. In practice, we often write $\mathcal{V}$
short for $\left(\rho,V\right)$ and treat $\mathcal{V}$ as an $\mathcal{A}$-module;
the action of $a\in\mathcal{A}$ on $v\in\mathcal{V}$ is then written
as $a\cdot v\coloneqq\rho\left(a\right)v$.

For a quasi-bialgebra $\left(\mathcal{A},\varDelta,\varepsilon\right)$,
a tensor category structure can be defined for $\mathcal{A}$\textsf{-Mod}.
Given any two representations $\mathcal{V}_{1}=\left(\rho_{1},V_{1}\right)$
and $\mathcal{V}_{2}=\left(\rho_{2},V_{2}\right)$, their tensor product
is $\mathcal{V}_{1}\otimes\mathcal{V}_{2}=\left(\rho_{12},V_{1}\otimes V_{2}\right)$
with
\begin{equation}
\rho_{12}\coloneqq\left(\rho_{1}\otimes\rho_{2}\right)\circ\varDelta,
\end{equation}
which is also a representation of $\mathcal{A}$. The tensor product
of morphisms is the standard tensor product of linear maps. 

The unit object is the trivial representation $\left(\varepsilon,\mathbb{C}\right)$.
The following intertwiners 
\begin{equation}
\begin{array}{ccccc}
\mathbb{C}\otimes V & \cong & V & \cong & V\otimes\mathbb{C}\\
1\otimes v & \mapsto & v & \mapsfrom & v\otimes1
\end{array}
\end{equation}
are isomorphisms and are called the left and right unitors of $\mathcal{A}$\textsf{-Mod}.

Given three representations $\mathcal{V}_{j}=\left(\rho_{j},V_{j}\right),j=1,2,3$,
we can construct two representations $\left(\mathcal{V}_{1}\otimes\mathcal{V}_{2}\right)\otimes\mathcal{V}_{3}$
and $\mathcal{V}_{1}\otimes\left(\mathcal{V}_{2}\otimes\mathcal{V}_{3}\right)$,
which are the same vector space but not necessarily identical as an
$\mathcal{A}$-module. They are isomorphic by the intertwiner 
\begin{equation}
\rho_{1}\otimes\rho_{2}\otimes\rho_{3}\left(\phi\right):\left(V_{1}\otimes V_{2}\right)\otimes V_{3}\rightarrow V_{1}\otimes\left(V_{2}\otimes V_{3}\right),
\end{equation}
which is called the associator of $\mathcal{A}$\textsf{-Mod}.

In case that $\left(\mathcal{A},\varDelta,\varepsilon\right)$ is
a quasi-Hopf algebra with an antipode $\left(S,\alpha,\beta\right)$,
given any representation $\mathcal{V}=\left(\rho,V\right)$ we can
construct a \emph{dual representation} $\mathcal{V}^{*}=\left(\rho^{*},V^{*}\right)$,
where $V^{*}\coloneqq\text{Hom}_{\mathbb{C}}\left(V,\mathbb{C}\right)$
and $\rho^{*}\left(a\right)=\rho\left(S\left(a\right)\right)^{\mathsf{T}}$
is the transpose of $\rho\left(S\left(a\right)\right)$ for any $a\in\mathcal{A}$.
Explicitly, the action of $\forall a\in A$ on $\forall f\in V^{*}$
is given by $\left(a\cdot f\right)\left(v\right)\coloneqq f\left(S\left(a\right)\cdot v\right),\forall v\in V$.
Using the properties of the antipode, delineated in Eqs.~(\ref{eq:antipode1}-\ref{eq:antipode4}),
we construct two intertwiners
\begin{align}
\alpha_{\mathcal{V}}:\mathcal{V}^{*}\otimes\mathcal{V}\rightarrow\mathbb{C}, & \;f\otimes v\mapsto f\left(\rho\left(\alpha\right)v\right),\label{eq:ev_alpha}\\
\beta_{\mathcal{V}}:\mathbb{C}\rightarrow\mathcal{V}\otimes\mathcal{V}^{*}, &\; 1\mapsto\rho\left(\beta\right)\in\mathcal{L}\left(\mathcal{V}\right)=\mathcal{V}\otimes\mathcal{V}^{*}\label{eq:coev_beta}
\end{align}
such that the compositions 
\begin{gather}
\mathcal{V}\xrightarrow{\beta_{\mathcal{V}}\otimes\text{id}_{\mathcal{V}}}\left(\mathcal{V}\otimes\mathcal{V}^{*}\right)\otimes\mathcal{V}\xrightarrow{\phi}\nonumber \\
\mathcal{V}\otimes\left(\mathcal{V}^{*}\otimes\mathcal{V}\right)\xrightarrow{\text{id}_{\mathcal{V}}\otimes\alpha_{\mathcal{V}}}\mathcal{V},\label{eq:zigzag1}\\
\mathcal{V}^{*}\xrightarrow{\text{id}_{\mathcal{V}^{*}}\otimes\beta_{\mathcal{V}}}\mathcal{V}^{*}\otimes\left(\mathcal{V}\otimes\mathcal{V}^{*}\right)\xrightarrow{\phi^{-1}}\nonumber \\
\left(\mathcal{V}^{*}\otimes\mathcal{V}\right)\otimes\mathcal{V}^{*}\xrightarrow{\alpha_{\mathcal{V}}\otimes\text{id}_{\mathcal{V}^{*}}}\mathcal{V}^{*}\label{eq:zigzag2}
\end{gather}
equal the identity maps $\text{id}_{\mathcal{V}}$ and $\text{id}_{\mathcal{V}^{*}}$
respectively. Thus, $\mathcal{V}^{*}$ is a \emph{left dual}
of $\mathcal{V}$ in the tensor category $\mathcal{A}$\textsf{-Mod}.
Another representation can be constructed on $V^{*}$ with $^{*}\rho=\rho\left(S^{-1}\left(a\right)\right)^{\mathsf{T}}$
and $^{*}\mathcal{V}=\left(^{*}\rho,V^{*}\right)$ is a \emph{right
dual} of $\mathcal{V}$. The notions of left dual and right dual can
be found in many references on tensor categories, such as Ref.~\cite{Etingof2015}.

In case that $\left(\mathcal{A},\varDelta,\varepsilon\right)$ is
quasi-triangular with $R=\sum_{j}r_{j}^{\left(1\right)}\otimes r_{j}^{\left(2\right)}$,
for any two objects $\mathcal{V}_{1}=\left(\rho_{1},V_{1}\right)$
and $\mathcal{V}_{2}=\left(\rho_{2},V_{2}\right)$ in $\mathcal{A}$\textsf{-Mod}
we can define a morphism $\mathcal{R}^{\mathcal{V}_{1},\mathcal{V}_{2}}:\mathcal{V}_{1}\otimes\mathcal{V}_{2}\rightarrow\mathcal{V}_{2}\otimes\mathcal{V}_{1}$
by 
\begin{equation}
\mathcal{R}^{\mathcal{V}_{1},\mathcal{V}_{2}}\left(v_{1}\otimes v_{2}\right)\coloneqq\sum_{j}\left(r_{j}^{\left(2\right)}\cdot v_{2}\right)\otimes\left(r_{j}^{\left(1\right)}\cdot v_{1}\right).\label{eq:Rab}
\end{equation}
The above works as a braiding for $\mathcal{A}$\textsf{-Mod}. For a quasi-triangular
quasi-Hopf algebra, it is guaranteed that $^{*}\mathcal{V}$ is equivalent
to $\mathcal{V}^{*}$ and that the double dual $\mathcal{V}^{**}$
is equivalent to $\mathcal{V}$~\cite{Altschuler1992}.

\subsection{Algebra structures of $\mathcal{D}^{\omega}\left(G\right)$}

\begin{figure}
\begin{minipage}[t]{0.5\columnwidth}%
\includegraphics[width=0.95\columnwidth]{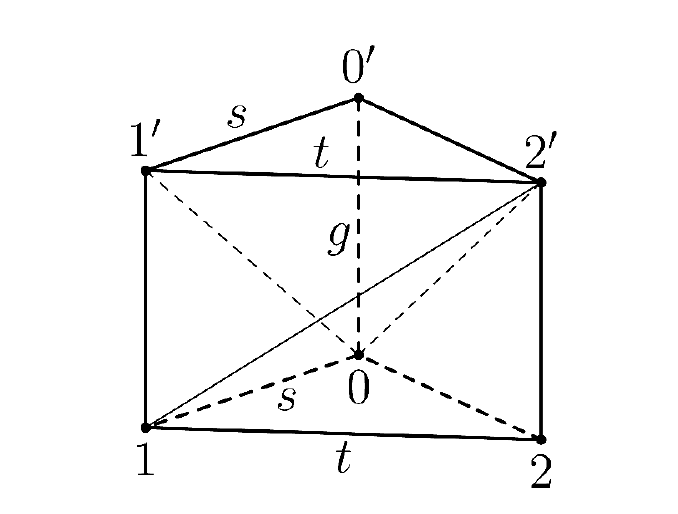}

(a) $\omega_{g}\left(s,t\right)$%
\end{minipage}%
\begin{minipage}[t]{0.5\columnwidth}%
\includegraphics[width=0.95\columnwidth]{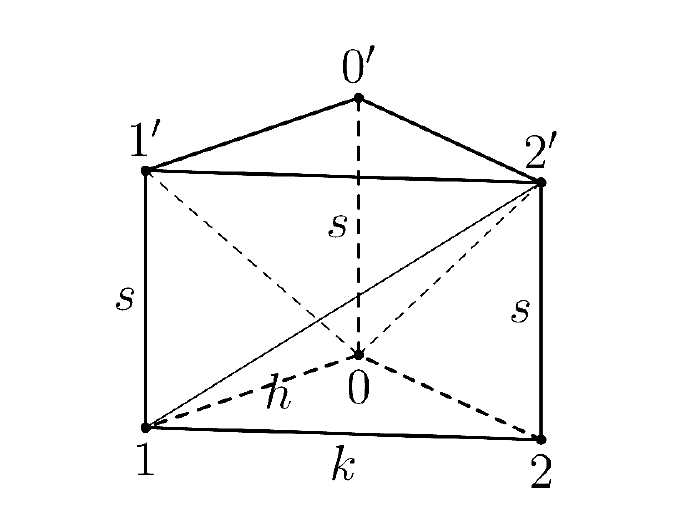}

(b) $\omega^{s}\left(h,k\right)$%
\end{minipage}

\caption{Graphic representation of $\omega_{g}\left(s,t\right)$ and $\omega^{s}\left(h,k\right)$.
The order of vertices is $0<0'<1<1'<2<2'$.}

\label{fig:prism}
\end{figure}

Given a finite group $G$, whose identity element is denoted by $e$,
and a normalized 3-cocycle $\omega\in Z^{3}\left(G,U\left(1\right)\right)$,
we can construct a quasi-triangular quasi-Hopf algebra $\left({\cal D}^{\omega}\left(G\right),\varDelta,\varepsilon,\phi,S,\alpha,\beta,R\right)$.
First of all, ${\cal D}^{\omega}\left(G\right)$ is a $\left|G\right|^{2}$-dimensional
vector space over $\mathbb{C}$ with a basis denoted as $\left\{ D_{g}^{s}\right\} _{g,s\in G}$.
The multiplication and comultiplication laws are given by 
\begin{align}
D_{g}^{s}D_{h}^{t} & =\delta_{g,shs^{-1}}\omega_{g}\left(s,t\right)D_{g}^{st},\\
\varDelta\left(D_{g}^{s}\right) & =\sum_{hk=g}\omega^{s}\left(h,k\right)D_{h}^{s}\otimes D_{k}^{s}.
\end{align}
Here $\omega_{g}\left(s,t\right)$ and $\omega^{s}\left(h,k\right)$
are phase factors defined as
\begin{align}
\omega_{g}\left(s,t\right) & \coloneqq\frac{\omega\left(g,s,t\right)\omega\left(s,t,\left(st\right)^{-1}gst\right)}{\omega\left(s,s^{-1}gs,t\right)},\label{eq:slant1}\\
\omega^{s}\left(h,k\right) & \coloneqq\frac{\omega\left(h,k,s\right)\omega\left(s,s^{-1}hs,s^{-1}ks\right)}{\omega\left(h,s,s^{-1}ks\right)}.\label{eq:slant2}
\end{align}
They correspond to the Dijkgraaf-Witten weights on the $\Delta$-complexes
with coloring shown in Fig.~\ref{fig:prism}. 

For convenience, we write
\begin{align}
D^{s} & \coloneqq\sum_{g}D_{g}^{s}.
\end{align}
It is evident that $D^{e}$ is the unit of the algebra, where $e$
is the identity element of $G$. In other words, $\mathbb{C}$ is
included in $\mathcal{D}^{\omega}\left(G\right)$ as $\mathbb{C}D^{e}$;
we often write $1$ instead of $D^{e}$ for the identity of $\mathcal{D}^{\omega}\left(G\right)$
for simplicity. Moreover, the counit is 
\begin{align}
\varepsilon:\mathcal{D}^{\omega}\left(G\right) & \rightarrow\mathbb{C}\nonumber \\
D_{g}^{s} & \mapsto\varepsilon\left(D_{g}^{s}\right)=\delta_{g,e}.
\end{align}
The Drinfeld associator is
\begin{equation}
\phi=\sum_{g,h,k\in G}\omega\left(g,h,k\right)^{-1}D_{g}^{e}\otimes D_{h}^{e}\otimes D_{k}^{e}.\label{eq:Drinfeld_DG}
\end{equation}

Further, $\left(\mathcal{D}^{\omega}\left(G\right),\varDelta,\varepsilon,\phi\right)$
is a quasi-Hopf algebra with an antipode $\left(S,\alpha,\beta\right)$
given by
\begin{align}
S\left(D_{g}^{s}\right) & =\frac{1}{\omega_{g^{-1}}\left(s,s^{-1}\right)\omega^{s}\left(g,g^{-1}\right)}D_{s^{-1}g^{-1}s}^{s^{-1}},\label{eq:S_DG}\\
\alpha & =1,\label{eq:S_alpha_DG}\\
\beta & =\sum_{g\in G}\omega\left(g,g^{-1},g\right)D_{g}^{e}.\label{eq:S_beta_DG}
\end{align}
It is also quasi-triangular with 
\begin{equation}
R=\sum_{g\in G}D_{g}^{e}\otimes D^{g}.
\end{equation}

In addition, $\mathcal{D}^{\omega}\left(G\right)$ is also a $*$-algebra.
The Hermitian conjugate of $D_{g}^{s}$ is given by
\begin{equation}
\left(D_{g}^{s}\right)^{\dagger}=\omega_{g}^{*}\left(s,s^{-1}\right)D_{s^{-1}gs}^{s^{-1}},
\end{equation}
where $\omega_{g}^{*}\left(s,s^{-1}\right)$ is the complex conjugate
of $\omega_{g}\left(s,s^{-1}\right)$. Moreover, the Hermitian
conjugate on $\mathcal{D}^{\omega}\left(G\right)\otimes\mathcal{D}^{\omega}\left(G\right)$
is given by 
\begin{equation}
\left(D_{g}^{s}\otimes D_{h}^{t}\right)^{\dagger}=\left(D_{g}^{s}\right)^{\dagger}\otimes\left(D_{h}^{t}\right)^{\dagger}.
\end{equation}
It can be checked that $\forall A\in\mathcal{D}^{\omega}\left(G\right)$,
\begin{align}
\varDelta\left(A^{\dagger}\right) & =\left(\varDelta\left(A\right)\right)^{\dagger},\\
\varepsilon\left(A^{\dagger}\right) & =\left(\varepsilon\left(A\right)\right)^{*}.
\end{align}
In the main text, $\mathcal{D}^{\omega}\left(G\right)$ is faithfully
represented, with the Hermitian conjugate respected, on a finite
Hilbert space.  So $\mathcal{D}^{\omega}\left(G\right)$ is in fact a $C^{*}$-algebra and is hence semisimple. 

\subsection{Representations of $\mathcal{D}^{\omega}\left(G\right)$}
\label{subsec:Reps_DG}

It is known~\cite{1991NuPhS..18...60D} that $\mathcal{D}^{\omega}\left(G\right)$
is semisimple: all its representations can be decomposed into irreducible
ones. Below, we construct all possible irreducible representations
$\left\{ {\cal V}_{\mathfrak{a}}\right\} _{\mathfrak{a}\in \mathfrak{Q}}$ of ${\cal D}^{\omega}\left(G\right)$.
The index set can be $\mathfrak{Q}=\left\{ \left(h,\varrho\right)|h\in J,\varrho\in\left(Z_{G}\left(h\right)\right)_{\text{ir}}^{\omega_{h}}\right\} $,
where $J$ is a subset of $G$ selecting a representative for each
conjugacy class and $\left(Z_{G}\left(h\right)\right)_{\text{ir}}^{\omega_{h}}$
selects a representative for each irreducible $\omega_{h}$-representation
isomorphism class of $Z_{G}\left(h\right)\coloneqq\left\{ g\in G|gh=hg\right\} $.
Here $Z_{G}\left(h\right)$ is called the centralizer of $h$ in $G$. 

In detail, a $\omega_{h}$-representation of $Z_{G}\left(h\right)$ is a vector space $V_{\varrho}$ equipped with a map $\ensuremath{\varrho}:Z_{G}\left(h\right)\rightarrow\text{GL}\left(V_{\varrho}\right)$ satisfying $\varrho\left(s\right)\varrho\left(t\right)=\omega_{h}\left(s,t\right)\varrho\left(st\right),\forall s,t\in G$, where $\text{GL}\left(V_{\varrho}\right)$ is the group of all invertible linear transformations of $V_{\varrho}$. 
Then we can define a representation $\varrho_{h}$ of ${\cal D}_{h}^{\omega}\left(G\right)$
on $V_{\varrho}$ by 
\begin{equation}
\varrho_{h}\left(D_{g}^{s}\right)=\delta_{g,h}\;\varrho\left(s\right),
\end{equation}
where ${\cal D}_{h}^{\omega}\left(G\right)$ is the subalgebra of
${\cal D}^{\omega}\left(G\right)$ spanned by $\left\{ D_{g}^{s}|g\in G,s\in Z_{G}\left(h\right)\right\} $.
Further, in short, 
\begin{equation}
{\cal V}_{\left(h,\varrho\right)}={\cal D}^{\omega}\left(G\right)\otimes_{{\cal D}_{h}^{\omega}\left(G\right)}V_{\varrho}.
\end{equation}
gives an explicit representation ${\cal V}_{\left(h,\varrho\right)}$ corresponding to $(h,\varrho)\in\mathfrak{Q}$.
Moreover, an inner product can be added to $\mathcal{V}_{\left(h,\varrho\right)}$
such that $\rho_{\left(h,\varrho\right)}\left(A^{\dagger}\right)=\left(\rho_{\left(h,\varrho\right)}\left(A\right)\right)^{\dagger},\forall A\in{\cal D}^{\omega}\left(G\right)$.

Explicitly, we pick a representative $q_{j}^{h}$ for
each left coset of $\nicefrac{G}{Z_{G}\left(h\right)}$. Since the
conjugacy class containing $h$ can be expressed as $\left[h\right]=\left\{ q_{j}^{h}h\left(q_{j}^{h}\right)^{-1}\right\} $,
the index $j$ goes from $1$ to $\left|\left[h\right]\right|$ (\emph{i.e.},
the cardinality of $\left[h\right]$). For convenience, we always
take $q_{1}^{h}=e$. To proceed, if $\left\{ \epsilon_{i}^{\varrho}\right\} _{i=1,2,\cdots,\deg\varrho}$
is a basis for $V_{\varrho}$, then so is $\left\{ \left|q_{j}^{h},\epsilon_{i}^{\varrho}\right\rangle \coloneqq A^{q_{j}^{h}}\otimes\left|\epsilon_{i}^{\varrho}\right\rangle \right\} _{j=1,2,\cdots,\left|\left[h\right]\right|}^{i=1,2,\cdots,\deg\varrho}$
for ${\cal V}_{\left(h,\varrho\right)}$. Then the representation
$\rho_{\left(h,\varrho\right)}$ on ${\cal V}_{\left(h,\varrho\right)}$
is given by
\begin{multline}
\rho_{\left(h,\varrho\right)}\left(D_{g}^{t}\right)\left|q_{j}^{h},\epsilon_{i}^{\varrho}\right\rangle =\\
\frac{\omega_{g}\left(t,q_{j}^{h}\right)}{\omega_{g}\left(q_{k}^{h},s\right)}\delta_{g,tq_{j}^{h}h\left(tq_{j}^{h}\right)^{-1}}\left|q_{k}^{h},\varrho\left(s\right)\epsilon_{i}^{\varrho}\right\rangle ,\label{eq:irrep-1}
\end{multline}
where $q_{k}^{h}$ and $s$ are specified by $tq_{j}^{h}=q_{k}^{h}s$
and $s\in Z_{G}\left(h\right)$. Two representations ${\cal V}_{\left(h,\varrho\right)}$
and ${\cal V}_{\left(h',\varrho'\right)}$ constructed this way are
equivalent if $h,h'$ are conjugate and $\varrho,\varrho'$ are equivalent. 

In addition, an inner product can be added on $\mathcal{V}_{\left(h,\varrho\right)}$
such that $\rho_{\left(h,\varrho\right)}\left(A^{\dagger}\right)=\left(\rho_{\left(h,\varrho\right)}\left(A\right)\right)^{\dagger},\forall A\in{\cal D}^{\omega}\left(G\right)$.
To see this, we first apply Weyl\textquoteright s unitarian trick to
$\left(\varrho,V_{\varrho}\right)\in\left(Z_{G}\left(h\right)\right)_{\text{ir}}^{\omega_{h}}$:
starting with any inner product $\left(\cdot,\cdot\right):V_{\varrho}\times V_{\varrho}\rightarrow\mathbb{C}$,
we can construct a new inner product by 
\begin{equation}
\left\langle v|w\right\rangle \coloneqq\frac{\sum_{g\in Z_{G}\left(h\right)}\left(\varrho\left(g\right)v,\varrho\left(g\right)w\right)}{\left|Z_{G}\left(h\right)\right|},\forall v,w\in V_{\varrho}.
\end{equation}
Since $\omega_{h}\left(f,g\right)\in U\left(1\right)$, it is straightforward to see that $\varrho$ is unitary under the new inner product $\left\langle \cdot|\cdot\right\rangle $,
\emph{i.e.}, $\forall v,w\in V_{\varrho},\forall g\in Z_{G}\left(h\right),$
\begin{equation}
\left\langle \varrho\left(g\right)v|\varrho\left(g\right)w\right\rangle =\left\langle v|w\right\rangle .
\end{equation}
Further, if $\left\{ \epsilon_{i}^{\varrho}\right\} _{i=1,2,\cdots,\deg\varrho}$
is an orthonormal basis for $V_{\varrho}$ with respect to $\left\langle \cdot|\cdot\right\rangle $,
then an inner product, also denoted $\left\langle \cdot|\cdot\right\rangle $,
on $\mathcal{V}_{\left(h,\varrho\right)}$ is given by requiring that $\left\{ \left|q_{j}^{h},\epsilon_{i}^{\varrho}\right\rangle \right\} _{j=1,2,\cdots,\left|\left[h\right]\right|}^{i=1,2,\cdots,\deg\varrho}$
is orthonormal as well. It can be checked that 
\begin{equation}
\rho_{\left(h,\varrho\right)}\left(A^{\dagger}\right)=\left(\rho_{\left(h,\varrho\right)}\left(A\right)\right)^{\dagger},\forall A\in{\cal D}^{\omega}\left(G\right),\label{eq:rep_dagger}
\end{equation}
where the two $\dagger$'s denote the Hermitian conjugates for ${\cal D}^{\omega}\left(G\right)$
and operators on $\left(\mathcal{V}_{d},\left\langle \cdot|\cdot\right\rangle \right)$
respectively.

Further, $\rho=\oplus_{\mathfrak{a}\in \mathfrak{Q}}\rho_{\mathfrak{a}}$ gives an isomorphism of algebras
\begin{equation}
\rho:{\cal D}^{\omega}\left(G\right)\simeq\bigoplus_{\mathfrak{a}\in \mathfrak{Q}}\mathcal{L}\left({\cal V}_{\mathfrak{a}}\right),\label{eq:wd-1}
\end{equation}
where $\mathcal{L}\left({\cal V}_{\mathfrak{a}}\right)$ is the algebra of all
linear operators on ${\cal V}_{\mathfrak{a}}$ and is isomorphic to the algebra
of all $\dim_{\mathbb{C}}{\cal V}_{\mathfrak{a}}\times\dim_{\mathbb{C}}{\cal V}_{\mathfrak{a}}$-matrices
$M_{\dim_{\mathbb{C}}{\cal V}_{\mathfrak{a}}}\left(\mathbb{C}\right)$. According
to the Artin-Wedderburn theorem, a finite dimensional algebra over
an algebraic closed field is semisimple if and only if such an isomorphism
exist. The inverse of $\rho$, denoted $\gamma$, can be specified
by its value on each basis vector $\left|q_{j}^{h},\epsilon_{i}^{\varrho}\right\rangle \left\langle q_{j'}^{h},\epsilon_{i'}^{\varrho}\right|\in\mathcal{L}\left({\cal V}_{\left(h,\varrho\right)}\right)$.
To keep notations compact, we may not write $\gamma$ explicitly as
long as the representation is clearly carried by a vector space.

To work out the details step by step, we start with the subalgebra
$\mathbb{C}\left[Z_{G}\left(h\right)\right]_{\omega_{h}}$ spanned
by $\left\{ D_{h}^{s}|s\in Z_{G}\left(h\right)\right\} $, which is
a twisted group algebra and hence semisimple~\cite{CHENG2015230}.
So we naturally write down
\begin{multline}
\gamma\left(\left|q_{1}^{h},\epsilon_{i}^{\varrho}\right\rangle \left\langle q_{1}^{h},\epsilon_{i'}^{\varrho}\right|\right)=\\
\sum_{s\in Z_{G}\left(h\right)}\frac{\deg\varrho}{\left|Z_{G}\left(h\right)\right|}\left\langle \epsilon_{i'}^{\varrho}\right|\varrho\left(s\right)^{-1}\left|\epsilon_{i}^{\varrho}\right\rangle D_{h}^{s}.\label{eq:basis-0-1}
\end{multline}
It can be checked that $\pi$ respects the action of $D_{h}^{s}$ for
any $s\in Z_{G}\left(h\right)$ and that it sends $\sum_{\varrho,i}\left|q_{1}^{h},\epsilon_{i}^{\varrho}\right\rangle \left\langle q_{1}^{h},\epsilon_{i}^{\varrho}\right|$
to $B_{h}$. Hence $\pi\circ\rho\left(D_{h}^{s}\right)=D_{h}^{s},\forall s\in Z_{G}\left(h\right)$.
Also, $\rho\circ\pi$ equals the identity on $\left|q_{1}^{h},\epsilon_{i}^{\varrho}\right\rangle \left\langle q_{1}^{h},\epsilon_{i'}^{\varrho}\right|$
by dimension counting. 

In addition, we notice that 
\begin{align}
\left\langle q_{j'}^{h},\epsilon_{i'}^{\varrho}\right| & =\left\langle q_{1}^{h},\epsilon_{i'}^{\varrho}\right|\left(D^{q_{j'}^{h}}\right)^{\dagger}=\frac{\left\langle q_{1}^{h},\epsilon_{i}^{\varrho}\right|D_{h}^{(q_{j'}^{h})^{-1}}}{\omega_{h}\left((q_{j'}^{h})^{-1},q_{j'}^{h}\right)}.
\end{align}
Therefore, the inverse of $\rho$ is given by
\begin{multline}
\gamma\left(\left|q_{j}^{h},\epsilon_{i}^{\varrho}\right\rangle \left\langle q_{j'}^{h},\epsilon_{i'}^{\varrho}\right|\right)=\gamma\left[D^{q_{j}^{h}}\frac{\left|q_{1}^{h},\epsilon_{i}^{\varrho}\right\rangle \left\langle q_{1}^{h},\epsilon_{i'}^{\varrho}\right|}{\omega_{h}\left((q_{j'}^{h})^{-1},q_{j'}^{h}\right)}D_{h}^{(q_{j'}^{h})^{-1}}\right]\\
=\sum_{s\in Z_{G}\left(h\right)}\frac{\deg\varrho}{\left|Z_{G}\left(h\right)\right|}D^{q_{j}^{h}}D_{h}^{s}\frac{\left\langle \epsilon_{i'}^{\varrho}\right|\varrho\left(s\right)^{-1}\left|\epsilon_{i}^{\varrho}\right\rangle }{\omega_{h}\left((q_{j'}^{h})^{-1},q_{j'}^{h}\right)}D_{h}^{(q_{j'}^{h})^{-1}}\\
=\sum_{s\in Z_{G}\left(h\right)}\frac{\deg\varrho}{\left|Z_{G}\left(h\right)\right|}\Gamma_{\varrho}^{i,j,i',j'}\left(s\right)D_{h_{j},}^{q_{j}^{h}s(q_{j'}^{h})^{-1}},
\end{multline}
where $h_{j}\coloneqq q_{j}^{h}h(q_{j}^{h})^{-1}$ and
\begin{multline}
\Gamma_{\varrho}^{i,j,i',j'}\left(s\right)\coloneqq\\
\frac{\left\langle \epsilon_{i'}^{\varrho}\right|\varrho\left(s\right)^{-1}\left|\epsilon_{i}^{\varrho}\right\rangle \omega_{h_{j}}\left(q_{j}^{h},s\right)\omega_{h_{j}}\left(q_{j}^{h}s,(q_{j'}^{h})^{-1}\right)}{\omega_{h}\left((q_{j'}^{h})^{-1},q_{j'}^{h}\right)}.
\end{multline}

In case that $G$ is Abelian, the basis vectors of ${\cal V}_{\left(h,\varrho\right)}$
can be written as $\left|h;\epsilon_{i}^{\varrho}\right\rangle \coloneqq\left|q_{1}^{h},\epsilon_{i}^{\varrho}\right\rangle $.
The representation Eq.~\eqref{eq:irrep-1} reduces to
\begin{equation}
\rho_{\left(h,\varrho\right)}\left(D_{g}^{s}\right)\left|h;\epsilon_{i}^{\varrho}\right\rangle =\delta_{g,h}\cdot\varrho\left(s\right)\left|h;\epsilon_{i}^{\varrho}\right\rangle ,\label{eq:rep_D_abelian}
\end{equation}
and the inverse of $\rho=\oplus\rho_{\left(h,\varrho\right)}$ reduces
to 
\begin{equation}
\gamma\left(\left|h;\epsilon_{i}^{\varrho}\right\rangle \left\langle h;\epsilon_{i'}^{\varrho}\right|\right)=\sum_{s\in G}\frac{\deg\varrho}{\left|G\right|}\left\langle \epsilon_{i'}^{\varrho}\right|\varrho\left(s\right)^{-1}\left|\epsilon_{i}^{\varrho}\right\rangle D_{h}^{s}.
\end{equation}

\subsection{The braided tensor category $\mathcal{D}^{\omega}\left(G\right)$\textsf{-Mod}}

Since $\left(\mathcal{D}^{\omega}\left(G\right),\varDelta,\varepsilon,\phi,S,\alpha,\beta,R\right)$
is a quasi-triangular quasi-Hopf algebra, its representation category
$\mathcal{D}^{\omega}\left(G\right)$\textsf{-Mod} is a braided tensor
category with duality. In addition, $\mathcal{D}^{\omega}\left(G\right)$\textsf{-Mod}
is semisimple with finitely many isomorphism classes of simple objects.
Let $\mathfrak{Q}$ label all the isomorphism classes of simple objects;
irreducible representations are simple objects in a representation
category. To be concrete, for each $\mathfrak{a}\in\mathfrak{Q}$,
we pick an explicit representation $\mathcal{V}_{a}$, such as the one
constructed in subsection~\ref{subsec:Reps_DG}. In particular,
the trivial representation is denoted by $\mathcal{V}_{\mathfrak{0}}$
and the corresponding element of $\mathfrak{Q}$ is denoted by $\mathfrak{0}$.
It can be checked that the dual representation $\mathcal{V}_{\mathfrak{a}}^{*}$
is also irreducible and works as both a right and left dual object
of $\mathfrak{a}$. Let $\overline{\mathfrak{a}}\in\mathfrak{Q}$ label
the isomorphism class of $\mathcal{V}_{\mathfrak{a}}^{*}$. By definition,
$\mathcal{V}_{\overline{\mathfrak{a}}}$ is isomorphic to $\mathcal{V}_{\mathfrak{a}}^{*}$. 

In the physics literature, a semisimple braided tensor category
is often specified by the following data (1) fusion rule $N_{\mathfrak{a}\mathfrak{b}}^{\mathfrak{c}}$,
(2) $6j$-symbols $F_{\mathfrak{d}\mathfrak{e}\mathfrak{f}}^{\mathfrak{a}\mathfrak{b}\mathfrak{c}}$
and (3) $\mathcal{R}$-symbols $\mathcal{R}_{\mathfrak{c}}^{\mathfrak{a}\mathfrak{b}}$,
where $\mathfrak{a},\mathfrak{b},\mathfrak{c},\mathfrak{d},\mathfrak{e},\mathfrak{f}\in\mathfrak{Q}$.
Below, let us work out the definitions of these data for $\mathcal{D}^{\omega}\left(G\right)$\textsf{-Mod}.
This can be done via the notion of a \emph{splitting space} $V_{\mathfrak{c}}^{\mathfrak{a}\mathfrak{b}}$,
defined as
\begin{align}
V_{\mathfrak{c}}^{\mathfrak{a}\mathfrak{b}} & \coloneqq\text{Hom}\left(\mathcal{V}_{\mathfrak{c}},\mathcal{V}_{\mathfrak{a}}\otimes\mathcal{V}_{\mathfrak{b}}\right),
\end{align}
where $\mathcal{V}_{\mathfrak{a}},\mathcal{V}_{\mathfrak{b}},\mathcal{V}_{\mathfrak{c}}$
are irreducible representations labeled by $\mathfrak{a},\mathfrak{b},\mathfrak{c}\in \mathfrak{Q}$.
In other words, $V_{\mathfrak{c}}^{\mathfrak{a}\mathfrak{b}}$ is
the vector space of intertwiners from $\mathcal{V}_{\mathfrak{c}}$
to $\mathcal{V}_{\mathfrak{a}}\otimes\mathcal{V}_{\mathfrak{b}}$. 

The fusion rule $N_{\mathfrak{a}\mathfrak{b}}^{\mathfrak{c}}$ is
just the dimension of $V_{\mathfrak{c}}^{\mathfrak{a}\mathfrak{b}}$,
\emph{i.e.},
\begin{equation}
N_{\mathfrak{a}\mathfrak{b}}^{\mathfrak{c}}\coloneqq\dim_{\mathbb{C}}V_{\mathfrak{c}}^{\mathfrak{a}\mathfrak{b}},
\end{equation}
for any $\mathfrak{a},\mathfrak{b},\mathfrak{c}\in \mathfrak{Q}$. It satisfies 
\begin{equation}
\sum_{\mathfrak{e}}N_{\mathfrak{ab}}^{\mathfrak{e}}N_{\mathfrak{ec}}^{\mathfrak{d}}=\sum_{\mathfrak{f}}N_{\mathfrak{af}}^{\mathfrak{d}}N_{\mathfrak{bc}}^{\mathfrak{f}},
\end{equation}
as a result of the isomorphism in Eq.~\eqref{eq:F_iso} below. The
fusion rule can be viewed as the multiplication rule of the Grothendieck
ring of $\mathcal{D}^{\omega}\left(G\right)$\textsf{-Mod}, and we
write 
\begin{equation}
\mathfrak{a}\times\mathfrak{b}=\sum_{\mathfrak{c}\in\mathfrak{Q}}N_{\mathfrak{a}\mathfrak{b}}^{\mathfrak{c}}\mathfrak{c}.
\end{equation}
Because of the isomorphism $\mathcal{R}^{\mathfrak{a}\mathfrak{b}}:\mathcal{V}_{\mathfrak{b}}\otimes\mathcal{V}_{\mathfrak{a}}\cong\mathcal{V}_{\mathfrak{a}}\otimes\mathcal{V}_{\mathfrak{b}}$,
we have 
\begin{equation}
N_{\mathfrak{a}\mathfrak{b}}^{\mathfrak{c}}=N_{\mathfrak{b}\mathfrak{a}}^{\mathfrak{c}}.
\end{equation}
In addition, for any semisimple braided tensor category with duality,
we notice that 
\begin{equation}
N_{\mathfrak{a}\mathfrak{b}}^{\mathfrak{0}}=\delta_{\overline{\mathfrak{a}}\mathfrak{b}},
\end{equation}
which identifies $\overline{\mathfrak{a}}$ from the fusion rule. 

To define $F_{\mathfrak{d}\mathfrak{e}\mathfrak{f}}^{\mathfrak{a}\mathfrak{b}\mathfrak{c}}$,
we observe the isomorphisms of vector spaces
\begin{align}
\bigoplus_{\mathfrak{e}\in\mathfrak{Q}}V_{\mathfrak{e}}^{\mathfrak{a}\mathfrak{b}}\otimes V_{\mathfrak{d}}^{\mathfrak{e}\mathfrak{c}} & \cong\text{Hom}\left(\mathcal{V}_{\mathfrak{d}},\left(\mathcal{V}_{\mathfrak{a}}\otimes\mathcal{V}_{\mathfrak{b}}\right)\otimes\mathcal{V}_{\mathfrak{c}}\right)\nonumber \\
\mu\otimes\nu & \mapsto\left(\mu\otimes\text{id}_{\mathfrak{c}}\right)\circ\nu,\label{eq:mn}
\end{align}
\begin{align}
\bigoplus_{\mathfrak{f}\in\mathfrak{Q}}V_{\mathfrak{d}}^{\mathfrak{a}\mathfrak{f}}\otimes V_{\mathfrak{f}}^{\mathfrak{b}\mathfrak{c}} & \cong\text{Hom}\left(\mathcal{V}_{\mathfrak{d}},\mathcal{V}_{\mathfrak{a}}\otimes\left(\mathcal{V}_{\mathfrak{b}}\otimes\mathcal{V}_{\mathfrak{c}}\right)\right)\nonumber \\
\kappa\otimes\lambda & \mapsto\left(\text{id}_{\mathfrak{a}}\otimes\lambda\right)\circ\kappa.\label{eq:klamba}
\end{align}
Because of $\phi:\left(\mathcal{V}_{\mathfrak{a}}\otimes\mathcal{V}_{\mathfrak{b}}\right)\otimes\mathcal{V}_{\mathfrak{c}}\cong\mathcal{V}_{\mathfrak{a}}\otimes\left(\mathcal{V}_{\mathfrak{b}}\otimes\mathcal{V}_{\mathfrak{c}}\right)$,
we have $\text{Hom}\left(\mathcal{V}_{\mathfrak{d}},\left(\mathcal{V}_{\mathfrak{a}}\otimes\mathcal{V}_{\mathfrak{b}}\right)\otimes\mathcal{V}_{\mathfrak{c}}\right)\cong\text{Hom}\left(\mathcal{V}_{\mathfrak{d}},\mathcal{V}_{\mathfrak{a}}\otimes\left(\mathcal{V}_{\mathfrak{b}}\otimes\mathcal{V}_{\mathfrak{c}}\right)\right).$
Thus, there is an isomorphism of vector spaces
\begin{equation}
\bigoplus_{\mathfrak{e}\in\mathfrak{Q}}V_{\mathfrak{e}}^{\mathfrak{a}\mathfrak{b}}\otimes V_{\mathfrak{d}}^{\mathfrak{e}\mathfrak{c}}\cong\bigoplus_{\mathfrak{f}\in\mathfrak{Q}}V_{\mathfrak{d}}^{\mathfrak{a}\mathfrak{f}}\otimes V_{\mathfrak{f}}^{\mathfrak{b}\mathfrak{c}}.\label{eq:F_iso}
\end{equation}
Restricting the isomorphism to the summand on the left hand side corresponding
to a given $\mathfrak{e}\in \mathfrak{Q}$ and projecting into the summand on the right
hand side corresponding to a given $\mathfrak{f}\in\mathfrak{Q}$,
we get the homomorphism 
\begin{equation}
F_{\mathfrak{d}\mathfrak{e}\mathfrak{f}}^{\mathfrak{a}\mathfrak{b}\mathfrak{c}}:V_{\mathfrak{e}}^{\mathfrak{a}\mathfrak{b}}\otimes V_{\mathfrak{d}}^{\mathfrak{e}\mathfrak{c}}\rightarrow V_{\mathfrak{d}}^{\mathfrak{a}\mathfrak{f}}\otimes V_{\mathfrak{f}}^{\mathfrak{b}\mathfrak{c}},
\end{equation}
which is called the $6j$-symbol for $\left(\mathfrak{a},\mathfrak{b},\mathfrak{c},\mathfrak{d},\mathfrak{e},\mathfrak{f}\right)\in\mathfrak{Q}^{6}$. 

The $R$-symbols $\mathcal{R}_{\mathfrak{c}}^{\mathfrak{a}\mathfrak{b}}$
are an isomorphism between $V_{\mathfrak{c}}^{\mathfrak{b}\mathfrak{a}}$
and $V_{\mathfrak{c}}^{\mathfrak{a}\mathfrak{b}}$; it is induced
by the braiding $\mathcal{R}^{\mathfrak{a}\mathfrak{b}}:\mathcal{V}_{\mathfrak{b}}\otimes\mathcal{V}_{\mathfrak{a}}\rightarrow\mathcal{V}_{\mathfrak{a}}\otimes\mathcal{V}_{\mathfrak{b}}$
in the following way
\begin{align}
\mathcal{R}_{\mathfrak{c}}^{\mathfrak{a}\mathfrak{b}}:V_{\mathfrak{c}}^{\mathfrak{b}\mathfrak{a}} & \rightarrow V_{\mathfrak{c}}^{\mathfrak{a}\mathfrak{b}}\nonumber \\
\mu & \mapsto\mathcal{R}^{\mathfrak{a}\mathfrak{b}}\circ\mu,\label{eq:Rabc}
\end{align}
for any $\mathfrak{a},\mathfrak{b},\mathfrak{c}\in\mathfrak{Q}$ and
any $\mu\in V_{\mathfrak{c}}^{\mathfrak{a}\mathfrak{b}}$. 

To give a matrix representation for the linear maps $F_{\mathfrak{d}\mathfrak{e}\mathfrak{f}}^{\mathfrak{a}\mathfrak{b}\mathfrak{c}}$
and $\mathcal{R}_{\mathfrak{c}}^{\mathfrak{a}\mathfrak{b}}$, we need
to pick a basis for each splitting space. Thus, one braided tensor
category may have different matrix representations of $F_{\mathfrak{d}\mathfrak{e}\mathfrak{f}}^{\mathfrak{a}\mathfrak{b}\mathfrak{c}}$
and $\mathcal{R}_{\mathfrak{c}}^{\mathfrak{a}\mathfrak{b}}$; they
are related by changes of basis, which are also referred to as \emph{gauge
transformations} sometimes in the physics literature. In order to distinguish
inequivalent braided tensor categories, we want to find some useful
quantities, invariant under changes of basis. The \emph{topological
spin} $\theta_{\mathfrak{a}}$ associated with each $\mathfrak{a}\in\mathfrak{Q}$
is an important one for this purpose; for $\mathcal{D}^{\omega}\left(G\right)$\textsf{-Mod},
it can be defined as 
\begin{equation}
\theta_{\mathfrak{a}}\coloneqq\sum_{\mathfrak{c}\in\mathfrak{Q}}\frac{\dim_{\mathbb{C}}\left(\mathcal{V}_{\mathfrak{c}}\right)}{\dim_{\mathbb{C}}\left(\mathcal{V}_{\mathfrak{a}}\right)}\text{tr}\left(\mathcal{R}_{\mathfrak{c}}^{\mathfrak{a}\mathfrak{a}}\right),\label{eq:topspin}
\end{equation}
which is a root of unity and satisfies $\theta_{\overline{\mathfrak{a}}}=\theta_{\mathfrak{a}}$.
More explicitly in terms of representations,
\begin{equation}
\theta_{\mathfrak{a}}=\frac{\text{tr}\left(\mathcal{R}^{\mathfrak{a}\mathfrak{a}}\right)}{\dim_{\mathbb{C}}\left(\mathcal{V}_{\mathfrak{a}}\right)}=\frac{\text{tr}\left(\wp R,\mathcal{V}_{\mathfrak{a}}\otimes\mathcal{V}_{\mathfrak{a}}\right)}{\dim_{\mathbb{C}}\left(\mathcal{V}_{\mathfrak{a}}\right)},\label{eq:topspin_irrep}
\end{equation}
where $\wp$ is the operator permuting the two factors $\mathcal{V}_{\mathfrak{a}}\otimes\mathcal{V}_{\mathfrak{a}}$
and $R$ is the universal $R$-matrix of $\mathcal{D}^{\omega}\left(G\right)$.
The topological spins are often collected into a matrix form $\mathcal{T}_{\mathfrak{a}\mathfrak{b}}=\theta_{\mathfrak{a}}\delta_{\mathfrak{a}\mathfrak{b}},\forall\mathfrak{a},\mathfrak{b}\in\mathfrak{Q}$,
which is called the \emph{topological $T$-matrix}. It is well-known
that the $R$-symbols satisfy the ``ribbon property''
\begin{equation}
\mathcal{R}_{\mathfrak{c}}^{\mathfrak{a}\mathfrak{b}}\mathcal{R}_{\mathfrak{c}}^{\mathfrak{b}\mathfrak{a}}=\frac{\theta_{\mathfrak{c}}}{\theta_{\mathfrak{a}}\theta_{\mathfrak{b}}}\text{id}_{V_{\mathfrak{c}}^{\mathfrak{a}\mathfrak{b}}}.\label{eq:mutualstat}
\end{equation}

The \emph{topological $S$-matrix} $\mathcal{S}=(\mathcal{S}_{\mathfrak{ab}})_{\mathfrak{a},\mathfrak{b}\in\mathfrak{Q}}$ is another important quantity.
For $\mathcal{D}^{\omega}\left(G\right)$\textsf{-Mod}, its matrix
element $\mathcal{S}_{\mathfrak{ab}}, \forall \mathfrak{a},\mathfrak{b}\in\mathfrak{Q}$ is defined as
\begin{align}
\mathcal{S}_{\mathfrak{a}\mathfrak{b}} & \coloneqq\frac{1}{\mathcal{D}}\sum_{\mathfrak{c}\in\mathfrak{Q}}\dim_{\mathbb{C}}\left(\mathcal{V}_{\mathfrak{c}}\right)\text{tr}\left(\mathcal{R}_{\mathfrak{c}}^{\overline{\mathfrak{a}}\mathfrak{b}}\mathcal{R}_{\mathfrak{c}}^{\mathfrak{b}\overline{\mathfrak{a}}}\right)\nonumber \\
 & =\frac{1}{\mathcal{D}}\sum_{\mathfrak{c}\in\mathfrak{Q}}N_{\overline{\mathfrak{a}}\mathfrak{b}}^{\mathfrak{c}}\frac{\theta_{\mathfrak{c}}}{\theta_{\mathfrak{a}}\theta_{\mathfrak{b}}}\dim_{\mathbb{C}}\left(\mathcal{V}_{\mathfrak{c}}\right),\label{eq:Sab}\\
{\cal D} & \coloneqq\sqrt{\sum_{\mathfrak{c}\in\mathfrak{Q}}\left[\dim_{\mathbb{C}}\left(\mathcal{V}_{\mathfrak{c}}\right)\right]^{2}}=\left|G\right|.
\end{align}
Given the topological
$S$-matrix, we can recover the fusion rule by the Verlinde formula~\cite{VERLINDE1988360,KITAEV20062,barkeshli}
\begin{equation}
N_{\mathfrak{a}\mathfrak{b}}^{\mathfrak{c}}=\sum_{\mathfrak{q}\in\mathfrak{Q}}\frac{\mathcal{S}_{\mathfrak{a}\mathfrak{q}}\mathcal{S}_{\mathfrak{b}\mathfrak{q}}\mathcal{S}_{\overline{\mathfrak{c}}\mathfrak{q}}}{\mathcal{S}_{\mathfrak{0}\mathfrak{q}}}.
\end{equation}
The topological $T$-matrix and $S$-matrix are also called the modular
invariants, as they are closely related to the modular transformations~\cite{dijkgraaf1990,hu13twisted}.

\subsection{Examples of $\mathcal{D}^{\omega}\left(G\right)$ and $\mathcal{D}^{\omega}\left(G\right)$\textsf{-Mod}}
\label{subsec:ExamplesBTC}

Below, let us study several concrete examples of $\mathcal{D}^{\omega}\left(G\right)$
and its representation category $\mathcal{D}^{\omega}\left(G\right)$\textsf{-Mod. }

\subsubsection{$\mathcal{D}\left(\mathbb{Z}_{2}\right)$}

Picking $G=\mathbb{Z}_{2}=\left\{ 0,1\right\} $ and $\omega$ trivial,
we get the quantum double algebra $\mathcal{D}\left(\mathbb{Z}_{2}\right)$.
Its four inequivalent irreducible representations, given by Eq.~\eqref{eq:rep_D_abelian},
are
\begin{equation}
\rho_{g}^{\lambda}\left(D_{h}^{s}\right)=\delta_{g,h}\cdot e^{i\pi\lambda s},
\end{equation}
labeled by $\left(g,\lambda\right)\in\mathbb{Z}_{2}\times\mathbb{Z}_{2}\equiv \mathfrak{Q}$,
all of which are one-dimensional. For example, 
\begin{equation}
\rho_{0}^{1}\left(D_{0}^{1}\right)=e^{i\pi\left(1\times1\right)}=-1.
\end{equation}
In the notation widely used in the toric code model, the four
simple objects $\left(0,0\right),\left(0,1\right),\left(1,0\right),\left(1,1\right)$
in $\mathcal{D}\left(\mathbb{Z}_{2}\right)$\textsf{-Mod} are denoted
by $\mathfrak{1},\mathfrak{e},\mathfrak{m},\boldsymbol{\varepsilon}$ respectively~\cite{KITAEV20032}.

The fusion rule is given by 
\begin{equation}
\left(g_{1},\lambda_{1}\right)\times\left(g_{2},\lambda_{2}\right)=\left(g_{1}+g_{2},\lambda_{1}+\lambda_{2}\right).
\end{equation}
In other words, 
\begin{gather}
\mathfrak{e}\times \mathfrak{e}=\mathfrak{m}\times \mathfrak{m}=\boldsymbol{\varepsilon}\times\boldsymbol{\varepsilon}=\mathfrak{1},\\
\mathfrak{e}\times \mathfrak{m}=\boldsymbol{\varepsilon},\;\mathfrak{e}\times\boldsymbol{\varepsilon}=\mathfrak{m},\;\mathfrak{m}\times\boldsymbol{\varepsilon}=\mathfrak{e}.
\end{gather}
The unit object is $\boldsymbol{1}\equiv\left(0,0\right)$ and $\overline{\mathfrak{a}}=\mathfrak{a},\forall \mathfrak{a}\in \mathfrak{Q}$. 

All $6j$-symbols, allowed by fusion rules, equal 1. The $R$-symbols
are given by
\begin{equation}
\mathcal{R}^{\left(g_{1},\lambda_{1}\right)\left(g_{2},\lambda_{2}\right)}=e^{i\pi\lambda_{1}g_{2}},
\end{equation}
where we omit $\mathfrak{c}$ in $\mathcal{R}_{\mathfrak{c}}^{\mathfrak{ab}}$ since $\mathfrak{c}$ is uniquely
determined by $\mathfrak{a}$ and $\mathfrak{b}$. Then Eq.~\eqref{eq:topspin}
gives the topological spins 
\begin{align}
\left(\theta_{\mathfrak{1}},\theta_{\mathfrak{e}},\theta_{\mathfrak{m}},\theta_{\boldsymbol{\varepsilon}}\right) & =\left(1,1,1,-1\right).
\end{align}
By Eq.~\eqref{eq:Sab}, the topological $S$-matrix is given by
\begin{equation}
\mathcal{S}_{\mathfrak{a}\mathfrak{b}}=\frac{1}{2}\frac{\theta_{\mathfrak{a}\times\mathfrak{b}}}{\theta_{\mathfrak{a}}\theta_{\mathfrak{b}}}.
\end{equation}

In general, $\mathcal{R}^{\mathfrak{a}\mathfrak{b}}\mathcal{R}^{\mathfrak{b}\mathfrak{a}}$
is a scalar multiplication by $\frac{\theta_{\mathfrak{a}\times\mathfrak{b}}}{\theta_{\mathfrak{a}}\theta_{\mathfrak{b}}}$.
For $\mathcal{D}^{\omega}\left(\mathbb{Z}_{2}\right)$\textsf{-Mod},
we notice that $\mathcal{R}^{\mathfrak{a}\mathfrak{a}}$ is a scalar
multiplication by $\theta_{\mathfrak{a}}$. 

\subsubsection{$\mathcal{D}^{\omega}\left(\mathbb{Z}_{2}\right)$ with $\omega\left(1,1,1\right)=-1$}
\label{subsec:semion}

The nontrivial element of $H^{3}\left(G,U\left(1\right)\right)=\mathbb{Z}_{2} $
for $G=\mathbb{Z}_2=\{0,1\}$ is represented by the normalized $3$-cocycle
\begin{equation}
\omega\left(g,h,k\right)=\begin{cases}
-1, & g=h=k=1,\\
1, & \text{otherwise}.
\end{cases}
\end{equation}
Different from $\mathcal{D}\left(\mathbb{Z}_{2}\right)$, in $\mathcal{D}^{\omega}\left(\mathbb{Z}_{2}\right)$,
we have $D_{1}^{1}D_{1}^{1}=-D_{1}^{0}$ because $\omega_{1}\left(1,1\right)=-1$
by the definition in Eq.~\eqref{eq:slant1}. 

The irreducible representations of $\mathcal{D}^{\omega}\left(\mathbb{Z}_{2}\right)$
are 
\begin{equation}
\rho_{g}^{\lambda}\left(D_{h}^{s}\right)=\delta_{g,h}\cdot i^{g}\cdot e^{i\pi\lambda s},
\end{equation}
labeled by $\left(g,\lambda\right)\in\mathbb{Z}_{2}\times\mathbb{Z}_{2}\equiv \mathfrak{Q}$.
For example, we have 
\begin{equation}
\rho_{1}^{0}\left(D_{h}^{s}\right)=i\delta_{1,h}.
\end{equation}
The fusion rule is still
\begin{equation}
\left(g_{1},\lambda_{1}\right)\times\left(g_{2},\lambda_{2}\right)=\left(g_{1}+g_{2},\lambda_{1}+\lambda_{2}\right).
\end{equation}
The unit object is $\left(0,0\right)$ and $\overline{\left(g,\lambda\right)}=\left(g,\lambda\right)$.

Suppose that $\mathcal{V}_{\left(g,\lambda\right)}$ is spanned by
$e_{g}^{\lambda}$. Then $e_{g_{1}+g_{2}}^{\lambda_{1}+\lambda_{2}}\mapsto e_{g_{1}}^{\lambda_{1}}\otimes e_{g_{2}}^{\lambda_{2}}$
spans $V_{\left(g_{1}+g_{2},\lambda_{1}+\lambda_{2}\right)}^{\left(g_{1},\lambda_{1}\right)\left(g_{2},\lambda_{2}\right)}$.
Using such a basis for each splitting space and noticing that 
\begin{equation}
\rho_{g_{1}}^{\lambda_{1}}\otimes\rho_{g_{2}}^{\lambda_{2}}\otimes\rho_{g_{3}}^{\lambda_{3}}\left(\phi\right)=\begin{cases}
-1, & g_{1}=g_{2}=g_{3}=1,\\
1, & \text{otherwise},
\end{cases}
\end{equation}
we have 
\begin{equation}
F^{\left(g_{1},\lambda_{1}\right)\left(g_{2},\lambda_{2}\right)\left(g_{3},\lambda_{3}\right)}=\begin{cases}
-1, & g_{1}=g_{2}=g_{3}=1,\\
1, & \text{otherwise},
\end{cases}
\end{equation}
where $\mathfrak{e},\mathfrak{d},\mathfrak{f}$ are omitted in $F_{\mathfrak{edf}}^{\mathfrak{abc}}$
as they are uniquely determined by $\mathfrak{a},\mathfrak{b},\mathfrak{c}$. 

By Eq.~\eqref{eq:Rab}, we directly read the $R$-matrix 
\begin{align}
\mathcal{R}^{\left(g_{1},\lambda_{1}\right)\left(g_{2},\lambda_{2}\right)} & =\sum_{g}\rho_{g_{2}}^{\lambda_{2}}\left(D^{g}\right)\otimes\rho_{g_{1}}^{\lambda_{1}}\left(D_{g}^{e}\right)\nonumber \\
 & =i^{g_{2}}\cdot e^{i\pi\lambda_{2}g_{1}}
\end{align}
from the universal $R$-matrix $R=\sum_{g}D_{g}^{e}\otimes D^{g}$ with $D^{s}  \coloneqq\sum_{g}D_{g}^{s}$.
Then Eq.~\eqref{eq:topspin} gives the topological spins
\begin{align}
\left(\theta_{\left(0,0\right)},\theta_{\left(0,1\right)},\theta_{\left(1,0\right)},\theta_{\left(1,1\right)}\right) & =\left(1,1,i,-i\right).
\end{align}
Hence, the simple objects $\left(1,0\right)$ and $\left(1,1\right)$
are often called semions. In addition, Eq.~\eqref{eq:Sab} gives
the topological $S$-matrix 
\begin{equation}
\mathcal{S}_{\mathfrak{a}\mathfrak{b}}=\frac{1}{2}\frac{\theta_{\mathfrak{a}\times\mathfrak{b}}}{\theta_{\mathfrak{a}}\theta_{\mathfrak{b}}}.
\end{equation}

In general,$\mathcal{R}^{\mathfrak{a}\mathfrak{b}}\mathcal{R}^{\mathfrak{b}\mathfrak{a}}$
is a scalar multiplication by $\frac{\theta_{\mathfrak{a}\times\mathfrak{b}}}{\theta_{\mathfrak{a}}\theta_{\mathfrak{b}}}$.
For $\mathcal{D}^{\omega}\left(\mathbb{Z}_{2}\right)$\textsf{-Mod},
we notice that $\mathcal{R}^{\mathfrak{aa}}$ is a scalar multiplication
by $\theta_{\mathfrak{a}}$. 

\subsubsection{$\mathcal{D}^{\omega}\left(\mathbb{Z}_{2}^{3}\right)$ with $\omega\left(g,h,k\right)=e^{i\pi g^{\left(1\right)}h^{\left(2\right)}k^{\left(3\right)}}$}

This gives an example in which not all irreducible representations are
one-dimensional, even though $G$ itself is Abelian. The three $\mathbb{Z}_{2}$
components of $g\in G=\mathbb{Z}_{2}^{3}=\mathbb{Z}_{2}\times\mathbb{Z}_{2}\times\mathbb{Z}_{2}$
are denoted by $g^{\left(1\right)},g^{\left(2\right)},g^{\left(3\right)}$.
To keep notations compact, we will use $100$ short for $\left(1,0,0\right)\in G$;
thus, the eight group elements are also denoted $000, 100, 010, 001, 110, 011, 101, 111$. Moreover, we write $0$ short for $000$ when it is clear from context.
As $G$ is Abelian, Eqs.~\eqref{eq:slant1} and \eqref{eq:slant2}
define a normalized $2$-cocycle $\omega_{g}=\omega^{g}\in Z^{2}\left(G,U\left(1\right)\right)$
for each fixed $g\in G$. Explicitly,
\begin{equation}
\omega_{g}\left(h,k\right)=e^{i\pi\left(g^{\left(1\right)}h^{\left(2\right)}k^{\left(3\right)}-h^{\left(1\right)}k^{\left(2\right)}g^{\left(3\right)}+k^{\left(1\right)}g^{\left(2\right)}h^{\left(3\right)}\right)}.
\end{equation}
Whenever $g\neq0$, we notice $\left[\omega_{g}\right]$
corresponds to a non-trivial element of $H^{2}\left(G,U\left(1\right)\right)$.

In total, ${\cal D}^{\omega}\left(G\right)$ has 22 inequivalent irreducible
representations. Eight of them are one-dimensional, labeled as $\rho_{0}^{\lambda}$
with $\lambda=\left(\lambda^{\left(1\right)},\lambda^{\left(2\right)},\lambda^{\left(3\right)}\right)\in\mathbb{Z}_{2}^{3}$;
explicitly, 
\begin{equation}
\rho_{0}^{\lambda}\left(D_{h}^{s}\right)=\delta_{0,h}\,e^{i\pi\lambda \cdot s},
\end{equation}
where $\lambda \cdot s \coloneqq\lambda^{\left(1\right)}s^{\left(1\right)}+\lambda^{\left(2\right)}s^{\left(2\right)}+\lambda^{\left(3\right)}s^{\left(3\right)}\pmod2$.
The other fourteen irreducible representations are two-dimensional,
labeled as $\rho_{g}^{+}$ and $\rho_{g}^{-}$
with $g\neq0$ in $\mathbb{Z}_{2}^{3}$;
they can be specified by the action of $D_{h}^{100},D_{h}^{010},D_{h}^{001}$
as shown in Table~\ref{table:irrep}. The corresponding representation
isomorphism classes are denoted by $\left(0,\lambda\right)$
for $\lambda\in\mathbb{Z}_{2}^{3}$ and $\left(g,\pm\right)\equiv\left(g,\pm1\right)$
for $g\in\mathbb{Z}_{2}^{3}$, $g\neq0$.

The dual of each simple object is isomorphic to itself; in particular,
each two dimensional irreducible representation fuses with itself as
\begin{align}
\left(100,\pm\right)^{2} & =\mathfrak{0}+\left(0,010\right)+\left(0,001\right)+\left(0,011\right),\label{eq:fusion_100}\\
\left(010,\pm\right)^{2} & =\mathfrak{0}+\left(0,001\right)+\left(0,100\right)+\left(0,101\right),\label{eq:fusion_010}\\
\left(001,\pm\right)^{2} & =\mathfrak{0}+\left(0,100\right)+\left(0,010\right)+\left(0,110\right),\label{eq:fusion_001}\\
\left(011,\pm\right)^{2} & =\mathfrak{0}+\left(0,100\right)+\left(0,011\right)+\left(0,111\right),\label{eq:fusion_011}\\
\left(101,\pm\right)^{2} & =\mathfrak{0}+\left(0,010\right)+\left(0,101\right)+\left(0,111\right),\label{eq:fusion_101}\\
\left(110,\pm\right)^{2} & =\mathfrak{0}+\left(0,001\right)+\left(0,110\right)+\left(0,111\right),\label{eq:fusion_110}\\
\left(111,\pm\right)^{2} & =\mathfrak{0}+\left(0,110\right)+\left(0,101\right)+\left(0,011\right),\label{eq:fusion_111}
\end{align}
where $\mathfrak{0}=\left(0,0\right)$ denotes
the unit object. We notice that Eqs.~(\ref{eq:fusion_100}-\ref{eq:fusion_001})
and Eqs.~(\ref{eq:fusion_011}-\ref{eq:fusion_110}) are related
by permuting components of $g,\lambda\in\mathbb{Z}_{2}^{3}$.
In addition, $\left(g,\pm\right)\times\left(g,\mp\right)$
equals the sum of one-dimensional representations that do not appear
in $\left({g},\pm\right)\times\left({g},\pm\right)$.
For example,
\begin{multline}
\left(100,\pm\right)\times\left(100,\mp\right)=\\
\left(0,100\right)+\left(0,110\right)+\left(0,101\right)+\left(0,111\right).
\end{multline}
The rest of the fusion rules are
\begin{align}
\left(0,{\lambda}\right)\times\left(0,{\mu}\right) & =\left(0,{\lambda}+{\mu}\right),\\
\left(0,{\lambda}\right)\times\left({g},\pm\right) & =\left({g},\pm e^{i\pi{\lambda}\cdot{g}}\right),\label{eq:fusion_0g}\\
\left({g},\kappa\right)\times\left({h},\kappa'\right) & =\left({g}+{h},+\right)+\left({g}+{h},-\right)
\end{align}
for ${g},{h}\neq0$, ${g}\neq{h}$
in $G$ and $\kappa,\kappa'=\pm$. 

For $6j$-symbols, let us compute $F_{\mathfrak{aef}}^{\mathfrak{aaa}}$
with $\mathfrak{a}=\left(111,\pm\right)$ as an example. In the current
category, all allowed splitting spaces are one-dimensional. We pick
a basis for each splitting space relevant here as follows
\begin{gather}
\mu_{0}\coloneqq\frac{1}{\sqrt{2}}\left(0,1,-1,0\right)^{T}\in V_{\left(0,000\right)}^{\mathfrak{aa}},\label{eq:u0}\\
\mu_{1}\coloneqq\frac{1}{\sqrt{2}}\left(1,0,0,-1\right)^{T}\in V_{\left(0,011\right)}^{\mathfrak{aa}},\label{eq:u1}\\
\mu_{2}\coloneqq\frac{1}{\sqrt{2}}\left(1,0,0,1\right)^{T}\in V_{\left(0,101\right)}^{\mathfrak{aa}},\label{eq:u2}\\
\mu_{3}\coloneqq\frac{1}{\sqrt{2}}\left(0,1,1,0\right)^{T}\in V_{\left(0,110\right)}^{\mathfrak{aa}},\label{eq:u3}\\
\sigma_{0}=\left(\begin{array}{cc}
1 & 0\\
0 & 1
\end{array}\right)\in V_{\mathfrak{a}}^{\left(0,000\right)\mathfrak{a}},V_{\mathfrak{a}}^{\mathfrak{a}\left(0,000\right)},\\
\sigma_{1}=\left(\begin{array}{cc}
0 & 1\\
1 & 0
\end{array}\right)\in V_{\mathfrak{a}}^{\left(0,011\right)\mathfrak{a}},V_{\mathfrak{a}}^{\mathfrak{a}\left(0,011\right)},\\
\sigma_{2}=\left(\begin{array}{cc}
0 & -i\\
i & 0
\end{array}\right)\in V_{\mathfrak{a}}^{\left(0,101\right)\mathfrak{a}},V_{\mathfrak{a}}^{\mathfrak{a}\left(0,101\right)},\\
\sigma_{3}=\left(\begin{array}{cc}
1 & 0\\
0 & -1
\end{array}\right)\in V_{\mathfrak{a}}^{\left(0,110\right)\mathfrak{a}},V_{\mathfrak{a}}^{\mathfrak{a}\left(0,110\right)},
\end{gather}
where the intertwiners are presented in matrix form. From Eq.~\eqref{eq:mn},
we get that $\left\{ \left(\mu_{j}\otimes\text{id}_{\mathfrak{a}}\right)\sigma_{j}\right\} _{j=0,1,2,3}$
forms a basis of $\text{Hom}\left(\mathcal{V}_{\mathfrak{a}},\left(\mathcal{V}_{\mathfrak{a}}\otimes\mathcal{V}_{\mathfrak{a}}\right)\otimes\mathcal{V}_{\mathfrak{a}}\right)$.
By the Drinfeld associator $\left(\rho_{111}^{\pm}\otimes\rho_{111}^{\pm}\otimes\rho_{111}^{\pm}\right)\left(\phi\right)=-1$,
the basis is converted into $\left\{ \psi_{j}\coloneqq-\left(\mu_{j}\otimes\text{id}_{\mathfrak{a}}\right)\sigma_{j}\right\} _{j=0,1,2,3}$
as a basis of $\text{Hom}\left(\mathcal{V}_{\mathfrak{a}},\mathcal{V}_{\mathfrak{a}}\otimes\left(\mathcal{V}_{\mathfrak{a}}\otimes\mathcal{V}_{\mathfrak{a}}\right)\right)$.
On the other hand, $\left\{ \varphi_{j}\coloneqq\left(\text{id}_{\mathfrak{a}}\otimes\mu_{j}\right)\sigma_{j}\right\}_{j=0,1,2,3}$
forms another basis of $\text{Hom}\left(\mathcal{V}_{\mathfrak{a}},\mathcal{V}_{\mathfrak{a}}\otimes\left(\mathcal{V}_{\mathfrak{a}}\otimes\mathcal{V}_{\mathfrak{a}}\right)\right)$
from Eq.~\eqref{eq:klamba}. It is straightforward to check 
\begin{align}
\psi_{j}^{\dagger}\psi_{j'} & =\delta_{jj'}\text{id}_{\mathfrak{a}},\forall j,j'=0,1,2,3.\\
\varphi_{j}^{\dagger}\varphi_{j'} & =\delta_{jj'}\text{id}_{\mathfrak{a}},\forall j,j'=0,1,2,3.
\end{align}
Then, the $4\times4$ matrix $\left(F_{\mathfrak{a}}^{\mathfrak{aaa}}\right)_{\mathfrak{e},\mathfrak{f}}\coloneqq F_{\mathfrak{aef}}^{\mathfrak{aaa}}$,
for $\mathfrak{e},\mathfrak{f}=\left(0,000\right),\left(0,011\right),\left(0,101\right),\left(0,110\right)$,
describes the basis transformation; thus,
\begin{align}
\left(F_{\mathfrak{a}}^{\mathfrak{aaa}}\right)_{\mathfrak{e},\mathfrak{f}} & =\left(\left\langle \varphi_{j}|\psi_{j'}\right\rangle \right)_{j,j'}\nonumber \\
 & =\frac{1}{2}\left(\begin{array}{cccc}
1 & -1 & -i & 1\\
1 & -1 & i & -1\\
-i & -i & -1 & i\\
-1 & -1 & -i & -1
\end{array}\right),
\end{align}
where $\left\langle \varphi_{j}|\psi_{j'}\right\rangle $ is defined
by $\varphi_{j}^{\dagger}\psi_{j'}=\left\langle \varphi_{j}|\psi_{j'}\right\rangle \text{id}_{\mathfrak{a}}$.
For example, $F_{\mathfrak{a}\left(0,000\right)\left(0,011\right)}^{\mathfrak{aaa}}=-\frac{1}{2}$.
All the other $6j$ symbols can be computed in this way.

\begin{table}
	\begin{tabular}{|c|c|c|c||c|}
		\hline 
		& $D_{{h}}^{100}$ & $D_{{h}}^{010}$ & $D_{{h}}^{001}$ & $\theta$\tabularnewline
		\hline 
		\hline 
		$\rho_{100}^{\pm}$ & $\pm\delta_{100,{h}}\cdot\sigma_{0}$ & $\delta_{100,{h}}\cdot\sigma_{1}$ & $\delta_{100,{h}}\cdot\sigma_{3}$ & $\pm1$\tabularnewline
		\hline 
		$\rho_{010}^{\pm}$ & $\delta_{010,{h}}\cdot\sigma_{3}$ & $\pm\delta_{010,{h}}\cdot\sigma_{0}$ & $\delta_{010,{h}}\cdot\sigma_{1}$ & $\pm1$\tabularnewline
		\hline 
		$\rho_{001}^{\pm}$ & $\delta_{001,{h}}\cdot\sigma_{1}$ & $\delta_{001,{h}}\cdot\sigma_{3}$ & $\pm\delta_{001,{h}}\cdot\sigma_{0}$ & $\pm1$\tabularnewline
		\hline 
		$\rho_{011}^{\pm}$ & $\delta_{011,{h}}\cdot\sigma_{3}$ & $\delta_{011,{h}}\cdot\sigma_{1}$ & $\pm\delta_{011,{h}}\cdot\sigma_{1}$ & $\pm1$\tabularnewline
		\hline 
		$\rho_{101}^{\pm}$ & $\pm\delta_{101,{h}}\cdot\sigma_{1}$ & $\delta_{101,{h}}\cdot\sigma_{3}$ & $\delta_{101,{h}}\cdot\sigma_{1}$ & $\pm1$\tabularnewline
		\hline 
		$\rho_{110}^{\pm}$ & $\delta_{110,{h}}\cdot\sigma_{1}$ & $\pm\delta_{110,{h}}\cdot\sigma_{1}$ & $\delta_{110,{h}}\cdot\sigma_{3}$ & $\pm1$\tabularnewline
		\hline 
		$\rho_{111}^{\pm}$ & $\pm\delta_{111,{h}}\cdot\sigma_{1}$ & $\pm\delta_{111,{h}}\cdot\sigma_{2}$ & $\pm\delta_{111,{h}}\cdot\sigma_{3}$ & $\mp i$\tabularnewline
		\hline 
	\end{tabular}
	
	\caption{Two-dimensional irreducible representations of $\mathcal{D}^{\omega}\left(\mathbb{Z}_{2}^{3}\right)$
		with $\omega\left({g},{h},{k}\right)=e^{i\pi g^{\left(1\right)}h^{\left(2\right)}k^{\left(3\right)}}$,
		specified by the action of a set of generators. The $2\times2$ identity
		matrix and the Pauli matrices are denoted by $\sigma_{0},\sigma_{1},\sigma_{2},\sigma_{3}$
		respectively. The topological spin $\theta$ of each irreducible representation
		is listed in the last column. \textcolor{red}{}}
	
	\label{table:irrep}
\end{table}

The universal $R$-matrix $R=\sum_{g}D_{g}^{e}\otimes D^{g}$ is elegant
enough to describe braidings; we will not list all $\mathcal{R}_{\mathfrak{c}}^{\mathfrak{ab}}$,
which can be obtained by a well-defined but tedious computation from $R$. To illustrate
the computation, let us calculate $\mathcal{R}_{\mathfrak{c}}^{\mathfrak{aa}}$
with $\mathfrak{a}=\left(0,111\right)$ as an example.
Given by Eq.~\eqref{eq:Rab}, the matrix form of $\mathcal{R}^{\mathfrak{aa}}:\mathcal{V}_{\mathfrak{a}}\otimes\mathcal{V}_{\mathfrak{a}}\rightarrow\mathcal{V}_{\mathfrak{a}}\otimes\mathcal{V}_{\mathfrak{a}}$
is 
\begin{align}
\mathcal{R}^{\mathfrak{aa}} & =\left(\rho_{111}^{\pm}\left(D^{111}\right)\otimes\text{id}_{\mathfrak{a}}\right)\wp,\nonumber \\
 & =\mp i\left(\begin{array}{cccc}
1 & 0 & 0 & 0\\
0 & 0 & 1 & 0\\
0 & 1 & 0 & 0\\
0 & 0 & 0 & 1
\end{array}\right),
\end{align}
where $\wp:\mathcal{V}_{\mathfrak{a}}\otimes\mathcal{V}_{\mathfrak{a}}\rightarrow\mathcal{V}_{\mathfrak{a}}\otimes\mathcal{V}_{\mathfrak{a}}$
is the exchange of the two factors and 
\begin{multline}
\rho_{111}^{\pm}\left(D^{111}\right)=\rho_{111}^{\pm}\left(D_{111}^{111}\right)=\rho_{111}^{\pm}\left(D_{111}^{100}D_{111}^{011}\right)\\
=\rho_{111}^{\pm}\left(-D_{111}^{100}D_{111}^{010}D_{111}^{001}\right)=\mp\sigma_{1}\sigma_{2}\sigma_{3}=\mp i.
\end{multline}
Using Eqs.~(\ref{eq:u0}-\ref{eq:u3}) and Eq.~\eqref{eq:Rabc},
we get 
\begin{align}
\mathcal{R}_{\mathfrak{0}}^{\mathfrak{aa}}=\pm i\cdot\text{id}_{V_{\mathfrak{0}}^{\mathfrak{aa}}}, & \quad\mathcal{R}_{\mathfrak{c}}^{\mathfrak{aa}}=\mp i\cdot\text{id}_{V_{\mathfrak{c}}^{\mathfrak{aa}}},
\end{align}
where $\mathfrak{c}=\left(0,011\right),\left(0,101\right)$
or $\left(0,110\right)$. Thus, the topological spin
of $\mathfrak{a}=\left(111,\pm\right)$, defined by Eq.~\eqref{eq:topspin},
is 
\begin{equation}
\theta_{\left(111,\pm\right)}=\mp i.
\end{equation}
The topological spins of all two-dimensional irreducible representations
are listed in Table~\ref{table:irrep}, while all the topological
spins of one-dimensional representations are $1$. Given the topological
spins and the fusion rules, we can read off the topological $S$-matrix
from Eq.~\eqref{eq:Sab}.

\newpage{}

\bibliography{TwistedFracton}

\end{document}